\documentclass[12pt,a4paper]{book}
\usepackage{amsfonts,amsbsy,graphicx,color,latexsym}
\setkeys{Gin}{keepaspectratio=true}
\newfont{\peter}{cmss17 scaled 2084}

\usepackage{macros}

\begin{document}
\frontmatter

\begin{titlepage}

\vspace*{\fill}


\hspace{.04\textwidth}\begin{minipage}{.96\textwidth}
\begin{center}

{\peter Quantum Systems Exactly}

\vspace{.5cm}

{\huge\sffamily Habilitation}

\vspace{1.1cm}

{\large\sffamily Peter Schupp

\vspace{.5cm}

\sl\sffamily
Fakult\"at f\"ur Physik\\ 
Ludwig-Maximiliams-Universit\"at M\"unchen

\vspace{1.1cm}

\includegraphics[bb=30 200 590 590, width = .9\textwidth]{%
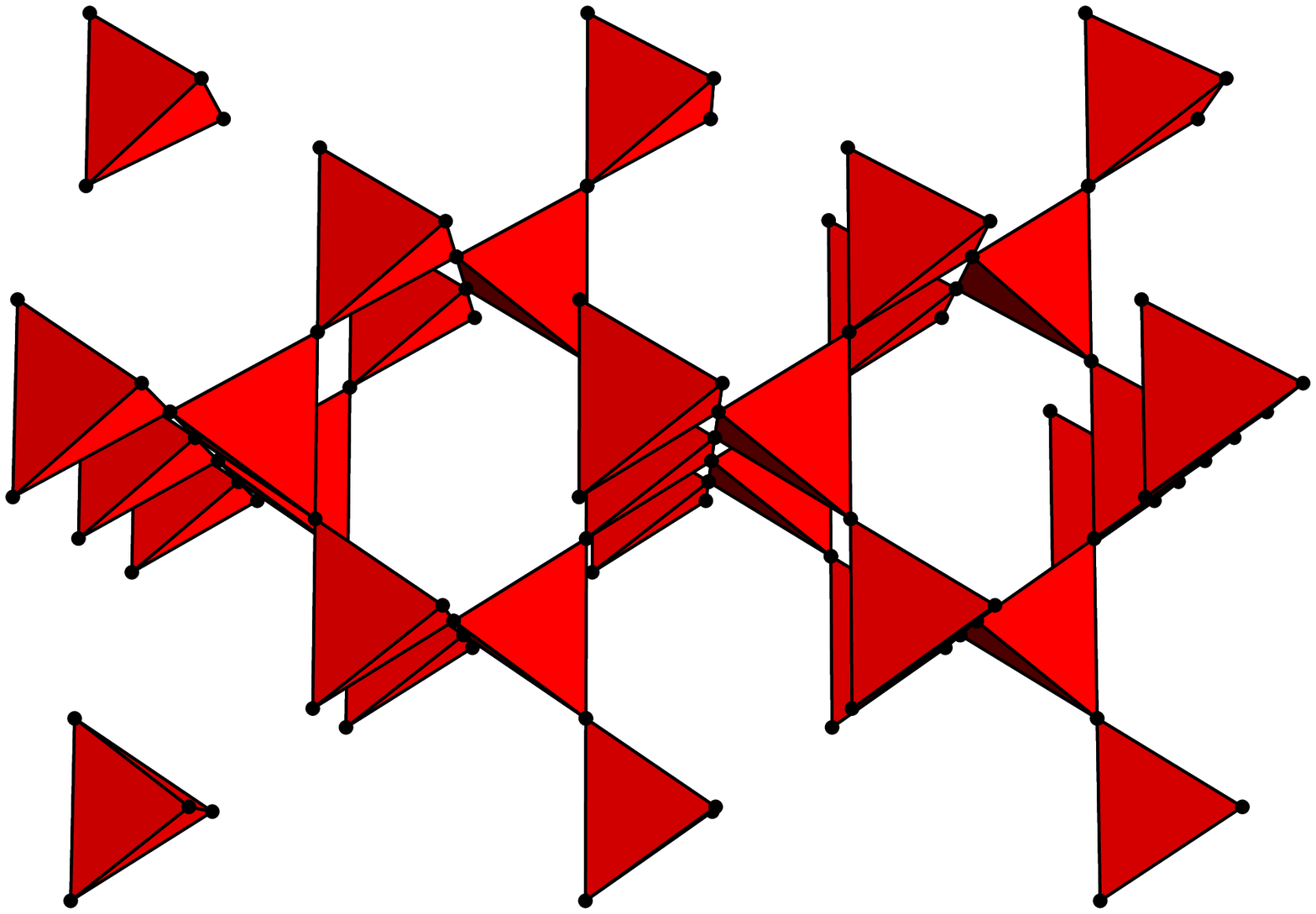}

\vspace{1.1cm}

April 2001}

\vspace*{1cm}
\vspace*{\fill}

\end{center}
\end{minipage}

\pagebreak

\mbox{}
\vfill
\noindent cover picture: pyrochlore lattice viewed
from a direction that reveals the
hidden two-dimensional kagome lattice substructure

\end{titlepage}

\tableofcontents

\chapter[{\sffamily Preface}]{{\sffamily Preface}}

There are two facets to research in theoretical/mathematical physics that 
go hand in hand with each other and that are both, whenever technically 
feasible, guided and ultimately confronted by experiment: There is the quest 
to find the underlying theories that govern natural phenomena, and then 
there is the equally complex and important task to use these fundamental 
building blocks to understand actual physical systems and to solve given 
problems. Here we will be mainly concerned with the latter. 

This thesis contains a selection of works in theoretical/mathematical 
physics that are all in one way or the other concerned with exact results 
for quantum systems. The problems are taken from various fields of physics, 
ranging from concrete quantum systems, in particular spin and strongly 
correlated electron systems, over abstract quantum integrable systems to gauge 
theories on quantum spaces. 

Since we are interested in mathematically rigorous results one would a 
priori expect the need to invent a novel approach for each given problem. 
However, it turns out that one can often use similar ideas in many different 
fields of physics -- the methods are transferable. There are several 
recurring themes in this thesis, the most obvious being the use of symmetries. 
Then there is the trick to simplify a system by enlarging it. In the work 
on frustrated spin systems and on quantum integrable systems we construct 
doubles; in the first case by adjoining a suitable mirror image, yielding a 
re positive whole, in the second case by adjoining the dual of configuration 
space and thereby linearizing the equations of motion. In the work on 
nonabelian noncommutative gauge theories we enlarge space-time by extra 
internal dimensions to reduce the problem to the abelian case. 

The last chapter on noncommutative gauge theories and star products 
is the focus of my current research. At first sight this topic does not seem 
to fit in the central theme of this thesis since it frequently employs formal 
power series, i.e., something inherently perturbative, or in other words, not 
``exact'' However, the framework of deformation quantization allows us to 
separate algebraic questions from difficult problems regarding representation 
theory and convergence, thereby enabling, e.g., a rigorous proof by explicit 
construction of the existence of a Seiberg-Witten map. 

This thesis is based on a series of publications [1]--[7]. The original 
presentation has been streamlined and illustrations have been added. New 
unpublished results are scattered throughout the text. Each chapter starts 
with an introductory overview including an indication of the original sources 
and, where appropriate, a discussion of cross relations between the chapters. 
These introductory sections starting on pages 1, 21, 37, 49, 83 and the 
historical remarks on the Lieb-Mattis theorem starting on page 12 can be also 
read independently of the main text as a brief summary. 

\subsection*{Acknowledgments}

I am indebted to Julius Wess and Elliott Lieb for letting me join their 
respective groups in Munich and Princeton where most of the research of this 
thesis was done. \\[1em]
I thank Branislav Jur\v co, Elliott Lieb and Julius Wess for fruitful and 
enjoyable collaboration on parts of the material included in this thesis 
as well as Bianca Cerchiai, John Madore, Lutz M\"oller, Stefan Schraml 
and Wolfgang Spitzer for joint work on closely related topics [8]--[12].\\[1em]
Many helpful and inspiring discussions with Michael Aizenman, Paolo 
Aschieri, Almut Burchard, Rainer Dick, Gaetano Fiore, Dirk Hundertmark, 
Michael Loss, Roderich Moessner, Nicolai Reshetikhin, Michael Schlieker, 
Andrzej Sitarz, Harold Steinacker, Stefan Theisen, Lawrence Thomas and 
Bruno Zumino have also contributed much to this work.

\mainmatter


\chapter[{\sffamily Frustrated quantum spin systems}]{Frustrated 
quantum spin systems}

\label{chap:spin}

\begin{figure}[!htb]
\begin{center}
\includegraphics[bb=-10 145 585 710, height=.16\textwidth]{%
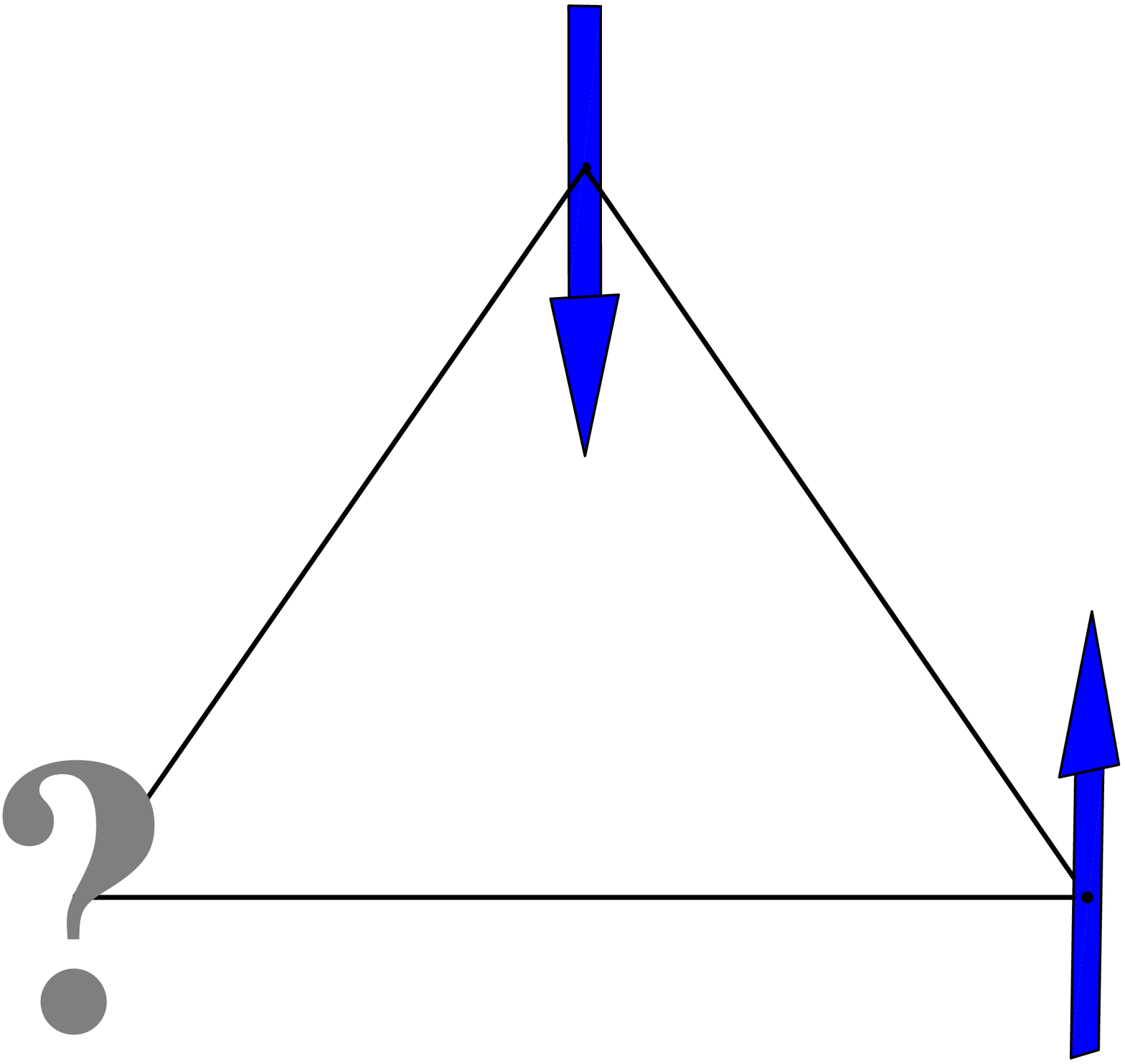}\hspace{3em}
\includegraphics[bb=20 80 580 730, height=.18\textwidth]{%
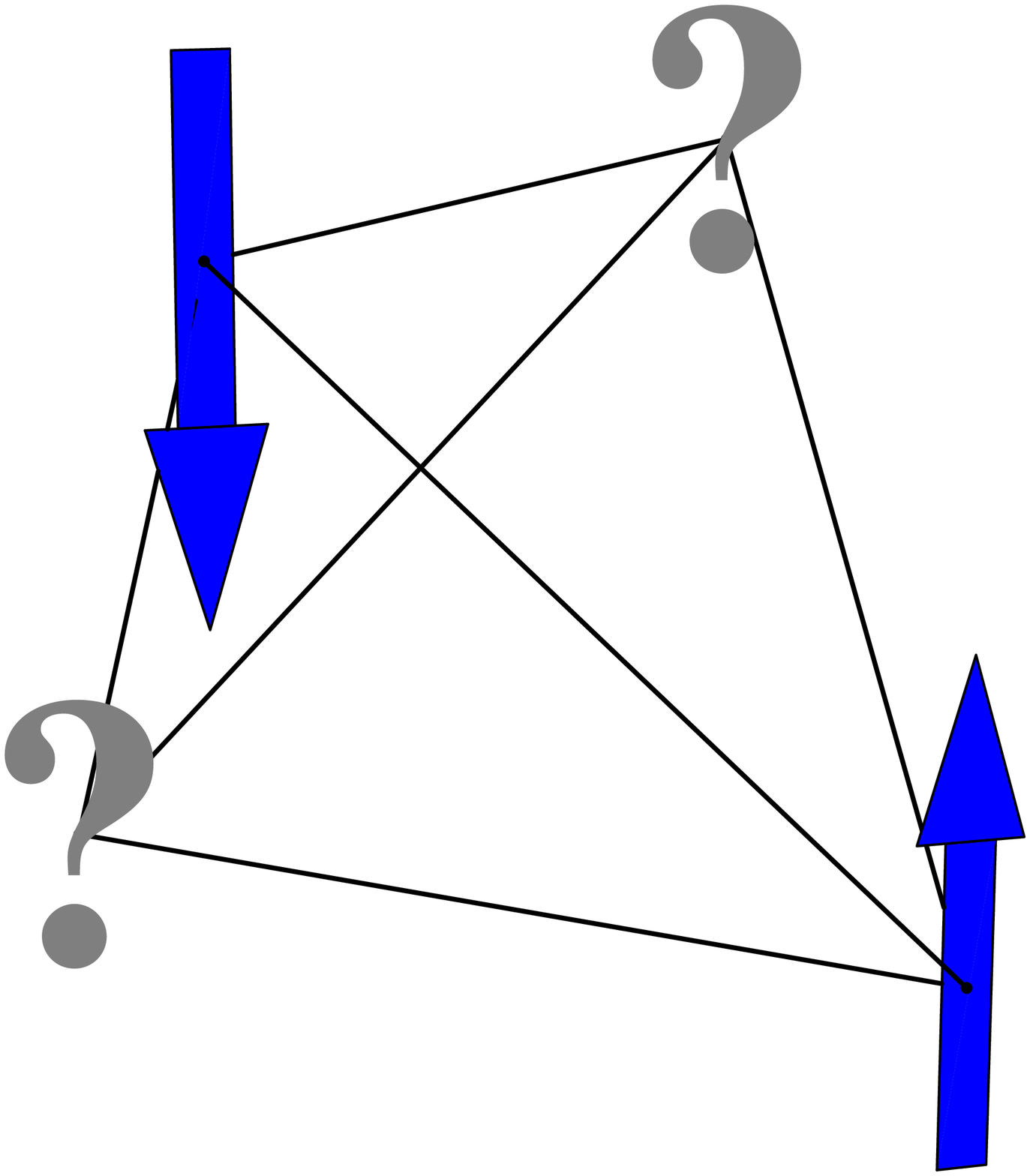}
\end{center} 
\caption{Triangular and tetrahedral frustrated units.} \label{fig:tri}
\end{figure}
\begin{figure}[htb]
\begin{center}
\includegraphics[bb=125 170 550 640, width=.66\textwidth]{
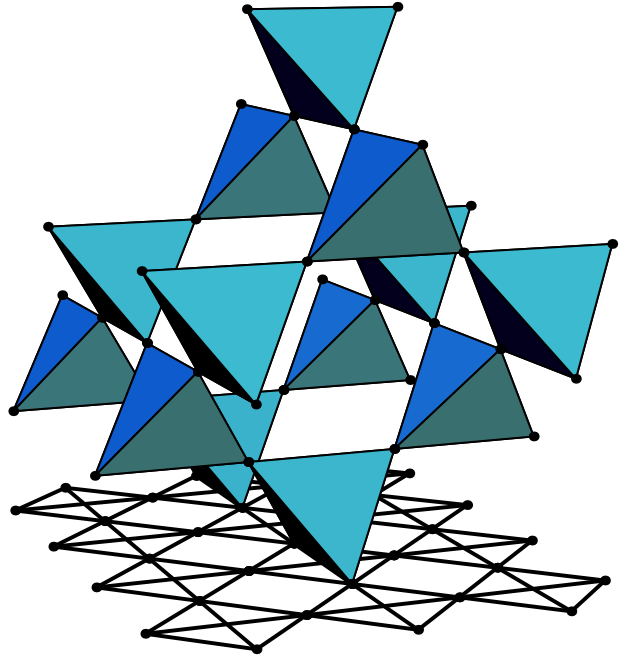}
\end{center}
\caption{Pyrochlore lattice over checkerboard lattice.} \label{pyro}
\end{figure}

Geometric frustration occurs in spin systems with 
interactions that favor anti-align\-ment and involve fully connected units 
of three or more spins, that can obviously not all be 
mutually anti-align\-ed (Fig.~\ref{fig:tri}).
The kagome lattice is an example of a frustrated spin system
with site-sharing triangular units, the pyrochlore lattice
and its two-dimensional version, the pyrochlore checkerboard are examples with
site-sharing tetrahedra (Fig.~\ref{pyro}). 
Common to all these systems is the richness of
classical ground states which make them very interesting but
unfortunately also very hard to understand, especially in the quantum
case.
We were able to obtain some exact results (among the first in this field) for 
the fully frustrated quantum antiferromagnet on a pyrochlore checkerboard:
With the help of the reflection symmetry of this two-dimensional lattice
we have established rigorously that 
there is always a ground state with total spin zero (i.e., a singlet), 
furthermore, 
in the periodic case
all ground states (if there is more than one) 
are singlets and the spin-expectation vanishes for
each frustrated unit (this is the quantum analog of the 
``{\em ice rule}'' for the corresponding Ising
system). With the same methods we also found explicit upper bounds on the 
susceptibility,
$\chi(T) \leq \frac{1}{8}$ (in natural units), 
both for the
ground state and at finite temperature. This bound becomes exact in the
classical high spin limit.
The pyrochlore checkerboard is a system in which frustration effects are
especially strong, so it is very exciting that a door to a more rigorous
understanding of such systems has been opened. 

Our approach combines old ideas of reflection positivity with 
more recent methods of Lieb and myself that we had originally tried to use
with the  Hubbard model (see chapter~\ref{chap:electron}).
Figure~\ref{refpos} illustrates the basic idea
for the simpler case of a classical antiferromagnetic spin system:
\begin{figure}[htb]
\begin{center}
\includegraphics[bb=40 230 565 585, width=.7\textwidth]{%
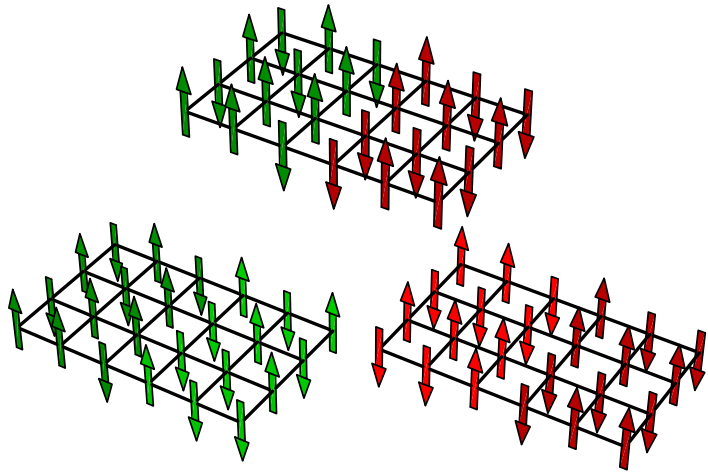}
\end{center}
\caption{Reflection positivity for a classical antiferromagnet.} \label{refpos}
\end{figure}
The lattice on the top shows a given state of the spin system,
the two lattices on the bottom show derived states. They are constructed from
the original state by reflecting the left (resp.\ right) half over to the
right (resp.\ left) half while simultaneously flipping all the spins.  This 
leads in either case to a state with a lower energy. (Recall that
we are considering \emph{anti}-ferromagnetic interaction.) If the original 
state was a ground state then the derived states must also be ground 
states of the system. In the quantum case things get more involved since
states are in general superpositions of configurations like the ones in
Figure~\ref{refpos}. Mathematically the extra complexity can be handled by
trace inequalities.  Both in the classical and quantum case
we gain a lot of useful information about ground
states of such spin systems -- the results mentioned above are some of the 
consequences. The method and some of the results generalize to
other reflection symmetric lattices and hold in arbitrary
dimensions. In section~\ref{sec:spin} we shall in particular establish
the following exact properties of reflection symmetric
spin systems with antiferromagnetic crossing bonds: 
At least one ground state has total spin zero
and a positive semidefinite coefficient matrix and
the crossing bonds obey an ice rule.
This augments some previous results which were limited to bipartite spin systems 
and is of particular interest for frustrated spin systems.

There are many open
questions and there is hope that much remains to be learned with the new
approach which, by the way, has its origin in old work by Osterwalder on quantum
field theory.
This work was done in collaboration with E.\ Lieb at Princeton University
and is published in~\cite{prl,physica}. (The following two sections are based
on these publications.)

\label{end:spin}


\section{The Pyrochlore checkerboard}

\label{sec:pyro}

\subsection{Geometric frustration}

Geometrically frustrated spin systems are known to have many interesting
properties that are quite unlike those of conventional magnetic systems
or spin glasses~\cite{HTD}. 
Most results are for classical systems.
The first frustrated system, for which the richness
of classical ground states was noted, was the triangular lattice \cite{WH}.
Subsequently, the pyrochlore lattice,
which consists of tetrahedra that share sites, was identified as a
lattice on which the frustration effects are especially strong~\cite{PWA}.
Unusual low-energy properties -- in particular the absence of ordering at
any temperature, was predicted both for discrete \cite{PWA} 
and continuous \cite{JV} classical
spin sytems. The ground state and low energy properties of the classical 
pyrochlore antiferromagnet
-- whose quantum version we are interested in --
has been extensivly studied in \cite{MC}.
Both the interest and difficulty in studying frustrated spin systems stem
from the large ground-state degeneracy, which precludes most perturbative
approaches. 

As is the case for most other strong interacting systems in
more than one dimension, very little is known exactly about the ground states of
frustrated quantum spin systems. Most of the present knowledge has been
obtained by numerics or clever approximations.
Quantum fluctuations
have been studied in the limits of large-$S$ \cite{SL},
where a tendency towards lifting the ground-state degeneracy in favor of
an ordered state (``quantum order by disorder") was detected. In the
opposite limit -- $S=1/2$, where quantum fluctuations are much stronger --
the pyrochlore antiferromagnet has been identified as a candidate for a quantum disordered
magnet (``quantum spin liquid") \cite{CL}, and it has also been discussed in
terms of a resonating valence bond approach \cite{RVB}.
However, there are no exact results 
against which to test
the reliability of the results in this limit. In contrast to this, for
conventional -- bipartite -- 
antiferromagnetic spin systems it is well known, for
example, that the energy levels are ordered in a natural
way according to spin, starting from spin zero \cite{LM}. 
Geometrically frustrated systems are not
bipartite and thus this otherwise quite general
theorem does not apply.

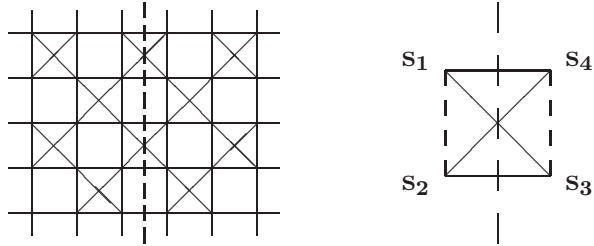
\begin{figure}[htb]
\begin{center}
\unitlength 1.00mm
\linethickness{0.4pt}
\begin{picture}(77.00,33.00)
\put(6.00,2.00){\line(0,1){30.00}}
\put(12.00,32.00){\line(0,-1){30.00}}
\put(18.00,2.00){\line(0,1){30.00}}
\put(24.00,32.00){\line(0,-1){30.00}}
\put(30.00,2.00){\line(0,1){30.00}}
\put(36.00,32.00){\line(0,-1){30.00}}
\put(3.00,5.00){\line(1,0){36.00}}
\put(39.00,11.00){\line(-1,0){36.00}}
\put(3.00,17.00){\line(1,0){36.00}}
\put(39.00,23.00){\line(-1,0){36.00}}
\put(3.00,29.00){\line(1,0){36.00}}
\put(6.00,29.00){\line(1,-1){24.00}}
\put(6.00,17.00){\line(1,-1){12.00}}
\put(18.00,29.00){\line(1,-1){18.00}}
\put(30.00,29.00){\line(1,-1){6.00}}
\put(6.00,23.00){\line(1,1){6.00}}
\put(6.00,11.00){\line(1,1){18.00}}
\put(12.00,5.00){\line(1,1){24.00}}
\put(24.00,5.00){\line(1,1){12.00}}
\put(21.00,1.00){\line(0,1){2.00}}
\put(21.00,4.00){\line(0,1){2.00}}
\put(21.00,7.00){\line(0,1){2.00}}
\put(21.00,10.00){\line(0,1){2.00}}
\put(21.00,13.00){\line(0,1){2.00}}
\put(21.00,16.00){\line(0,1){2.00}}
\put(21.00,19.00){\line(0,1){2.00}}
\put(21.00,22.00){\line(0,1){2.00}}
\put(21.00,25.00){\line(0,1){2.00}}
\put(21.00,28.00){\line(0,1){2.00}}
\put(21.00,31.00){\line(0,1){2.00}}
\put(75.00,24.00){\line(-1,0){14.00}}
\put(61.00,10.00){\line(1,0){14.00}}
\put(75.00,10.00){\line(-1,1){14.00}}
\put(61.00,10.00){\line(1,1){14.00}}
\put(68.00,15.00){\line(0,1){4.00}}
\put(68.00,22.00){\line(0,1){4.00}}
\put(68.00,29.00){\line(0,1){4.00}}
\put(68.00,5.00){\line(0,-1){4.00}}
\put(61.00,10.00){\line(0,1){2.00}}
\put(61.00,14.00){\line(0,1){2.00}}
\put(61.00,18.00){\line(0,1){2.00}}
\put(61.00,22.00){\line(0,1){2.00}}
\put(75.00,10.00){\line(0,1){2.00}}
\put(75.00,14.00){\line(0,1){2.00}}
\put(75.00,18.00){\line(0,1){2.00}}
\put(75.00,22.00){\line(0,1){2.00}}
\put(59.00,24.00){\makebox(0,0)[rb]{$\bf s_1$}}
\put(59.00,10.00){\makebox(0,0)[rt]{$\bf s_2$}}
\put(77.00,10.00){\makebox(0,0)[lt]{$\bf s_3$}}
\put(77.00,24.00){\makebox(0,0)[lb]{$\bf s_4$}}
\put(68.00,12.00){\line(0,-1){4.00}}
\end{picture}
\caption{(a) pyrochlore checkerboard, reflection symmetric 
about dashed line
(b)~frustrated unit with crossing bonds}
\label{fig:che1}
\end{center}
\end{figure}

In the following we shall focus on
the pyrochlore checkerboard: this is a 
two dimensional array of site-sharing tetrahedra, whose projection 
onto a plane is a square lattice with two
extra diagonal bonds on every other square, see figures~\ref{pyro},~\ref{fig:che1}(a).
(The regular pyrochlore lattice is
a three-dimensional array of site-sharing tetrahedra; it coincides with the checkerboard
if suitable periodic boundary conditions are imposed.)
The tetrahedra -- or squares with extra diagonal bonds -- are 
the frustrated units and will henceforth be called {\em boxes}.

The hamiltonian of a quantum Heisenberg
antiferromagnet  on a general lattice is (in natural units)
\eq
H_{\mbox{\tiny AF}} = \sum_{\la i, j\ra} {\bf s_i} \cdot {\bf s_j} ,
\en
where the sum is over bonds $\la i, j \ra$ that connect sites $i$ and $j$
and ${\bf s} = (s^1, s^2, s^3)$ are spin operators in the spin-$s$
representation, where $s$ can be anything.
For the checkerboard lattice the hamiltonian is
up to a constant equal to half the sum 
of the total spin squared of all boxes (labelled by $x$)
\eq
H = \frac{1}{2}\sum_x ({\bf s_1} +{\bf s_2} + {\bf s_3}+ {\bf s_4})_x^2.
\label{checker}
\en

A 3 $\times$ 3 checkerboard with periodic boundary
conditions, i.e., with four independent sites,
provides the simplest example. It has
a hamiltonian that is (up to a constant) the total spin
squared of one box and the energy levels, degeneracies, and
eigenstates follow from the decomposition of the Hilbert space of four spin-$s$
particles into components of total spin; all ground states have total
spin zero and there are $2s +1$ of them.

\subsection{Reflection symmetry}

A checkerboard lattice of arbitrary size,
with or without
periodic boundary conditions but with an even number of independent
sites, has the property
that it can be split into two equal parts
that are mirror images of one another about a line
that cuts bonds, as indicated in figure~\ref{fig:che1}, and that
contains no sites. We shall now show that such a system has at least
one spin-zero ground state.
It is actually not important, for the following argument,
what the lattice looks
like on the left or right; these sublattices neither need to
be checkerboards nor do they have to be purely antiferromagnetic 
(as long as total spin is a good quantum number).
What is impotant is, that the whole system is reflection symmetric about the
line that separates left and right and that
the crossing bonds are of checkerboard type. 
(For a system with periodic
boundary conditions in one direction there will 
actually be two such lines, but we
emphasize that periodic boundary conditions are
not needed here even though it is needed in the usual
reflection positivity applications; see \cite{DLS} and references therein.)
A key observation is that these crossing bonds (solid lines in
figure~\ref{fig:che1}(b)) form antiferromagnetic bonds
$\bf S_L \cdot S_R$ between pairs of spins $\bf S_L = s_1 + s_2$ and 
$\bf S_R = s_3 + s_4$ of each box on the symmetry line.

The hamiltonian is $H = H_L + H_R + H_C$, where $H_L$ and $H_R$
act solely in the Hilbert spaces of the left, respectively right, subsystem
and $H_C$
contains the crossing bonds.
For the checkerboard
$H_C = \sum_y (({\bf s_1 + s_2})\cdot ({\bf s_3 + s_4}))_y$,
with the sum over boxes $y$ that are bisected by the
symmetry line.
$H_L$ and $H_R$ are completely arbitrary as long as they commute with
the total spin operator. 
We will, however, assume here that they are real in the $S^3$ basis.
Any state of the system can be written in terms of a 
matrix $c$ as
\eq
\psi = \sum_{\al, \be} c_{\al\be} 
\psi^L_\al \ot (\psi^R_\be)_{\mbox{\tiny rot}} , \label{psi}
\en
where the $\psi^L_\al$ form a real orthonormal basis of $S^3$ eigenstates
for the left subsystem and the $(\psi^R_\be)_{\mbox{\tiny rot}}$ are
the corresponding states for the right subsystem, 
but rotated by an angle $\pi$
around the 2-direction in spin space. This rotation takes $\uparrow$ into
$\downarrow$, $\downarrow$ into $-\uparrow$, and more generally
$|s,m\ra$ into $(-)^{s-m}|s,-m\ra$. It reverses the signs of the
operators $S^1$ and $S^3$, while it keeps $S^2$ unchanged.
The eigenvalue problem $H \psi = E \psi$ is now a matrix
equation
\eq
h_L c + c (h_R)^T -\sum_{i=1}^3 \sum_y t_y^{(i)} c (t_y^{(i)})^T = E c,
\label{ev}
\en
where $(h_L)_{\al\be}$ and $(h_R)_{\al\be}$ are real, symmetric matrix
elements of the corrosponding terms
in the hamiltonian and the $t_y^{(i)}$
are the \emph{real} matrices defined for the
spin operators $\bf s_1$ and $\bf s_2$
in box $y$ by
$t^{(1,3)}_{\al\be} = \la\psi_\al^L|s^{(1,3)}_1 + s^{(1,3)}_2|\psi_\be^L\ra$
and $t^{(2)}_{\al\be} = i \la\psi_\al^L|s^{(2)}_1 + s^{(2)}_2|\psi_\be^L\ra$.
Note the overall minus sign of the crossing term in (\ref{ev}):
replacing $\bf s_1 + s_2$ by $\bf s_3 + s_4$ and $\psi^L$ by
$(\psi^R)_{\mbox{\tiny rot}}$ gives a change of sign for directions 1 and 3,
while the $i$ in the definition of $t^{(2)}$ gives the minus sign 
for direction~2.

Consider now the energy expectation in terms of $c$:
\eq
\la\psi| H |\psi\ra =\tr c c^\dagger h_L + \tr c^\dagger c h_R
-\sum_{i,y} \tr c^\dagger t_y^{(i)} c (t_y^{(i)})^\dagger .
\label{ee}
\en
Since $H$ is left-right symmetric and by assumption real, we find that for an
eigenstate of $H$ with coefficient matrix $c$,
there is also an eigenstate with matrix $c^\dagger$ and, by linearity,
with $c+ c^\dagger$ and $i(c - c^\dagger)$. Without loss of generality we
may, therefore, write eigenstates of
$H$ in terms of Hermitean
$c = c^\dagger$.
We shall also take $\psi$ to be normalized:
$\la\psi|\psi\ra = \tr c^\dagger c = 1$.
Following \cite{L}, 
let us write the trace in the last term of (\ref{ee}) in the 
diagonal basis of $c$:
$-\tr c^\dagger t_y^{(i)} c (t_y^{(i)})^\dagger =
-\sum_{k,l} c_k c_l |(t_y^{(i)})_{kl}|^2$.
This expression clearly does not increase if we replace all the $c_k$
by their absolute values $|c_k|$, i.e., if we replace the matrix
$c$ by the positive semidefinite matrix $|c| = \sqrt{c^2}$.
The first two terms in (\ref{ee}) and the norm of $\psi$
remain unchanged under this operation.
We conclude that if $c$ is a ground state than so is $|c|$. Since
$c = c^+ - c^-$ and $|c| = c^+ + c^-$, with positive semidefinite (p.s.d.)
$c^+$ and $c^-$, we may, in fact, chose a basis of ground states with p.s.d.\ coefficient
matrices.

\subsection{Singlets and magnetization}

Next, we will show that the state $\psi_0$ with the unit matrix as
coefficient matrix (in the $S^3$ eigenbasis) has total spin zero.
Since the overlap of $\psi_0$ with a state with matrix $c$ is simply the
trace of $c$, which is neccessarily non-zero for states with a p.s.d.\
matrix, and because spin is a good quantum number of the problem, this will
imply that there is a least one ground state with total spin zero.
First consider a spin-1/2 system. In the $S^3$ eigenbasis
every site has then either spin up or down. The state with unit coefficient
matrix is a tensor product of singlets on corresponding 
pairs of sites $i \in L$,
$i' \in R$ of the two sublattices (see Fig.~\ref{fig:spin0}):
\begin{figure}[tb]
\begin{center}
\includegraphics[bb=50 160 580 680, width=.3\textwidth]{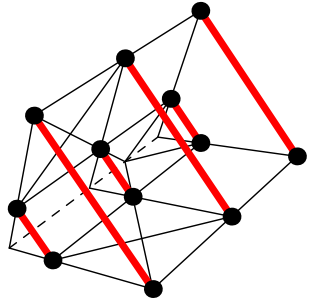}
\caption{Canonical spin-zero state -- the pyrochlore lattice is shown
folded along the symmetry axis, thick lines represent singlets.}
 \label{fig:spin0}
 \end{center}
\end{figure}
\eq
\psi_0 = \bigotimes_{i \in L} \left(\uparrow (\uparrow)_{\mbox{\tiny rot}} + 
\downarrow (\downarrow)_{\mbox{\tiny rot}}\right)_{i i'} =
\bigotimes_{i \in L} \left(\uparrow \downarrow 
- \downarrow \uparrow\right)_{i i'} .
\en
The analogous state for a system with arbitrary spins,
\eq
\psi_0 
=\bigotimes_{i\in L} \sum_{m=-s}^s 
\left((-)^{s-m}|s,m\ra\otimes|s,-m\ra\right)_{i i'} ,
\en 
is also a tensor product of spin-zero states.

Finally, we would like to show that
the projection onto the spin zero part of a state with p.s.d.\
coefficient matrix preserves its positivity. This is only of academic
interest here, but it is non-trivial and may very well be important for
other physical questions.
To find the projection onto spin zero we
need to decompose the whole Hilbert space into tensor products of the
spin components
$[j]_k \ot [j']_{k'}$ of the two subsystems;
here $k$, $k'$ are additional quantum numbers that
distinguish multiple multiplets with the same spin $j$. Only tensor
products with $j = j'$ can have a spin zero subspace, which is
unique, in fact, and generated by the spin zero state
$\displaystyle\sum_{m=-j}^j |j,m,k\ra\ot (-)^{j-m} |j,-m,k'\ra$.
Noting that $(-)^{j-m}|j,-m,k'\ra$ is precisely the spin-rotated state
$(|j,m,k'\ra)_{\mbox{\tiny rot}}$, we convince
ourselves that the projection onto spin
zero amounts to a partial trace over $m$ in a suitably parametrized
matrix $c$.
This operation preserves positive semidefiniteness, so we actually proved that
the checkerboard has at least one ground state that has
both total spin zero \emph{and} a p.s.d. coefficient matrix $c$.

We do not know how many ground states there are.
To determine the spin of any remaining ground states
we add an external field to the hamiltonian and study the resulting
magnetization. We will see that the spontaneous magnetization of every box
on the symmetry line vanishes for all ground states, 
and thus if we have periodic
boundary conditions in at least one direction, 
the total magnetization vanishes. 
Since $S^3_{\mbox{\tiny tot}}$ is a good quantum number and
$S^\pm_{\mbox{\tiny tot}}$ commute with
the hamiltonian, this will imply that all ground states in such a system
have total spin zero. 
Let us thus modify the original hamiltonian (\ref{checker}) by  
replacing the term
$(s_1^{(3)} +s_2^{(3)} +s_3^{(3)} +s_4^{(3)})^2_z$
for a single box, $z$, on the symmetry line
by $(s_1^{(3)} +s_2^{(3)} +s_3^{(3)} +s_4^{(3)} - b)^2_z$, i.e.,
effectively adding a field $b$ to the spins in box $z$ 
and a constant term $\frac{1}{2}b^2$
to the hamiltonian. We want to distribute the resulting
$b$-terms $(s_1^{(3)} +s_2^{(3)} - b/2)^2$, $(s_3^{(3)} +s_4^{(3)} - b/2)^2$,
and $2 (s_1^{(3)} +s_2^{(3)} - b/2)(s_3^{(3)} +s_4^{(3)} - b/2)$
to $H_L$, $H_R$, $H_C$ respectively.
We cannot use the spin rotation as before, because
the crossing terms in the hamiltonian would no longer
be left-right symmetric in the basis (\ref{psi}). To avoid
this problem we will, instead, expand eigenstates $\psi$ 
in the same basis on the left and on the right:
\eq
\psi = \sum_{\al, \be} \tilde c_{\al\be} 
\psi^L_\al \ot \psi^R_\be.
\en
In this basis the hamiltonian is left-right symmetric and
we may assume, as before, that $\tilde c = \tilde c^\dagger$.
Except for the presence of $b$ in box $z$
the energy expectations on the left and right
are as before. The energy expectation of the crossing terms of box $z$
in the diagonal
basis of $\tilde c$ is now
$$
\sum_{k,l} \tilde c_k \tilde c_l \left(|(t_z^{(1)})_{kl}|^2
- |(t_z^{(2)})_{kl}|^2
+ |(t_z^{(3)})_{kl} - b/2|^2\right) .
$$
This expression clearly does not increase if we replace
the $c_k$ by their absolute value $|c_k|$ \emph{and}
change the signs of the first and last terms. The sign change
can be achieved by simultaneously performing a spin rotation
and changing the sign of the field $b$ in the right subsystem.
This actually completly removes $b$ from
the hamiltonian. We have thus shown that the ground state energies
of the systems $H_b$ with and $H_0$ without the $b$-terms satisfy the
inequality $E_b \geq E_0$. Let $|b\ra$ be a ground state of $H_b$
and $|0\ra$ a ground state of $H_0$. It follows from the variational
principle, that $\la 0|H_b|0\ra \geq \la b|H_b|b\ra = E_b \geq E_0$.
Expressed in terms of spin operators this reads
$E_0 - 2 \la 0|b (s_1^{(3)}+s_2^{(3)}+s_3^{(3)}+s_4^{(3)})_z|0 \ra
+ b^2 \geq E_0$.
Recalling that we are free to choose both the sign and
the magnitude of $b$ we find that the ground state magnetization of
box $z$ must be zero:
\eq
\la 0|(s_1^{(3)}+s_2^{(3)}+s_3^{(3)}+s_4^{(3)})_z|0\ra = 0.
\en
This quantum analog of the ``\emph{ice rule}''
is true for any box on the symmetry line and it holds 
for all three spin components. 
In a system with periodic
boundary conditions and an even number of sites in at least one direction
we can choose the symmetry line(s) to intersect any given box, so in such
a system the magnetization is zero both for every single 
box separately and also
for the whole system: $\la 0|S^{(3)}_{\mbox{\tiny tot}}|0\ra = 0$.
As mentioned previously this
implies that the total spin is zero for 
all ground states of such a system.

\subsection{Susceptibility}

Let us return to the inequality $E_b \geq E_0$. It
implies a bound on the local susceptibility of the system: Let
$E(b) \equiv \la b|H_0 - b S^{(3)}_{\mbox{\tiny box}}|b\ra$ be the 
ground state energy of the periodic
pyrochlore checkerboard with a single box immersed in an external field
$b$. Recalling 
$H_b = H_0 - b S^{(3)}_{\mbox{\tiny box}} + \frac{1}{2} b^2$, we see
that the above inequality becomes $E(b) + \frac{1}{2} b^2 \geq E(0)$ and,
assuming differentiability, implies an
upper bound on the susceptibility at zero field for single-box magnetization
\eq
\chi_{\mbox{\tiny loc}} = 
-\frac{1}{\lambda} \left.\frac{\partial^2 E(b)}{\partial b^2}\right|_{b=0} 
\leq \frac{1}{4},
\en
where $\lambda = 4$ is the number of spins in a box. 
(The susceptibility is given
in natural units in which we have absorbed the g-factor 
and Bohr magneton in the definition 
of the field $b$.)

We would like to get more detailed information about the response of the
spin system to a global field $\{b_x\}$ in a hamiltonian
$H_{\{b_x\}}$ which is identical to (\ref{checker}), 
except for the terms for the third
spin component, which are replaced by 
$(s_1^{(3)}+s_2^{(3)}+s_3^{(3)}+s_4^{(3)} - b_x)_x^2$.
From what we have seen so far, it is apparent that the corresponding
ground state energy $E_{\{b_x\}}$
is extremal for $b_x =0$. With the help of a more sophisticated trace 
inequality \cite{KLS,physica}, that becomes relevant 
whenever the matrix $c$ in (\ref{psi})
cannot be diagonalized, one can actually show that 
$E_{\{b_x\}}$ has an absolute minimum at
$b_x =0$:
\eq
E_{\{b_x\}} \geq E_0 . \label{ineq}
\en
Note that we had to put the field on the boxes 
for this result to hold; not every field 
on the individual spins can be written this way.
The special choice $b_x =B/2$ corresponds to a global homogenous field $B$ on
all spins. (The factor $1/2$ adjusts for the fact that every spin is shared by
two boxes.) If 
$E(B) = \la B|H_0 - B S^{(3)}_{\mbox{\tiny tot}}|B\ra$ is the ground
state energy of the periodic pyrochlore checkerboard 
in the external field $B$, 
then (\ref{ineq}) implies $E(B) + \frac{\Lambda}{16} B^2 \geq E(0)$, 
and thus an upper
bound on the susceptibility per site at zero field (in natural units)
\eq
\chi = 
-\frac{1}{\Lambda}\left.\frac{\partial^2 E(B)}{\partial B^2}\right|_{B=0} 
\leq \frac{1}{8} ,
\en
where $\Lambda$ is the number of independent sites,
which equals twice the number of boxes.

All these results continue to hold at finite temperature. 
The analog of (\ref{ineq})
holds also for the partition function corresponding to $H_{\{b_x\}}$:
\eq
Z_{\{b_x\}} \leq Z_0 , \label{ineqz}
\en
as can be shown by a straightforward application of lemma~4.1 
in section 4 of \cite{DLS}
to the pyrochlore checkerboard. The physically relevant partition function
for the periodic pyrochlore checkerboard at finite temperature in a homogenous
external field, 
$Z(B) \equiv \tr e^{-\beta(H_0 - B S^{(3)}_{\mbox{\tiny tot}})}$,
differs from $Z_{\{b_x\}}$, where $b_x = B/2$, 
only by a factor corresponding to the 
constant term in $H_{\{b_x\}}$. Due to (\ref{ineqz}), the free energy 
$F(B)= -\beta^{-1}\ln Z(B)$ satisfies
\eq
F(B) + \frac{\Lambda}{16} B^2 \geq F(0).
\en
This implies (i) that the magnetization at zero field is still zero at finite
temperature,
\eq
M_T = 
-\frac{1}{\Lambda}\left.\frac{\partial F(B)}{\partial B}\right|_{B=0} = 0,
\en
and, more interestingly, (ii) the same upper bound for the susceptibility 
per site at zero field as we had for the ground state:
\eq
\chi_T = 
-\frac{1}{\Lambda}\left.\frac{\partial^2 F(B)}{\partial B^2}\right|_{B=0} 
\leq \frac{1}{8} .
\en
The bounds on the susceptibility hold for arbitrary intrinsic spin-$s$
and agree very well with the results of \cite{MB} for the classical
pyrochlore antiferromagnet in the un-diluted case.

\begin{figure}[htb]
\begin{center}
\includegraphics[bb=30 60 590 720, width=.45\textwidth]{%
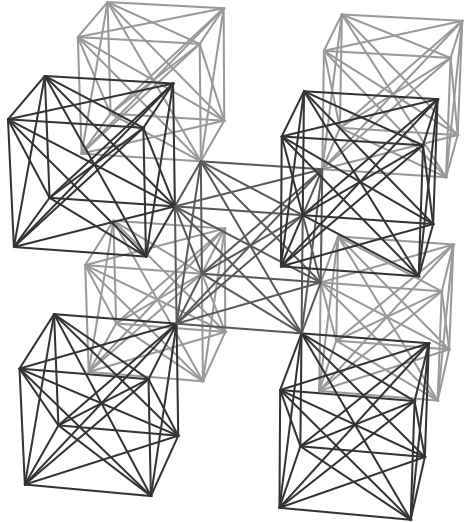}
\caption{A 3-dimensional checkerboard: Every cube has 8~vertices -- each
shared among two cubes. This lattice exhibits even
stronger frustration than the pyrochlore checkerboard.
The expectation value of the total spin vector of each cube is zero,
all ground states are singlets, the susceptibility is bounded above
by 1/16.}\label{3dpyro}
\end{center} 
\end{figure}

It is not essential for our method that only every other square of the
pyrochlore checkerboard is a frustrated unit, 
only the reflection symmmetry and the
antiferromagnetic crossing bonds are important. We could, e.g., have
diagonal bonds on \emph{every} square, but then the horizontal and/or
vertical bonds must have twice the coupling strength.
Our results also apply to  various 3-dimensional cubic
versions of the checkerboard, e.g., 
with diagonal crossing bonds in every other
cube, see figure~\ref{3dpyro}.
While the method does not directly work for the 3D pyrochlore lattice
because its geometry is too complicated, it has been seen in \cite{MC}
that classically this system has similar properties to the pyrochlore
checkerboard, which is also fully frustrated, and has the added advantage
of being more easily visualizeable.


\section{Singlets in reflection symmetric spin
systems}
\label{sec:spin}

\subsection{Ordering energy levels according to spin}

Total spin is often a useful quantum number to classify energy eigenstates of spin
systems. An example is the antiferromagnetic Heisenberg Hamiltonian on a
bipartite lattice, whose energy levels plotted versus total spin form towers of
states. The spin-zero tower extends furthest down the energy scale, the
spin-one tower has the next higher base, and so on, all the way up the spin
ladder: $E(S+1) > E(S)$, where $E(S)$ denotes the lowest energy eigenvalue for
total spin $S$ \cite{LM}. The ground state, in particular, has total
spin zero; it is a singlet. This fact had been suspected for a long time, but
the first rigorous proof was probably given by Marshall~\cite{M} for a 
one-dimensional antiferromagnetic chain with an even number of sites, each with
intrinsic spin-1/2 and with periodic boundary conditions. 
This system is bipartite, it can be split into two
subsystems, each of which contains only every other site, so that all
antiferromagnet bonds are \emph{between} these subsystems.
Marshall bases his proof on a theorem
that he attributes to Peierls: Any ground state of the system, expanded
in terms of $S^{(3)}$-eigenstates has coefficients with alternating signs that
depend on the $S^{(3)}$-eigenvalue of one of the subsystems. After a 
canonical transformation, consisting of a rotation of one of the subsystems
by $\pi$ around the 2-axis in spin space, the theorem simply states that all
coefficients of a ground state can be chosen to be positive. To show that this
implies zero total spin, Marshall
works in a subspace with $S^{(3)}$-eigenvalue $M=0$ and uses translation
invariance. His argument easily generalizes to higher dimensions
and higher intrinsic spin. Lieb, Schultz and Mattis~\cite{LSM}
point out that translational invariance is not really necessary,
only reflection symmetry is needed to relate the two subsystems, and the
ground state is unique in the connected case.
Lieb and Mattis~\cite{LM} ultimately remove the requirement of translation invariance
or reflection symmetry and apply the $M$-subspace method to classify excited
states. Like Peierls they use a Perron-Frobenius type argument to prove that in
the $S^{(3)}$-basis the ground state wave function for the connected case is a
positive vector and it is unique. Comparing this wave function with the 
positive wave function of a simple soluble model in an appropriate
$M$-subspace~\cite{L} they conclude that the ground state has total spin
$S = |S_A - S_B|$, where $S_A$ and $S_B$ are the maximum possible spins
of the two subsystems. (In the antiferromagnetic case $S_A = S_B$ and the ground
state has total spin zero.)
We shall now reintroduce
reflection symmetry, but for other reasons: we want to exploit methods and
ideas of ``reflection positivity" (see \cite{DLS} and references therein.)
We do not require bipartiteness. 
The main application is to frustrated spin systems similar to the pyrochlore
lattices discussed in the previous section.
\label{ends:spin}

\subsection{Reflection symmetric spin system}

We would like to  consider a spin system that consists of 
two subsystems that are mirror images of one another, except for a rotation by
$\pi$ around the \mbox{2-axis} in spin-space, 
and that has antiferromagnetic crossing bonds 
between corresponding sets of sites of the two subsystems. 
The spin Hamiltonian is
\eq
H = H_L + H_R + H_C ,
\en
and it acts on a tensor product of two identical copies of a Hilbert space that 
carries a representation of SU(2). ``$H_L = \tilde H_R$'' in the
sense that $H_L = h \ot 1$ and
$H_R = 1 \ot \tilde h$, where the tilde shall henceforth denote the rotation
by $\pi$ around the 2-axis in spin-space. We make no further assumptions
about the nature of $H_L$ and $H_R$, in particular we do \emph{not} assume that
these subsystems are antiferromagnetic.
The crossing bonds are of anti-ferromagnetic type in the sense that
$H_C = \sum_A \vec S_A \cdot \vec S_{A'}$, with
$\vec S_A = \sum_{i \in A} j_i \,
\vec s_i$ and $\vec S_{A'} = \sum_{i' \in A'} j_i \, \vec s_{i'}$, where
$A$ is a set of sites in the left subsystem, $A'$ is the corresponding
set of sites in the right subsystem, and $j_i$ are \emph{real} coefficients. 
The intrinsic spins $s_i$ are arbitrary
and can vary from site to site, as long as the whole system is reflection symmetric.
We shall state explicitly when we make further assumptions, e.g., that
the whole system is invariant under spin-rotations.
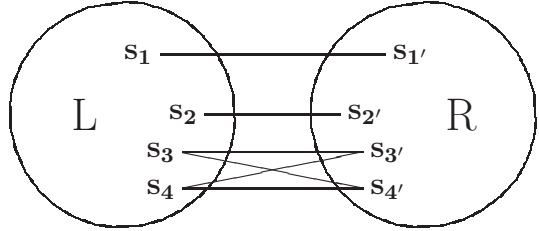
\begin{figure}[htb]
\begin{center}
\unitlength 1.00mm
\linethickness{0.4pt}
\begin{picture}(75.00,30.00)
\multiput(20.00,30.00)(0.99,-0.10){3}{\line(1,0){0.99}}
\multiput(22.97,29.70)(0.36,-0.11){8}{\line(1,0){0.36}}
\multiput(25.83,28.82)(0.22,-0.12){12}{\line(1,0){0.22}}
\multiput(28.45,27.39)(0.13,-0.11){17}{\line(1,0){0.13}}
\multiput(30.74,25.47)(0.12,-0.15){16}{\line(0,-1){0.15}}
\multiput(32.60,23.14)(0.11,-0.22){12}{\line(0,-1){0.22}}
\multiput(33.96,20.48)(0.12,-0.41){7}{\line(0,-1){0.41}}
\multiput(34.77,17.60)(0.11,-1.49){2}{\line(0,-1){1.49}}
\multiput(35.00,14.63)(-0.09,-0.74){4}{\line(0,-1){0.74}}
\multiput(34.62,11.66)(-0.12,-0.35){8}{\line(0,-1){0.35}}
\multiput(33.67,8.83)(-0.11,-0.20){13}{\line(0,-1){0.20}}
\multiput(32.18,6.24)(-0.12,-0.13){17}{\line(0,-1){0.13}}
\multiput(30.20,4.00)(-0.15,-0.11){16}{\line(-1,0){0.15}}
\multiput(27.82,2.20)(-0.24,-0.12){11}{\line(-1,0){0.24}}
\multiput(25.13,0.90)(-0.41,-0.11){7}{\line(-1,0){0.41}}
\multiput(22.24,0.17)(-1.49,-0.07){2}{\line(-1,0){1.49}}
\multiput(19.25,0.02)(-0.74,0.11){4}{\line(-1,0){0.74}}
\multiput(16.30,0.46)(-0.31,0.11){9}{\line(-1,0){0.31}}
\multiput(13.49,1.49)(-0.20,0.12){13}{\line(-1,0){0.20}}
\multiput(10.94,3.04)(-0.13,0.12){17}{\line(-1,0){0.13}}
\multiput(8.75,5.07)(-0.12,0.16){15}{\line(0,1){0.16}}
\multiput(7.01,7.50)(-0.11,0.25){11}{\line(0,1){0.25}}
\multiput(5.78,10.22)(-0.11,0.49){6}{\line(0,1){0.49}}
\put(5.12,13.13){\line(0,1){2.99}}
\multiput(5.04,16.12)(0.10,0.59){5}{\line(0,1){0.59}}
\multiput(5.56,19.06)(0.11,0.28){10}{\line(0,1){0.28}}
\multiput(6.65,21.84)(0.12,0.18){14}{\line(0,1){0.18}}
\multiput(8.27,24.35)(0.12,0.12){18}{\line(0,1){0.12}}
\multiput(10.36,26.49)(0.16,0.11){15}{\line(1,0){0.16}}
\multiput(12.83,28.17)(0.28,0.12){10}{\line(1,0){0.28}}
\multiput(15.58,29.33)(0.74,0.11){6}{\line(1,0){0.74}}
\multiput(60.00,30.00)(0.99,-0.10){3}{\line(1,0){0.99}}
\multiput(62.97,29.70)(0.36,-0.11){8}{\line(1,0){0.36}}
\multiput(65.83,28.82)(0.22,-0.12){12}{\line(1,0){0.22}}
\multiput(68.45,27.39)(0.13,-0.11){17}{\line(1,0){0.13}}
\multiput(70.74,25.47)(0.12,-0.15){16}{\line(0,-1){0.15}}
\multiput(72.60,23.14)(0.11,-0.22){12}{\line(0,-1){0.22}}
\multiput(73.96,20.48)(0.12,-0.41){7}{\line(0,-1){0.41}}
\multiput(74.77,17.60)(0.11,-1.49){2}{\line(0,-1){1.49}}
\multiput(75.00,14.63)(-0.09,-0.74){4}{\line(0,-1){0.74}}
\multiput(74.62,11.66)(-0.12,-0.35){8}{\line(0,-1){0.35}}
\multiput(73.67,8.83)(-0.11,-0.20){13}{\line(0,-1){0.20}}
\multiput(72.18,6.24)(-0.12,-0.13){17}{\line(0,-1){0.13}}
\multiput(70.20,4.00)(-0.15,-0.11){16}{\line(-1,0){0.15}}
\multiput(67.82,2.20)(-0.24,-0.12){11}{\line(-1,0){0.24}}
\multiput(65.13,0.90)(-0.41,-0.11){7}{\line(-1,0){0.41}}
\multiput(62.24,0.17)(-1.49,-0.07){2}{\line(-1,0){1.49}}
\multiput(59.25,0.02)(-0.74,0.11){4}{\line(-1,0){0.74}}
\multiput(56.30,0.46)(-0.31,0.11){9}{\line(-1,0){0.31}}
\multiput(53.49,1.49)(-0.20,0.12){13}{\line(-1,0){0.20}}
\multiput(50.94,3.04)(-0.13,0.12){17}{\line(-1,0){0.13}}
\multiput(48.75,5.07)(-0.12,0.16){15}{\line(0,1){0.16}}
\multiput(47.01,7.50)(-0.11,0.25){11}{\line(0,1){0.25}}
\multiput(45.78,10.22)(-0.11,0.49){6}{\line(0,1){0.49}}
\put(45.12,13.13){\line(0,1){2.99}}
\multiput(45.04,16.12)(0.10,0.59){5}{\line(0,1){0.59}}
\multiput(45.56,19.06)(0.11,0.28){10}{\line(0,1){0.28}}
\multiput(46.65,21.84)(0.12,0.18){14}{\line(0,1){0.18}}
\multiput(48.27,24.35)(0.12,0.12){18}{\line(0,1){0.12}}
\multiput(50.36,26.49)(0.16,0.11){15}{\line(1,0){0.16}}
\multiput(52.83,28.17)(0.28,0.12){10}{\line(1,0){0.28}}
\multiput(55.58,29.33)(0.74,0.11){6}{\line(1,0){0.74}}
\put(55.00,23.00){\line(-1,0){30.00}}
\put(31.00,15.00){\line(1,0){18.00}}
\put(52.00,10.00){\line(-1,0){24.00}}
\put(28.00,10.00){\line(5,-1){24.00}}
\put(52.00,5.20){\line(-1,0){24.00}}
\put(28.00,5.20){\line(5,1){24.00}}
\put(15.00,15.00){\makebox(0,0)[cc]{{\Large L}}}
\put(65.00,15.00){\makebox(0,0)[cc]{{\Large R}}}
\put(24.00,23.00){\makebox(0,0)[rc]{$\bf s_1$}}
\put(56.00,23.00){\makebox(0,0)[lc]{$\bf s_{1'}$}}
\put(50.00,15.00){\makebox(0,0)[lc]{$\bf s_{2'}$}}
\put(30.00,15.00){\makebox(0,0)[rc]{$\bf s_2$}}
\put(27.00,10.00){\makebox(0,0)[rc]{$\bf s_3$}}
\put(53.00,10.00){\makebox(0,0)[lc]{$\bf s_{3'}$}}
\put(53.00,5.00){\makebox(0,0)[lc]{$\bf s_{4'}$}}
\put(27.00,5.00){\makebox(0,0)[rc]{$\bf s_4$}}
\end{picture}
\end{center}
\caption{Some possible crossing bonds.}
\end{figure}

Any state of the system can be expanded in terms of a square matrix
$c$,
\eq
\psi = \sum_{\al,\be} c_{\al \be}  \psi_\al \ot \widetilde \psi_\be ,
\label{psi2}
\en
where $\{\psi_\al\}$ is a basis of $S^{(3)}$-eigenstates. (The
indices $\al$, $\be$ may contain additional non-spin 
quantum numbers, as needed, and the
tilde on the second tensor factor denotes the spin rotation.)
We shall assume that the state is normalized:
$\la \psi | \psi \ra = \tr c c^\dagger = 1$.
The energy expectation in terms of $c$ is a matrix expression
\eq
\la\psi|H|\psi\ra = \tr c c^\dagger h + \tr (c^\dagger c)^T h
- \sum_A \sum_{a=1}^3  \tr c^\dagger S_A^{(a)} c (S_A^{(a)})^\dagger ,
\label{ee2}
\en
here $(h)_{\al\be} = \la\psi_\al|H_L|\psi_\be\ra 
= \la\widetilde\psi_\al|H_R|\widetilde\psi_\be\ra$,
$(S_A^{(a)})_{\al\be}  
=\la\psi_\al| \sum_{i \in A} j_i \,s^{(a)}_i  |\psi_\be\ra$,
and we have used $(\widetilde{S_A^{(a)}})^T = -(S_A^{(a)})^\dagger$.
(For $a = 1, 3$ the minus sign comes from the spin rotation, for $a=2$
it comes from complex conjugation. This can be seen by writing
$S^{(1)}$ and $S^{(2)}$ in terms of the real matrices $S^+$ and $S^-$.)
Note, that we do not assume $(h)_{\al\be}$ to be real or symmetric, otherwise
the following considerations would simplify considerably~\cite{prl}.

We see, by inspection, that the energy expectation value remains unchanged if
we replace $c$ by its transpose $c^T$, and, by linearity, if we replace it
by $c+c^T$ or $c-c^T$. So, if $c$ corresponds to a ground state, then we
might as well assume for convenience that $c$ is either symmetric or
antisymmetric. 
Note, that in either case we have $(c_R)^T = c_L$, where $c_L \equiv 
\sqrt{c c^\dagger}$ and $c_R \equiv \sqrt{c^\dagger c}$.
(Proof:
$(c_R^2)^T = (c^\dagger c)^T = c c^\dagger = c_L^2$, if $c^T = \pm c$;
now take the unique square root of this.) Using this we see that the 
first two terms in the energy
expectation equal $2 \tr c_L^2 h $ and thus depend on $c$ only through 
the positive semidefinite matrix $c_L$.
With the help of a trace inequality we will show that
the third term does not increase if we replace $c$ by the positive semidefinite
matrix $c_L$.

\subsection[Trace inequality and spin rings]{Trace inequality}

For any square matrices $c$, $M$, $N$ it is true that \cite{KLS}
\eq
|\tr c^\dagger M c N^\dagger| \leq \frac{1}{2}\left( \tr c_L M c_L M^\dagger
+ \tr c_R N c_R N^\dagger \right) , \label{trinequality}
\en
where $c_R = \sqrt{c^\dagger c}$, $c_L = \sqrt{c c^\dagger}$ are the
unique square roots of the positive semidefinite matrices $c^\dagger c$
and $c c^\dagger$. Here is a proof: By the polar decomposition theorem
$c = u c_R$ with a unitary matrix $u$ and
$(u c_R u^\dagger)^2 = u c^\dagger c u^\dagger = c c^\dagger = c_L^2$, 
so by the uniqueness of the square root
$u c_R u^\dagger = c_L$. Similarly,
for any function $f$ on the non-negative
real line $u f(c_R) u^\dagger$ = $f(c_L)$, and in particular
$u \sqrt{c_R} = \sqrt{c_L} u$
and thus $c = \sqrt{c_L} u \sqrt{c_R}$.
Let $P \equiv u^\dagger \sqrt{c_L} M \sqrt{c_L} u$ and
$Q \equiv \sqrt{c_R} N^\dagger \sqrt{c_R}$, then
\eqa
|\tr c^\dagger M c N^\dagger| & = & |\tr P Q| 
\, \leq \, \frac{1}{2}(\tr P P^\dagger + \tr Q Q^\dagger) \nn
& = & \frac{1}{2}\left( \tr c_L M c_L M^\dagger
+ \tr c_R N c_R N^\dagger \right) ,
\ena
where the inequality is simply the geometric arithmetic
mean inequality for matrices
$$
|\tr P Q| = |\sum_{i,j} P_{ij} Q_{ji}| \leq 
\frac{1}{2} \sum_{i,j} (|P_{ij}|^2 +  |Q_{ji}|^2)
= \frac{1}{2}(\tr P P^\dagger + \tr Q Q^\dagger) .
$$

\subsubsection{Spin rings -- with or without reflection symmetry}

\begin{figure}[htb]
\begin{center}
\includegraphics[bb=40 90 580 700, width=.28\textwidth]{%
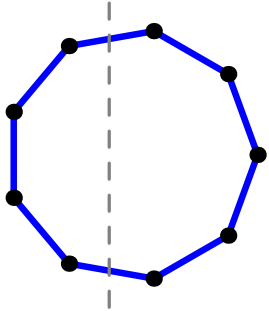}\vspace{-1em}
\caption{Ring with $n+m$ spins. Reflection along the
dottet line leads to two rings with $2n$ and $2m$ spins
respectively. The trace inequality and (\ref{ee2})
imply an inequality for the  ground state energies:
$E_{n+m} \geq \frac{1}{2}(E_{2n} + E_{2m})$.}\label{fig:nmspins}
\end{center}
\end{figure}

\begin{figure}[htb]
\begin{center}
\includegraphics[bb=160 80 450 710, angle=90, width=.3\textwidth]{%
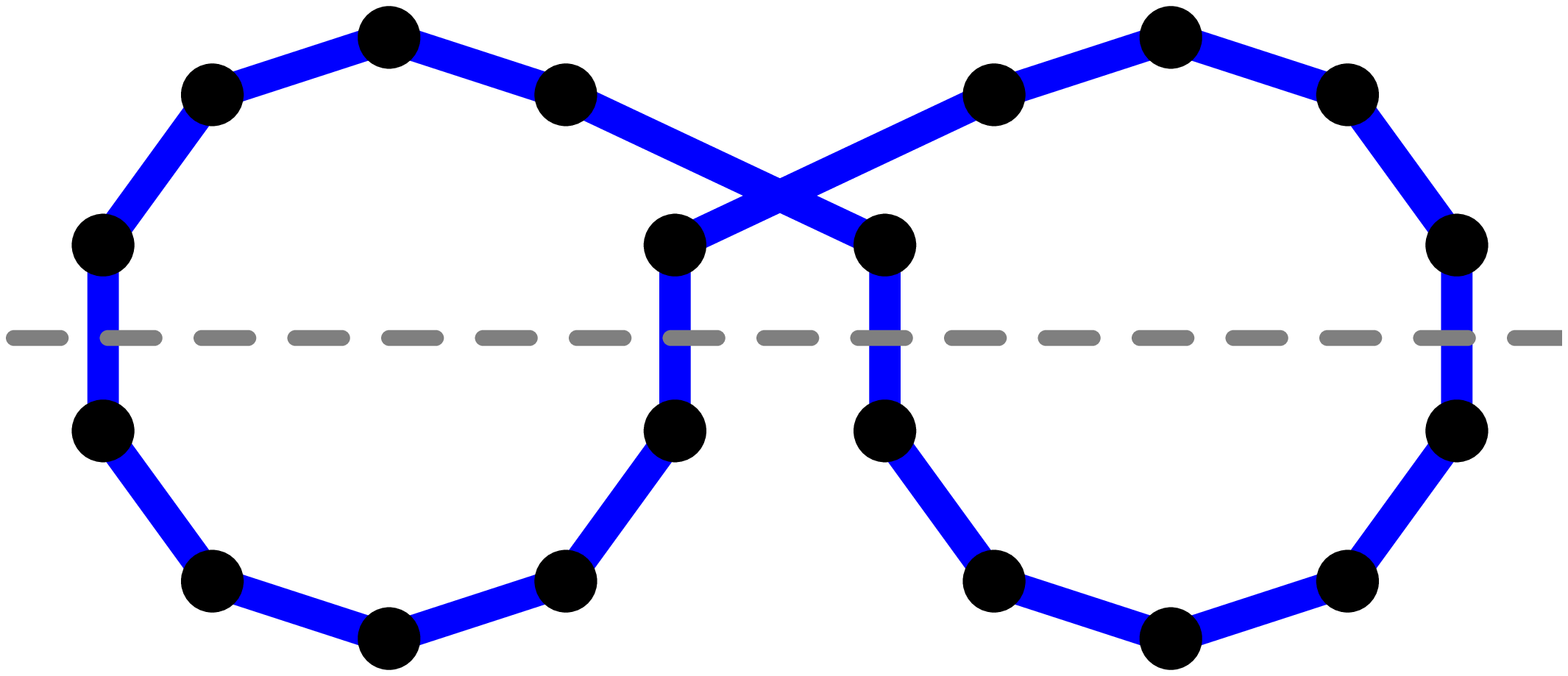}%
\parbox{.07\textwidth}{\raisebox{4.5em}{{\large$\;\rightarrow$}}}%
\includegraphics[bb=160 80 450 710, angle=90, width=.3\textwidth]{%
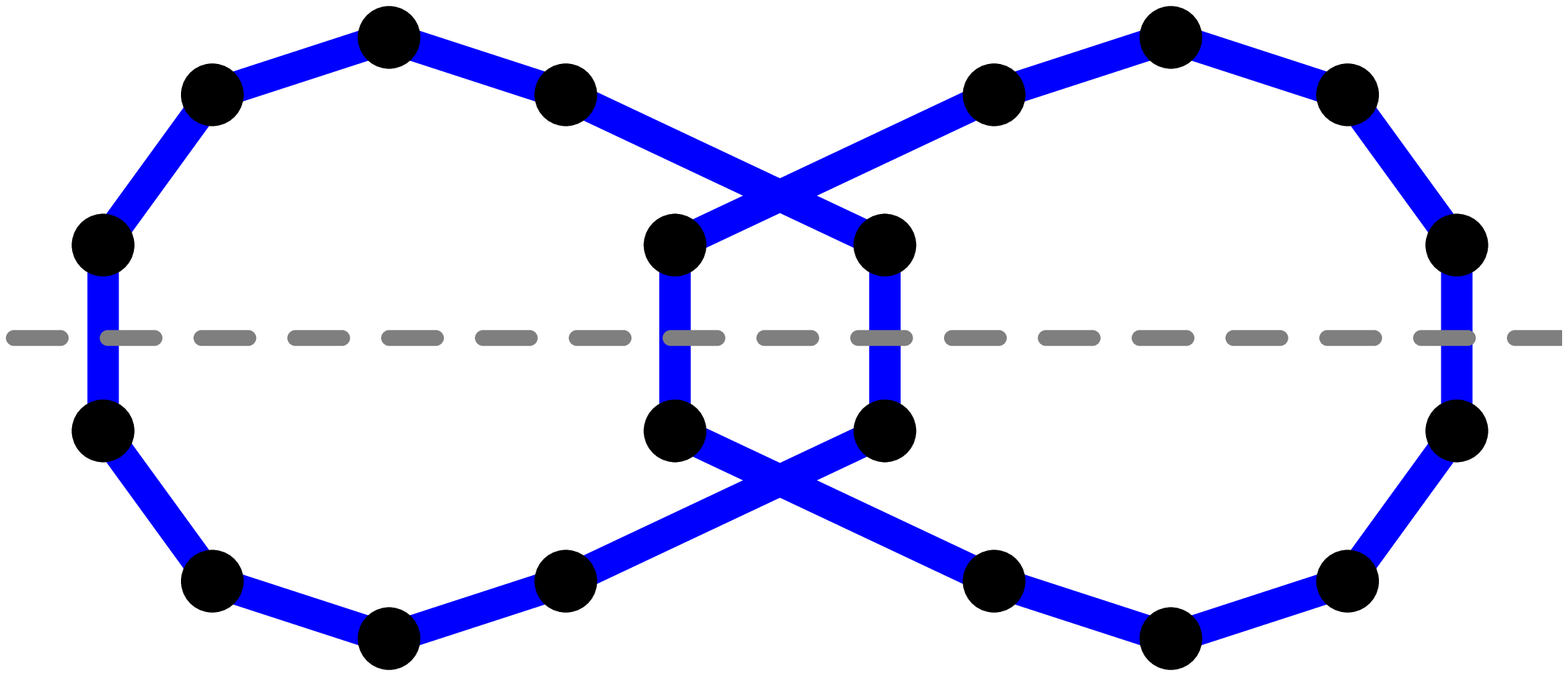}%
\parbox{.03\textwidth}{\raisebox{4.5em}{{\large$\,,$}}}%
\includegraphics[bb=160 80 450 710, angle=90, width=.3\textwidth]{%
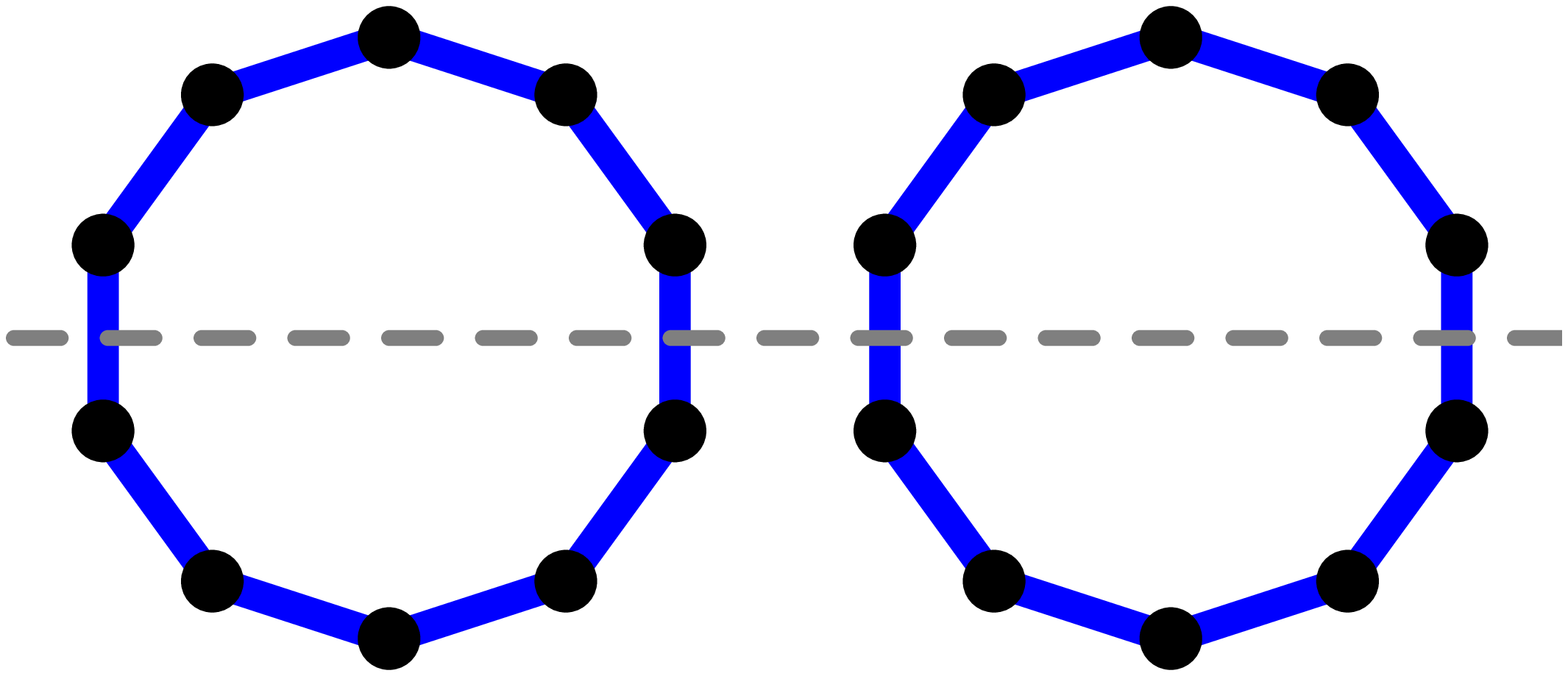} \vspace{-2em}
\caption{Figure-8 ring with $2n$ spins ($n$ even). Reflection
along the dotted line leads to two pairs of rings with
$n$ spins each. The trace inequality and (\ref{ee2})
imply an inequality for the ground state energy \emph{per spin}:
$e_{2n} \geq e_n$.}\label{fig:2nspins}
\end{center}
\end{figure}

Inequality (\ref{trinequality}) also
holds for rectangular matrices:
If $c$ is an $m\times n$ matrix then $M$ and $N$ are $m\times m$ and
$n\times n$-matrices respectively, $u$ is a partial isometry and
$c_L$ and $c_R$ are positive $m\times m$ and $n\times n$ matrices
respectively.
This was originally of interest for the ordering
of energy levels in the Hubbard model (see chapter~\ref{chap:electron})
and for Bose-Einstein
condensation in the hard core lattice approximation away from half-filling, 
but may also be used to find rigorous
inequalities for the 
ground state energy of 1-dimensional periodic spin systems
(spin rings) with an even or odd number of spins,
as illustrated in figures~\ref{fig:nmspins} and~\ref{fig:2nspins}.

\subsection{Positive spin-zero ground state}
\subsubsection{Existence of a positive ground state}
\label{positive}

Consider any ground state of the system with coefficient matrix 
$c = \pm c^T$
and apply the trace inequality to the terms in $\la\psi|H_C|\psi\ra$:
$$
-\tr c^\dagger S_A^{(a)} c (S_A^{(a)})^\dagger
\geq
-\frac{1}{2} \left(\tr c_L S_A^{(a)} c_L (S_A^{(a)})^\dagger
+ \tr c_R S_A^{(a)} c_R (S_A^{(a)})^\dagger\right) ,
$$
but $(c_R S_A^{(a)} c_R (S_A^{(a)})^\dagger)^T 
= ((S_A^{(a)})^T)^\dagger c_L (S_A^{(a)})^T c_L
= (S_A^{(a)})^\dagger c_L (S_A^{(a)}) c_L$, so in fact
$$
-\tr c^\dagger S_A^{(a)} c (S_A^{(a)})^\dagger 
\geq -\tr c_L S_A^{(a)} c_L (S_A^{(a)})^\dagger .
$$
Since the normalization of the state and 
the other terms in (\ref{ee}) are unchanged if we replace $c$ by 
$c_L = \sqrt{c c^\dagger}$, and because we have assumed that $c$ is 
the coefficient matrix of a ground state, it follows that the 
positive semidefinite matrix $c_L$ must also be the coefficient 
matrix of a ground state.

\subsubsection{Overlap with canonical spin zero state}

Consider the (not normalized) canonical state with coefficient matrix given by
the identity matrix in a basis of $S^{(3)}$-eigenstates of either
subsystem
\eqa
\Xi & = & \displaystyle \sum_{k,k'}   
\sum_{j}\sum_{m = -j}^j \psi_{(j,m,k)} 
\ot \widetilde \psi_{(j,m, k')} \nn
& = & \displaystyle \sum_{k,k'}       
\sum_{j}\sum_{\tilde m = -j}^j \psi_{(j,m,k)} 
\ot(-)^{j-m} \psi_{(j,-m, k')}.
\ena
The states are labeled by the usual spin quantum numbers $j$, $m$ and an 
additional symbolic quantum number $k$ to lift remaining ambiguities. 
The state $\Xi$ has total spin zero because of the spin rotation in 
the right subsystem:
Its $S_{\text{tot}}^{(3)}$-eigenvalue is zero and acting with either 
$S^+_{\text{tot}}$ or $S^-_{\text{tot}}$ on it gives zero.
The overlap of any state with coefficient matrix $c$ with the canonical
state $\Xi$ is simply the trace of $c$. 
In the previous section we found that the reflection symmetric spin
system necessarily has a ground state with positive semidefinite, non-zero 
coefficient matrix, which, by definition, has a (non-zero) positive trace. 
Since the trace is proportional to the overlap with the canonical spin-zero
state, we have now shown that there is always a ground state that
contains a spin-zero part. Provided that total spin is a good quantum
number, we can conclude further that our system always has a ground
state with total spin zero, i.e., a singlet. 

\subsubsection{Projection onto spin zero}

Consider any state
$\psi = \sum c_{\al \be} \psi_\al \ot \widetilde \psi_\be$
with positive semidefinite $c = |c|$. We have seen that this implies
that $\psi$ has a spin-zero component. If total spin is a good quantum
number it is interesting to ask what happens to $c$ when we project
$\psi$ onto its spin zero part 
\eq
\psi^0 = \sum c^0_{\al \be} \psi_\al \ot \widetilde \psi_\be .
\en
We shall show that the
coefficient matrix $c^0$ of $\psi^0$ is a partial trace of $c$ and thus
still positive semidefinite.  A convenient parametrisation of the 
$S^3$ eigenstates $\psi_\al$ for this task is, as before, 
$\al = (j,m,k)$, where $k$ labels spin-$j$
multiplets $[j]_k$ in the decomposition of the Hilbert space
of one subsystem into components of total spin.
Note that $[j]_k \ot [j']_{k'} = [j+j']  \op \ldots \op [|j-j'|]$, so
$[j]_k \ot [j']_{k'}$ contains a spin zero subspace only if $j = j'$, and for
each $k$, $k'$ that subspace is unique and generated by the normalized
spin zero state
\eq
\xi_{k,k'} = (2j+1)^{-\frac{1}{2}} 
\sum_{\tilde m = -j}^j \psi_{(j,\tilde m,k)} 
\ot \widetilde \psi_{(j,\tilde m, k')} .
\en
(Recall that
$\widetilde \psi_{(j,\tilde m, k')} 
= (-)^{j-\tilde m} \psi_{(j,-\tilde m, k')}$
is the rotation of $\psi_{(j,\tilde m, k')}$ by $\pi$ around the 2-axis in
spin space.)
The projection of $\psi$ onto spin zero is thus amounts to replacing $c$ with
$c^0$, where
\eq
c^0_{(j,m,k)(j',m',k')} = \left\{
\begin{array}{l}
0 \quad\mbox{if $j \neq j'$ or $m \neq m'$} \\
\displaystyle \frac{N}{2j + 1} \sum_{\tilde m = -j}^j 
c_{(j,\tilde m, k)(j,\tilde m, k')} \quad\mbox{else} .
\end{array}
\right. \label{czero}
\en
($N$ is a overall normalisation constant, independent of $j$, $m$, $k$.)
Let us now show that this partial trace preserves positivity, i.e.,
$(v, c^0 \; v) \geq 0$ for any vector $v=\left(v_{(j,m,k)}\right)$ 
of complex numbers.
If we decompose $v$ into a sum of vectors $v_{j m}$
with definite $j$, $m$ and use (\ref{czero}), we see
\eq
(v , c^0 \, v)  =  \sum_{j,m} (v_{j m} , c^0 \, v_{j m})
=  \sum_{j,m,\tilde m} (\omega_{j \tilde m}^m ,c \, \omega_{j \tilde m}^m)
\geq 0 ,
\en
where the $\omega_{j \tilde m}^m$ are new vectors with components
$\omega_{(j,\tilde m,k)}^m = v_{(j,m,k)}$, independent of $\tilde m$. 
Every term in the last sum
is non-negative because $c$ is positive semidefinite by assumption.
This result implies in particular that a reflection symmetric spin
system always has a ground state with total spin zero \emph{and} positive
semidefinite coefficient matrix -- provided that total spin is a good
quantum number.

\subsection{Ice rule for crossing bonds}

The expectation of the third spin component
of the sites involved in each crossing bond $B$, weighted by their coefficients
$j_i$, vanishes for \emph{any} ground state $\psi_0$,
\eq
\la \psi_0 |\sum_{i \in B} j_i (s_i^{(3)} + s_{i'}^{(3)})|
\psi_0\ra = 0,
\label{ice}
\en
provided that either the left
and right subsystems are invariant under the spin rotation,
$h = \tilde h$, or that their matrix elements are real (the latter
is equivalent to the assumption $h = h^T$, since we know
that $h = h^\dagger$ or otherwise the whole spin Hamiltonian would not be 
Hermitean).
By symmetry (\ref{ice}) will also be true for the first spin component and, if 
we are dealing with a spin Hamiltonian that is invariant under spin rotations, 
it is also true for the second spin component.
For ground states with symmetric or antisymmetric coefficient matrix
we automatically have $\la s_i^{(3)} + s_{i'}^{(3)} \ra = 0$ for
any pair of sites $i$ and $i'$, so in that case (\ref{ice}) is trivial.

For the proof we introduce a real parameter $b$ in the spin Hamiltonian:
$H(b) \equiv H - b(S^{(3)}_B + S^{(3)}_{B'}) + b^2/2$, where $B$ is one
of the sets of sites involved in the crossing bonds of the original
Hamiltonian $H$. Let $E_b$ be the 
ground state energy of $H(b)$ and $E_0$ the ground state energy of $H$.
One can show that $E_b \geq E_0$ and
(\ref{ice}) follows then by a variational argument:
\eq
\la \psi_0 | H(b) | \psi_0 \ra \geq E_b \geq E_0 = \la \psi_0 | H | \psi_0 \ra,
\en
or, $\la\psi_0| b(S^{(3)}_B + S^{(3)}_{B'})|\psi_0\ra \leq b^2/2$, which
implies (\ref{ice}). Note, that we did not make any assumptions about the symmetry
or antisymmetry of the coefficient matrix of $\psi_0$ here.

Sketch of the proof of $E_b \geq E_0$ (see also \cite{KLS,prl}): 
$H(b) = H_L(b) + H_R(b) + H_C(b) + b^2/4$ with 
$H_{L,R}(b) = H_{L,R} - b/2 \cdot S^{(3)}_B$ and 
$H_C(b)$ equal to $H_C$ except for the term $S^{(3)}_B \cdot S^{(3)}_{B'}$, which
is replaced by $(S^{(3)}_B - b/2)\cdot(S^{(3)}_{B'} - b/2)$.
If we now write the ground state energy expectation of $H(b)$ as 
a matrix expression like (\ref{ee}) and apply the trace inequality to
it, we will find an equal or lower energy expectation
not of $H(b)$, but rather of $H$: The trace inequality effectively removes the 
parameter $b$ from the Hamiltonian. By the variational principle the true
ground state energy of $H$ is even lower and we
conclude that $E_b \geq E_0$. Role of the technical assumptions mentioned above:
If $h = h^T$, then the transpose in the second term in (\ref{ee}) vanishes,
the matrix expression is symmetric in $c_L$ and $c_R$ (except for the sign
of the parameter $b$), and the trace inequality gives
$\la H(b) \ra_c \geq \frac{1}{2}\left\{\la H \ra_{c_L} + \la H \ra_{c_R}\right\}$.
If $h = \tilde h$, then we should drop the spin rotation  on the
second term of the analog of expression (\ref{psi2}) for $\psi_b$. 
The matrix expression for $\la H(b) \ra$ is then symmetric in $c$ and $c^T$ 
and we may assume $c = \pm c^T$ to prove $E_b \geq E_0$. The calculation is
similar to the one in section~\ref{positive}. Note, that $c = \pm c^T$ only
enters the proof of $E_b \geq E_0$, we still do not need to assume that the coefficient
matrix of $\psi_0$ in (\ref{ice}) has that property.

The preferred configurations of
four spins with antiferromagnetic crossed bonds in a classical
Ising system are very similar to the configurations of the 
four hydrogen atoms that surround each oxygen
atom in ice (Fig.~\ref{fig:ice}):
\begin{figure}
\begin{center}
\includegraphics[bb=100 200 600 650,clip,width=.32\textwidth]{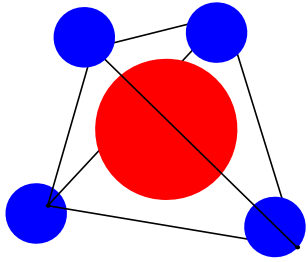}
\caption{Pauli's ice rule for hydrogen atoms around an oxigen atom:
Two hydrogen atoms are closer to and two are further away from
each oxygen atom. In the analogous Ising system this would correspond
to the magnetic quantum number $M = 0$. We have found such an ice rule
for the spins involved in each crossing bond.}
\label{fig:ice}
\end{center}
\end{figure}
There are always two hydrogen atoms close and two further
away from each oxygen atom, and there are always
two spins ``up'' and two ``down'', i.e. $M = 0$, in the Ising system.
Equation $(\ref{ice})$
is a (generalized) quantum mechanical version of this --
that is why we use the term ``ice rule''. This phrase
is also used in the context of \emph{ferromagnetic} pyrochlore
with Ising anisotropy (``spin ice'') \cite{HBMZG} and we hope that does not cause
confusion.

\subsection{Comparison with previous results}

We would like to discuss similarities between our method and 
previous work, in particular the approach of \cite{LM} for
the bipartite antiferromagnet:
There, the spin Hamiltonian splits into two parts $H = H_0 + H_1$. 
The expectation value of $H_0$ with respect to a state
$\psi = \sum f_\al \phi_\al$, expanded in an appropriate basis
$\{\phi_\al\}$, depends only on $|f_\al|$, 
and the expectation of $H_1$ does not increase under the transformation 
$f_\al \rightarrow |f_\al|$. The variational principle then implies that there must be 
a ground state with only non-negative coefficients $|f_\al|$.
The present setup is very similar, except that 
we use coefficient matrices $(c_{\al \be})$ to expand states,
since we work on a tensor product of Hilbert spaces. In our case
the expectation value of
$H_0 = H_L + H_R$ depends only on $c$ via the positive matrices
$c_L$ and $c_R$, and the expectation value of $H_C$ increases
if we ``replace" $c$ by these positive matrices. The similarity is even
more apparent if $h$ has real matrix elements: In that case we may
assume that $c$ is diagonalisable and its eigenvalues play the
role of the coefficients $f_\al$. 
The spin of a positive ground state is established in all cases
from the overlap with a state of known spin that is also positive.
In a system with sufficient symmetry we can, however, also use the
``ice rule'' to prove that \emph{all} ground states have
total spin zero \cite{prl}. (E.g., in a system with constant coefficients
$j_i$ and enough translational invariance, so that every spin can
be considered to be involved in a crossing bond and thus in an ice rule,
we would conclude that all ground states have $S^{(3)}_{\text{tot}} = 0$
and, assuming rotational invariance in spin space,
$S_{\text{tot}} = 0$.)
It is not clear, if $M$-subspace methods can be used in the present setting
to get information about excited states.
An important point in the our work is that we 
consider not only antiferromagnetic
bonds between single sites but also bonds between sets of sites. This frees us
from the requirement of bipartiteness and even allows some ferromagnetic
crossing bonds, for example in $(s_1 - s_2)(s_{1'} - s_{2'})$.
There is no doubt that the scheme can be further generalized, e.g., to other
groups or more abstract ``crossing bonds''. In the present form the most interesting
applications are in the field of frustrated spin systems \cite{prl}.

We did not address the question of the degeneracy of ground states. Classically
a characteristic feature of frustrated systems is 
their large ground state degeneracy. For frustrated quantum spin systems this is an
important open problem.

\chapter{{\sffamily Wehrl entropy of Bloch coherent states}}
\label{chap:wehrl}

There is a hybrid between the quantum mechanical entropy,
$-\mbox{tr} \rho \ln \rho$, and the entropy of the 
corresponding classical system: Wehrl proposed to use the expectation 
$\langle z|\rho|z\rangle$ 
of the
density matrix between coherent states 
as probability density for the
Shannon entropy;
$S_{W} = -\int d z \; \langle z|\rho|z\rangle \ln
\langle z|\rho|z\rangle$.
It turns out that this entropy has nice properties, some of which either of 
its `parents' lack: The Wehrl entropy is sub-additive, 
monotone and always positive -- even 
for pure states. In fact, it is always larger than the quantum 
entropy. 

Wehrl conjectured that the minimum of $S_{W}$ for a 
single particle on the line is reached only for density matrices that are
projectors onto coherent states. This was proved by Lieb using 
advanced theorems in Fourier analysis. \mbox{While} it is well-known that 
entropy considerations can lead to non-trivial inequalities (they 
e.g.\ improve on Heisenberg's uncertainty), it was still quite 
surprising that the proof of Wehrl's conjecture was so hard.
In an effort to shed more light on this, Lieb was led to a related 
conjecture for
Bloch coherent states. Even though many attempts have been made to find a proof
of the latter, there had been virtually no progress for the
last twenty years.
Using a geometric representation of spin states 
we will see in the following sections how to compute
the Wehrl entropy explicitly and will settle
the conjecture for cases of low spin. We will also give a group 
theoretic proof for all spin of a related inequality.

Sharp inequalities that stem from entropy considerations have in the
past been seen to be very useful in mathematical physics.
In this particular case there is a way to use Lieb's conjecture
to get much better approximations to probability distributions than are
in use today. There is also a direct physical interpretation:\
The Wehrl Entropy for states of spin $j$ is the 
entropy of a point vortex on the sphere in the background of $2 j$
fixed point vortices with the usual vortex-interaction.
We find that a proof of Lieb's conjecture for low spins can be reduced
to some beautiful spherical geometry, 
but the unreasonable difficulty of a complete proof
is still a great puzzle; its resolution may very well lead to
interesting mathematics and perhaps physics.

The results have been published in~\cite{bloch}, the remainder of
this chapter is based on that publication. Much of the early work
was done in collaboration with Wolfgang Spitzer who 
\label{end:wehrl}

\section{Conjectures of Wehrl and Lieb}

For a quantum mechanical system with density matrix $\rho$, Hilbert space
$\Hil$,
and a family of normalized
coherent states $|z\ra$,
parametrized symbolically by
$z$ and satisfying
$\int dz \, |z\ra\la z| = {\bf 1}$ (resolution of identity),
the Wehrl entropy~\cite{W} is
\eq
S_W(\rho)  = -\int dz \, \la z|\rho|z\ra \ln \la z|\rho|z\ra , \label{W0}
\en
i.e., this new entropy is the ordinary Shannon entropy of
the probability density provided by the lower
symbol of the density matrix.
If we are dealing with a tensor product $\Hil_1\ot\Hil_2$
of Hilbert spaces we can either consider the
total entropy $S_{12}$ directly, or we can use partial traces
to compute reduced density matrices $\rho_1$, $\rho_2$
and the associated entropies $S_1$ and $S_2$. It would
be physically desirable to have inequalities
\eq
S_1 \leq S_{12} \leq S_1 + S_2
\en
but it turns out that while subadditivity $S_{12} \leq S_1 + S_2$
is no problem, monotonicity $S_1 \leq S_1 + S_2$ in general fails
for quantum entropy and for classical continuous entropy. For
the Wehrl entropy, however, both inequalities are valid. Furthermore
it also satisfies concavity in $\rho$ and strong subadditivity~\cite{B,E}.

Like quantum mechanical entropy,
$S_Q = -\tr \rho\ln\rho $,
Wehrl entropy is always non-negative, in fact
$S_W > S_Q \geq 0$.
In view of this inequality it is interesting to ask for the minimum
of $S_W$ and the corresponding minimizing density matrix.
It follows from concavity of $-x \ln x$ that a minimizing
density matrix must be a pure state, \ie, $\rho = |\psi\ra\la\psi|$
for a normalized vector $|\psi\ra \in \Hil$ \cite{A}. (Note that
$S_W(|\psi\ra\la\psi|)$ depends on $|\psi\ra$ and is non-zero,
unlike the quantum entropy which is of course zero for pure states.)

For Glauber coherent states  Wehrl conjectured \cite{W} and Lieb proved
\cite{A}
that the minimizing state $|\psi\ra$ is again a coherent
state. It turns out that all Glauber coherent states
have Wehrl entropy one, so Wehrl's conjecture can be written as follows:
\begin{thm}[\rm Lieb] \label{thmWL}
The minimum of $S_W(\rho)$ for states in $\Hil = L^2(\RE )$
is one,
\eq
S_W(|\psi\ra\la\psi|)
= -\int dz \, |\la\psi|z\ra|^2 \ln |\la\psi|z\ra|^2 \geq 1 , \label{W1}
\en
with equality if and only if $|\psi\ra$ is a coherent state.
\end{thm}
To prove this, Lieb used a clever combination of the sharp
Hausdorff-Young inequality \cite{Y1,Y3,LL}
and the sharp Young inequality \cite{Y2,Y1,Y3,LL}
to show that
\eq
s \int dz \, |\la z|\psi\ra|^{2 s} \leq 1, \quad s \geq 1, \label{W2}
\en
again with equality if and only if $|\psi\ra$ is a coherent state.
Wehrl's conjecture follows from this in the limit $s \rightarrow 1$
essentially because
(\ref{W1}) is the derivative of (\ref{W2}) with respect to $s$ at $s=1$.
All this easily generalizes to $L^2(\RE^n)$ \cite{A,Y4}.

The lower bound on the Wehrl entropy is related to
Heisenberg's uncertainty principle \cite{AH,G} and it has been speculated that
$S_W$ can be used to measure uncertainty due to both quantum and thermal
fluctuations \cite{G}.

It is very surprising that `heavy artillery' like the sharp constants
in the mentioned inequalities are needed in Lieb's proof. To elucidate
this situation, Lieb suggested \cite{A} studying the analog of Wehrl's
conjecture for Bloch coherent states $|\Omega\ra$, where one should
expect significant 
simplification since these are finite dimensional Hilbert spaces.
However, no progress has been made, not even for a single spin, even though
many attempts have been made \cite{B}. Attempts to proceed again 
along the lines of Lieb's original proof have failed to provide a sharp
inequality and the direct computation of the entropy and related integrals,
even numerically, was unsuccessful \cite{S}.

The key idea turns out to be a geometric representation of a
state of spin~$j$ as $2j$ points on a sphere.
In this representation the expression
$|\la\Omega|\psi\ra|^2$ factorizes into a product
of $2j$ functions $f_i$ on the sphere,  
which measure the square chordal distance
from the antipode of the point parametrized by
$\Omega$ to each of the $2j$ points on the sphere.
Lieb's conjecture,
in a generalized form analogous to (\ref{W2}),
then looks like the quotient of two H\"older inequalities
\eq
\frac{|\!|f_1 \cdots f_{2j}|\!|_s}{|\!|f_1 \cdots f_{2j}|\!|_1}
\leq 
\frac{\prod_{i=1}^{2j}|\!|f_i|\!|_{2js}}{\prod_{i=1}^{2j} |\!|f_i|\!|_{2j}},
\label{holder}
\en
with the one with the higher power winning against the other one.
We shall give a group theoretic proof of this inequality
for the special case $s \in \N$ in theorem~\ref{natural}.

In the geometric representation the 
Wehrl entropy of spin states finds a direct physical
interpretation: It is the classical entropy of a single particle on a sphere
interacting via Coulomb potential with $2j$ fixed sources; $s$ plays the
role of inverse temperature.

The entropy integral (\ref{W0}) can now be done because 
$|\la\Omega|\psi\ra|^2$ factorizes
and one finds a formula for the Wehrl entropy of any state.
When we evaluate the entropy explicitly for states of spin 1,
3/2, and 2 we find surprisingly simple expressions  solely in
terms of the square chordal distances between the points on the
sphere that define the given state. 

A different, more group theoretic approach seems to point 
to a connection between
Lieb's conjecture and the norm of certain spin $j s$ states with
$1 \leq s \in \RE$ \cite{J}.
So far, however, this has only been useful for proving the analog
of inequality (\ref{W2}) for $s \in \N$.

\subsection{Bloch coherent spin states}

Glauber coherent states 
\eq
|z\ra = \pi^{-\frac{1}{4}} e^{-(x-q)^2/2} e^{ipx},
\en
parametrized by $z = (q + i p)/\sqrt{2}$ and equiped with a measure
$dz = dp dq/2\pi$,
are usually introduced as
eigenvectors of the annihilation operator
$a = (\hat x + i \hat p)/\sqrt{2}$, 
$a|z\ra = z |z\ra$,
but the same states can also be
obtained by the action of the Heisenberg-Weyl group
$H_4 = \{a^\dagger a, a^\dagger, a, I\}$
on the extremal state
$|0\ra = \pi^{-\frac{1}{4}} e^{-x^2/2}$.
Glauber coherent states are thus elements of the coset
space of the Heisenberg-Weyl group
modulo the stability subgroup $U(1)\ot U(1)$ that leaves the extremal state
invariant. (See \eg\ \cite{C1} and references therein.)
This construction easily generalizes
to other groups, in particular to SU(2), where it gives
the Bloch coherent spin states \cite{BC} that we are interested in:
Here the Hilbert space can be any one of the finite dimensional spin-$j$
representations $[j] \equiv \C^{2j+1}$ of SU(2), 
$j = {1\over 2}, 1, \frac{3}{2}, \ldots$,
and
the extremal state for each $[j]$ is the
highest weight vector $|j,j\ra$. The stability subgroup is U(1)
and the coherent states are thus 
elements of the sphere $S_2 = $SU(2)/U(1);
they can be labeled by
$\Omega = (\theta,\phi)$ and are
obtained from $|j,j\ra$ by rotation:
\eq
|\Omega\ra_j = \R_j(\Omega) |j,j\ra. \label{Om}
\en
For spin $j = \half$ we find
\eq
|\omega\ra = p^{\half} e^{-i{\phi\over 2}} |\U\ra 
           + (1-p)^{\half} e^{i{\phi\over 2}} |\D\ra ,  \label{coh}
\en
with $p \equiv \cos^2\frac{\theta}{2}$.
(Here and in the following $|\omega\ra$ is short for the spin-$\half$ coherent
state $|\Omega\ra_\half$; $\omega = \Omega = (\theta,\phi)$. 
$|\U\ra \equiv |\half,\half\ra$ and
$|\D\ra \equiv |\half,-\half\ra$.)
An important observation for what follows is that
the product of two coherent states for the same $\Omega$
is again a coherent state:
\eqa 
|\Omega\ra_j \ot |\Omega\ra_{j'} 
	& = & (\R_j \ot \R_{j'})\, (|j,j\ra \ot |j',j'\ra)  \nn
	& = & \R_{j+j'}\, |j+j',j+j'\ra 	           
	\; = \; |\Omega\ra_{j+j'} .	     
\ena
Coherent states are in fact the only states for which
the product of a spin-$j$ state with a spin-$j'$ state is
a spin-$(j+j')$ state and not a more general element of
$[j+j'] \oplus \ldots \oplus [\,|j - j'|\,]$.
From this key property
an explicit representation for Bloch coherent states of higher spin 
can be easily derived:
\eqa
|\Omega\ra_j & = & \left(|\omega\ra\right)^{\ot 2j} 
             \; = \;  \left(p^{\half} e^{-i{\phi\over 2}} |\U\ra 
               + (1-p)^{\half} e^{i{\phi\over 2}} |\D\ra\right)^{\ot 2j} \nn
	     & = & \sum_{m=-j}^j {2 j \choose j + m}^{\half}
	     p^{j+m\over 2}  (1-p)^{j-m\over 2} 
	     e^{-i m {\phi\over 2}} |j,m\ra.  \label{Coh}
\ena
(The same expression can also be obtained directly from (\ref{Om}), see
\eg\ \cite[chapter 4]{C2}.)
The coherent states as given are normalized $\la\Om|\Om\ra_j =1$
and satisfy
\eq
(2j+1) \int\dOm \, |\Om\ra_j\la\Om|_j = P_j , \qquad \mbox{(resolution of
identity)} \label{project}
\en
where $P_j = \sum |j,m\ra\la j,m|$ is the projector onto $[j]$.
It is not hard to compute the Wehrl entropy for a coherent state
$|\Om'\ra$: Since the integral over the sphere is invariant under rotations
it is enough to consider the coherent state $|j,j\ra$; then use
$|\la j,j|\Om\ra|^2 = |\la\U\!\!|\om\ra|^{2\cdot 2j} = p^{2j}$
and $d\Om/4\pi = -dp\,d\phi/2\pi$, where 
$p =\cos^2 \frac{\theta}{2}$ as above, to obtain
\eqa
S_W(|\Om'\ra\la\Om'|) 
& = & -(2j+1) \int\dOm \, |\la\Om|\Om'\ra|^2 \ln |\la\Om|\Om'\ra|^2 \nn
& = & -(2j+1) \int_0^1 dp \, p^{2j} \, 2j \ln p \, = \, \frac{2j}{2j+1}.
\ena
Similarly, for later use,
\eq
(2js+1) \int \dOm |\la\Om'|\Om\ra|^{2s} = (2js+1) \int_0^1 dp \, p^{2js} = 1.
\en
As before the density matrix that minimizes $S_W$
must be a pure state $|\psi\ra\la\psi|$. 
The analog of theorem~\ref{thmWL} for spin states is:
\begin{conj}[\rm Lieb] \label{conject1}
The minimum of $S_W$ for states in $\Hil = \C^{2j+1}$
is $2j/(2j+1)$,
\eq
S_W(|\psi\ra\la\psi|) 
= -(2j+1) \int\dOm \, |\la\Om|\psi\ra|^2 \ln |\la\Om|\psi\ra|^2
\geq \frac{2j}{2j+1},
\en
with equality if and only if $|\psi\ra$ is a coherent state.
\end{conj}

\noindent
\emph{Remark:} For spin 1/2 this is an identity because
all spin~1/2 states are coherent states. The first non-trivial case
is spin $j=1$.

\section{Proof of Lieb's conjecture for low spin}

In this section we shall geometrize the description of spin states, use this
to solve the entropy integrals
for all spin and prove Lieb's conjecture for low spin by actual computation
of the entropy.

\begin{lemma}
States of spin $j$ are in one to one correspondence to $2j$ points
on the sphere $S_2$: With $2j$ points,
parametrized by $\om_k = (\theta_k, \phi_k)$, $k = 1, \ldots , 2j$,
we can associate a state
\eq
|\psi\ra = c^\half P_j (|\om_1\ra \ot \ldots \ot |\om_{2j}\ra) \; \in \; [j] ,
\label{psiprod}
\en
and every state $|\psi\ra \in [j]$ is of that form. (The spin-$\half$ states
$|\om_k\ra$ are given by (\ref{coh}),
$c^\half \neq 0$ fixes the
normalization of $|\psi\ra$, and $P_j$ is the projector onto spin~$j$.)
\label{sphere}
\end{lemma}

\noindent \emph{Remark:} 
Some or all of the points may coincide.
Coherent states are exactly
those states for which all points on the sphere coincide. 
$c^\half \in \C$ may contain an (unimportant) phase
that we can safely ignore in the following.
This representation is unique up to permutation
of the $|\om_k\ra$. The $\om_k$ may be found by looking at
$\la\Om|\psi\ra$ as a function of $\Om = (\theta,\psi)$:
they are the antipodal points to the zeroes of this function. 

\noindent {\sc Proof}:
Rewrite (\ref{Coh}) in complex coordinates for $\theta\neq 0$
\eq
z = \left(\frac{p}{1-p}\right)^\half e^{i\phi} = \cot\frac{\theta}{2} e^{i\phi}
\en
(stereographic projection)
and contract it with $|\psi\ra$ to find
\eq
\la\Omega|\psi\ra = \frac{e^{-ij\phi}}{(1+z\bar z)^j} 
\sum_{m=-j}^{j_{\mbox{\tiny max}}} 
	   {2 j \choose j + m}^{\half} z^{j+m} \psi_m , \label{poly}
\en
where $j_{\mbox{\scriptsize max}}$ is the largest value of $m$ for which
$\psi_m$ in the expansion
$|\psi\ra = \sum\psi_m |m\ra$
is nonzero. This is a polynomial of degree
$j+j_{\mbox{\scriptsize max}}$ in $z \in \C$ and can thus be factorized:
\eq
\la\Omega|\psi\ra = 
\frac{ e^{-ij\phi} \psi_{j_{\mbox{\tiny max}}} }{ (1+z\bar z)^j } 
\prod_{k=1}^{j+j_{\mbox{\tiny max}}} (z - z_k) . \label{fact}
\en
Consider now the spin~$\half$ states
$|\om_k\ra = (1+z_k \bar z_k)^{-\half}(|\U\ra - z_k |\D\ra)$
for $1 \leq k \leq j+j_{\mbox{\tiny max}}$ and
$|\om_m\ra = |\D\ra$ for $j+j_{\mbox{\tiny max}} < m \leq 2j$. According
to (\ref{poly}):
\eq
\la\om|\om_k\ra = \frac{e^{-\frac{i\phi}{2}}}{(1+z\bar z)^\half 
(1+ z_k\bar z_k)^\half}(z - z_k) , 
\qquad \la\om|\om_m\ra = 
\frac{e^{-\frac{i\phi}{2}}}{(1+z\bar z)^\half},
\en
so by comparison with (\ref{fact}) and with an appropriate constant
$c$
\eq
\la\Om|\psi\ra 
= c^\half \la\om|\om_1\ra \cdots \la\om|\om_{2j}\ra
= c^\half \la\Om|\om_1\ot\ldots\ot\om_{2j}\ra.
\label{fac}
\en
By inspection we see that this expression is still valid when
$\theta = 0$ and with the help of (\ref{project}) we can complete the
proof the lemma.\hfill $\mybox$\\[1em]
We see that the geometric representation of spin states leads to a
factorization of $\la\Om|\psi\ra|^2$. In this representation we can
now do the entropy integrals, essentially because the logarithm becomes a
simple sum.

\begin{thm}   \label{theorem}
Consider any state $|\psi\ra$ of spin $j$. According to
lemma~\ref{sphere}, it can be written as
$|\psi\ra = c^\half P_j (|\om_1\ra \ot \ldots \ot |\om_{2j}\ra).$
Let $\R_i$ be the rotation that turns $\om_i$ to the `north pole',
$\R_i|\om_i\ra = |\U\ra$, let $|\psi^{(i)}\ra = \R_i|\psi\ra$,
and let $\psi_m^{(i)}$ be the coefficient of $|j,m\ra$ in the expansion
of $|\psi^{(i)}\ra$,
then the Wehrl entropy is:
\eq
S_W(|\psi\ra\la\psi|) =
\sum_{i=1}^{2j} \sum_{m=-j}^{j} \left(\sum_{n=0}^{j-m}
\frac{1}{2j+1-n}\right) |\psi_m^{(i)}|^2  -  \ln c . \label{formula}
\en
\end{thm}

\noindent \emph{Remark:}
This formula reduces the computation of the Wehrl entropy of any
spin state to its factorization in the sense of lemma~\ref{sphere},
which in general requires the solution of a
$2j$'th order algebraic equation. This may explain why previous 
attempts to do the entropy integrals have failed.
The $n=0$ terms in the expression for the entropy
sum up to $2j/(2j+1)$, the entropy of a coherent state, 
and Lieb's conjecture can be thus be written
\eq
\ln c \leq \sum_{i=1}^{2j} \sum_{m=-j+1}^{j-1} \left(\sum_{n=1}^{j-m}
\frac{1}{2j+1-n}\right) |\psi_m^{(i)}|^2.
\en
Note that $\psi^{(i)}_{-j} = 0$ by construction of
$|\psi^{(i)}\ra$: $\psi^{(i)}_{-j}$ contains a factor
$\la\downarrow|\U\ra$.\\
A similar calculation gives
\eq
\ln c = 2j + \int\dOm \, \ln|\la\Om|\psi\ra|^2 .
\en

\noindent
{\sc Proof}:
Using lemma~\ref{sphere}, (\ref{project}),
the rotational invariance of the
measure and the inverse Fourier transform in $\phi$ we find
\eqa
\lefteqn{S_W(|\psi\ra\la\psi|) \; = \;
{ -(2j+1)}
\int\dOm |\la\Om|\psi\ra|^2 \sum_{i=1}^{2j} \ln |\la\om|\om_i\ra|^2
- \ln c} \nn
&& = { -(2j+1)} \sum_{i=1}^{2j} \int\dOm
|\la\Om|\psi^{(i)}\ra|^2  \ln |\la\om|\U\ra|^2  - \ln c \nn
&& = {\scriptstyle -(2j+1)} \sum_{i=1}^{2j}\sum_{m=-j}^j |\psi^{(i)}_m|^2
{\scriptstyle {2j \choose j + m}}
\int_0^1 dp \, 
 p^{j+m} (1 \! - \! p)^{j-m} \ln p - \ln c.
\ena
It is now easy to do the remaining $p$-integral by partial integration
to proof the theorem.\hfill $\mybox$\\[1em]
Lieb's conjecture for low spin can be proved with the help of
formula (\ref{formula}). For spin 1/2 there is
nothing to prove, since all states  of spin 1/2 are coherent states.
The first nontrivial case is spin 1:

\begin{cor}[\rm spin 1]
Consider an arbitrary state of spin 1. Let
$\mu$ be the square of the
chordal distance between the two points on the sphere of radius~$\half$
that represent this state. It's Wehrl entropy is given by
\eq
S_W(\mu) = \frac{2}{3} + c\cdot\left(\frac{\mu}{2} + \frac{1}{c} \ln
\frac{1}{c}\right) , \label{entropy1}
\en
with
\eq
\frac{1}{c} = 1 - \frac{\mu}{2}.
\en
Lieb's conjecture holds for all states of spin 1:
$S_W(\mu) \geq 2/3 = 2j/(2j+1)$ with equality for $\mu = 0$, \ie\ for
coherent states.
\end{cor}

\noindent {\sc Proof}:
Because of rotational invariance we can assume without loss of generality
that the first point is at the `north pole' of the sphere and that
the second point is parametrized as $\om_2 = (\tilde\theta, \tilde\phi = 0)$,
so that
$\mu = \sin^2\frac{\tilde\theta}{2}$ . Up to normalization (and an irrelevant
phase)
\eq
|\tilde\psi\ra = P_{j=1}|\U\ot\tilde\om\ra
\en
is the state of interest. But from (\ref{coh})
\eq
|\U\ot\tilde\om\ra = (1-\mu)^\half|\U\;\U\ra + \mu^\half|\U\;\D\ra.
\en
Projecting onto spin 1 and inserting the normalization constant $c^\half$
we find
\eq
|\psi\ra = c^\half\left((1-\mu)^\half |1,1\ra + \mu^\half
\frac{1}{\sqrt{2}}|1,0\ra\right). \label{state}
\en
This gives (ignoring a possible phase) 
\eq
1 = \la\psi|\psi\ra = c\left(1 - \mu + \frac{\mu}{2}\right) = c\left(1 -
\frac{\mu}{2}\right)   \label{cvalue}
\en
and so $1/c = 1 - \mu/2$. Now we need to compute the components
of $|\psi^{(1)}\ra$ and $|\psi^{(2)}\ra$. Note that
$|\psi^{(1)}\ra = |\psi\ra$ because $\om_1$ is already pointing to
the `north pole'. To obtain $|\psi^{(2)}\ra$ we need to rotate point 2
to the `north pole'. We can use the remaining rotational freedom
to effectively exchange the two points, thereby recovering the original
state $|\psi\ra$. The components of 
both $|\psi^{(1)}\ra$ and $|\psi^{(2)}\ra$ can thus be read off (\ref{state}):
\eq
\psi_1^{(1)} = \psi_1^{(2)} = c^\half(1-\mu)^\half  ,
\qquad \psi_0^{(1)} = \psi_0^{(2)} = c^\half \mu^\half/\sqrt{2}.
\en
Inserting now $c$, $|\psi_1^{(1)}|^2 = |\psi_1^{(2)}|^2 = c(1-\mu)$,
and $|\psi_0^{(1)}|^2 = |\psi_0^{(2)}|^2 = c \mu/2$ into (\ref{formula})
gives the stated entropy.

To prove Lieb's conjecture for states of spin~1 we use
(\ref{cvalue}) to show that
the second term in (\ref{entropy1}) is always non-negative and zero only for
$\mu = 0$, \ie\ for a coherent state. This follows from
\eq
\frac{c \mu}{2} - \ln c \geq \frac{c \mu}{2} + 1 - c = 0
\en
with equality for $c=1$ which is equivalent to $\mu=0$ .\hfill 
$\mybox$
\vspace{2ex}
\begin{figure}
\begin{center}
\unitlength 1.00mm
\linethickness{0.4pt}
\begin{picture}(30.00,21.00)(10,5)
\put(10.00,0.00){\line(5,1){25.00}}
\put(35.00,5.00){\line(-4,3){20.00}}
\put(15.00,20.00){\line(-1,-4){5.00}}
\put(9.00,0.00){\makebox(0,0)[rc]{$1$}}
\put(15.00,21.00){\makebox(0,0)[cb]{$3$}}
\put(36.00,5.00){\makebox(0,0)[lc]{$2$}}
\put(11.00,10.00){\makebox(0,0)[rc]{$\mu$}}
\put(26.00,13.00){\makebox(0,0)[lb]{$\epsilon$}}
\put(27.00,2.00){\makebox(0,0)[ct]{$\nu$}}
\end{picture}
\end{center}
\caption{Spin~3/2}
\label{three}
\end{figure}
\begin{cor}[\rm spin 3/2]
Consider an arbitrary state of spin 3/2. Let $\ep$, $\mu$, $\nu$
be the squares of the chordal distances between the three points on
the sphere of radius $\half$ that represent this state (see figure~\ref{three}). 
It's Wehrl entropy
is given by
\eq
S_W(\ep,\mu,\nu) = \frac{3}{4} + 
c\cdot\left(\frac{\ep+\mu+\nu}{3} - \frac{\ep\mu + \ep\nu + \mu\nu}{6}
+\frac{1}{c} \ln\frac{1}{c} \right)  \label{spin32}
\en
with 
\eq
\frac{1}{c} = 1 - \frac{\ep+\mu+\nu}{3} .
\en
Lieb's conjecture holds for all states of spin 3/2:
$S_W(\ep,\mu,\nu) \geq 3/4 = 2j/(2j+1)$ with equality for 
$\ep = \mu = \nu = 0$, \ie\ for
coherent states.
\end{cor}

\noindent {\sc Proof}: The proof is similar to the spin~1 case, but the
geometry and algebra is more involved.
Consider a sphere of radius ${1\over 2}$, with points 1, 2, 3 on its surface,
and two planes through its center; the first plane
containing
points 1 and 3, the second plane containing points 2 and 3. The intersection
angle $\phi$
of these two planes satisfies
\eq
 2\cos\phi \sqrt{\ep\mu(1-\ep)(1-\mu)} = \ep + \mu - \nu - 2\ep\mu .
\label{phi}
\en
$\phi$ is the azimuthal angle of point 2, if point 3 is at the `north pole' of
the sphere and point 1 is assigned zero azimuthal angle.

The states $|\psi^{(1)}\ra$,
$|\psi^{(2)}\ra$, and $|\psi^{(3)}\ra$ all have one point at the
north pole of the sphere. It is enough to compute the values of
$|\psi_m^{(i)}|^2$ for
one $i$, the other values can be found by appropriate permutation of
$\ep$, $\mu$, $\nu$. (Note that we make no restriction on
the parameters $0\leq \ep$, $\mu$, $\nu \leq 1$ other than that they are
square chordal distances between three points on a sphere of
radius $\half$.)
We shall start with $i = 3$: Without loss of generality
the three points can be parametrized as $\om^{(3)}_1 = (\tilde\theta,0)$,
$\om^{(3)}_2 = (\theta,\phi)$, and $\om^{(3)}_3 = (0,0)$
with $\mu = \sin^2{\tilde\theta\over 2}$ and $\ep = \sin^2{\theta\over 2}$.
Corresponding spin-$\half$ states are
\eqa
|\om^{(3)}_1\ra & = & (1-\mu)^\half|\U\ra + \mu^\half|\D\ra ,\label{om1}\\
|\om^{(3)}_2\ra & = & (1-\ep)^\half e^{-i\phi\over 2}|\U\ra 
+ \ep^\half e^{i\phi\over 2}|\D\ra ,  \label{om2}\\
|\om^{(3)}_3\ra & = & |\U\ra ,      \label{om3} 
\ena
and up to normalization, the state of interest is
\eqa
|\tilde\psi^{(3)}\ra 
& = & P_{j=3/2} |\om^{(3)}_1\ot\om^{(3)}_2\ot\om^{(3)}_3\ra  \nn
& = & (1-\ep)^\half (1-\mu)^\half e^{-i\phi\over 2}
      |{3\over 2},{3\over 2}\ra \nn
&&    + \left( (1-\mu)^\half \ep^\half e^{i\phi\over 2} 
      + \mu^\half (1-\ep)^\half e^{-i\phi\over 2} \right) 
      { {1 \over \sqrt{3}}} |{3\over 2},{1\over 2}\ra \nn
&&    + \mu^\half \ep^\half e^{i\phi\over 2} 
      { {1 \over \sqrt{3}}} |{3\over 2},-{1\over 2}\ra .
\ena
This gives 
\eqa
|\tilde\psi^{(3)}_{3\over 2}|^2 & = & (1-\ep)(1-\mu),\\
|\tilde\psi^{(3)}_{1\over 2}|^2 & = & {1 \over 3}\left(
\ep(1-\mu) + \mu(1 - \ep) + 2 \sqrt{\ep\mu(1-\mu)(1-\ep)} \cos\phi\right) \nn
& = & {2\over 3}\ep(1-\mu) + {2\over 3}\mu(1 - \ep) -{\nu\over 3}, \\
|\tilde\psi^{(3)}_{-{1\over 2}}|^2 & = & {\ep \mu\over 3},
\ena
and
$|\tilde\psi^{(3)}_{-{3\over 2}}|^2  = 0$. The sum of these expressions
is
\eq
{1\over c} = \la\tilde\psi|\tilde\psi\ra =
1 - {\ep + \mu + \nu \over 3} ,
\en
with $0 < 1/c \leq 1$.
The case $i=1$ is found by exchanging $\mu \leftrightarrow \nu$ (and also
$3 \leftrightarrow 1$, $\phi \leftrightarrow -\phi$).
The case $i=2$ is found by permuting
$\ep\rightarrow\mu\rightarrow\nu\rightarrow\ep$ (and also $1 \rightarrow 3
\rightarrow 2 \rightarrow 1$).
Using (\ref{formula}) then gives the stated entropy.

To complete the proof Lieb's conjecture for all states of 
spin~$3/2$ we need to show
that the second term in (\ref{spin32}) is always non-negative and zero
only for $\ep=\mu=\nu=0$.
From the inequality $(1-x)\ln(1-x) \geq -x + x^2/2$ for $0 \leq x < 1$,
we find
\eq
{1\over c}\ln{1\over c} \geq -{\ep+\mu+\nu\over 3} + {1\over 2}\left(
{\ep+\mu+\nu\over 3}\right)^2 ,
\en
with equality for $c=1$. Using the inequality between algebraic and geometric
mean it is not hard to see that
\eq
\left({\ep+\mu+\nu\over 3}\right)^2 \geq {\ep\mu + \nu\ep + \mu\nu \over 3}
\en
with equality for $\ep=\mu=\nu$. Putting everything together and inserting
it into (\ref{spin32}) we have, as desired, $S_W \geq 3/4$ with equality
for $\ep=\mu=\nu=0$, \ie\ for coherent states.
\hfill $\mybox$\\[1em]
\begin{figure}
\begin{center}
\unitlength 1.00mm
\linethickness{0.4pt}
\begin{picture}(30.00,36.00)(10,7)
\put(10.00,15.00){\line(5,1){25.00}}
\put(35.00,20.00){\line(-4,3){20.00}}
\put(15.00,35.00){\line(-1,-4){5.00}}
\put(10.00,15.00){\line(3,-2){15.00}}
\put(25.00,5.00){\line(2,3){10.00}}
\put(9.00,15.00){\makebox(0,0)[rc]{$1$}}
\put(15.00,36.00){\makebox(0,0)[cb]{$3$}}
\put(36.00,20.00){\makebox(0,0)[lc]{$2$}}
\put(25.00,4.00){\makebox(0,0)[ct]{$4$}}
\put(11.00,25.00){\makebox(0,0)[rc]{$\mu$}}
\put(26.00,28.00){\makebox(0,0)[lb]{$\epsilon$}}
\put(19.00,24.00){\makebox(0,0)[lb]{$\gamma$}}
\put(27.00,17.00){\makebox(0,0)[ct]{$\nu$}}
\put(17.00,9.00){\makebox(0,0)[rt]{$\alpha$}}
\put(31.00,11.00){\makebox(0,0)[rt]{$\beta$}}
\put(15.00,35.00){\line(1,-3){10.00}}
\end{picture}
\end{center}
\caption{Spin~2}
\end{figure}
\begin{cor}[\rm spin 2]
Consider an arbitrary state of spin 2. Let $\ep$, $\mu$, $\nu$, $\al$, $\be$,
$\ga$
be the squares of the chordal distances between the four points on
the sphere of radius $\half$ that represent this state
(see figure). It's Wehrl entropy
is given by
\eq
S_W(\ep,\mu,\nu,\al,\be) = \frac{4}{5} + c \cdot \left( \sigma + \frac{1}{c}
\ln\frac{1}{c}\right), \label{S2}
\en
where
\eq
\frac{1}{c} = 1 - \frac{1}{4}\sum\lipic
+\frac{1}{12}\sum\papic \label{c2}
\en
and
\eq
\sigma = \frac{1}{12}\left(-\frac{1}{2}\sum\trpic
-\frac{5}{3}\sum\papic-\sum\wepic+3\sum\lipic
\right)
\en
with
\eq
\sum\trpic \equiv \al\mu\nu+\ep\be\nu+\ep\mu\ga+\al\be\ga,
\en
\eq
\sum\papic \equiv \al\ep+\be\mu+\ga\nu,
\qquad
\sum\lipic \equiv \al+\be+\ga+\mu+\nu+\ep,
\en
\eq
\sum\wepic \equiv \al\mu+\al\nu+\mu\nu+\be\ep
+\be\nu+\ep\nu+\ep\ga+\ep\mu+\mu\ga
+\al\be+\al\ga+\be\ga.
\en \label{spin2}
\end{cor}

\noindent \emph{Remark:} The fact that the four points lie on the surface
of a sphere imposes a complicated constraint on the parameters
$\ep$, $\mu$, $\nu$, $\al$, $\be$,
$\ga$. Although we have convincing numerical evidence
for Lieb's conjecture for spin~2,
so far a rigorous proof has been limited to
certain symmetric configurations
like equilateral triangles with centered fourth point ($\ep=\mu=\nu$ and
$\al=\be=\ga$), and squares ($\al=\be=\ep=\mu$ and
$\ga=\nu$). It is not hard to find values of the parameters
that give values of $S_W$ below the entropy for coherent states, but they
do \emph{not} correspond to any configuration of points on the sphere,
so in contrast to spin 1 and spin 3/2
the constraint is now important.
$S_W$ is concave in each of the parameters $\ep$, $\mu$, $\nu$, $\al$, $\be$,
$\ga$.

\noindent {\sc Proof}: The proof is analogous to the spin~1 and spin~3/2
cases but the geometry and algebra are considerably more complicated,
so we will just give a sketch. Pick four points on the sphere,
without loss of generality parametrized as $\om_1^{(3)} =(\tilde\theta,0)$,
$\om_2^{(3)} =(\theta,\phi)$, $\om_3^{(3)} = (0,0)$,
and $\om_4^{(3)} =(\bar\theta,\bar\phi)$. Corresponding spin $\half$ states
are $|\om_1^{(3)}\ra$, $|\om_2^{(3)}\ra$, $|\om_3^{(3)}\ra$,
as given in (\ref{om1}), (\ref{om2}), (\ref{om3}), and
\eq
|\om_4^{(3)}\ra = (1-\ga)^\half e^{-i\bar\phi \over 2} |\U\ra
+ \ga^\half e^{i\bar\phi \over 2} |\D\ra.
\en
Up to normalization, the state of interest is
\eq
|\tilde\psi^{(3)}\ra =
P_{j=2} |\om_1^{(3)} \ot \om_2^{(3)} \ot\om_3^{(3)} \ot\om_4^{(3)}\ra.
\en
In the computation of $|\tilde\psi^{(3)}_m|^2$ we encounter
again the angle $\phi$, compare (\ref{phi}),and two new
angles $\bar\phi$ and
$\bar\phi -\phi$.
Luckily both can again be expressed as angles between planes that
intersect the circle's center and we have
\eqa
2 \cos\bar\phi\sqrt{\mu\ga(1-\mu)(1-\ga)} & = & 
\mu + \ga - \al - 2\mu\ga, \\
2 \cos(\bar\phi-\phi)\sqrt{\ep\ga(1-\ep)(1-\ga)}
& = & \ga + \ep - \be - 2\ga\ep,
\ena
and find $1/c = \sum_m |\tilde\psi^{(3)}_m|^2$ as given in (\ref{c2}).
By permuting the parameters $\ep$, $\mu$, $\nu$, $\al$, $\be$,
$\ga$ appropriately we can derive expressions for the remaining
$|\tilde\psi^{(i)}_m|^2$'s and then compute $S_W$ (\ref{S2}) with the
help of $(\ref{formula})$.\hfill $\mybox$

\subsection{Higher spin}

The construction outlined in the proof of corollary~\ref{spin2}
can in principle also be applied to states of higher spin, but
the expressions pretty quickly become quite unwieldy.
It is, however, possible to use theorem~\ref{theorem} to show that
the entropy is extremal for coherent states:

\begin{cor}[\rm spin $j$]
Consider the state of spin $j$ characterized by $2j -1$ coinciding
points on the sphere and a $2j$'th point, a small (chordal) distance
$\ep^\half$ away from them. The Wehrl entropy of this small deviation
from a coherent state, up to third order in $\ep$, is
\eq
S_W(\ep) = {2j\over 2j+1} + {c \over 8 j^2} \ep^2 \quad + {\cal O}[\ep^4] ,
\en
with
\eq
{1 \over c} = 1 - {2j - 1 \over 2j} \ep \quad \mbox{(exact)} .
\en
\end{cor}

A generalized version of Lieb's conjecture, analogous to (\ref{W2}), 
is \cite{A}
\begin{conj} \label{conject2}
Let $|\psi\ra$ be a normalized state of spin $j$, then
\eq
(2j s + 1) \int\dOm \, |\la\Om|\psi\ra|^{2s} \leq 1 , \quad s > 1 ,
\label{norms}
\en
with equality if and only if $|\psi\ra$ is a coherent state.
\end{conj}

\noindent \emph{Remark:} 
This conjecture is equivalent to the ``quotient of two H\"older inequalities"
(\ref{holder}).
The original conjecture~\ref{conject1} 
follows from it in the limit $s \rightarrow 1$.
For $s=1$ we simply get the norm of the spin $j$ state $|\psi\ra$,
\eq
(2j + 1) \int\dOm \, |\la\Om|\psi\ra_j|^2 = | P_j | \psi\ra|^2 ,
\label{norm}
\en
where $P_j$ is the projector onto spin $j$.
We have numerical evidence for low spin
that an analog of conjecture~\ref{conject2}
holds in fact for a much larger class of convex functions than
$x^s$ or $x \ln x$.

For $s \in \N$ there is a surprisingly simple group theoretic argument
based on (\ref{norm}):
\begin{thm}
Conjecture~\ref{conject2} holds for $s \in \N$. \label{natural}
\end{thm}

\noindent \emph{Remark:} For spin 1 and spin 3/2  (at $s=2$) this was
first shown by Wolfgang Spitzer by direct computation of the
integral.

\noindent {\sc Proof}: Let us consider
$s=2$, $|\psi\ra \in [j]$ with $||\psi\ra|^2 = 1$,
rewrite (\ref{norms}) as follows
and use (\ref{norm})
\eqa
\lefteqn{(2j\cdot 2 + 1) \int\dOm \, |\la\Om|\psi\ra|^{2\cdot 2}} \nn
&& = (2(2j)+ 1) \int\dOm \, |\la\Om\ot\Om|\psi\ot\psi\ra|^2 
=  |P_{2j} |\psi\ot\psi\ra |^2.
\ena
But $|\psi\ra\ot|\psi\ra \in [j]\ot[j] = [2j]\oplus[2j-1]\oplus\ldots
\oplus[0]$, so $|P_{2j} |\psi\ot\psi\ra |^2 < ||\psi\ot\psi\ra |^2 = 1$
unless $|\psi\ra$ is a coherent state, in which case
$|\psi\ra\ot|\psi\ra \in [2j]$ and we have equality. The proof for
all other $s \in \N$ is completely analogous.
\hfill $\mybox$\\[1em]
It seems that there should also be a similar group theoretic
proof for all real, positive $s$ related to (infinite dimensional)
spin~$js$ representations of su(2) (more precisely: sl(2)). 
There has been some progress and it is now clear that there will
not be an argument as simple as the one given above \cite{J}.
Coherent states of the form discussed in \cite{C3} (for the hydrogen atom)
could be of importance here, since they easily generalize to non-integer
`spin'.

Theorem~\ref{natural} provides a quick, crude, lower limit on the
entropy:

\begin{cor}
For states of spin $j$
\eq
S_W(|\psi\ra\la\psi|) \geq \ln{4j+1\over 2j+1} > 0.
\en
\end{cor}

\noindent {\sc Proof}: This follows from Jensen's inequality and
concavity of $\ln x$:
\eqa
S_W(|\psi\ra\la\psi|) & = &
-{\textstyle (2j+1)}\int\dOm\, |\la\Om|\psi\ra|^2 \ln |\la\Om|\psi\ra|^2 \nn
& \geq & -\ln\left({\textstyle (2j+1)} 
\int\dOm\, |\la\Om|\psi\ra|^{2\cdot 2}\right) \nn
& \geq & -\ln{2j+1 \over 4j + 1} .
\ena
In the last step we have used theorem~\ref{natural}.\hfill 
$\mybox$\\[1em]
We hope to have provided enough evidence to convince the
reader that it is reasonable to expect that
Lieb's conjecture is indeed true for all spin. All cases
listed in Lieb's original article, $1/2$, $1$, $3/2$, are now
settled --
it would be nice if someone
could take care of the remaining ``dot, dot, dot" \ldots


\chapter{{\sffamily Symmetries of the Hubbard model}}
\label{chap:electron}

Research on high temperature superconductivity in cuprates 
has greatly revived interest in the Hubbard model \cite{H} as a model of
strongly correlated electron systems \cite{An,AS,Da}.
Despite its formal simplicity this model continues to resist complete
analytical or numerical understanding. Symmetries of the Hubbard
Hamiltonian play a major role in the reduction of the problem. They have
for instance been used  to construct eigenstates of the
Hamiltonian with off-diagonal long range order \cite{Y}, to
simplify numerical diagonalization\cite{BA} and
to show completeness of the solution \cite{LW} to
the one-dimensional model \cite{EKS}.

In addition to the regular spin su(2) symmetry, the Hubbard model has
a second su(2) symmetry at half filling, which is called 
pseudo-spin~ \cite{HL,Y,YZ}.
This is a consequence of the particle hole symmetry at half filling,
which maps su(2)-spin into su(2)-pseudo-spin and vice versa. 
According to Yang the generators of the pseudo-spin symmetry indicate
off-diagonal long range order.
The closest thing to a rigorous prove of long range order at half
filling is Lieb's proof that the ground state of both the 
repulsive and the attractive
Hubbard model is a singlet in this case. In the physically 
interesting
repulsive case I was able to extend this result to show 
that in fact all energy levels are arranged
according to spin at half-filling. For this I used reflection positivity
arguments to prove that the Hubbard states have non-vanishing
overlap with special
states of known spin. A similar result has been independently
found by Tasaki using arguments related to the prove of
the Lieb-Mattis theorem.
 
Away from half-filling
the symmetry of the Hubbard model and also that
of related models is deformed to a special
kind of quantum group symmetry, as we will discuss 
in the following sections. 
This suggests that even away from half-filling
the energy levels of the Hubbard model should be arranged according
to their spin. There are simple qualitative arguments in the
limits of strong and weak on-site interaction: Consider
the attractive case. In the limit of weak interaction the kinetic
energy (hopping term) dominates and energy levels will fill up
pair-wise with spin singlets until only spins of one type are
left over. If there are $n$ spin ``up'' and $m$ spin ``down'' electrons
then the ground state will have total spin $1/2\cdot |m -n|$. In the
limit of strong interaction the result is the same. Electrons will again
try to minimize the ground state energy by arranging themselfs in
pairs (this time in position space instead of momentum space)
until only electrons of one type are left over. The conjecture is thus
that this is also true for all intermediate coupling strengths. 
A corresponding conjecture
for the physically more interesting repulsive case can be found by
a particle-hole transformation on every other site. 

In his proof Lieb expanded states of the Hubbard model in terms
of a basis of electron states with spin ``up'' and a dual basis
of electron states with spin ``down'',
$$\psi = \sum c_{ij} \psi^i_\uparrow \otimes \psi^j_\downarrow,$$
and then used trace-inequalities for the coefficient matrices $c_{ij}$
as is familiar to us from chapter~\ref{chap:spin}.
Lieb worked at half-filling, so the $c_{ij}$ were square matrices 
in his case
and the required trace-inequalities were already known from older 
work on reflection positivity. 
Away from half-filling (repulsive
case) or for $S_z \neq 0$ (attractive case) the coefficient matrices 
are rectangular and  appropriate
trace inequalities that would prove the conjecture are not known.
It is possible,
however, to find inequalities that relate rectangular to square 
matrices. These also give interesting information about the 
ground state of the Hubbard model and are in fact the same inequalities that 
let to the breakthrough in the mathematically closely related 
work on frustrated quantum spin systems (see chapter~\ref{chap:spin}).


In the following we will discuss so-called
superconducting quantum 
symmetries in extended one-band one-dimensional Hubbard models.
We will see that they  originate from classical 
(pseudo-)spin symmetries
of a class of models of which the standard
Hubbard model is a special case. Motivated by this
we give the Hamiltonian of the most general extended Hubbard model
with spin and pseudo-spin symmetry. (As far as we know this
result has not appeared in the literature before.)
The quantum symmetric models provide extra parameters, which
makes them interesting for phenomenology, but they are 
restricted to one dimension. The equivalent new models with
classical symmetries do not have that drawback.
Especially notworthy is that the filling factor is one of the
free parameters. (Unlike in the case of the standard Hubbard
model which has the full (pseudo-)spin symmetry only at
half-filling.
All models that we will discuss
are related by generalized Lang-Firsov transformations.
This work has been published in~\cite{unravel}. 

\section{The Hubbard model}

Originally introduced as a simplistic description of narrow d-bands in
transition metals, the Hubbard model combines band-like and atomic behavior.
In the standard Hubbard Hamiltonian
\beq
\hhub =  u \sum_i n_{i \1} n_{i \0} - \mu \sum_{i,\s} n_{i \s}
+ t \sum_{\la i,j \ra \s} a^\dag_{j \s } a_{i \s},\label{hhub}
\eeq
this is achieved by a local Coulomb term and a competing 
non-local hopping term.
Here $a^\dag_{i \s }$, $a_{i \s}$ are creation and annihilation
operators\footnote{We will use the convention that
operators at different sites {\em commute}. On a bipartite lattice
one can easily switch to anticommutators without
changing any of our results.} for electrons of
spin $\s\in\{\1, \0\}$ at site $i$ of a $D$-dimensional lattice, 
$\la i,j \ra$ denotes nearest neighbor sites and
$n_{i \s} \equiv a^\dag_{i \s }a_{i \s}$.
The average number of electrons $\la \sum_{i,\s} n_{i \s} \ra$
is fixed by the chemical potential $\mu$.
 
\subsection{Classical symmetries} 
 
The {\em standard Hubbard model\/} has a $\su2\times\su2/\zet_2$ symmetry
at $\mu = u/2$, the value of $\mu$ corresponding 
to half filling in the
band-like limit. This symmetry is the product of a {\em magnetic\/}
$\su2_m$ (spin) with local generators
\beq
\xp_m = a^\dag_\1 a_\0,\quad \xm_m = a^\dag_\0 a_\1, 
\quad H_m = n_\1 - n_\0 ,
\label{mag}
\eeq
and a {\em superconducting\/} $\su2_s$
(pseudo-spin) with local generators
\beq
\xp_s = a^\dag_\1 a^\dag_\0, \quad \xm_s = a_\0 a_\1,
\quad H_s = n_\1 + n_\0 - 1 ,
\label{suco}
\eeq
modulo a $\zet_2$, generated by the unitary transformation
($a_\0 \leftrightarrow a^\dag_\0$) that interchanges the
two sets of local generators.
The mutually orthogonal
algebras generated by (\ref{mag}) and (\ref{suco})
are isomorphic to the algebra generated by the Pauli matrices
and have unit elements
$1_s = H_s^2$, $1_m = H_m^2$ with $1_s + 1_m = 1$.
The superconducting generators commute
with each term of the local part $\hloc$ (first two terms) 
of the Hubbard Hamiltonian (\ref{hhub})
provided that
$\mu = u/2$.
This can either be seen by direct 
computation or by studying the action of the
generators on the four possible electron states at each site.
It is also easily seen that the magnetic generators
commute with each term of $\hloc$; in the following
we will however
focus predominantly on the superconducting symmetry.

To check the symmetry of the non-local hopping term
we have to consider global generators ${\cal O}$:
These generators are here simply
given by the sum $\sum {\cal O}_i$ of the 
local generators for all sites $i$.
The rule that governs the combination of
representations for more than one lattice site is abstractly given by
the diagonal map
or coproduct $\Delta$ of $U(su(2))$.
Generators for two sites are directly obtained from the coproduct
$$
\Delta(X^\pm)  =  X^\pm \otimes 1 + 1 \otimes X^\pm, \quad
\Delta(H)  =  H \otimes 1 + 1 \otimes H,
$$
while generators for $N$ sites require 
$(N-1)$-fold iterative application of
$\Delta$. Coassociativity of $\Delta$ ensures that it does not matter which
tensor factor is split up at each step.
Another distinguishing property of this {\em classical\/} coproduct is its
symmetry (cocommutativity). This property and coassociativity ensure that we
can arrange that the two factors of the last coproduct coincide with any
given pair $\la i,j \ra$ of next-neighbor sites; see Fig.~\ref{coass}. 
It is hence enough to study
symmetry of a single next-neighbor term of the Hamiltonian
to prove global symmetry.
\begin{figure}[tbp]
\begin{center}
\unitlength 1.00mm
\linethickness{0.4pt}
\begin{picture}(71.00,9.00)
\put(0.00,4.00){\circle*{2.00}}
\put(20.00,4.00){\circle*{2.00}}
\put(30.00,4.00){\circle*{2.00}}
\put(40.00,4.00){\circle*{2.00}}
\put(50.00,4.00){\circle*{2.00}}
\put(70.00,4.00){\circle*{2.00}}
\put(60.00,4.00){\makebox(0,0)[cc]{$\cdots$}}
\put(10.00,4.00){\makebox(0,0)[cc]{$\cdots$}}
\put(0.00,9.00){\makebox(0,0)[cc]{$(\id$}}
\put(70.00,9.00){\makebox(0,0)[cc]{$\id)$}}
\put(10.00,9.00){\makebox(0,0)[cc]{$\cdots$}}
\put(60.00,9.00){\makebox(0,0)[cc]{$\cdots$}}
\put(5.00,9.00){\makebox(0,0)[cc]{$\otimes$}}
\put(15.00,9.00){\makebox(0,0)[cc]{$\otimes$}}
\put(25.00,9.00){\makebox(0,0)[cc]{$\otimes$}}
\put(45.00,9.00){\makebox(0,0)[cc]{$\otimes$}}
\put(55.00,9.00){\makebox(0,0)[cc]{$\otimes$}}
\put(65.00,9.00){\makebox(0,0)[cc]{$\otimes$}}
\put(20.00,9.00){\makebox(0,0)[cc]{$\id$}}
\put(50.00,9.00){\makebox(0,0)[cc]{$\id$}}
\put(35.00,9.00){\makebox(0,0)[cc]{$\Delta$}}
\put(30.00,4.00){\line(1,0){10.00}}
\put(30.00,1.00){\makebox(0,0)[cc]{$i$}}
\put(40.00,1.00){\makebox(0,0)[cc]{$j$}}
\end{picture}
\end{center}
\caption{Coassociativity of $\Delta$ reduces global symmetry to symmetry of
next-neighbor terms $\la i,j \ra$ if $D = 1$.} 
\label{coass}
\end{figure}

\subsection{Quantum symmetries}

The (pseudo-)spin symmetries
of the standard Hubbard model are restricted to
the case of an average of one electron per site  
(half-filling), so recent speculations \cite{MR} about 
extended Hubbard models with generalized
(quantum group) symmetries away from half-filling attracted some attention.
A careful analysis of the new models reveals that this quantum symmetry
exists only on one-dimensional lattices and in an appropriate approximation
seems still to be restricted to
half-filling. Despite these shortcomings 
the existence of novel symmetries in
Hubbard models is very interesting and worth investigating.
Quantum symmetries of the Hubbard model were first investigated in the form of
Yangians \cite{UK}; quantum supersymmetries of Hubbard models have
also been considered \cite{GHLZ}.
In the following we shall investigate the origin of quantum symmetries in
extended Hubbard models. We will find a one-to-one correspondence 
between Hamiltonians with quantum and classical symmetries.
Guided by our results we will 
then be able to identify models whose
symmetries are neither restricted to one-dimensional lattices nor to half
filling.

The search for quantum group symmetries in the Hubbard model is motivated by
the observation that the local generators $\xp_s$, $\xm_s$ and $H_s$ in the
superconducting representation of $\su2$ also
satisfy the $\suq2$ algebra as given in the Jimbo-Drinfel'd basis \cite{Ji}
\beq
\left[\xp , \xm\right] = {q^H - q^{-H} \over q - q^{-1}}, \qquad \left[H ,
X^\pm\right] = \pm 2 X^\pm.
\eeq
(The proof uses $H_s^3 = H_s$.)
It immediately follows that $\hloc$ has a local quantum symmetry.
As is, this is a trivial
statement because we did not yet consider global quantum symmetries.
Global generators are now defined via the deformed coproduct of $\suq2$,
$q \in {\bf R}\backslash\{0\}$
\bea
\Delta_q(X^\pm) & = & X^\pm \ot q^{-H/2} + q^{H/2} \ot X^\pm ,\nonumber\\
\Delta_q(H) & = & H \ot 1 + 1 \ot H.
\eea
The local symmetry can be extended to a non-trivial global quantum
symmetry by a modification of the Hubbard Hamiltonian. The idea of
\cite{MR} was to achieve this by including phonons.
Before we proceed to study the resulting extended Hubbard Hamiltonian 
$\hext$,
we would like to make two remarks: (i) We call a Hamiltonian quantum symmetric
if it commutes with all global generators. This implies invariance under the
quantum adjoint action and vice versa.
(ii) Coproducts of quantum groups are coassociative but not cocommutative. This
means that the reduction of global symmetry to that of next-neighbor terms
holds only for one-dimensional lattices. The practical implication is an
absence of quantum symmetries for higher-dimensional lattices.
(For a triangular lattice this is illustrated in Fig.~\ref{tri}.)
\begin{figure}[tbp]
\begin{center}\hspace{-7em}%
\unitlength 0.75mm
\linethickness{0.4pt}
\begin{picture}(65.00,13.00)
\put(15.00,5.00){\circle*{2.00}}
\put(5.00,5.00){\circle*{2.00}}
\put(10.00,12.00){\circle*{2.00}}
\put(25.00,5.00){\circle*{2.00}}
\put(35.00,5.00){\circle*{2.00}}
\put(45.00,5.00){\circle*{2.00}}
\put(55.00,5.00){\circle*{2.00}}
\put(50.00,12.00){\circle*{2.00}}
\put(30.00,12.00){\circle*{2.00}}
\put(5.00,5.00){\line(1,0){10.00}}
\put(35.00,5.00){\line(-3,4){5.25}}
\put(45.00,5.00){\line(3,4){5.25}}
\put(20.00,8.00){\makebox(0,0)[cc]{$\sim$}}
\put(40.00,8.00){\makebox(0,0)[cc]{$\sim$}}
\put(60.00,8.00){\makebox(0,0)[cc]{$\Leftrightarrow$}}
\put(65.00,8.00){\makebox(0,0)[lc]{$\Delta$ is cocommutative}}
\end{picture}
\caption{In $D \neq 1$ symmetry of next-neighbor terms
implies global symmetry only if
$\Delta$ is classical.} 
\label{tri}
\end{center}
\end{figure}

\subsection{Extended Hubbard model with phonons}

The {\em extended Hubbard model} of~\cite{MR} 
(with some modifications \cite{CS}) 
introduces Einstein oscillators (parameters: $M$,
$\omega$) and electron-phonon couplings (local: $\vec\lambda$-term,
non-local: via $T_{ij\s}$):
\bea
\hext & = & u \sum_i n_{i \1} n_{i \0} - \mu \sum_{i,\s} n_{i \s}
            - \vec\lambda\cdot\sum_{i \s} n_{i \s} \vec x_i  + {}\nonumber\\
	 && + \sum_i\left({\vec {p_i}^2 \over 2 M} + {1 \over 2} M \omega^2 
	      \vec x_i^2\right) 
            + \sum_{\la i,j \ra \s} a^\dag_{j \s } a_{i \s} T_{ij\sigma},
\eea
with hopping amplitude
\beq
T_{ij\sigma} = T_{ji\sigma}^\dag = 
t \exp(\zeta \hat e_{ij}\cdot(\vec x_i - \vec x_j) 
+ i \kappa\cdot(\vec p_i - \vec p_j)).
\eeq
The displacements $\vec x_i$ of the ions from their rest positions
and the corresponding momenta $\vec p_i$ satisfy
canonical commutation relations. The
$\hat e_{ij}$ are unit vectors from site $i$ to site $j$.
For $\vec\kappa = 0$ the model reduces to the Hubbard model with
phonons and atomic orbitals
$\psi(r) \sim \exp(-\zeta r)$
in $s$-wave approximation \cite{CS}.

The local part of $\hext$
commutes with the generators of $\suq2_s$ 
iff
\beq 
\mu = {u \over 2} - {\vec\lambda^2 \over M \omega^2}.
\eeq
(For technical reasons one needs to use modified generators
$\tilde X^\pm_s \equiv e^{\mp 2 i \vec\kappa\cdot\vec p} X^\pm_s$ here
that however still satisfy the $\suq2$ algebra.)\\
The nonlocal part of $\hext$ and thereby the whole extended
Hubbard Hamiltonian commutes with the global generators
iff
\beq
\vec\lambda = \hbar M \omega^2 \vec\kappa , \qquad
q = \exp(2 \kappa \zeta \hbar),
\eeq
where $\kappa \equiv - \hat e_{ij}\cdot\vec\kappa$
for $i,j$ ordered next neighbour sites. 
\emph{For $q \neq 1$ the symmetry is restricted to models
given on a 1-dimensional lattice with naturally ordered sites.}

From what we have seen so far we could be let to the premature conclusion
that the quantum symmetry is due to phonons and that we have found symmetry
away from half filling because $\mu \neq u/2$. However: the pure Hubbard
model with phonons has $\vec\kappa = 0$ and hence a classical symmetry $(q=1)$.
Furthermore $\vec\lambda \neq 0$ implies non-vanishing local electron-phonon
coupling so that a mean field approximation cannot be performed and we simply
do not know how to compute the actual filling.
Luckily there is an equivalent model that is not plagued with this problem:
A {\em Lang-Firsov transformation} 
with unitary operator $U = \exp(i \vec\kappa\cdot
\sum_j \vec p_j n_{j \s})$.
leads to the Hamiltonian 
\beq
\hqsym = U \hext U^{-1} = \hext(\vec\lambda',u',\mu',{T'}_{ij\sigma}),
\label{LF}
\eeq
of what we shall call the {\em quantum symmetric
Hubbard model}. It has the same form as $\hext$, but
with a new set of parameters 
\bea
\vec\lambda' & = & \vec\lambda - M \omega^2 \hbar \vec\kappa\\
u' & = & u - 2 \hbar \vec\lambda\cdot\vec\kappa + M \omega^2 \hbar^2 \kappa^2\\ 
\mu' & = & \mu + \hbar \vec\lambda\cdot\vec\kappa 
- {1 \over 2}M \omega^2 \hbar^2 \kappa^2
\eea 
and a modified hopping amplitude
\beq
T'_{ij,-\s} = \tilde t_{ij} (1 + 
(q^{{\hat e_{ji} \over 2}} - 1) n_{i \s})(1 + 
(q^{{\hat e_{ij} \over 2}} -1) n_{j \s})
\eeq
where $\tilde t_{ij} = t \exp(\zeta \hat e_{ij}\cdot (\vec x_i - \vec x_j))$. 
The condition for symmetry expressed in terms of the new parameters is 
\beq
\vec\lambda' =0,\qquad \mu' = {u' \over 2},
\eeq
\ie requires vanishing local phonon coupling
and corresponds to {\em half filling}!
$\tilde t_{ij}$ may also be turned
into a (temperature-dependent) constant via a mean field approximation. This
approximation is admissible for the quantum symmetric Hubbard model because
$\vec\lambda' = 0$.

\begin{figure}[htb]
\begin{center}
\unitlength 3.50pt
\linethickness{0.4pt}
\begin{picture}(61.00,56.00)
\put(26.00,26.00){\circle*{2.00}}
\put(27.00,27.00){\line(1,1){3.00}}
\put(31.00,31.00){\circle{2.00}}
\put(32.00,32.00){\line(1,1){3.00}}
\put(36.00,36.00){\circle*{2.00}}
\put(27.00,26.00){\line(1,0){5.00}}
\put(33.00,26.00){\circle{2.00}}
\put(34.00,26.00){\line(1,0){5.00}}
\put(40.00,26.00){\circle*{2.00}}
\put(41.00,27.00){\line(1,1){3.00}}
\put(45.00,31.00){\circle{2.00}}
\put(37.00,36.00){\line(1,0){5.00}}
\put(43.00,36.00){\circle{2.00}}
\put(46.00,32.00){\line(1,1){3.00}}
\put(50.00,36.00){\circle*{2.00}}
\put(44.00,36.00){\line(1,0){5.00}}
\put(12.00,26.00){\circle*{2.00}}
\put(13.00,27.00){\line(1,1){3.00}}
\put(17.00,31.00){\circle{2.00}}
\put(18.00,32.00){\line(1,1){3.00}}
\put(22.00,36.00){\circle*{2.00}}
\put(13.00,26.00){\line(1,0){5.00}}
\put(19.00,26.00){\circle{2.00}}
\put(20.00,26.00){\line(1,0){5.00}}
\put(23.00,36.00){\line(1,0){5.00}}
\put(29.00,36.00){\circle{2.00}}
\put(30.00,36.00){\line(1,0){5.00}}
\put(37.00,37.00){\line(1,1){3.00}}
\put(41.00,41.00){\circle{2.00}}
\put(42.00,42.00){\line(1,1){3.00}}
\put(46.00,46.00){\circle*{2.00}}
\put(51.00,37.00){\line(1,1){3.00}}
\put(55.00,41.00){\circle{2.00}}
\put(47.00,46.00){\line(1,0){5.00}}
\put(53.00,46.00){\circle{2.00}}
\put(56.00,42.00){\line(1,1){3.00}}
\put(60.00,46.00){\circle*{2.00}}
\put(54.00,46.00){\line(1,0){5.00}}
\put(23.00,37.00){\line(1,1){3.00}}
\put(27.00,41.00){\circle{2.00}}
\put(28.00,42.00){\line(1,1){3.00}}
\put(32.00,46.00){\circle*{2.00}}
\put(33.00,46.00){\line(1,0){5.00}}
\put(39.00,46.00){\circle{2.00}}
\put(40.00,46.00){\line(1,0){5.00}}
\put(36.00,37.00){\line(0,1){5.00}}
\put(36.00,43.00){\circle{2.00}}
\put(36.00,35.00){\line(0,-1){5.00}}
\put(36.00,29.00){\circle{2.00}}
\put(29.41,34.97){\line(2,-5){1.18}}
\put(29.63,36.86){\line(1,1){5.53}}
\put(29.99,36.39){\line(5,2){10.07}}
\put(40.06,40.42){\line(0,0){0.00}}
\put(36.97,42.60){\line(5,-2){3.05}}
\put(36.79,42.31){\line(1,-1){5.53}}
\put(41.62,40.13){\line(1,-3){1.05}}
\put(32.03,31.19){\line(5,2){10.18}}
\put(29.85,35.37){\line(1,-1){5.53}}
\put(40.60,40.06){\line(-2,-5){4.06}}
\put(35.50,42.07){\line(-2,-5){4.04}}
\put(31.92,30.49){\line(5,-2){3.08}}
\put(36.87,29.59){\line(1,1){5.57}}
\put(36.00,44.00){\line(0,1){5.00}}
\put(36.00,28.00){\line(0,-1){5.00}}
\put(36.00,22.00){\circle*{2.83}}
\put(36.00,50.00){\circle*{2.83}}
\put(36.00,51.00){\line(0,1){5.00}}
\put(36.00,21.00){\line(0,-1){5.00}}
\put(36.00,16.00){\line(0,0){0.00}}
\put(36.00,16.00){\line(0,0){0.00}}
\put(36.00,16.00){\line(0,0){0.00}}
\put(24.00,28.00){\line(0,1){6.00}}
\put(24.00,27.00){\circle*{2.83}}
\put(24.00,35.00){\circle*{2.83}}
\put(24.00,36.00){\line(0,1){5.00}}
\put(24.00,26.00){\line(0,-1){5.00}}
\put(14.00,3.00){\circle*{2.00}}
\put(15.00,4.00){\line(1,1){3.00}}
\put(19.00,8.00){\circle{2.00}}
\put(20.00,9.00){\line(1,1){3.00}}
\put(24.00,13.00){\circle*{2.00}}
\put(15.00,3.00){\line(1,0){5.00}}
\put(21.00,3.00){\circle{2.00}}
\put(22.00,3.00){\line(1,0){5.00}}
\put(28.00,3.00){\circle*{2.00}}
\put(29.00,4.00){\line(1,1){3.00}}
\put(33.00,8.00){\circle{2.00}}
\put(25.00,13.00){\line(1,0){5.00}}
\put(31.00,13.00){\circle{2.00}}
\put(34.00,9.00){\line(1,1){3.00}}
\put(38.00,13.00){\circle*{2.00}}
\put(32.00,13.00){\line(1,0){5.00}}
\put(0.00,3.00){\circle*{2.00}}
\put(1.00,4.00){\line(1,1){3.00}}
\put(5.00,8.00){\circle{2.00}}
\put(6.00,9.00){\line(1,1){3.00}}
\put(10.00,13.00){\circle*{2.00}}
\put(1.00,3.00){\line(1,0){5.00}}
\put(7.00,3.00){\circle{2.00}}
\put(8.00,3.00){\line(1,0){5.00}}
\put(11.00,13.00){\line(1,0){5.00}}
\put(17.00,13.00){\circle{2.00}}
\put(18.00,13.00){\line(1,0){5.00}}
\put(25.00,14.00){\line(1,1){3.00}}
\put(29.00,18.00){\circle{2.00}}
\put(30.00,19.00){\line(1,1){3.00}}
\put(34.00,23.00){\circle*{2.00}}
\put(39.00,14.00){\line(1,1){3.00}}
\put(43.00,18.00){\circle{2.00}}
\put(35.00,23.00){\line(1,0){5.00}}
\put(41.00,23.00){\circle{2.00}}
\put(44.00,19.00){\line(1,1){3.00}}
\put(48.00,23.00){\circle*{2.00}}
\put(42.00,23.00){\line(1,0){5.00}}
\put(11.00,14.00){\line(1,1){3.00}}
\put(15.00,18.00){\circle{2.00}}
\put(16.00,19.00){\line(1,1){3.00}}
\put(20.00,23.00){\circle*{2.00}}
\put(21.00,23.00){\line(1,0){5.00}}
\put(27.00,23.00){\circle{2.00}}
\put(28.00,23.00){\line(1,0){5.00}}
\put(24.00,14.00){\line(0,1){5.00}}
\put(24.00,20.00){\circle{2.00}}
\put(24.00,12.00){\line(0,-1){5.00}}
\put(24.00,6.00){\circle{2.00}}
\put(17.41,11.97){\line(2,-5){1.18}}
\put(17.63,13.86){\line(1,1){5.53}}
\put(17.99,13.39){\line(5,2){10.07}}
\put(28.06,17.42){\line(0,0){0.00}}
\put(24.97,19.60){\line(5,-2){3.05}}
\put(24.79,19.31){\line(1,-1){5.53}}
\put(29.62,17.13){\line(1,-3){1.05}}
\put(20.03,8.19){\line(5,2){10.18}}
\put(17.85,12.37){\line(1,-1){5.53}}
\put(28.60,17.06){\line(-2,-5){4.06}}
\put(23.50,19.07){\line(-2,-5){4.04}}
\put(19.92,7.49){\line(5,-2){3.08}}
\put(24.87,6.59){\line(1,1){5.57}}
\put(24.00,5.00){\line(0,-1){5.00}}
\put(35.00,-5.00){\line(0,0){0.00}}
\put(35.00,-5.00){\line(0,0){0.00}}
\put(35.00,-5.00){\line(0,0){0.00}}
\put(36.00,15.00){\circle*{2.83}}
\put(36.00,14.00){\line(0,-1){5.00}}
\end{picture}
\end{center}
\caption{Typical cuprate superconductor 
with CuO${}_2$ conduction planes.} \label{cuprate}
\end{figure}
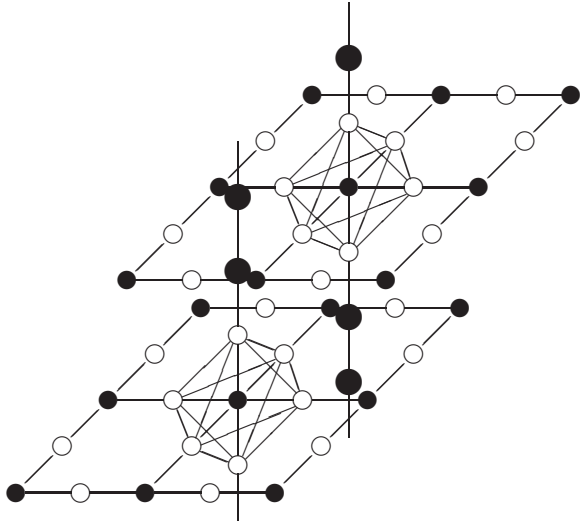

\section{Quantum symmetry unraveled}

We have so far identified several quantum group symmetric models (with and
without phonons) and have achieved a better understanding of $\hext$'s
superconducting quantum symmetry. There are however still open questions:
(i) Does a new model exist that is equivalent to $\hqsym$ in 1-D but can
also be formulated on higher dimensional lattices without breaking the
symmetry? This would be important for realistic models, see
Fig.~\ref{cuprate}.
(ii) Are there models with symmetry away from half-filling?
(iii) What is the precise relation between models with
classical and quantum symmetry in
this setting?

As we shall see the answer to the last question also leads to the resolution
of the first two. Without loss of generality (see argument given above)
we will focus on one pair of next-neighbor sites in the following.
We shall present two approaches that supplement each other:

\subsection{Generalized Lang-Firsov transformation}

We recall that the Hubbard model with phonons (with classical symmetry)
can be transformed into the standard Hubbard
Hamiltonian in two steps: A Lang-Firsov transformation changes the
model to one with vanishing local phonon coupling and a mean field
approximation removes the phonon operators from the model by averaging
over Einstein oscillator eigenstates \cite{RMR}. There
exists a similar transformation that relates the extended Hubbard model
(with quantum symmetry) to the standard Hubbard model:
$$
\hext \longleftrightarrow \hqsym \longleftrightarrow \hhub.
$$
(We have already
seen the first step of this transformation above in (\ref{LF}).)
It is easy to see that the hopping terms of $\hqsym$ and $\hhub$ have
different spectrum so the transformation that we are looking for
cannot be an equivalence
transformation. There exists however an invertible operator $M$, with 
$M M^* = 1 +  (\alpha^2 -1) \xi$, 
$\xi^2 = \xi$ (\ie similar to a partial isometry), that transforms the
coproducts of the classical 
Chevalley generators into their
Jimbo-Drinfel'd quantum counterparts
\bea
M \Delta_c(X^\pm)_s M^* & = & \Delta_q(X^\pm)_s \nonumber \\
M \Delta_c(H)_s M^* & = & \Delta_q(H)_s, \label{copxh}
\eea
and the standard Hubbard Hamiltonian into $\hqsym$
\beq
M  \hhub M^* = \hqsym  .
\eeq
This operator $M$ is
\beq
M = 1 \ot 1 + (\alpha -1) \xi + \beta f,
\eeq
with $f = X^-_s \ot X^+_s - X^+_s \ot X^-_s$,
$\xi = - f^2 = {1 \over 2}(H_s^2 \ot H_s^2 - H_s \ot H_s)$
and $\alpha \pm \beta = q^{\pm {1 \over 2}}$.
With this knowledge the proof of the quantum symmetry of $\hext$ is greatly
simplified.

\subsection{Quantum vs.\ Classical Groups---Twists}

A systematic way to study the relation of quantum and classical symmetries
was given by Drinfel'd \cite{D}. 
He argues that the classical U(g) and $q$-deformed
U$_q$(g) universal enveloping algebras are isomorphic {\em as algebras}.
The relation of the Hopf algebra structures is slightly more involved:
the undeformed universal enveloping algebra U(g) of a
Lie algebra, interpreted as a quasi-associative Hopf
algebra whose coassociator is an invariant 3-tensor,
is twist-equivalent to the Hopf algebra U$_q$(g) (over $[[\ln q]]$).

All we need to know here is that
classical ($\Delta_c$) and quantum ($\Delta_q$)
coproducts are related via conjugation (``twist'')
by the so-called 
universal ${\cal F} \in \mbox{U}_q(\mbox{su}_2)^{\hat\ot 2}$:
\beq
\Delta_q (x) = {\cal F} \Delta_c(x) {\cal F}^{-1}. 
\label{copt}
\eeq
(For notational simplicity we did not explicitly write the map that describes
the algebra isomorphism of $\mbox{U}(\mbox{su}_2)$ and 
$\mbox{U}_q(\mbox{su}_2)$ but
we should not be fooled by the apparent similarity between
(\ref{copxh}) and (\ref{copt}): The algebra isomorphism does not
map Chevalley generators to Jimbo-Drinfel'd generators  
and $M$ is not a representation of ${\cal F}$.)

The fundamental matrix representation of the universal ${\cal F}$
for SU(N) is an orthogonal matrix 
\bea
\rho^{\ot 2}({\cal F}) 
& = & \sum_i e_{ii} \ot e_{ii} 
      + \cos\varphi \sum_{i \neq j} e_{ii} \ot e_{jj} \nonumber \\
& + & \sin\varphi \sum_{i < j} \left( e_{ij} \ot e_{ji} - e_{ji} \ot
      e_{ij}\right),
\eea
where $\cos\varphi \pm \sin\varphi = \sqrt{2 q^{\pm 1}/(q + q^{-1})}$ ,
$i,j = 1\ldots N$ and $e_{ij}$ are N$\times$N matrices with lone
``1'' at position $(i,j)$.
The universal ${\cal F}$ in the superconducting
spin-${1\over 2}$ representation, {\em i.e.} essentially
the $N = 2$ case with the Pauli
matrices replaced by (\ref{suco}), is
\beq
F_s = \exp(\varphi f)_s = \tilde\xi + \cos\varphi \,\xi 
      + \sin\varphi \,f
\eeq
and $\tilde\xi + \xi = 1_s \ot 1_s$.
We are interested in a representation of the universal
${\cal F}$ on the 16-dimensional Hilbert
space of states of two sites:
\bea
F & = & (\epsilon_m \oplus \rho_s)^{\otimes 2}({\cal F}) 
        = \exp(\varphi f) \nonumber\\
  & = & 1 \ot 1 - 1_s \ot 1_s + F_s.
\eea
Note that the trivial magnetic representation $\epsilon_m$ enters here
even though we decided to study only deformations of the superconducting
symmetry---$F_s$ alone would have been
identically zero on the hopping term
and would hence have lead to a trivial model.

We now face a puzzle: By construction $F^{-1} \hqsym F$ should
commute with the (global)
generators of $\suq2_s$ just like $\hhub$.
But $F^{-1} \hqsym F$ obviously has the same spectrum
as $\hqsym$ so it cannot be equal to $\hhub$.
There must be other models with the same symmetries.
In fact we find a six-parameter family of classically 
symmetric models in any dimension.
In the one-dimensional case twist-equivalent quantum symmetric models
can be constructed as deformations of each of these classical models.
$\hhub$ and $\hqsym$ are not a twist-equivalent pair but
all models mentioned are related by 
generalized Lang-Firsov transformations.

To close we would like to present the most general Hamiltonian
with $\su2\times\su2/\zet_2$ symmetry and symmetric next-neighbor
terms.
(The group-theoretical derivation and detailed description of this
model is quite involved and will be given elsewhere.)
The Hamiltonian is written with eight real 
parameters ($\mu$, $r$, $s$, $t$, $u$, $v$, Re($z$), Im($z$)):
\bea
\hsym \:
&& =  u \sum_i n_{i\1} n_{i\0} \quad - \: \mu \sum_{i,\s} n_{i\s}
      \quad + \: t \sum_{\la i,j \ra \s} a^\dag_{i\s} a_{j\s}\nonumber\\
&& +  \: r \sum_{\la i,j \ra \s} n_{i\s} n_{j -\s}
      \quad +  \: s \sum_{\la i,j \ra \s} n_{i\s} n_{j \s} \nonumber\\    
&& +  \: {2 \mu - u \over e} \sum_{\la i,j \ra \s}
      a^\dag_{i\1} a^\dag_{i\0} a_{j\1} a_{j\0} \nonumber\\ 
&& +  \: (s - r) \sum_{\la i,j \ra \s} a^\dag_{i\s} a_{i -\s} a^\dag_{j -\s}
      a_{j \s} \nonumber\\
&& +  \: v \sum_{\la i,j \ra \s} 
      \left(n_{i\1} n_{i\0} n_{j\1} n_{j\0}
      - n_{i\1} n_{i\0} n_{j\s} 
      - n_{i\s} n_{j\1} n_{j\0}\right) \nonumber\\ 
&& +  \sum_{\la i,j \ra \s} a^\dag_{i -\s} a_{j -\s}
      \left(z(n_{i\s} - 1) n_{j\s} + z^* n_{i\s} (n_{j\s} -1)\right) 
      \nonumber\\    
&&    \mbox{\sc + h.c.} 
\eea
For symmetry $v = r + s + u - 2 \mu$ must hold.
One parameter can be absorbed into an overall multiplicative
constant, so we have six free parameters.
The first three terms comprise the standard Hubbard model but now
without the restriction to half-filling.
The filling factor is fixed by the coefficient of the
pair hopping term (6$^{th}$ term). The number $e$ in
the denominator of this coefficient 
is the number of edges per site. For a single pair of
sites $e = 1$, for a one-dimensional chain $e = 2$, for a honeycomb lattice
$e = 3$, for a square lattice $e = 4$, for a triangular lattice $e = 6$
and for a $D$-dimensional hypercube $e = 2 D$.
For a model on a general graph $e$ will vary with the site.
The 4$^{th}$ and 5$^{th}$ term describe density-density interaction 
for anti-parallel and parallel spins respectively.
The balance of these two interactions is governed by the
coefficient of the spin-wave term (7$^{th}$ term).
The last term is a modified hopping term that is reminiscent of the
hopping term in the $t$-$J$ model with hopping strength depending
on the occupation of the sites;
after deformation this term is the origin of the 
non-trivial quantum symmetries
of $\hsym$.

The known and many new quantum symmetric Hubbard models can be derived 
from $\hsym$ by twisting as described above.
While the deformation provides up to two extra parameters for the
quantum symmetric models the 
advantage of the corresponding classical
models is that they are not restricted to one dimension.
There are both classically and quantum symmetric
models with symmetries away from half filling.

The way the filling and the spin-spin interactions appear as coefficients
of the pair-hopping and spin-wave terms respectively looks quite
promising for a physical interpretation.
Due to its symmetries $\hsym$ should share some of the
nice analytical properties of the standard Hubbard Hamiltonian and
could hence be of interest in its own right.


\chapter[{\sffamily Quantum integrable systems}]{{\sffamily Solution by 
factorization of quantum integrable systems}}
\label{TQLE}

Adler, Kostant and Symes have introduced a beautiful unifying approach 
to the construction and solution by factorization of classical integrable 
systems in the framework of Lie and Lie-Poisson groups. In collaboration with 
B. Jur\v co an analogous quantum mechanical construction was found with the 
Heisenberg equations of motion in quantum Lax form. Like in the classical 
theory, the equations of motion are linearized and solved in an enlarged space 
(the Heisenberg double); the solution to the original problem is then found 
by factorization. For this project a formalism based on quantum groups was 
introduced that can handle the dual non-commutativities (quantum commutators 
and group structure). The method can be further generalized to 
the case of face Hopf algebras (elliptic quantum groups with dynamical 
R-matrices.) We give the construction of twisted and untwisted quantum Lax 
equations associated with quantum groups and solve these equations using 
factorization properties of the corresponding quantum groups. We then give 
the construction of quantum Lax equations for IRF models and difference 
versions of Calogero-Moser-Sutherland models introduced by Ruijsenaars. 
We again solve the equations using factorization properties of the underlying 
elliptic quantum groups. So far the construction is quite abstract: we give 
a formal solution to the equations of motion and the theory should be 
further developed by studying explicit quantum factorization in specific models. 
(One can picture Hamilton-Jacobi theory as a useful analog from classical 
mechanics.) We are solving the quantum mechanical time evolution of quantum 
integrable systems, not the spectrum that is usually studied in quantum 
mechanics. In quantum optics, however, it is exactly the time-evolution that 
is at the center of interest and it would be interesting to see if our work can 
give some new insights in that field. 

The research was done in collaboration with Branislav Jur\v co
and is published in~\cite{twist,face}; the remainder of this
chapter is based on these publications.

In the next section we review the classical theory, giving a more detailed 
introduction to the problem that we want to solve.

\section{Twisted quantum Lax equations}

We would like to  understand quantum integrable systems related
to quasitriangular Hopf algebras in a way that generalizes the classical 
theory based on the construction of Adler \cite{Ad}, 
Kostant \cite{K}, Symes \cite{Symes} for
Lie groups and its subsequent generalization to the 
Lie-Poisson case due to Semenov-Tian-Shansky \cite{STSDress}. 
The classical theory which we briefly summarize in this section 
gives the construction and solution
of integrable systems possessing a (twisted) Lax pair and $r$-matrix
formulation. The rich structure of these integrable systems appears 
naturally as a consequence of the factorization
properties of the groups under consideration. 
Within this approach the fundamental methods (inverse scattering  
method \cite{Miura}, algebro-geometric methods of solution) 
and the fundamental  notions of the soliton theory, such as 
$\tau$-function \cite{JM} and  
Baker-Akhieser function \cite{DKN}, \cite{SeW}, found their unifying and 
natural group-theoretical explanation.
Here we are interested in the quantum case. 
The theory of integrable models in quantum mechanics and quantum field theory 
made remarkable progress with the quantum version of  
the inverse scattering method, which goes back to the seminal Bethe  
ansatz for the solution of the Heisenberg
spin chain. We refer the reader for a review of related topics to the  
books \cite{Gaudin}, \cite{BIK} and to the papers \cite{F}, \cite{KS},
\cite{Thacker}. This development suggested the introduction of 
quantum groups \cite{DrQG}, \cite{Jimbo} as  
algebraic objects that play in the quantum case a role analogous to that of  
Lie groups in the classical theory. However, we were still  
missing (with the exception of the quantum integrable systems with  
discrete time evolution \cite{Resh}) a quantum analogue of the  
factorization theorem for the solution of the Heisenberg equations of  
motion for quantum integrable systems. We have also to mention the 
remarkable paper \cite{Maillet} in this context.

The quantum systems we consider are quantum counterparts of 
those described by the classical factorization theorem. 
In~\cite{JSch2} 
a quantum version of the theory in the case without twisting
was formulated.
Our main result,
the quantum factorization theorem,
as well as the remaining discussion extends all constructions in the presence
of twisting.

\subsection{Classical integrable systems}

Here we briefly review the construction of integrable systems and
their solution by factorization which is due to
M. Semenov-Tian-Shansky \cite{STSDress} and
which generalizes to the case of the Poisson Lie groups
the construction of Adler, Kostant and Symes \cite{Ad}, \cite{K}, \cite{Symes}.
Let $G$ be a quasitriangular Poisson Lie group, which is for simplicity
assumed to be a matrix group. Let \gtg\ 
be the corresponding Lie algebra and
$r\in \gtg \otimes \gtg$ the classical $r$-matrix, a
solution to the classical Yang-Baxter equation.
In the following we will use a notation that does
not distinguish between
the universal element and its matrix representative.
We will denote by $G_r$ and $\gtgr$ 
the dual Poisson Lie group and its 
Lie algebra respectively. The pairing between \gtg\ and \gtgr\
is denoted by $\langle.,.\rangle$.
The Poisson structure on $G$ is given by the Sklyanin bracket
\begin{equation}
\{g_1,g_2\}=[r, g\otimes g]\,.\label{Sklyanin}
\end{equation}
Here we used the standard tensor notation:
$g_1=g\otimes 1$, $g_2=1\otimes g$, with
$g\in G$ being a group element (matrix) 
and $1$ the unit matrix. The commutator
on the right-hand side is the usual matrix 
commutator in $\gtg \otimes \gtg$.
With the universal $r$-matrix $r$ we can 
associate two mappings $r_{\pm}$:
$\gtgr \rightarrow \gtg$
\eq
r_+(X) = \langle X\otimes id, r\rangle\equiv X_+\,,\qquad
r_-(X) = -\langle id \otimes X, 
r\rangle\equiv X_- \label{maps}\,.
\en
These mappings are algebra homomorphism.
Let $\gtg\hspace{.5pt}_{\pm}=\mbox{Im}(r_{\pm})$
be the corresponding subalgebras in \gtg. 
Consider the combined mapping
$i_r = r_- \oplus r_+:\, \gtgr \rightarrow \gtg \oplus \gtg$.
Let us assume that the mapping
$r_+-r_-:\,\gtgr \rightarrow \gtg$ is an isomorphism of
linear spaces. In that case \gtg\ is called factorizable
and any $X\in \gtg$ has a unique decomposition
\eq 
X=X_+-X_-,
\en
with $(X_-, X_+) \in \mbox{Im}(i_r)$.
The map $i_r$ gives rise to a Lie group embedding 
$I_r:G_r=G_-\times G_+\rightarrow G\times G$. Followed 
by the group inversion in the first factor and a 
subsequent group multiplication of factors it 
defines a local homeomorphism on the Poisson Lie groups $G_r$ and $G$. 
This means that in the neighborhood of the group 
identity any group element $g\in G$ admits a unique decomposition
\eq
g=g_-^{-1}g_+,\label{cfac}
\en
with $(g_-,g_+)\in \mbox{Im}(I_r)$.

The group manifold equipped with Sklyanin bracket 
(\ref{Sklyanin}) plays the role of the phase space 
for a classical dynamical system governed by a 
Hamiltonian constructed in the following way. 
Let $\phi$ by an automorphism of \gtg\ which 
preserves the classical $r$-matrix
\eq
(\phi\otimes \phi)r=r
\en
and which defines an automorphism of $G$ denoted by the same symbol.
The Hamiltonian $h$ is taken as any smooth function on 
$G$ invariant with respect to the twisted conjugation. This means that
\eq
h(g)=h(g_1^{\phi} g (g_1)^{-1}),\,\,\, g^{\phi}\equiv\phi(g)\label{twconj}
\en
holds for any two group elements $g,g_1\in G$.
The functions satisfying (\ref{twconj}) are in involution 
with respect to the Sklyanin bracket, so they play the role
of integrals of motion for the 
dynamical system on $G$ just described above.
We shall denote this involutive 
subset of $C^{\infty}(G)$ as $I^{\phi}(G)$.

For any smooth function $f$ on $G$ let us introduce $D_f(g) \in \gtgr$
by the following equality
\eq
\langle D_f (g),X\rangle=(d/dt)_{t=0}f(ge^{tX}).
\en

We shall also use the symbol $\nabla_f(g)$ 
for the corresponding element of 
\gtg
\eq 
\nabla_f(g)=(D_f(g))_+-(D_f(g))_-
\en
and we shall refer to it as to the gradient of $f$.
If $h_1, h_2\in I^{\phi}(G)$ then the corresponding gradients commute.

Now we can formulate the main classical theorem: 
\begin{thm}[main classical theorem] \label{main}
(i) Functions \ $h\in I^{\phi}(G)$ are in involution with 
respect to the Sklyanin bracket on $G$. 
(ii) The equations of motion defined by Hamiltonians 
$h\in I^{\phi}(G)$ are of the Lax form
\eq 
dL/dt = \phi(M^{\pm})L -LM^{\pm}, \label{clax}
\en
with $M^{\pm}=(D_h)(L)_{\pm}$ and $L\in G$.
(iii) Let $g^{\pm}(t)$ be the solutions to the 
factorization problem \/\emph{(\ref{cfac})} 
with the left hand side given by
\eq
g(t)=\mbox {exp}(t\nabla_h(L(0))).
\en 
The integral curves of equation \/\emph{(\ref{clax})} are given by
\eq
L(t)=\phi(g_{\pm}(t))L(0)g_{\pm}(t)^{-1}.\label{csl}
\en
\end{thm}

An elegant  proof of this theorem as presented by
N.Y.\ Reshetikhin in lectures at the University of Munich
is  given in appendix~\ref{AppB}. Interested readers can also consult
\cite{STSDress}, \cite{ReySTS} for other  proofs. We would
like to mention that 
there is an easy direct proof
and a more conceptual one, which reflects the 
rich structure of the theory of Poisson Lie 
groups and the geometry related to the theory of 
integrable systems. The strategy of the second proof 
is to show that the Lax equations 
are obtained by a change of variables from the simplest 
$G_-\times G_+$-invariant Hamiltonian systems on the so-called 
classical Heisenberg double
(the Poisson Lie generalization of the of the 
cotangent bundle $T^*(G)$). 
This construction allows also, using the 
Poisson Lie variant of the symplectic reduction, 
to give one more description of the symplectic leaves 
of $G$ which are known 
to be the orbits of the dressing action of $G_r$ on $G$ \cite{STSDress}.

It is often useful to consider the 
Lax equations (\ref{clax}) corresponding to
different Hamiltonians from 
$I^{\phi}(G)$ simultaneously, using different
time parameters corresponding to the different Hamiltonians.
Usually there is given
a hierarchy (a complete set) of functionally independent
Hamiltonians $h_{\alpha}$, with $\alpha$ 
running over some label set $I$.
The corresponding time parameters are
$\{t_{\alpha}\}_{\alpha \in I}\equiv{\bf t}$
and we index by $\alpha$ also the corresponding gradients and matrices
$M^{\pm}$ entering the Lax equations. 
Then we obtain the following equations
for $L({\bf t})\in G$, $M^{\pm}_{\alpha}({\bf t})\in \gtg$
and $M_{\alpha}\equiv \nabla_{h_{\alpha}}\in \gtg$
\eq
\partial L/\partial t_{\alpha} 
= \phi(M_{\alpha}^{\pm})L -LM_{\alpha}^{\pm}, \label{1a}
\en
\eq
\partial M_{\alpha}/\partial t_{\beta} = [M^{\pm}_{\beta}, M_{\alpha}]  \label{1b}
\en
and
\eq
\partial M^{\pm}_{\alpha}/\partial t_{\beta} 
-\partial M^{\pm}_{\beta}/\partial t_{\alpha}
= [M^{\pm}_{\beta}, M^{\pm}_{\alpha}]. \label{1c}
\en
Here the integral curves $L({\bf t})$ corresponding 
to the commuting dynamical flows on $G$ are given as 
above by the twisted conjugation like in (\ref{csl})
with the factors of
\eq
g({\bf t})=\mbox {exp}(\sum_{\alpha}t_{\alpha}M_{\alpha}(0)). \label{1d}
\en
In the case of a concrete dynamical system, 
we pick up a proper symplectic leaf
on $G$. Usually it is taken in a way 
that $I^{\phi}(G)$ contains 
enough first integrals to ensure that the system is 
completely integrable (in a proper sense).
The group element $g({\bf t})$ can then be brought 
by a similarity transformation to the form
\eq g({\bf t})=\varphi(0)\mbox {exp}(\sum 
 t_{\alpha}X_{\alpha})\varphi(0)^{-1},
\en
where $X_{\alpha}$ are generators of some abelian subalgebra of $G$,
so that the group element $g({\bf t})$ describes an 
embedding of one of the commutative subgroups of $G$ 
into $G$ itself.

Let us now consider the 
quantization of the Lie-Poisson structure. 
The reader is referred to the existing monographs 
on quantum groups (e.g. \cite{ChP}) for the necessary 
information on quantum groups.

\subsection{Heisenberg equations of motion}

As quantized phase space we take the 
non-commutative Hopf algebra $F$
of functions on a quantum group, dual via the pairing
$\la.,.\ra:\, U \ot F \rightarrow k$ to a 
quasitriangular Hopf algebra $U$.
The quantum analog of Sklyanin's bracket is then \cite{FRT}
\eq
R T_1 T_2 = T_2 T_1 R. \label{RTT}
\en
The $R$-matrix can be expanded as 
$R = 1 + h r + {\cal O}(h^2)$, where $h$
is a deformation parameter that 
gives the correspondence to classical
mechanics:
\[
\frac{[f,g]}{h} = \{f,g\} \quad (\mbox{mod } h),
\]
such that for instance
$
0 = R T_1 T_2 - T_2 T_1 R 
= [T_1 , T_2] \quad (\mbox{mod } h)
$
and
\[
0 = \frac{R T_1 T_2 - T_2 T_1 R}{h} 
= \{T_1,T_2\} - [T_1 T_2, r] \quad (\mbox{mod } h) ,
\]
\emph{i.e.} 
$\{T_1,T_2\} = [T_1 T_2, r]$ is the
classical limit of (\ref{RTT}) as desired. 
(Note that (locally) all integrable systems have such a
classical $r$-matrix \cite{BaVi}).

In these and the following expressions $T$ 
may either be interpreted simply
as a Matrix $T \in M_n(F)$ or, much more general, 
as the canonical element of $U \ot F$ : \
Let $\{e_i\}$ and $\{f^i\}$
be dual linear bases of $U$ and $F$ respectively. 
It is very convenient to work with the canonical
element in $U \ot F$ (also called the universal $T$-matrix \cite{FRT}, 
for an elementary overview see
\cite{CSW}),
\eq
T = \sum_i e_i \ot f^i \quad \in \quad U \ot F,
\en
because equations expressed in terms of it will be
reminiscent of the familiar expressions
for matrix representation---but we still keep 
full Hopf algebraic generality.
For the same reason we will often write ``$T_1$'' in
place of ``$T_{12}$'' when we use a notation
that suppresses direct reference to the second tensor
space of $T$; multiplication in $F$ is understood
in that case. Example:
$R_{12} T_1 T_2 = T_2 T_1 R_{12}$
is short for
$R_{12} T_{13} T_{23} = T_{23} T_{13} R_{12}$.

The set of cocommutative elements of $F$ form a
commutative subalgebra $I \subset F$ \cite{DrQG}. 
If we choose a Hamiltonian
from this set, it will commute with all other cocommutative
elements, which will consequently be constants of motion.
This observation can be generalized to twisted cocommutative
elements: Let $\phi$ be an automorphism of $U$ that 
preserves the universal $R$-matrix,
\eq
(\phi \ot \phi)(R) = R. \label{phiR}
\en
($\phi$ is the quantum analog of an 
automorphism of a Poisson-Lie Group.)
Let $\phi^*$ be the pullback of $\phi$ to $F$, \ie\
$\langle x , \phi^*(f) \rangle = \langle \phi(x) , f
\rangle$.
The Hamiltonian $h \in F$ shall be a twisted cocommutative 
function,\footnote{Quantum traces are 
twisted cocommutative functions
with a non-trivial $\phi$ given by the square of the 
antipode, while ordinary traces are simply cocommutative.}
\ie
\eq
\Delta' h = (\phi^*\otimes\id)(\Delta h), \label{phiDelta}
\en
where $\Delta' = \tau \circ \Delta$ is the opposite
comultiplication. The set of twisted 
cocommutative functions also form a
commutative subalgebra $I^\phi \subset F$.

In the following we shall 
present the more general twisted case.
The untwisted formulation can be obtained by omitting ``$\phi$''
in all expressions.
The dynamics of our system is governed by the Heisenberg
equations of motion
\eq
i \dot f = [h , f], \quad \forall f \in F.
\en
These can equivalently be written in terms of the universal
$T$ as
\[i\dot T \equiv i\sum e_i \ot \dot f^i = [\id \ot h , T] 
= \langle T_{13} T_{23} - T_{23} T_{13} , 
h \ot \id^2 \rangle,\] 
or short
\eq
i \dot T_2 =  \langle [T_1 , T_2] , h \ot \id \rangle.
\label{tttheisenberg}
\en

\subsubsection{Strategy for solution}

\begin{figure}[thb]
\begin{center}
\begin{picture}(0,0)%
\includegraphics{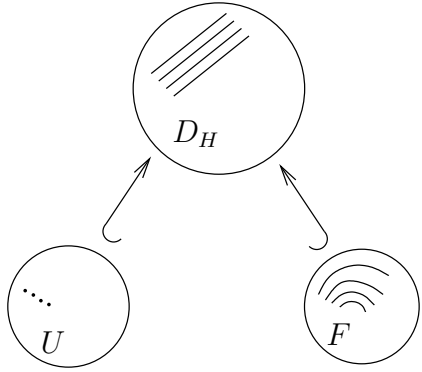}%
\end{picture}%
\setlength{\unitlength}{4144sp}%
\begingroup\makeatletter\ifx\SetFigFont\undefined%
\gdef\SetFigFont#1#2#3#4#5{%
  \reset@font\fontsize{#1}{#2pt}%
  \fontfamily{#3}\fontseries{#4}\fontshape{#5}%
  \selectfont}%
\fi\endgroup%
\begin{picture}(2477,2197)(3275,-3582)
\put(4411,-2221){\makebox(0,0)[b]{\smash{\SetFigFont{12}{14.4}{\familydefault}{\mddefault}{\updefault}
$D_H$
}}}
\put(3556,-3481){\makebox(0,0)[b]{\smash{\SetFigFont{12}{14.4}{\rmdefault}{\mddefault}{\updefault}
$U$
}}}
\put(5266,-3436){\makebox(0,0)[b]{\smash{\SetFigFont{12}{14.4}{\rmdefault}{\mddefault}{\updefault}
$F$
}}}
\end{picture}
\caption{Nonlinear evolution with Hamiltonian $h \in F$ becomes
linear after a nontrivial embeding  of $F$ into the Heisenberg 
double $D_H$ with $h \mapsto \tilde h$. This is due to the constant 
evolution that $\tilde h$ generates in the dual to $F$, 
$U \subset D_H$.} 
\label{ttdouble}
\end{center}
\end{figure}
Our strategy to solve equation (\ref{tttheisenberg})
will be to embed
the quantized phase-space
$F$ into a bigger algebra (the Heisenberg double, 
$D_H \approx F \ot U$ as a vector space),
where the equations take on a particularly simple form:
The image under this embedding of a 
Hamiltonian $h \in I^\phi$ 
is a casimir in $U$ which leads to trivial time evolution
in $U \subset D_H$ and simple (linear) evolution in 
$D_H$, see figure~\ref{ttdouble}. Projecting the solution 
back to $F$ 
we will find that Heisenberg's equations can be written
in (twisted) Lax form and our original problem
is solved by factorization just
like in the classical case.

\subsection{Heisenberg double with twist}

The Heisenberg Double of $F$ shall refer to the the semi-direct product
algebra $D_H = F \semi U$ \cite{Sweedler,AF,STS}. 
It is also known as the quantum
algebra of differential operators \cite{Zum,SWZ} or the quantum cotangent
bundle on $F$; see \emph{e.g.} \cite{Fr} for the corresponding classical
Heisenberg double. $D_H$ is isomorphic to $F \ot U$ as a vector
space; it inherits the product structures of $F$ and $U$;
mixed products are obtained from the left action of $U$ on $F$.
All relations can be conveniently summarized
in terms of the canonical element $T$ of $F \ot U$~\cite{CSW}:
\eq
T_{23} T_{12} = T_{12} T_{13} T_{23}. \label{HD} \qquad\mbox{\it
(Heisenberg Double)}\qquad
\en
This equation gives commutation relations for
elements  $x \in U$ with elements $f \in F$
that equip $F \ot U$ with an algebra structure:
$\: x\cdot f
= \la\id \ot \Delta x,\Delta f\ot \id \ra\:\in\: F \ot U$.
In the setting of interest to us, $F$ is a co-quasitriangular Hopf
algebra who's structure (\ref{RTT}) is determined by a universal
$R$-matrix. Following Drinfel'd's construction \cite{DrQG}
we shall assume that $U$ is itself the quantum double of a Hopf
algebra $U_+$; the universal $R$ arises then
as the canonical element in $U_- \ot U_+$, where
$U_- = U_+^{* \mbox{\tiny op} \Delta}$.
The Yang-Baxter equation
\eq R_{23} R_{13} R_{12} = R_{12} R_{13} R_{23} 
\qquad\mbox{\it
(Quantum Double)}\qquad \label{YBE}\en
plays the same role for
the quantum double $U = U_+ \Join U_-$
as (\ref{HD}) plays for the Heisenberg Double.
The spaces $U_+$ and $U_-$ are images of the
two mappings $R^\pm: F \rightarrow U_\pm$ associated with the universal 
$R \in U_- \ot U_+$:
\eq
R^+(f) = \la R_{21} , \id \ot f\ra, 
\qquad R^-(f) = \la R\inv_{12}, \id \ot f\ra.
\en

The twisted Heisenberg double was introduced in \cite{STS}.
The most convenient description for our purposes
of the dual Hopf algebra $U$ with twist is in terms of the universal 
invertible twisted-invariant 2-tensor $Y \in U \ot U$:
\eq
Y_1 R\p_{12} Y_2 R_{21} = R_{12} Y_2 R\p_{21} Y_1. \label{YY}
\en
Here as well as in the following the superscript ${}^\phi$ denotes
the application of the automorphism $\phi$ to the first tensor space:
$R^\phi \equiv (\phi\ot\id)(R)$.
Twisted invariance means:
\eq
Y_{12} T\p_1 T_2 =  T_1 T_2 Y_{12}.
\en 
The mixed relations
\eq
Y_1 T_2 = T_2 R_{21} Y_1 R\p_{12} \label{YT}
\en
complete the description of the Heisenberg double.
(In the case of a trivial twist $\phi = \id$ we may chose
$Y = R_{21} R_{12}$; equation (\ref{YY}) is then a consequence
of the Yang-Baxter equation (\ref{YBE}).)

Crucial for the construction that we are going to present 
is that $T$ factorizes in $U \ot F$ as \cite{FRT} 
\eq
T = \Lambda Z,\quad\mbox{with}\quad \Lambda \in U_- \ot F_-,\quad
Z \in U_+ \ot F_+ ,\quad F_\pm = (U_\pm)^* .
\en
We have $U = U_- \ot U_+$ as a linear space \emph{and coalgebra}
and $F = F_- \ot F_+$ as a linear space \emph{and algebra:}
\eq
(\Delta\ot\id)(\Lambda) = \Lambda_1 \Lambda_2, \quad
(\Delta\ot\id)(Z) = Z_1 Z_2, \quad
\Lambda_1 Z_2 = Z_2 \Lambda_1 .
\en
The universal elements $Z$ and $\Lambda$ of $U_\pm \ot F_\pm$
define projections $F \rightarrow F_\pm$ and
$U \rightarrow U_\pm$ that can be used to extract the $F_\pm$
parts of any element of $F$ and the $U_\pm$ parts of
$(-+)$-ordered expressions
in $U$. Let us denote by $(L^+)\inv \in U \ot U_+$ and $L^- \in U \ot U_-$
the corresponding images of $Y\inv$, such that $Y = L^+ (L^-)\inv$.
Using the maps based on $Z$ and $\Lambda$ we can derive a
host of relations between $Y$, $Z$, $\Lambda$ and $L^\pm$:
\begin{prop} The Heisenberg Double is defined by relations
(\ref{RTT}), (\ref{YY}) and (\ref{YT}).
The following relations are consequences of
these and (\ref{HD}):
\eq
Z_1 Y_1 Z_2 = Z_2 Z_1 Y_1 R^\phi_{12} \label{YZ}
\en
\eq
R_{12} Z_1 Z_2 = Z_2 Z_1 R_{12} \label{ZZ}
\en
\eq
R_{12} \Lambda_1 \Lambda_2 = \Lambda_2 \Lambda_1 R_{12}
\en
\eq
Z_1 L^+_1 Z_2 = Z_2 Z_1 L^+_1
\en
\eq
Z_1 L^+_1 \Lambda_2 = \Lambda_2 R_{21} Z_1 L^+_1
\en
\eq
L^-_1 Z_2 = Z_2 R\p_{12}{}\inv L^-_1
\en
\eq
L^-_1  \Lambda_2 = \Lambda_2 L^-_1
\en
\eq
R_{12} L^\pm_2 L^\pm_1 = L^\pm_1 L^\pm_2 R_{12}
\en
\eq
R\p_{12} L^+_2 L^-_1 = L^-_1 L^+_2 R_{12}
\en
\eq
Z_{23} \Lambda_{12} = \Lambda_{12} Z_{23}
\en
\eq
Z_{23} Z_{12} = Z_{12} Z_{13} Z_{23}.
\en
\end{prop}
\emph{Remark: } Similar relations involving $R$, $L^\pm$ and $T$ 
in the presence
of twisting were proposed in \cite{STS}.
Further relations involving $Y$ and $T$ can be easily obtained from
the ones given above. 
The apparent asymmetry between relations involving $Z$ versus those
involving $\Lambda$ is due to our choice of factorizing $T$ as $\Lambda Z$.
We could have also based our analysis on $T = S V$ with $S \in U_+ \ot F_+$
and $V \in U_- \ot F_-$; this would 
restore the $+$/$-$-symmetry.\\[1ex]
{\sc Proof: } We shall only proof relation (\ref{YZ}) that we are
going to use extensively in the next sections.
(The other relations follow similarly; see also \cite{JSch2}
and the discussion in \cite{JSch}.)\\
The second tensor space of relation (\ref{YT}) is not $(U_-U_+)$-ordered
so we have to resort to a trick:
with the help of (\ref{RTT}) we can derive a new relation
\[
T_1 Y_1 T{}_{\stackrel{\scriptstyle 2}{\scriptscriptstyle -+}}
= R{}_{\stackrel{\scriptstyle 2}{\scriptscriptstyle -}}{}_1 
T{}_{\stackrel{\scriptstyle 2}{\scriptscriptstyle -+}} 
T_1 Y_1 R\p{}_1{}_{\stackrel{\scriptstyle 2}{\scriptscriptstyle +}}
\] 
whose second tensor
space is $(U_-U_+)$-ordered as is easily verified. Projecting to
$U_+$ we obtain $\: T_1 Y_1 Z_2 = Z_2 T_1 Y_1 R\p_{12}\:$ which
can be simplified using $T_1 = \Lambda_1 Z_1$ and $\Lambda_1 Z_2 = Z_2
\Lambda_1$ to yield (\ref{YZ}). \hfill $\mybox$

\subsubsection*{A remark on quantum traces and twisting}

We have argued that the cocommutative 
elements of $F$ are natural candidates
for Hamiltonians. Classically 
cocommutativity is equivalent to ad-invariance,
so it would also be natural to 
look for Hamiltonian functions in the
quantum case that are invariant under the
quantum adjoint coaction\footnote{This 
holds for instance for 
quantum traces.}
\eq
\Delta^{Ad}(h) \equiv h_{(2)} \ot S(h_{(1)}) h_{(3)}
= h \ot 1,\label{adinv}
\en
It turns out that both these and the cocommutative 
Hamiltonians are
treated on equal footing 
in the twisted formulation: Requirement (\ref{adinv}) is 
equivalent to
\eq
\Delta h = (\id \ot S^2)(\Delta' h),
\en
\ie\ corresponds to a twisted cocommutative function with the
pullback of the twist
$\phi^*$ given by the square of the antipode.
The twist $\phi = S^2$ is here generated via conjugation by 
an element $u \in U$:
\eq
S^2(x) = u x u^{-1},\quad\forall x \in U.
\en
It seems interesting to study the general case of a twist $\phi$ that
is given via conjugation by some element $\varphi$, \ie
\eq
\phi(x) = \varphi x \varphi\inv, \qquad (\varphi\ot\varphi) R = R 
(\varphi\ot\varphi).\label{varphi}
\en
If $\varphi \in U$ then
$\phi^*(f) = \la\varphi 
, f_{(1)}\ra f_{(2)} \la\varphi\inv , f_{(3)}\ra$
for all $f \in F$.
(Here we see by the way that $\phi^*(h) = h$ 
holds both for cocommutative
and twisted cocommutative $h$.)
Due to (\ref{varphi}), 
$f \mapsto \la \id \ot \varphi\inv , \Delta f\ra$ defines an
algebra isomorphism of $F$, that maps
$I \subset F$ to $I\p \subset F$, \ie\
cocommutative elements to twisted cocommutative
elements.

It is easily verified that all expressions 
containing $\phi$, \emph{e.g.}
(\ref{YY}), (\ref{YZ}), \emph{etc.} continue to hold 
if we omit $\phi$ \emph{and}
replace $Y$ by $Y\cdot(\varphi\ot 1)$. 
Examples:
\eq
R_{21} Y_1 \varphi_1 R_{12} Y_2 \varphi_2 
= Y_2 \varphi_2 R_{21}
Y_1 \varphi_1 R_{12},
\en
\eq
Z_1 Y_1 \varphi_1 Z_2 = Z_2 Z_1 Y_1 \varphi_1 R_{12}, \qquad etc.
\en
This gives a nice mnemonic for where
to put the $\phi$'s---\emph{even when an element $\varphi$ does
not exist in $U$}: First we write 
expressions without  $\phi$, then we
formally 
replace all $(L^-)^{-1}$'s by 
$(L^-)^{-1}\cdot(\varphi\ot 1)$ (and consequently 
$Y$ by $Y\cdot(\varphi\ot 1)$),
finally we
remove all $\varphi$'s from the 
expression with the help of
relation (\ref{varphi}).
\ \emph{Remark:}
$Y = L^+ (L^-)^{-1}$ but $Y \neq R_{21} R_{12}$ in the
twisted case. In case we know an element $\varphi$ that
satisfies (\ref{varphi}), we can realize $L^\pm$
in terms of the universal $R$ for instance as $L^+ = R_{21}$
and $L^- = \varphi_1 R\inv_{12}$. 
There is however some remaining ambiguity in this choice.

\subsection{Embedding the operator algebra into the double}
\label{s:embed}

Here we will show how to embed $F$ into $D_H$ 
in such a way that any (twisted)
cocommutative element of $F$ is mapped to a 
casimir operator of $U \subset
D_H$.
\begin{prop}
The following element of $U \ot D_H$ 
\eq
\widetilde T = \phi(Z) Y^{-1} Z^{-1},
\en
where $\phi(Z) \equiv (\phi\ot\id)(Z)$, satisfies 
\eq
R_{12} \widetilde T_1 \widetilde T_2 
= \widetilde T_1 \widetilde T_2 R_{12}
\label{RTTtilde}
\en
and thus defines an embedding of $F \hookrightarrow D_H:\:
f \mapsto \langle \widetilde T , f \ot \id \rangle$,
that is an algebra homomorphism. (The picture
of $F$ in $D_H$ by this embedding will be denoted $\widetilde F$.)
\end{prop}

\noindent {\sc Proof:} Start with (\ref{YY}) in form
$Y\inv_1 R^\phi_{21}{}\inv Y_2\inv 
= R_{21}\inv Y_2\inv R\p_{12}{}\inv Y_1\inv R_{12},$
multiply by $Z\inv_2 Z\inv_1$ from the right and use (\ref{ZZ}) to obtain
\[
Y\inv_1 Z\inv_1 Y\inv_2 Z\inv_2 = R_{21}\inv Y\inv_2 
\underline{R\p_{12}{}\inv Y\inv_1
Z\inv_1 Z\inv_2} R_{12}.
\]
Applying equation (\ref{YZ}) to the underlined part gives
\eq
Y\inv_1 Z\inv_1 Y\inv_2 Z\inv_2 = R_{21}\inv Y\inv_2 Z\inv_2 Y\inv_1 Z\inv_1 R_{12}
\label{zwischen}
\en
and as a corollary:
$R_{12} Z_2 Y_2 Z_1 Y_1 = Z_1 Y_1 Z_2 Y_2 R_{21}$. 
Now use equation (\ref{YZ}) twice: once in the form
$
Z_1 Y_1 \phi(Z_2) = \phi(Z_2) Z_1 Y_1 R_{12},
$
which follows from $(\phi\ot\phi)(R) =R$,
to replace $Y\inv_1 Z\inv_1$ on the LHS of (\ref{zwischen})
and once to replace $R_{21}{}\inv Y\inv_2 Z\inv_2$ on the RHS of (\ref{zwischen}).
Multiplying the resulting expression
by $\phi(Z_i)$ from the left and using (\ref{ZZ}) in the form
$
R_{12} \phi(Z_1) \phi(Z_2) = \phi(Z_1) \phi(Z_2) R_{12}
$
gives our result (\ref{RTTtilde}). \hfill$\mybox$

\begin{prop} 
The image $\widetilde h$ of the Hamiltonian $h$
under the embedding $F \rightarrow \widetilde F$ is a casimir in
$U \subset D_H$. We can find the following explicit expression:
\eq
\widetilde h = \la \widetilde T , h \ot \id \ra
  = \la u_1\inv Y_1\inv , h \ot \id \ra, \label{casimir}
\en
where\footnote{Here and in the following we
will use the following tensor notation: 
The second subscripts denote 
the \emph{order of
multiplication} in a given tensor space. 
Consider for example
$R = \sum_i \alpha_i \ot \beta^i$, then
$(S^2 \ot \id)(R)_{1_21_1}$ equals $\sum_i \beta^i S^2(\alpha_i)$ and lives in
tensor space 1.}
$u\inv = (S^2 \ot \id)(R)_{1_21_1}$ and
satisfies $u\inv x = S^2(x) u\inv, \ \forall x \in U$ .
\end{prop}

\noindent {\sc Proof:} We have to proof two things: 
1) $\widetilde h$ commutes with
all elements of $U$ and 2) $\widetilde h$ is an element of $U$ with the 
given expression.\\[1ex]
Ad 1): Here is a nice direct calculation that shows that $\widetilde h$ 
commutes with $Y^{-1}$
and hence (in the factorizable case) with all of $U$:\\[1ex]
Start with the twisted reflection equation (\ref{YY}) in the form
\[
Y\inv_1 \underline{R\p_{21}{}\inv Y\inv_2} 
= R_{21}\inv Y\inv_2 R\p_{12}{}\inv Y\inv_1 R_{12},
\]
apply (\ref{YZ}) with subscripts 1 and 2 exchanged 
to the underlined part,
rearrange and multiply by $\phi(Z_1)$ from the left to obtain:
\[
\phi(Z_1) Y\inv_1 Z\inv_1 Y\inv_2 
=\phi(Z_1) R_{21}\inv Y\inv_2 R\p_{12}{}\inv Y\inv_1 R_{12} Z\inv_2 Z\inv_1 Z_2.
\]
Now we use (\ref{YZ}) twice, first in the form
$\phi(Z_1) R_{21}\inv Y\inv_2 = Y\inv_2 Z\inv_2 \phi(Z_1) Z_2$ and then
in the form $Z_2 R\p_{12}{}\inv Y\inv_1 = Y\inv_1 Z\inv_1 Z_2 Z_1$, to 
remove two $R$'s from the RHS. The resulting expression,
simplified with the help of (\ref{ZZ}), is
\[
\phi(Z_1) Y\inv_1 Z\inv_1 Y\inv_2 
= Y\inv_2 Z\inv_2 \phi(Z_1) Y\inv_1 Z\inv_1 R_{12} Z_2.
\]
Contracting with $h$ in the first tensor space and using
$h_{(1)} \ot\ldots\ot h_{(4)} 
= h_{(2)}\ot\ldots\ot h_{(4)}\ot\phi^* h_{(1)}$,
which follows from the twisted 
cocommutativity of $h$, we can move $R_{12}$
three places to the left:
\[
\la\phi(Z_1) Y\inv_1 Z\inv_1 Y\inv_2, h \ot \id \ra
= \la Y\inv_2 Z\inv_2 R\p_{12} \phi(Z_1) 
\underline{Y\inv_1 Z\inv_1 Z_2},h \ot \id\ra.
\]
Applying (\ref{YZ}) once more to 
the underlined part and simplifying the
resulting expression
with the help of (\ref{ZZ}) written as $R\p_{12} \phi(Z_1) Z_2 
= Z_2 \phi(Z_1) R\p_{12}$, we finally obtain
$
\la\phi(Z_1) Y\inv_1 Z\inv_1 Y\inv_2, h \ot \id \ra
=\la Y\inv_2 \phi(Z_1) Y\inv_1 Z\inv_1, h \ot \id \ra,
$
\ie\
\eq
[1 \ot \widetilde h, Y\inv ] = 0 .
\en
\vspace{1ex}
Ad 2): 
Now we will derive the explicit expression for $\widetilde h$. 
(Using that expression
it is also possible to prove that $\widetilde h$ is a casimir
in $U$.)
We start with equation (\ref{YZ}), written as
$
Z_2 R\p_{12}{}\inv Y\inv_1 Z\inv_1 = Y\inv_1 Z\inv_1 Z_2,
$
and move the $R$ to the RHS with the help of
its opposite inverse 
$\bar R\p \equiv (S^2 \ot \id)(R\p)$, which satisfies 
$\bar R\p_{12_2} R\p_{12_1}{}\inv = 1 \ot 1$. We find
$
Z_2 Y\inv_1 Z\inv_1 = \bar R\p_{12_2} Y\inv_1 Z\inv_1 Z_{2_1}.
$
Let us now multiply tensor spaces 1 and 2 so that the
two $Z$'s on the RHS cancel
\[
Z_{1_3} Y\inv_{1_1} Z\inv_{1_2} 
= \bar R\p_{1_11_5} Y\inv_{1_2} Z\inv_{1_3} Z_{1_4}
= \bar R\p_{1_11_3} Y\inv_{1_2}.
\]
If we now contract this expression with $h$ in the first tensor space,
we can use the twisted cocommutativity of $h$ in the form
$h_{(1)} \ot h_{(2)} \ot h_{(3)} 
= h_{(2)} \ot h_{(3)} \ot \phi^* h_{(1)}$
to change the order of multiplication in the first tensor space
on both sides of the equation:
\[
\la \phi(Z_{1_1}) Y\inv_{1_2} Z\inv_{1_3},h \ot \id\ra
= \la \bar R_{1_21_1} Y\inv_{1_3},h \ot \id\ra,
\]
\ie\ 
$\la \widetilde T , h \ot \id \ra
  = \la u_1\inv Y_1\inv , h \ot \id \ra$.
This is precisely the expression (\ref{casimir}) 
that we wanted to prove.

We would like to briefly sketch how to prove that $\widetilde h$ is a
casimir starting from (\ref{casimir}):
$h' = \la u\inv \ot \id , \Delta(h)\ra$ is an element of $F$ which is
coinvariant with respect to the twisted adjoint action. 
 (This follows from twisted cocommutativity of $h$
and the fact that $u\inv$ generates $S^2$).
$Y$ on the other hand is a twisted invariant 2-tensor in $U \ot U$.
Being the contraction of $Y$ by $h'$, 
$\widetilde h$ is itself an ad-invariant
element of $U$ and hence a casimir operator: 
$T_2 \la Y\inv_{12} , h' \ot \id\ra
= \la T\p_1{}\inv Y\inv_{12} T_1 , h' \ot \id\ra T_2
= \la Y\inv_{12} , h' \ot \id\ra T_2$ .
\hfill$\mybox$

\subsection{Dynamics in the double}

Now that we have found the image of the Hamiltonian under the
embedding of the quantized phase space $F$ into the Heisenberg double
we can study Heisenberg's equations of motion in the double.
These are
\eq
i \dot\mathcal{O} 
= [\widetilde h , \mathcal{O}], \quad \forall \mathcal{O} \in D_H.
\label{heom}
\en
Time evolution in the $U$-part of $D_H$ 
is trivial (because $\widetilde h$ 
is central
in $U$)
\eq
i \dot x = [\widetilde h , x] = 0, 
\quad \forall x \in U \subset D_H.\label{uconst}
\en
In the $F$-part we find simple linear equations
\eq
i \dot T = [1\ot\widetilde h , T] 
= T (\Delta \widetilde h - 1 \ot \widetilde h) =: T \xi
\label{Tdot}
\en
that are solved by exponentiation because $\xi$ 
is an element of $U\ot U \subset U \ot D_H$ and hence
time-independent, see
(\ref{uconst}),
\eq
T(t) = T(0) e^{-i t \xi}.
\en
Here are some alternative useful expressions for 
$\xi = \Delta \widetilde h - 1 \ot \widetilde h$:
Equation (\ref{Tdot}) slightly rewritten gives
\eq
T_2 \xi_2 = \la\widetilde T_1 T_2 - T_2 \widetilde T_1 , h \ot \id\ra.
\label{txi}
\en
Starting from (\ref{YT})
one can derive
\eq
\xi = \left\la u_1\inv(R\p_{12}{}\inv Y\inv_1 R_{21}\inv - Y\inv_1) 
              , h \ot id\right\ra
\en
and
\eq
\xi = \left\la \left((S\inv \circ \phi \ot \id)(R_{12} R_{21})
      - 1 \right) u\inv_1 Y\inv_1 , h \ot \id\right\ra.
\en	  
We have thus far been able to give the explicit time evolution in the
Heisenberg double. In section~\ref{s:qlax} we will come closer
to the solution to
the original problem---Heisenberg's equations of motion---via
explicit expressions for the evolution of $\widetilde T(t)$.

\subsection{Quantum Lax equation}
\label{s:qlax}

We will now derive an explicit expression for the time evolution 
of $\widetilde T$.
Using the time-independence of 
$Y\inv \in U \ot U \subset U \ot D_H$ we find
\eq
\widetilde T(t) = \phi(Z(t)) \, Y\inv \, Z\inv(t) = 
\phi(Z(t) Z\inv(0)) \, \widetilde T(0) \, Z(0) Z\inv(t). \label{Ttfirst}
\en
If we had started with an alternative $\widetilde T$ expressed
in terms of $\Lambda$ and $Y$ 
we would have found an expression involving $\Lambda$ instead of $Z$.
Such considerations lead to the following proposition:
\begin{prop}
Let \hfill$\widetilde g_+(t) = Z(t) Z(0)\inv$,\hfill
$\widetilde g_-(t) = \Lambda\inv(t) \Lambda(0)$ \hfill and
$\widetilde M^\pm(t) = i \,\dot{\widetilde g}_\pm(t) 
\widetilde g_\pm\inv(t)$.
The time-evolution of $\widetilde T$ is given via conjugation
by
\eq
\widetilde g_\pm(t) = \exp(-it(1 \ot \widetilde h)) 
\exp(it(1 \ot \widetilde h - \widetilde M_\pm(0))) : \label{gpt}
\en
\eq
\widetilde T(t) = \phi(\widetilde g_+(t)) \widetilde T(0)
                  \widetilde g_+(t)\inv 
                = \phi(\widetilde g_-(t)) \widetilde T(0)
                  \widetilde g_-(t)\inv 
\label{Tt}
\en
and Heisenberg's equation of motion can be written in Lax form
\eq
i \,\frac{d}{dt}\widetilde T = \phi(\widetilde M^+) \widetilde T 
- \widetilde T \widetilde M^+
= \phi(\widetilde M^-) \widetilde T 
- \widetilde T \widetilde M^- . \label{mttm}
\en
\end{prop}
\noindent {\sc Proof}: The definition of $\widetilde M^\pm(t)$
can be used to express $\widetilde g_\pm(t)$
in terms of $\widetilde M^\pm(t)$. From (\ref{heom}) we have
\[
i \,\frac{d}{dt}\widetilde g_\pm(t) 
=\widetilde M_\pm(t) \widetilde g_\pm(t) 
= e^{-it(1 \ot \widetilde h)} \widetilde M_\pm(0) e^{i t(1 \ot
\widetilde h)} \widetilde g_\pm(t) ;
\]
this can be integrated
with the initial condition $\widetilde g_\pm(0) = 1$ to give
equation~(\ref{gpt}). 
If we differentiate (\ref{Tt}), we find equation (\ref{mttm}).
What is left to proof is equation (\ref{Tt}).
$
\widetilde T(t) = \phi(\widetilde g_+(t)) \widetilde T(0)
\widetilde g_+(t)\inv 
$
is simply (\ref{Ttfirst})
expressed in terms of $\widetilde g_+(t)$.
The time evolution in $D_H$ is an algebra 
homomorphism and so we can decompose
$T(t)$ = $\Lambda(t) Z(t)$ with $\Lambda(t) \in U_- \ot F_-(t)$ and 
$Z(t) \in U_+ \ot F_+(t)$.
It is now easy to see that 
\[
\phi(\widetilde g_+(t)) \widetilde T(0) \widetilde g_+(t)\inv 
= \phi(\widetilde g_-(t)) \widetilde T(0) \widetilde g_-(t)\inv
\]   
is equivalent to          
\[
\phi(T(t)) Y\inv T\inv(t) = \phi(T(0)) Y\inv T\inv(0),
\] \ie, we need to show
that $\phi(T) Y\inv T\inv$ is time-independent: From (\ref{Tdot})
we get
\[
i \, \frac{d}{dt} \left(\phi(T) Y\inv T\inv\right)
= \phi(T) \left(\phi(\xi)  Y\inv - Y\inv \xi\right) T\inv =0.
\] 
(That this is zero
can be seen from the explicit expression
$\xi = \Delta\widetilde h - 1 \ot \widetilde h$: \quad
$(\phi\ot\id)(\Delta\widetilde h) Y\inv - Y\inv \Delta\widetilde h = 0$ 
because $Y\inv$ is a twisted invariant 2-tensor and 
$[1 \ot \widetilde h , Y\inv] = 0$ because 
$\widetilde h$ is a casimir operator.)
\hfill$\mybox$\\[1ex]
We will now proceed to derive explicit expressions for 
$\widetilde M^\pm$ in terms of $h$. 
We will not use the expressions for $\xi$
but rather work directly with $Z$ and $\Lambda$. 
First we prove the following lemma:
\begin{lemma} 
The following two relations hold in $U \ot U \ot D_H$:
\eq
\Lambda\inv_2 \widetilde T_1 \Lambda_2 
= \widetilde T_1 R_{21}\inv, \label{first}
\en
\eq
Z_2 \widetilde T_1 Z_2\inv = R\p_{12} \widetilde T_1. \label{second}
\en
\end{lemma}
\noindent {\sc Proof}:
We need to use 
\eq
Z_1 Y_1 \Lambda_2 = \Lambda_2 R_{21} Z_1 Y_1 ,
\en
\begin{flushleft}
   which follows with $T = \Lambda Z$ from (\ref{YT}) and (\ref{YZ}).
   We have $\Lambda\inv_2 \widetilde T_1 \Lambda_2 =
   \Lambda\inv_2 \phi(Z_1) 
   \underline{Y_1\inv Z\inv_1 \Lambda_2} 
    = \Lambda\inv_2 \phi(Z_1) \Lambda_2 Y\inv_1 Z\inv_1
   R_{21}\inv = \widetilde T_1 R_{21}\inv$ 
   which proves (\ref{first}).
   Similarly:
   $Z_2 \widetilde T_1 Z_2\inv = Z_2 \phi(Z_1) 
   \underline{Y\inv_1 Z\inv_1 Z\inv_2} =
   \underline{Z_2 \phi(Z_1) R\p_{12}} Z_2\inv Y_1\inv Z_1\inv 
    = R\p_{12} \widetilde T_1$,
   which proves (\ref{second}). \hfill$\mybox$
\end{flushleft}
\begin{prop} 
It holds that
\begin{eqnarray}
\widetilde M^+(t) & = & 1 \ot \widetilde h 
- \la\widetilde T_1(t) R_{12} , h \ot \id\ra\quad\in\quad U_+ \ot
\widetilde F , \\
\widetilde M^-(t) & = & 1 \ot \widetilde h 
- \la \widetilde T_1(t) R\inv_{21}  , h \ot \id\ra \quad\in\quad U_- \ot
\widetilde F.\label{mminus}
\end{eqnarray}
\end{prop}
\begin{flushleft}{\sc Proof}:
   $\widetilde M^+(t) = i \dot Z(t) Z(t)\inv =
   \displaystyle{
      {d \over d t}\left(e^{-i t (1 \ot \widetilde h)}
      Z(0) e^{i t (1 \ot \widetilde h)}\right) Z(t)\inv
   }$
   $=\displaystyle{
        e^{-i t (1 \ot \widetilde h)}
        \left(1 \ot\widetilde h 
        - Z(0)(1 \ot \widetilde h)Z(0)\inv\right)
        e^{i t (1 \ot \widetilde h)}
   }$
   $=\displaystyle{1 \ot \widetilde h - 
   e^{-i t (1 \ot \widetilde h)} 
   \la (Z_2 \widetilde T_1 Z_2\inv)(0),h \ot \id\ra
   e^{i t (1 \ot \widetilde h)}
   }$
   $=1 \ot \widetilde h - \la R\p_{12} \widetilde T_1(t) , h \ot \id\ra$
   $=1 \ot \widetilde h 
   - \la\widetilde T_1(t) R_{12} , h \ot \id\ra$,\\
   where we have used (\ref{second}) and $(\phi^*\ot\id)(\Delta h) =
   \Delta'h$.\\ The completely analogous proof
   of (\ref{mminus}) is based on (\ref{first}). 
   \hfill$\mybox$
\end{flushleft}
Just like $T$ factorizes as $T(t) = \Lambda(t) Z(t)$, we shall think of
$\widetilde g_\pm \in U_\pm \ot D_H$ as factors
of an element of $U \ot D_H$:
\eq
\widetilde g(t) = \widetilde g_-\inv(t) \widetilde g_+(t) .
\en
\begin{prop}
$\widetilde g(t) \equiv 
\widetilde g_-\inv(t) \widetilde g_+(t)$ and its factors $g_-(t)$
and $g_+(t)$ are in fact elements of 
$U \ot \widetilde F \subset U \ot D_H$ as
is apparent from the following expression:
\eq
\widetilde g(t) = Z(0) \exp(-i t \xi) Z\inv(0) = \exp(-i t \widetilde M(0)),
\en
where
$
\widetilde M \equiv \widetilde M_+ - \widetilde M_-
= \la \widetilde T_1 (R\inv_{21} - R_{12}), h \ot \id\ra
\in U \ot \widetilde F.
$
\end{prop}
\begin{flushleft}{\sc Proof: } $\widetilde g(t) = 
\widetilde g\inv_-(t) \widetilde g_+(t) = \Lambda\inv(0) T(t) Z\inv(0)
= Z(0) \exp(-i t \xi) Z\inv(0)$.
{}From equation (\ref{txi}) and 
$T = \Lambda Z$:
$Z \xi Z\inv = \Lambda\inv_2 \la \widetilde T_1 T_2 
- T_2 \widetilde T_1 , h \ot \id \ra
Z_2\inv = \la \Lambda\inv_2 \widetilde T_1 \Lambda_2 
- Z_2 \widetilde T_1 Z_2\inv , h \ot \id\ra
= \la \widetilde T_1 R\inv_{21} - R\p_{12} \widetilde T_1 , h \ot \id\ra = 
\widetilde M_+ - \widetilde M_-$ and hence 
$Z \exp(-i t \xi) Z\inv
= \exp(-i t (\widetilde M_+ - \widetilde M_-)) 
= \exp(-i t \widetilde M)$.
\hfill$\mybox$
\end{flushleft}
So far we have learned a great deal about the 
equations of motion in the Heisenberg double and their solution. 
We are now ready to go back to our original
problem, \ie\ the formulation of the equations of motion in the quantized
phase space $F$ in terms of quantum Lax equations and their solution by
factorization, thus generalizing what has become known as the ``Main Theorem''
to the realm of quantum mechanics. Let us mention that the Lax equations 
presented in this section formalize and generalize the concrete examples 
known for particular integrable models \cite{IK}, \cite{Skly}, \cite{KS}, 
\cite{Maillet}, \cite{Zhang}, \cite{SoWa}.

\subsection{Solution by factorization}

Using the fact that the embedding via $\widetilde T$ of $F$
into $D_H$ is
an algebra homomorphism we can drop all the $\widetilde{~~}$'s in the
previous section, thereby projecting the solution of the time evolution of
$\widetilde F$ back to $F$. The result can be summarized in a quantum
mechanical analog of theorem~\ref{main}:

\begin{thm}[main quantum theorem] \label{mainquantum}
\mbox{ }
\begin{enumerate}
\item[(i)] The set of twisted cocommutative functions
$I^{\phi}$ is a commutative subalgebra of $F$. 
\item[(ii)] The equations of motion defined by Hamiltonians
$h\in I^{\phi}$ are of the Lax form
\eq 
i\frac{dT}{dt} = \phi(M^{\pm}) T -T M^{\pm}, \label{qlax}
\en
with 
$\;M^\pm  = \la T_1 (1 - R^\pm)_{21},h \ot\id \ra \in U_\pm \ot F\;$,
 $R^+ \equiv R_{21}$, $R^- \equiv R\inv_{12}$
and $T \in U \ot F$.
\item[(iii)] Let $g_{\pm}(t) \in U_\pm \ot F$ be the solutions to the 
factorization problem
\eq
g_-\inv(t) g_+(t) =  \exp(-i t M(0))\;\in\; U \ot F , \label{facprob}
\en 
where $M(0) = M^+ - M^-$, then
\eq
T(t)=\phi(g_{\pm}(t)) T(0) g_{\pm}(t)^{-1} \label{qsoln}
\en
solves the Lax equation {\rm (\ref{qlax})}; $\:g_\pm(t)$ are given by 
\eq
g_\pm(t) = \exp(-it(1\ot h))\,\exp(it(1\ot h - M^\pm(0))
\en
and are the solutions to the differential equation
\eq
i  \frac{d}{dt}g_\pm(t) =
M^\pm(t) g_\pm(t), \qquad  g_\pm(0) = 1 . \label{gpm}
\en
\end{enumerate}
\end{thm}
This theorem follows from the geometric construction given in the previous
sections, but we shall also present a direct proof:
\begin{flushleft}{\sc Proof:}\\
(i) Let $f,g \in I\p \subset F$, then $f g \in I\p$. 
Using (\ref{RTT}), (\ref{phiDelta}) 
and (\ref{phiR})
we can show that $f$ and $g$ commute:
\begin{quote}
$f g = \la T_1 T_2, f \ot g\ra
= \la R\inv_{12} T_2 T_1 R_{12}, f \ot g\ra
= \la R\inv_{13} T_3 T_1 R_{24}, \Delta f \ot \Delta g\ra
= \la R\inv_{13} T_3 T_1 \:(\phi \ot \phi)(R_{24}), 
\Delta' f \ot \Delta' g\ra
= \la R\inv_{13} T_3 T_1 R_{24}, 
\Delta' f \ot \Delta' g\ra
= \la R_{12} R\inv_{12} T_2 T_1, f \ot g\ra = g f$ .
\end{quote}
(ii) From $R^\pm_{21} T_1 T_2 = T_2 T_1 R^\pm_{21}$, (\ref{phiDelta}) 
and (\ref{phiR})  it follows that $\la
T_1\:(\id\ot\phi)(R^\pm_{21}) T_2, h\ot\id\ra 
= \la T_2 T_1 R^\pm_{21} , h\ot\id\ra$
and as a consequence  all terms that contain
$R^\pm_{21}$ cancel on the RHS of 
equation (\ref{qlax}); we are left with
$\phi(M^\pm) T - T M^\pm = \la T_1 T_2 - T_2 T_1 , h \ot \id\ra 
= [ h , T ] = i dT/dt$.\\[1ex]
(iii) The Lax equation (\ref{qlax}) 
follows immediately from (\ref{qsoln}) and
(\ref{gpm}).\\ 
The rest can be proven in three steps:\\ 
a) Let $m_\pm = 1 \ot h - M_\pm$; \quad
$g_\pm(t) = e^{-i t (1\ot h)} e^{i t m_\pm}$ 
are elements of $U_\pm\ot F$ and 
are the solutions to (\ref{gpm}) as can be checked by differentiation: 
$i \frac{d}{dt} g_\pm(t) = 
e^{-i t (1\ot h)}(1\ot h -  m_\pm) e^{i t m_\pm}
= M_\pm(t) e^{i t (1 \ot h)} e^{i t m_\pm} = M_\pm(t) g_\pm(t)$, 
and $g_\pm(0) = 1$\\[1mm]
b) $m_+ = \la T_1 R_{12},h\ot\id\ra$ and 
$m_- = \la T_1 R\inv_{21} , h \ot id\ra$ commute:\\
Using (\ref{RTT}), (\ref{YBE}), and
(\ref{phiDelta}) we find
\begin{quote}
$m_+ m_- = \la T_1 R_{13} T_2 R\inv_{32}, h \ot h \ot \id\ra
= \la R\inv_{12} T_2 T_1 R_{12} R_{13} R\inv_{32} , h \ot h \ot \id\ra
= \la R\inv_{12} T_2 R\inv_{32} T_1 R_{13} R_{12} , h \ot h \ot \id\ra
= \la T_2 R\inv_{32} T_1 R_{13} R_{12} R\inv_{12} , h \ot h \ot \id\ra
= m_- m_+$
\end{quote}  
{}From a) and b) follows\\[1mm] 
c) $g_-\inv(t) g_+(t) =  e^{-i t m_-} e^{i t m_+} = e^{i t (m_+ - m_-)}
= e^{- i t M(0)}$, \ie\ the $g_\pm(t)$ of (\ref{gpm}) solve the
factorization problem (\ref{facprob}). \hfill $\mybox$
\end{flushleft}
{\em Remark: } If we replace $R^\pm$ in the definition of $M^\pm$ in the
previous theorem by $(L^\mp)^{-1}$, then the $\phi$'s in (\ref{qlax}) 
and (\ref{qsoln}) will not appear explicitly anymore.

\subsubsection*{Dressing transformations}

We have found two (identical) solutions for the time-evolution
in $F$:
\[
f(t) = \la T(t), f(0) \ot \id\ra , \quad f(0) \in F
\]
with $T(t)$ given in (\ref{qsoln}).
Let us verify that
\[
\phi(g_+(t)) T(0) g_+(t)\inv = \phi(g_-(t)) T(0) g_-(t)\inv .
\]
Let  $g(t) = g_-(t)\inv g_+(t) = \exp(-it M(0))$;  we have to show
that
\eq
\phi(g(t)) = T(0) g(t) T(0)\inv  \label{phigt}
\en
which is implied by:
\begin{flushleft}
\begin{quote}
$T M T\inv = \la T_2 T_1 (R_{21}\inv - R_{12}) T_2\inv , h \ot \id\ra
= \la (R_{21}\inv - R_{12}) T_1 , h \ot \id\ra
= \la T_1 (\id \ot \phi)(R_{21}\inv - R_{12}) , h \ot \id\ra =  \phi(M) .$
\end{quote}
\end{flushleft}
With (\ref{phigt}) we can re-express (\ref{qsoln}) as
\eq
T(t) = \Big( T(0) g(t) T(0)\inv \Big)_\pm T(0) \:\Big(g(t)\Big)_\pm\inv 
\en
and thus find that the time-evolution in $F$ has the form of a dressing
transformation.
More precisely 
we can identify elements of $F$ 
with elements of $U \ot F$ (via the factorization map):
\eq
e^{i t h} 
\:\mapsto \: e^{i t m_\pm} = R^\pm \cdot e^{i t (1 \ot h)}\cdot (R^\pm)\inv 
\:\mapsto \: g = e^{-i t m_-} e^{i t m_+} 
\en
and hence have a map
\eq
F \ni e^{ith}: \: U \ot F \rightarrow U \ot F : \;\; T(0) \mapsto T(t).
\en
Let us choose the same $\pm$-conventions for
$T = \Lambda Z$ and $Y\inv = L^- (L^+)\inv$ 
as we did for $g = g_-\inv g_+$, \ie\ 
$(T)_- = \Lambda\inv$, $(T)_+ = Z$,  \emph{etc}.
 We can then write the
embedding of section~\ref{s:embed} in a way that parallels the classical theory:
\begin{eqnarray}
\lefteqn{\phi(T L^+)_- \cdot  (\phi(T) L^-) \cdot (T L^+)_+{}\inv }\nonumber\\ 
&& = \phi\left((\Lambda\inv)\inv Z L^+\right)_- 
      \cdot  (\phi(T) L^-) \cdot  \left((\Lambda\inv)\inv Z L^+\right)_+ \nonumber\\
&& = \phi(\Lambda\inv) \cdot  (\phi(T) L^-) \cdot   (L^+)\inv Z\inv\nonumber\\
&& = \phi(Z) Y\inv Z\inv .
\end{eqnarray}
(``$\pm$'' refers to the first tensor space.)\\
\emph{Remark 1: } Note that the multiplication
``$\cdot$'' is
taken in $U \ot D_H$ rather than $U \ot (F \ot U)$; this and the 
form in which the dressing transformations appear here are
somewhat non-standard.\\  
\emph{Remark 2: } The formal factorization problem in the case of $U$ 
being factorizable \cite{ReshSTS} remains the same as in the untwisted case.
See Appendix 2 in reference \cite{JSch2}.\\
\emph{Remark 3: }
Also in the quantum case we can consider Lax equations corresponding 
to different twisted cocommutative Hamiltonians simultaneously. 
The equations 
(\ref{1a}), (\ref{1b}), (\ref{1c}) and (\ref{1d}) are still valid.

\section{Face algebras and Ruijsenaars models}

Face Hopf algebras \cite{Hay} have been found to be the
algebraic structure that 
underlies some particularly interesting
integrable models of statistical physics. They generalize Hopf algebras
and quantum groups. Another closely related
generalization of quantum groups, the
so-called elliptic quantum groups, were introduced by Felder \cite{Fel}
in the context of IRF (face) models
\cite{Bax}. 
Face Hopf algebras and elliptic quantum groups
play the same role for face models as quantum groups do for vertex models.
The R-matrices are now replaced by dynamical R-matrices,
which first appeared in the context of Liouville string field theory
\cite{Ger}; they can
be understood as a reformulation of the Boltzmann weights in
Baxters solutions of the face-type Yang-Baxter equation.
A partial classification of the dynamical R-matrices is given in 
\cite{EV}.
Recently another type of integrable quantum systems, known as Ruijsenaars
models \cite{Rui}, 
and their various limiting cases have been shown to be connected to
quantum groups and elliptic quantum groups
\cite{Has,FV,ABB,ACF,EK}
through different approaches.

The Calogero-Moser-Sutherland class of integrable 
models describe the motion of
particles on a one-dimensional line or circle interacting
via pairwise potentials that are given by 
Weierstrass elliptic functions and their various degenerations.
The simplest case is an inverse $r^2$ potential.
The Ruijsenaars-Schneider model is a relativistic generalization, whose
quantum mechanical version, the Ruijsenaars model, is the model that
we are interested in here.

The Hamiltonian of the Ruijsenaars model 
for two particles with coordinates $x_1$ and $x_2$ has the form
\[
\ha = \left\{\frac{\theta(\frac{c \eta}{2} - \lambda)}{\theta(-\lambda)}
t^{(\lambda)}_1
+ \frac{\theta(\frac{c \eta}{2} + \lambda)}{\theta(\lambda)} 
t^{(\lambda)}_2\right\} , 
\]
where $\lambda = x_1 - x_2$. Here $c \in \C$ is the coupling 
constant, $\eta$ is the
relativistic deformation parameter, the $\theta$-function is 
given in (\ref{theta}); 
we have set $\hbar =1$. The Hamiltonian acts  on a wave function as
\[
\ha \,\psi(\lambda) = 
\frac{\theta(\frac{c \eta}{2} 
- \lambda)}{\theta(-\lambda)}\psi(\lambda - \eta) 
+ \frac{\theta(\frac{c \eta}{2} 
+ \lambda)}{\theta(\lambda)}\psi(\lambda + \eta),
\]
the $t_i^{(\lambda)}$ that appear in the Hamiltonian are hence
shift-operators in the variable 
$\lambda$; in the present case of
two particles they generate a one-dimensional
graph:
\begin{center}
\unitlength 0.50mm
\linethickness{0.4pt}
\thicklines
\begin{picture}(164.00,20.00)
\put(40.00,10.00){\circle*{2.00}}
\put(100.00,10.00){\circle*{2.00}}
\put(73.00,10.00){\vector(1,0){24.00}}
\put(67.00,10.00){\vector(-1,0){24.00}}
\put(70.00,7.00){\makebox(0,0)[ct]{$\lambda$}}
\put(84.00,11.00){\makebox(0,0)[cb]{$t_1$}}
\put(55.00,11.00){\makebox(0,0)[cb]{$t_2$}}
\put(10.00,10.00){\circle*{2.00}}
\put(130.00,10.00){\circle*{2.00}}
\put(70.00,10.00){\circle*{2.00}}
\put(70.00,10.00){\makebox(0,0)[cc]{$\times$}}
\thinlines
\put(0.00,10.00){\line(1,0){140.00}}
\put(142.00,10.00){\line(1,0){2.00}}
\put(146.00,10.00){\line(1,0){2.00}}
\put(150.00,10.00){\line(1,0){2.00}}
\put(-12.00,10.00){\line(1,0){2.00}}
\put(-8.00,10.00){\line(1,0){2.00}}
\put(-4.00,10.00){\line(1,0){2.00}}
\end{picture}
\end{center}
Relative to a fixed vertex $\lambda$ the
vertices of this graph 
are at points $\eta\cdot\Z \in \RE$, the (ordered) paths 
connect neighboring vertices and are hence
intervals in $\RE$. 
There are two paths per interval,
one in positive, one in negative direction.

Now consider a $N$-particle Ruijsenaars model: 
The graph relative to a fixed $\lambda$ is
then a $N -1$ dimensional hyper-cubic lattice. 
Relative to
$\lambda \in \RE^{N-1}$ the vertices are at points 
$\left(\eta \cdot \Z\right)^{\times (N-1)}$.
This picture can obviously be generalized and in the following we
shall consider an arbitrary ordered graph.

\paragraph{\it Remark:} For the Ruijsenaars system 
$\lambda$ can take any value in 
$\RE^{N-1}$, so we
are a priori let to a huge graph that 
consists of a continuous
family of disconnected graphs. Since the graphs are 
disconnected it is
sufficient to consider one.
Later we will in fact 
write all expressions with respect to one fixed vertex 
$\lambda$. This
will lead to equations containing explicit 
shifts on the graph and thus to dynamical $R$-matrices.
Apart from the physical interpretation there is a priori no reason
to restrict the values of particle-coordinates  at the level 
of the shift operator to $\RE$. We can and shall everywhere
in the following take $\C$ instead of $\RE$. 

\subsection{Face Hopf algebras}
 
The Hilbert space of the model 
will be build from vector spaces on paths of fixed length.
As we shall argue in the following, the operators
on this Hilbert space can be chosen to be elements of
a Face Algebra $F$ \cite{Hay} (or weak $C^*$-Hopf
algebras \cite{Vol,Boe}, which is essentially the same with a $*$-structure).

There are two commuting projection operators $e^i, e_i \in F$ 
for each
vertex of the graph (projectors 
onto bra's and ket's corresponding to vertex $i$):
\eq
e_i e_j = \delta_{ij} e_i, \quad e^i e^j = \delta_{ij} e^i, \quad
{\textstyle\sum} e_i = {\textstyle\sum} e^i = 1 .
\en
$F$ shall  be equipped with a coalgebra structure such that the
combination $e^i_j \equiv e^i e_j = e_j e^i$ is a
corepresentation:
\eq
\Delta(e^i_j) = {\textstyle\sum}_k e^i_k \ot e^k_j,
\quad \epsilon(e^i_j) = \delta_{ij}.
\en
It follows that $\Delta(1) = \sum_k e_k \ot e^k \neq 1 \ot 1$ 
(unless the graph has only a single vertex) 
-- this is a key feature of face algebras (and weak 
$C^*$-Hopf
algebras).

\begin{figure}[htb]
\begin{center}
\unitlength 1.00mm
\linethickness{0.4pt}
\begin{picture}(40.00,40.00)
\thinlines
\put(0.00,0.00){\line(0,1){40.00}}
\put(0.00,40.00){\line(1,0){40.00}}
\put(40.00,40.00){\line(0,-1){40.00}}
\put(40.00,0.00){\line(-1,0){40.00}}
\put(0.00,10.00){\line(1,0){40.00}}
\put(40.00,20.00){\line(-1,0){40.00}}
\put(0.00,30.00){\line(1,0){40.00}}
\put(10.00,40.00){\line(0,-1){40.00}}
\put(20.00,40.00){\line(0,-1){40.00}}
\put(30.00,40.00){\line(0,-1){40.00}}
\thicklines \color{green}
\put(0.00,40.00){\line(0,-1){30.00}}
\put(10.00,0.00){\line(1,0){10.00}}
\put(20.00,-0.30){\line(0,1){10.30}}
\put(10.00,30.00){\line(1,0){10.30}}
\put(20.00,30.00){\line(0,-1){10.00}}
\put(20.00,20.00){\line(1,0){10.00}}
\put(30.00,40.00){\line(0,-1){10.00}}
\put(30.00,30.00){\line(1,0){10.00}}
\put(40.00,30.00){\line(0,1){10.00}}
\put(10.00,0.00){\line(0,1){0.50}}
\put(10.00,0.60){\line(1,0){9.60}}
\put(0.20,40.00){\line(0,-1){30.00}}
\put(10.00,-0.20){\line(1,0){10.00}}
\put(20.20,-0.30){\line(0,1){10.30}}
\put(10.00,30.20){\line(1,0){10.30}}
\put(20.20,30.00){\line(0,-1){10.00}}
\put(20.00,20.20){\line(1,0){10.00}}
\put(30.20,40.00){\line(0,-1){10.00}}
\put(30.00,30.20){\line(1,0){10.00}}
\put(39.80,30.00){\line(0,1){10.00}}
\put(10.20,-0.30){\line(0,1){1.20}}
\put(10.00,0.80){\line(1,0){9.60}}
\end{picture}
\end{center}
\caption{Some paths of length 3 on a graph that is
a square lattice, \eg\ on part of a graph corresponding to a 
particular Ruijsenaars system.}
\label{fig:box}
\end{figure}
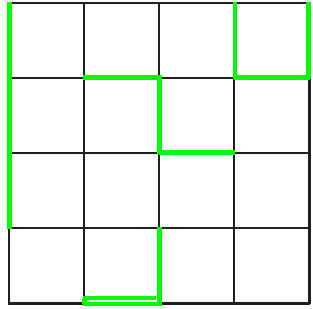

So far we have considered matrices with indices that are vertices,
\ie\ paths of length zero. In the given setting it is natural to also
allow paths of fixed length on a finite
oriented graph $\cal G$
as matrix indices. This is illustrated in
figure~\ref{fig:box}. 
We shall use capital letters 
to label paths. A path $P$ has an origin (source) $\cdot P$,
an end (range) $P \cdot$ and a length 
$\no P$. 
%
%
Two paths $Q$, $P$ can be concatenated to form a new path
$Q \cdot P$, if the end of the first path coincides with the start of
the second path, \ie\ if 
$Q \cdot = \cdot P$ (this explains our choice
of notation).

The important point is, that the symbols
$T^A_B$, where $\no A = \no B \geq 0$,
with relations
\eq
\Delta\left( T^A_B \right) = \sum_{A'} T^A_{A'} \ot T^{A'}_B  \label{delta}
\quad (\no A = \no A' = \no B) 
\en
\eq
\epsilon( T^A_B ) = \delta_{A B} \label{eps}
\en
\eq
T^A_B T^C_D = \delta_{A\cdot,\cdot C} \delta_{B\cdot,\cdot D} T^{A \cdot
C}_{B \cdot D} \label{tt}
\en
span an object that obeys the axioms of a face algebra.
Relations (\ref{delta}) and (\ref{eps}) make $T^A_B$ a
corepresentation; (\ref{tt}) is the rule for combining
representations.
The axioms of a face algebra can be found in \cite{Hay,Hay1}.
\paragraph{\it Pictorial representation:}
$$T^A_B \,\sim\;
\unitlength1mm
\begin{picture}(10,10)(0,4)\small
\put(0,0){\line(1,0){10}}
\put(0,10){\line(1,0){10}}
\multiput(0,0)(0,1){10}{\line(0,1){0.5}}
\multiput(10,0)(0,1){10}{\line(0,1){0.5}}
\put(0,5){\vector(0,1){1}}
\put(10,5){\vector(0,1){1}}
\put(5,0){\vector(1,0){1}}
\put(5,10){\vector(1,0){1}}
\put(5,-1){\makebox(0,0)[ct]{$B$}}
\put(5,11){\makebox(0,0)[cb]{$A$}}
\end{picture}
\quad
T^A_B T^C_D \,\sim\;
\begin{picture}(20,10)(0,4)\small
\put(0,0){\line(1,0){10}}
\put(0,10){\line(1,0){10}}
\multiput(0,0)(0,1){10}{\line(0,1){0.5}}
\multiput(10,0)(0,1){10}{\line(0,1){0.5}}
\put(0,5){\vector(0,1){1}}
\put(10,5){\vector(0,1){1}}
\put(5,0){\vector(1,0){1}}
\put(5,10){\vector(1,0){1}}
\put(5,-1){\makebox(0,0)[ct]{$B$}}
\put(5,11){\makebox(0,0)[cb]{$A$}}
\put(10,0){\line(1,0){10}}
\put(10,10){\line(1,0){10}}
\multiput(20,0)(0,1){10}{\line(0,1){0.5}}
\put(20,5){\vector(0,1){1}}
\put(15,0){\vector(1,0){1}}
\put(15,10){\vector(1,0){1}}
\put(15,-1){\makebox(0,0)[ct]{$D$}}
\put(15,11){\makebox(0,0)[cb]{$C$}}
\end{picture}\quad
\Delta T^A_B = \sum_{A'} T^A_{A'} \otimes T^{A'}_B \,\sim\;
\begin{picture}(15,10)(0,9)\small
\put(0,20){\line(1,0){10}}
\multiput(0,10)(0,1){10}{\line(0,1){0.5}}
\multiput(10,10)(0,1){10}{\line(0,1){0.5}}
\put(0,15){\vector(0,1){1}}
\put(10,15){\vector(0,1){1}}
\put(5,20){\vector(1,0){1}}
\put(5,21){\makebox(0,0)[cb]{$A$}}
\put(0,0){\line(1,0){10}}
\put(0,10){\line(1,0){10}}
\multiput(0,0)(0,1){10}{\line(0,1){0.5}}
\multiput(10,0)(0,1){10}{\line(0,1){0.5}}
\put(0,5){\vector(0,1){1}}
\put(10,5){\vector(0,1){1}}
\put(5,0){\vector(1,0){1}}
\put(5,10){\vector(1,0){1}}
\put(5,-1){\makebox(0,0)[ct]{$B$}}
\end{picture}\vspace{5mm}
$$
The dashed paths indicate the $F$-space(s), their orientation is from lower to
upper ``index''. Inner paths are summed over.

Often it is convenient not to consider a particular representation, \ie\
$T$-matrices corresponding to paths of a given fixed length but rather an
abstract, universal $T$.
Heuristically one can think of the universal $T$ as an abstract matrix or
group element, but it does in fact simply provide an alternative to the
usual Hopf algebra notation:
$T$, $T_1 \ot T_1$, $T_1 T_2$, \eg\ correspond to the identity map,
coproduct map and 
multiplication map of $F$ respectively. (More details are given in the
appendix.)
Keeping this in mind we shall---without loss of generality---nethertheless 
use a formal
notation that treats $T$ as if it was the
canonical element $T_{12}$ of $U \ot F$, 
where $U$ is the dual of $F$ via the
pairing $\la \: , \: \ra$.\footnote{Here 
and in the following we 
shall frequently suppress the
the second index of $T$; it corresponds to the $F$-space.
The displayed expressions are
short for $\la T_{12},f\ot \id\ra = f \in F$,
$\la T_{12} T_{13}, f \ot \id \ot \id\ra 
= \Delta\, f \in F \ot F$
and $\la T_{13} T_{23}, f\ot g\ot \id\ra = f g \in F$.}
\[ 
\la T , f \ra = f, \quad \la T_1 \ot T_1, f \ra = \Delta\,f, \quad
\la T_1 T_2 , f \ot g \ra = f g ; \quad f , g \in F
\]

A face \emph{Hopf} algebra has an anti-algebra and anti-coalgebra endomorphism,
called the antipode and denoted by $S$ or---in the universal tensor
formalism---by $\tilde T$: $\la \tilde T, f\ra = S(f)$. The antipode
satisfies some compatibility conditions with the
coproduct that are given in the appendix.
\paragraph{\it Remark:}
In the limit of a graph with a single vertex a face Hopf algebra
is the same as a Hopf algebra. Ordinary matrix indices correspond
to closed loops in that case.

\subsubsection{Universal-$T$ formalism}

For many reasons it is very convenient to use a formalism based
on the so-called universal $T$. The expressions formally resemble those in
a matrix representation but give nevertheless general face Hopf algebra 
statements. This greatly simplifies notation but also
the interpretation and application of the resulting expressions.
When we are dealing with quantizations of the functions on a group
we need to keep track both of the non-commutativity of the
quantized functions and the residue of the underlying group structure.
Both these structures can be encoded in algebraic relations for the
universal $T$ which easily allows to control
two non-commutative structures. 
In fact $T$ can be regarded as a universal group element.
Universal tensor expressions can \emph{formally} be read in two ways:
Either as ``group'' operations (or rather operations in $U$)
or as the corresponding pull-back maps in the
dual space. This simplifies the heuristics of ``dualizing and reversing
arrows'' and allows us to keep track of the classical limit.
Example: Multiplication in $U$, $x \ot y  \mapsto x y$: The corresponding
pull-back map in the dual space $F$ is the coproduct $\Delta$. Both operations
are summarized in the same universal expression $T_{12} T_{13}$.

Sometimes $T$ can be realized as the canonical element  $U \ot F$, but we
do not need to limit ourselves to these cases an will instead define $T$
as the identity map from $F$ to itself and will
use the same symbol for the identity map $U \rightarrow U$. 

For the application of this identity
map to an element of $F$ we shall nevertheless use the same bracket
notation as we would for a true
canonical element in the finite case, \emph{i.e.} 
$f \equiv \la \id , f \ra = \la T_{12} , f \ot \id\ra$.
This notation is very convenient---like inserting the unity in quantum
mechanics.
The identity map on a product is given by $a b \mapsto a \cdot b$ or
\[ \la T_{12} , a b \ot \id\ra = \la T_{13} T_{23}, a \ot b \ot \id\ra 
= a \cdot b\]
We shall write
$\Delta_1 T_{12} = T_{13} T_{23}$
to express this fact---this is hence the uni\-ver\-sal-T notation for the
multiplication map in $F$ (and the coproduct map in $U$).
The 
coproduct map in $F$ (multiplication map in $U$) is
$T_{12} T_{13}$:
\[f \mapsto \la T_{12} T_{13} , f \ot \id \ot \id\ra = \Delta(f)\]
The antipode map is $\tilde T$: $f \mapsto S(f)$, the
contraction map $x \ot 1$  with $x \in U = F^*$ maps $f \in F$ to
$1 \la x , f \ra$, the
counit map is $1 \ot 1$, \emph{etc.}. 
For face algebras the counit is not an algebra
homomorphism but rather $\Delta 1_{F} = \sum_k e_k \ot e^k$ and 
$\Delta 1_{U} = \sum_k E_k \ot E^k$,
where $\la E_k , f \ra = \epsilon(f e_k)$ and
$\la E^k , f \ra = \epsilon(e^k f)$. Therefore
$\epsilon(a b) = \sum_k \epsilon(a e_k) \epsilon(e^k b)$.
This is one of the face algebra axioms. 
It differs from the corresponding ordinary Hopf algebra axiom
(there: $\epsilon(a b) = \epsilon(a) \epsilon(b)$).
Some important relations involving the coproduct and antipode maps are
\[ \tilde T T = \sum_i E_i \ot e_i, 
\quad T \tilde T = \sum_i E^i \ot e^i,
\quad \tilde T T \tilde T = \tilde T ,
\quad T \tilde T T = T .\]
In the pictorial representation this relations imply that the vertices of the
paths in $U$ and $F$ that form $T$ match to give a closed ``square''.
They also give us two ways to fix the four vertices of $T$;
\[ (1 \ot e^i_k) T (1 \ot e^j_l) \; = \; (E^i_j \ot 1) T (E^k_l \ot 1)
\quad\sim\;
\unitlength1mm
\begin{picture}(15,6)(-12,4)\scriptsize
\put(0,0){\line(1,0){10}}
\put(0,10){\line(1,0){10}}
\multiput(0,0)(0,1){10}{\line(0,1){0.5}}
\multiput(10,0)(0,1){10}{\line(0,1){0.5}}
\put(0,5){\vector(0,1){1}}
\put(10,5){\vector(0,1){1}}
\put(5,0){\vector(1,0){1}}
\put(5,10){\vector(1,0){1}}
\put(0,0){\makebox(0,0)[cc]{$\times$}}
\put(10,0){\makebox(0,0)[cc]{$\times$}}
\put(0,10){\makebox(0,0)[cc]{$\times$}}
\put(10,10){\makebox(0,0)[cc]{$\times$}}
\put(-1,0){\makebox(0,0)[rt]{$k$}}
\put(11,0){\makebox(0,0)[lt]{$l$}}
\put(-1,10){\makebox(0,0)[rb]{$i$}}
\put(11,10){\makebox(0,0)[lb]{$j$}}
\end{picture}\vspace{4mm}
\]
Further useful relations can be obtained from this by summing over some
of the indices and using 
$\sum_i e^i_j = e_j$, $\sum_j e_j =1$, \emph{etc.} .

\subsection{Boltzmann weights}

By dualization we can describe a
coquasitriangular structure of $F$ by giving
a quasitriangular structure  for $U$.
The axioms \cite{Hay} for a quasitriangular face algebra are similar to
those of a quasitriangular Hopf algebra; there is a universal
$R \in U \ot U$ that controls the non-cocommutativity of the coproduct in $U$
and the non-commutativity of the
product in $F$,
\eq
R T_1 T_2 = T_2 T_1 R , \quad \tilde R T_2 T_1 = T_1 T_2 \tilde R,
\quad \tilde R \equiv (S \ot \id)(R),
\label{rtt}
\en
however the antipode of $R$ is
no longer inverse of $R$ but rather 
\eq 
\tilde R R = \Delta(1) , \quad R \tilde R = \Delta'(1) .
\en
The numerical 
``$R$-matrix'' obtained by contracting $R$ with two face corepresentations
is given by the face Boltzmann weight $W$:
\[ 
\la \R , T^A_B \ot T^C_D \ra \; = \; R^{A C}_{B D}  
\;\equiv\; W\Big( {C {B \atop A} D} \Big)
\quad\sim\qquad
\unitlength1mm
\begin{picture}(10,7)(-2,4)\small
\put(0,0){\line(1,0){10}}
\put(0,0){\line(0,1){10}}
\put(10,10){\line(0,-1){10}}
\put(10,10){\line(-1,0){10}}
\put(0,5){\vector(0,-1){1}}
\put(10,5){\vector(0,-1){1}}
\put(5,0){\vector(1,0){1}}
\put(5,10){\vector(1,0){1}}
\put(5,-1){\makebox(0,0)[ct]{$A$}}
\put(5,11){\makebox(0,0)[cb]{$B$}}
\put(-1,5){\makebox(0,0)[rc]{$C$}}
\put(11,5){\makebox(0,0)[lc]{$D$}}
\end{picture}\]\vspace{4mm}
The pictorial representation makes sense since 
the Boltzmann weight is zero unless $C\cdot A$ and $B\cdot D$
are valid paths with common source and range
as will be discussed in more detail
below.
Also note that $T^A_B \mapsto \la \R , T^A_B \ot T^C_D \ra$ is a
\emph{representation} of the matrix elements of $(T^A_B)$ while
$T^C_D \mapsto \la \R , T^A_B \ot T^C_D \ra$ is an \emph{anti-representation}.
Consistent with our 
pictorial representation for the $T$-matrices
we see that the orientation of the
paths in $F$-space remain the same for the first case but are reversed for the
latter:
\[\unitlength1mm
\begin{picture}(10,8)(0,4)\small
\put(0,0){\line(1,0){10}}
\put(0,10){\line(1,0){10}}
\multiput(0,0)(0,1){10}{\line(0,1){0.5}}
\multiput(10,0)(0,1){10}{\line(0,1){0.5}}
\put(0,5){\vector(0,1){1}}
\put(10,5){\vector(0,1){1}}
\put(5,0){\vector(1,0){1}}
\put(5,10){\vector(1,0){1}}
\put(5,-1){\makebox(0,0)[ct]{$B$}}
\put(5,11){\makebox(0,0)[cb]{$A$}}
\end{picture}\; \sim \;
\begin{picture}(10,10)(0,4)\small
\put(0,0){\line(1,0){10}}
\put(0,10){\line(1,0){10}}
\multiput(0,0)(0,1){10}{\line(0,1){0.5}}
\multiput(10,0)(0,1){10}{\line(0,1){0.5}}
\put(0,5){\vector(0,-1){1}}
\put(10,5){\vector(0,-1){1}}
\put(5,0){\vector(1,0){1}}
\put(5,10){\vector(1,0){1}}
\put(5,-1){\makebox(0,0)[ct]{$A$}}
\put(5,11){\makebox(0,0)[cb]{$B$}}
\end{picture}\quad
\stackrel{\it rep.}{\longrightarrow}\quad
\begin{picture}(10,10)(-2,4)\small
\put(0,0){\line(1,0){10}}
\put(0,0){\line(0,1){10}}
\put(10,10){\line(0,-1){10}}
\put(10,10){\line(-1,0){10}}
\put(0,5){\vector(0,-1){1}}
\put(10,5){\vector(0,-1){1}}
\put(5,0){\vector(1,0){1}}
\put(5,10){\vector(1,0){1}}
\put(5,-1){\makebox(0,0)[ct]{$A$}}
\put(5,11){\makebox(0,0)[cb]{$B$}}
\put(-1,5){\makebox(0,0)[rc]{$C$}}
\put(11,5){\makebox(0,0)[lc]{$D$}}
\end{picture}\qquad
\stackrel{\it anti-rep.}{\longleftarrow}\quad
\begin{picture}(10,10)(-2,4)\small
\put(0,0){\line(0,1){10}}
\put(10,10){\line(0,-1){10}}
\multiput(10,10)(-1,0){10}{\line(-1,0){0.5}}
\multiput(10,0)(-1,0){10}{\line(-1,0){0.5}}
\put(0,5){\vector(0,-1){1}}
\put(10,5){\vector(0,-1){1}}
\put(5,0){\vector(-1,0){1}}
\put(5,10){\vector(-1,0){1}}
\put(-1,5){\makebox(0,0)[rc]{$C$}}
\put(11,5){\makebox(0,0)[lc]{$D$}}
\end{picture}
\quad\;\; \sim \;
\begin{picture}(10,10)(0,4)\small
\put(0,0){\line(1,0){10}}
\put(0,10){\line(1,0){10}}
\multiput(0,0)(0,1){10}{\line(0,1){0.5}}
\multiput(10,0)(0,1){10}{\line(0,1){0.5}}
\put(0,5){\vector(0,1){1}}
\put(10,5){\vector(0,1){1}}
\put(5,0){\vector(1,0){1}}
\put(5,10){\vector(1,0){1}}
\put(5,-1){\makebox(0,0)[ct]{$D$}}
\put(5,11){\makebox(0,0)[cb]{$C$}}
\end{picture}\]\vspace{0mm}
\paragraph{\it Definition:} For $f \in F$ we can define two algebra homomorphisms
$F \rightarrow U$:
\eq
R^+(f)  = \la R, f \ot \id\ra ,\qquad R^-(f) = \la \tilde R, \id \ot f\ra \label{rplus}
\en
\paragraph{\it Yang-Baxter Equation.}
As a consequence of the axioms of a quasitriangular face Hopf algebra $R$
satisfies the Yang-Baxter Equation
\eq
R_{12} R_{13} R_{23} = R_{23} R_{13} R_{12} \quad \in \quad U \ot U \ot U \; .
\en
Contracted 
with $T^A_B \ot T^C_D \ot T^E_F$ this expression yields
a numerical Yang-Baxter equation with the following pictorial representation
\cite{Bax}:
\begin{center}
\unitlength 0.50mm
\linethickness{0.4pt}
\begin{picture}(145.00,46.00)
\put(0.00,20.00){\line(3,-4){15.00}}
\put(15.00,0.00){\line(1,0){25.00}}
\put(40.00,0.00){\line(3,4){15.00}}
\put(55.00,20.00){\line(-3,4){15.00}}
\put(40.00,40.00){\line(-1,0){25.00}}
\put(15.00,40.00){\line(-3,-4){15.00}}
\put(90.00,20.00){\line(3,-4){15.00}}
\put(105.00,0.00){\line(1,0){25.00}}
\put(130.00,0.00){\line(3,4){15.00}}
\put(145.00,20.00){\line(-3,4){15.00}}
\put(130.00,40.00){\line(-1,0){25.00}}
\put(105.00,40.00){\line(-3,-4){15.00}}
\put(0.00,20.00){\line(1,0){25.00}}
\put(25.00,20.00){\line(3,4){15.00}}
\put(25.00,20.00){\line(3,-4){15.00}}
\put(105.00,40.00){\line(3,-4){15.00}}
\put(120.00,20.00){\line(-3,-4){15.00}}
\put(120.00,20.00){\line(1,0){25.00}}
\put(15.00,40.00){\vector(-3,-4){9.00}}
\put(0.00,20.00){\vector(3,-4){9.00}}
\put(15.00,40.00){\vector(1,0){15.00}}
\put(0.00,20.00){\vector(1,0){15.00}}
\put(15.00,0.00){\vector(1,0){15.00}}
\put(25.00,20.00){\vector(3,-4){9.00}}
\put(40.00,40.00){\vector(-3,-4){9.00}}
\put(40.00,40.00){\vector(3,-4){9.00}}
\put(55.00,20.00){\vector(-3,-4){9.00}}
\put(90.00,20.00){\vector(3,-4){9.00}}
\put(105.00,40.00){\vector(-3,-4){9.00}}
\put(105.00,40.00){\vector(3,-4){9.00}}
\put(120.00,20.00){\vector(-3,-4){9.00}}
\put(130.00,40.00){\vector(3,-4){9.00}}
\put(145.00,20.00){\vector(-3,-4){9.00}}
\put(105.00,40.00){\vector(1,0){15.00}}
\put(120.00,20.00){\vector(1,0){15.00}}
\put(105.00,0.00){\vector(1,0){15.00}}
\put(20.00,30.00){\makebox(0,0)[cc]{$R_{13}$}}
\put(40.00,20.00){\makebox(0,0)[cc]{$R_{23}$}}
\put(20.00,10.00){\makebox(0,0)[cc]{$R_{12}$}}
\put(105.00,20.00){\makebox(0,0)[cc]{$R_{23}$}}
\put(125.00,10.00){\makebox(0,0)[cc]{$R_{13}$}}
\put(125.00,30.00){\makebox(0,0)[cc]{$R_{12}$}}
\put(72.00,20.00){\makebox(0,0)[cc]{$=$}}
\put(27.00,42.00){\makebox(0,0)[cb]{$B$}}
\put(27.00,-2.00){\makebox(0,0)[ct]{$A$}}
\put(5.00,30.00){\makebox(0,0)[rb]{$E$}}
\put(5.00,10.00){\makebox(0,0)[rt]{$C$}}
\put(50.00,10.00){\makebox(0,0)[lt]{$F$}}
\put(50.00,30.00){\makebox(0,0)[lb]{$D$}}
\put(117.00,42.00){\makebox(0,0)[cb]{$B$}}
\put(117.00,-2.00){\makebox(0,0)[ct]{$A$}}
\put(95.00,30.00){\makebox(0,0)[rb]{$E$}}
\put(95.00,10.00){\makebox(0,0)[rt]{$C$}}
\put(140.00,10.00){\makebox(0,0)[lt]{$F$}}
\put(140.00,30.00){\makebox(0,0)[lb]{$D$}}
\end{picture}
\end{center}
\vspace{1ex}
The inner edges are paths that are summed over. 
Moving along the outer edges of the hexagon 
we will later
read of the shifts in the Yang-Baxter equation for 
dynamical $R$-matrices.

So far we have argued heuristically that the Ruijsenaars system naturally leads to 
graphs and face algebras.
The formal relation between face Hopf algebras and oriented graphs
it is established by a lemma of Hayashi 
(Lemma 3.1 of \cite{Hay1}) which says
that any right or left comodule $M$ of a face 
Hopf algebra $F$ decomposes as a linear space
to a direct sum $M=\oplus_{i,j}M_{ij}$ with indices $i,j$ running
over all vertices. So we can naturally associate paths
from 
$j$ to $i$ to any pair of indices such that $M_{ij}\neq \emptyset$. 
So we may and in fact will speak of the 
vectors of a comodule or a module as of paths.
As the dual object $U$ to a face Hopf algebra is again a face Hopf algebra
characterized by the same set of vertices \cite{Hay1}, 
the same applies to its 
comodules. It is convenient to choose the orientation of paths
appearing in the decomposition of a comodule of $U$ 
(and hence in the module of $F$)
opposite to the convention that one uses in the case of $F$.

Particularly we have for any matrix corepresentation
$T^A_B$ of $F$ with symbols $A, B$ used to label some 
linear basis in $M$, the linear span $\langle T^A_B \rangle$ of all $T^A_B$ decomposes as 
linear space (bicomodule
of the face Hopf algebra) as a direct sum
\[
\bigoplus_{i,j,k,l}\langle e^ie_jT^A_B e^ke_l \rangle = \langle T^A_B \rangle 
\]
i.e. a sum over paths with fixed starting
and ending vertices. The upper indices $i$ and $k$ fix the beginning and the end
of the path $A$ and the lower indices $j$ and $l$ fix the beginning and the end 
of the path $B$.

Let us assume that
the matrix elements $T^A_B$ of a corepresentation of $F$
act in a module of paths that we shall label by greek characters 
$\alpha$, $\beta$, etc.
The definition of the dual face Hopf algebra implies
that the matrix element $(T^A_B)^{\alpha}_{\beta}$ is nonzero only if 
$\cdot \alpha = \cdot B$, $\cdot A=\alpha \cdot$, $ B\cdot=\cdot \beta$ and
$A\cdot = \beta \cdot $, \ie\ if paths $\alpha\cdot A$ and $B\cdot\beta$ have
common starting and endpoints. This justifies the pictorial 
representation used
in this chapter.
It also follows immediately that in the case of a coquasitriangular
Face Hopf algebra 
$
\la \R , T^A_B \ot T^C_D \ra = R^{A C}_{B D}  
\equiv W\Big( {\mbox{\scriptsize$C$} {B \atop A} \mbox{\scriptsize$D$}} \Big)
$
is zero unless $\cdot C = \cdot B$, $B\cdot = \cdot D$,
$C\cdot = \cdot A$ and $A\cdot = D\cdot$.

To make a contact with the Ruijsenaars type of models, 
we have to assume that 
the (coquasitriangular) face Hopf algebra $F$ is 
generated by the matrix elements of some fundamental 
corepresentation of it. We shall postulate the paths
of the corresponding corepresentation to be of length~$1$.
The paths belonging to the $n$-fold tensor product of the 
fundamental corepresentation are then by definition of length~$n$. 
Taking an infinite tensor
product of the fundamental corepresentation we get a graph that corresponds
to the one generated by the shift operators of the related integrable model.

In the next section we are going to formulate a quantum version of the
so-called Main Theorem which gives the solution by factorization of the
Heisenberg equations of motion. For this construction $F$ needs to have a
coquasitriangular structure -- this will  
also fix its algebra structure.

\subsection{Quantum factorization}

The cocommutative functions in $F$ are of particular interest to us
since they form a set of mutually commutative operators. We shall
pick a Hamiltonian from this set. Cocommutative
means that the result of an application of comultiplication $\Delta$ 
is invariant under exchange of the two resulting factors. The typical
example is a trace of the T-matrix. The following theorem gives for
the case of face Hopf algebras what has become known as the
``Main theorem'' for the solution by factorization of the equations
of motion \cite{Adl,Kost,Sym,RS,Res}. The following theorem 
is a direct generalization of our
previous results for Hopf algebras/Quantum Groups \cite{JSch2,twist}:

\begin{thm}[Main theorem for face algebras]
\mbox{ }

\begin{enumerate}
\item[(i)] The set of cocommutative functions, denoted $I$, is a commutative
subalgebra of $F$.
\item[(ii)] The Heisenberg 
equations of motion defined by a Hamiltonian $\ha \in I$
are of Lax form
\eq
i \frac{d T}{dt} = \left[ M^\pm , T \right], \label{laxeqn}
\en
with $M^\pm = 1 \ot \ha - m_\pm \in U_\pm \ot F$, 
$m_\pm = R^\pm(\ha_{(2)}) \ot \ha_{(1)}$; see (\ref{rplus}).
\item[(iii)] Let $g_\pm(t) \in U_\pm \ot F$
be the solutions to the factorization problem
\eq
g_-^{-1}(t) g_+(t) = \exp( i t (m_+ - m_-)) \: \in \: U \ot F ,
\en
then 
\eq
T(t) = g_\pm(t) T(0) g_\pm(t)^{-1} 
\en
solves the Lax equation {\rm (\ref{laxeqn})}; $\:g_\pm(t)$ are given by 
\eq
g_\pm(t) = \exp(-it(1\ot h))\,\exp(it(1\ot h - M^\pm(0))
\en
and are the solutions to the differential equation
\eq
i  \frac{d}{dt}g_\pm(t) =
M^\pm(t) g_\pm(t), \qquad  g_\pm(0) = 1 . 
\en
\end{enumerate}
\end{thm}
\noindent {\sc Proof:}
The proof is similar to the one given in \cite{twist} for factorizable
quasitriangular Hopf algebras. Here we shall only emphasize the points
that are different because we are now dealing with face algebras.
An important relation that we shall use several times in the proof
is
\eq
\Delta(1) T_1 T_2 = T_1 T_2 = T_1 T_2 \Delta(1).
\en
(Note that the generalization is not trivial since
now $\Delta(1) \neq 1 \ot 1$ in general.)
\begin{enumerate}
\item[(i)] Let $f , g \in I \subset F$, then $f g \in I$. Let us show that
$f$ and $g$ commute:
\begin{quote}
$\displaystyle f g = \la T_1 T_2 , f \ot g \ra = \la \Delta(1) T_1 T_2 , f
\ot g \ra = \la \tilde R R T_1 T_2 , f \ot g \ra = \la \tilde R T_2 T_1 R, f
\ot g \ra = \la T_2 T_1 R \tilde R, f \ot g \ra = \la T_2 T_1 \Delta'(1), f
\ot g \ra = \la T_2 T_1, f \ot g \ra = g f. $
\end{quote}
($\tilde R \equiv (S \ot \id)(R)$ can be commuted with $T_2 T_1 R$ in the
fifth step because $f$ and $g$ are both cocommutative.)
\item[(ii)] We need to show that 
$\left[ R^\pm(\ha_{(2)}) \ot \ha_{(1)} , T \right] = 0$. 
This follows from the cocommutativity of $\ha$ and (\ref{rtt}):
\begin{quote}
$\displaystyle \la T_1 R^\pm_{21} T_2 , \ha \ot \id \ra
= \la R^\pm_{21} T_1 T_2 , \ha \ot \id \ra
= \la T_2 T_1 R^\pm_{21} , \ha \ot \id \ra .$
\end{quote}
\item[(iii)] We need to show that $[ m_+ , m_- ] = 0 $; then the proof of 
\cite{twist} applies.\\[1ex]
$\displaystyle m_+ m_- = 
\la T_1 R_{13} T_2 \tilde R_{32} , \ha \ot \ha \ot id\ra
\equiv \la T_1 T_2 R_{13} \tilde R_{32} , \ha \ot \ha \ot \id \ra \\
= \la \tilde R_{12} T_2 T_1 R_{12} R_{13} \tilde R_{32}, 
\ha \ot \ha \ot \id\ra
= \la \tilde R_{12} T_2 T_1 \tilde R_{32} R_{13} R_{12}, 
\ha \ot \ha \ot \id \ra \\
= \la R_{12} \tilde R_{12} T_2 T_1 \tilde R_{32} R_{13} , 
\ha \ot \ha \ot \id \ra
= \la \Delta'(1) T_2 T_1 \tilde R_{32} R_{13} , 
\ha \ot \ha \ot \id \ra \\ 
= \la T_2 \tilde R_{32} T_1 R_{13} , \ha \ot \ha \ot \id \ra
= m_- m_+ . $
\end{enumerate}

\paragraph{\it Remark:} The objects in this theorem
($M^\pm, m_\pm, g_\pm(t), T(t)$)
can be interpreted (a) as elements of
$U \ot F$, (b) as maps $F \rightarrow F$ or (c), 
when a representation of $U$ is
considered, as matrices with $F$-valued
matrix elements.

\subsection{Dynamical operators}

In the main theorem we dealt with expressions that live in $U \ot F$
(and should be understood as maps from $F$ into itself). In this section
we want to write expressions with respect to one fixed vertex. Like we
mentioned in the introduction we are particularly interested in the action of
the Hamiltonian with respect to a fixed vertex.
Since the Hamiltonian is an element of $F$, we shall initially fix the
vertex in this space; due to the definition of the dual face Hopf algebra $U$
this will also fix a corresponding vertex in that space. 
We shall proceed as follows: We will fix a vertex in the
$T$-matrix with the help of $e^\lambda, e_\lambda \in F$ and
will also introduce the corresponding universal $T(\lambda)$.
Next we will contstruct dynamical $R$-matrices
$R^\pm(\lambda)$ as $R^\pm(T(\lambda))$ -- this is
an algebra homomorphism -- and will give the Yang-Baxter and $RTT$ 
equations with
shifts as an illustration.  Finally we shall plug everything into the main
theorem.

Convention for corepresentation $T$ with respect to a fixed vertex:
$T(\lambda)^A_B$ is zero unless  the range (end) of path $A$ is equal to the
fixed vertex $\lambda$. Such a $T$ will map the vector space spanned by
vectors $v^A$ with $A\cdot = \lambda$ fixed to itself. With the help of the
projection operator $e^\lambda \in F$ we can give the following explicit
expression:
\eq
T(\lambda)^A_B \; = \; T^A_B \, e^\lambda
\quad\sim\quad
\unitlength1mm
\begin{picture}(10,8)(0,4)\small
\put(0,0){\line(1,0){10}}
\put(0,10){\line(1,0){10}}
\multiput(0,0)(0,1){10}{\line(0,1){0.5}}
\multiput(10,0)(0,1){10}{\line(0,1){0.5}}
\put(0,5){\vector(0,1){1}}
\put(10,5){\vector(0,1){1}}
\put(5,0){\vector(1,0){1}}
\put(5,10){\vector(1,0){1}}
\put(10,10){\makebox(0,0)[cc]{$\times$}}
\put(11,10){\makebox(0,0)[lb]{$\lambda$}}
\put(5,-1){\makebox(0,0)[ct]{$B$}}
\put(5,11){\makebox(0,0)[cb]{$A$}}
\end{picture}\vspace{3ex}
\en
The universal $T(\lambda)$ is an abstraction of this an is defined
analogously.
\eq
T_{12}(\lambda) \; = \; T_{12} \,(e^\lambda)_2\quad\sim\quad
\unitlength1mm
\begin{picture}(10,10)(0,4)\small
\put(0,0){\line(1,0){10}}
\put(0,10){\line(1,0){10}}
\multiput(0,0)(0,1){10}{\line(0,1){0.5}}
\multiput(10,0)(0,1){10}{\line(0,1){0.5}}
\put(0,5){\vector(0,1){1}}
\put(10,5){\vector(0,1){1}}
\put(5,0){\vector(1,0){1}}
\put(5,10){\vector(1,0){1}}
\put(10,10){\makebox(0,0)[cc]{$\times$}}
\put(11,10){\makebox(0,0)[lb]{$\lambda$}}
\end{picture}
\en
\paragraph{\it Coproducts of $\,T$.} Expressions for the coproducts 
$\Delta_1 T(\lambda)$ and $\Delta_2 T(\lambda)$ follow either directly from
the definition of $T(\lambda)$ or can be read of the corresponding pictorial
representations.\\
(i)
Coproduct in $F$-space \cite{Fel}: \vspace{-4ex}
\eq 
\Delta_2 T(\lambda) \; = \; T_{12}(\lambda) T_{13}(\lambda - h_2)
\quad
\sim
\quad
\unitlength 1mm
\begin{picture}(15,10)(0,9)\small
\put(10,20){\makebox(0,0)[cc]{$\times$}}
\put(12,20){\makebox(0,0)[lc]{$\lambda$}}
\put(10,10){\makebox(0,0)[cc]{$\times$}}
\put(12,10){\makebox(0,0)[lc]{$\lambda - h_2$}}
\put(0,20){\line(1,0){10}}
\multiput(0,10)(0,1){10}{\line(0,1){0.5}}
\multiput(10,10)(0,1){10}{\line(0,1){0.5}}
\put(0,15){\vector(0,1){1}}
\put(10,15){\vector(0,1){1}}
\put(5,20){\vector(1,0){1}}
\put(0,0){\line(1,0){10}}
\put(0,10){\line(1,0){10}}
\multiput(0,0)(0,1){10}{\line(0,1){0.5}}
\multiput(10,0)(0,1){10}{\line(0,1){0.5}}
\put(0,5){\vector(0,1){1}}
\put(10,5){\vector(0,1){1}}
\put(5,0){\vector(1,0){1}}
\put(5,10){\vector(1,0){1}}
\end{picture} \vspace{4ex}
\en
The shift operator $h$ in $F$-space that appears here is
\eq
h_{(F)} = \sum_{\eta,\mu} (\mu - \eta) \, e^\mu_\eta \; \in \; F,
\en 
where we assume some appropriate (local) embedding of the vertices of the underlying
graph in $\C^n$ so that the difference of vertices makes sense.\\
{\sc Proof:} 
$\Delta_2 T_{12} (e^\lambda)_2 
= \sum_\eta T_{12} T_{13} (e_\eta^\lambda)_2 (e^\eta)_3
= \sum_{\eta,\mu} T_{12} (e^\lambda e^\mu)_2 (e_\eta)_2 T_{13} (e^\eta)_3
= \sum_{\eta,\mu} T_{12}(\lambda) (e^\mu_\eta)_2 T_{13}(\lambda + \eta - \mu)
= T_{12}(\lambda) T_{13}(\lambda - h_2)$. In the  second and third step we
used $e^\lambda e^\mu \propto \delta_{\lambda,\mu}$.\\[1mm]
(ii) Coproduct in $U$-space: (gives multiplication
in $F$)
\eq
\Delta_1 T(\lambda) \; = \; T_{13}(\lambda - h_2) T_{23}(\lambda)
\quad \sim \quad
\unitlength 1mm
\begin{picture}(20,10)(0,4)\small
\put(0,0){\line(1,0){10}}
\put(0,10){\line(1,0){10}}
\multiput(0,0)(0,1){10}{\line(0,1){0.5}}
\multiput(10,0)(0,1){10}{\line(0,1){0.5}}
\put(0,5){\vector(0,1){1}}
\put(10,5){\vector(0,1){1}}
\put(5,0){\vector(1,0){1}}
\put(5,10){\vector(1,0){1}}
\put(10,0){\line(1,0){10}}
\put(10,10){\line(1,0){10}}
\multiput(20,0)(0,1){10}{\line(0,1){0.5}}
\put(20,5){\vector(0,1){1}}
\put(15,0){\vector(1,0){1}}
\put(15,10){\vector(1,0){1}}
\put(20,10){\makebox(0,0)[cc]{$\times$}}
\put(20,12){\makebox(0,0)[cb]{\small $\lambda$}}
\put(10,10){\makebox(0,0)[cc]{$\times$}}
\put(10,12){\makebox(0,0)[cb]{\small $\lambda - h_2$}}
\end{picture}\quad\vspace{4mm}
\en 
This time we have a shift operator in $U$-space:
\eq
h_{(U)} = \sum_{\eta,\mu} (\mu - \eta) \, E_\mu E^\eta \; \in \; U .
\en
The proof is a little more involved than the one given for the $F$-case above
and uses $(e^\eta)_2 T_{12} = (E^\eta)_1 T_{12}$ and
$T_{12} (e^\mu)_2 = (E_\mu)_1 T_{12}$ which follow from 
$T \tilde T T = T$, $\tilde T T \tilde T = \tilde T$ 
and $T \tilde T = \sum_\xi E^\xi \ot
e^\xi$.
\paragraph{\it Dynamical $R$-matrix.}
Using the fact that
$f \mapsto R^+(f) \equiv \la R , f \ot \id \ra$
is an algebra-homomorphism we define
\eq
R_{12}(\lambda) \equiv  \la R , T_1(\lambda) \ot T_2 \ra
\en
The numerical $R$-matrix is defined analogously:
$R(\lambda)^{AC}_{BD} = \la R, T(\lambda)^A_B \ot T^C_D\ra$.
In the pictorial representation this fixes the one vertex of $R$
that is only
endpoint to paths.

\paragraph{\it Dynamical $RTT$-equation.}
{}From the coproduct $\Delta_1 T$ and $R \Delta(x) = \Delta'(x) R$ for all $x
\in U$ follows \cite{Fel}: 
\eq
R_{12}(\lambda) T_{1}(\lambda - h_2) T_{2}(\lambda)
= T_{2}(\lambda - h_1) T_{1}(\lambda) R_{12}(\lambda - h_3).
\en
Shifts $h_1$, $h_2$ are in $U$-space, shift $h_3$ is in $F$-space.
Twice contracted with $R$ the dynamical $RTT$-equation yields the dynamical
Yang-Baxter equation:
\paragraph{\it Dynamical Yang-Baxter equation \cite{Ger}} 
\eq
R_{12}(\lambda) R_{13}(\lambda - h_2) R_{23}(\lambda)
= R_{23}(\lambda - h_1) R_{13}(\lambda) R_{12}(\lambda - h_3).
\en
\paragraph{\it Hamiltonian and Lax operators in dynamical setting}
\mbox{ }\\
The Hamiltonian $\ha$ should be a cocomutative element of $F$. In the case of the
Ruijsenaars model it can be chosen to be the trace of a $T$-matrix, \ie\ the
$U$-trace of $T$ in an appropriate representation $\rho$: 
$\ha = \mbox{tr}^{(\rho)}_1 T_1$.
This can be written as a sum over vertices $\lambda$ of operators that act in
the respective subspaces corresponding to paths ending in the vertex
$\lambda$:
\eq
\ha = \sum_{Q, \no Q \,{\rm fixed}} T^Q_Q = \sum_\lambda \ha(\lambda)
\en
with
$\ha(\lambda) = \ha e^\lambda = \sum T(\lambda)^Q_Q$.
The pictorial representation of the Hamiltonian is two closed dashed
paths ($F$-space) connected by paths $Q$ of fixed length
that are summed over. In $\ha(\lambda)$ the end of path
$Q$ is fixed. When we look at a representation on Hilbert space the
paths $Q$ with endpoint $\lambda$ that appear in the component
$\ha(\lambda)$ of the Hamiltonian $\ha$ will shift the argument
of a state $\psi(\lambda)$
corresponding to the vertex $\lambda$ to a new vertex corresponding to the
starting point of the path $Q$. 
In the next section we will see in detail
how this construction is applied to the Ruijsenaars system.
For this we will have to take a representation of the face Hopf algebra $F$. We
shall then denote the resulting Lax operator by $L$. The Hamiltonian
will contain a sum 
(coming from the trace) over shift
operators.

For the Lax operators it is convenient to fix two vertices 
$\lambda$ and $\mu$
corresponding
to paths from $\mu$ to $\lambda$ . (We did not need to do this for the
Hamiltonian since it acts only on closed paths.)
The $T$ that appears in the Lax equation (\ref{laxeqn})
becomes
\[T_{12}(\lambda,\mu) = T_{12} (e_\mu e^\lambda)_{2} =  
(E_\lambda)_1 T_{12} (E_\mu)_1 ;\]
it should be taken in some representation of $F$ on the appropriate
Hilbert space. On space $U$ we are interested in a finite-dimensional
representation that is going to give us matrices.
Both is done in the next section and 
we shall call the resulting operator $L(\lambda,\mu)$. 
Similarly we proceed with the other two Lax operators $M^\pm$:
\[
M^\pm(\lambda,\mu) \equiv (E_\lambda \ot 1)M^\pm(E_\mu\ot 1)
= E_\lambda \delta_{\lambda,\mu}\ot \ha
- m_\pm(\lambda,\mu).
\]
(Recall that $E_\lambda E_\mu = E_\lambda \delta_{\lambda,\mu}$.)
If we define $M^\pm_{012} = T_{02} (1 - R^\pm_{01})$ then $M^\pm_{12}
= \la M^\pm_{012} , \ha \ot \id^2 \ra$ and we can write dynamical Lax
equations in an obvious notation as
\begin{eqnarray}
i \frac{d T(\lambda,\mu)}{d t} 
& = & \Big\la M^\pm_{012}(\lambda, \lambda - h_0)
T_{12}(\lambda - h_0,\mu) \\
& & - T_{12}(\lambda, \mu - h_0)
M^\pm_{012}(\mu - h_0,\mu) \, , \, \ha \ot \id^2 \Big\ra \nonumber
\end{eqnarray}
where $h_0 = \sum_{\alpha,\beta} (\alpha - \beta) E_\alpha^\beta$ is the shift
operator in the space contracted by the Hamiltonian. If the Hamiltonian is a 
trace we shall find a sum over shifts.

\paragraph{\it Remark:} There is another possible choice of conventions for
the fixed vertex. We could worked with $T_{12}\{\nu\} = (e_\nu)_2 T_{12}$ 
instead
of $T_{12}(\lambda) = T_{12} (e^\lambda)_2$. This would have fixed the lower
left vertex in the pictorial representation of $T^A_B$ and the vertex that is
\emph{starting}-point for all paths in $R\{\nu\}$. The dynamical equations
would of course look a little different from the ones that we have given.

\subsection{The Lax Pair}

Here we give as an example the Lax pair for the case of the $N$-particle
quantum Ruijsenaars model.
We may think of the face Hopf algebra $F$ as of the elliptic 
quantum group associated 
to $sl(N)$ introduced by Felder \cite{Fel}. We will continue to use the
the symbol $F$ for it. In that case there is an additional
spectral parameter entering all relations in the same way as it is in the case
ordinary quantum groups.

Let $h$ be the Cartan subalgebra of $sl(N)$ and $h^*$ its dual.
The actual graph related to the elliptic quantum group $F$ is 
$h^*\sim \C^{N-1}$.
This is the huge graph of the remark made in the introduction. 
However it decomposes into a continuous family of disconnected graphs, 
each one isomorphic to $-\eta . \Lambda$, the  
$-\eta\in \C$ multiple of the 
weight lattice $\Lambda$ of $sl(N)$, and we can restrict ourselves to this 
one component for simplicity. Correspondingly the shifts in all formulas
are rescaled by a factor $-\eta$.
The space $h^*\sim \C^{N-1}$
itself will be considered as the orthogonal complement of $\C^{N}=
\oplus_{i=1,...,N}\C\varepsilon_i$, 
$\langle \varepsilon_i, \varepsilon_j \rangle
=\delta_{ij}$ with respect to $\sum_{i=1,...,N}\varepsilon_i$.
We write the orthogonal projection
$\epsilon_i=\varepsilon_i - \frac{1}{N}\sum_k \varepsilon_k$ 
for the generator of $h^*$; $\la \epsilon_i , \epsilon_j \ra 
= \delta_{ij} - 1/N$.
The points of the lattice $\eta . \Lambda$ will be denoted by greek
characters $\lambda$, $\mu$, etc. 

The elliptic quantum group is defined by the matrix elements of the 
``fundamental corepresentation''.
This is described with the help of the paths of length
$1$ in the following way:
Let us
associate a one-dimensional linear space 
$V_{\rho,\lambda} \sim \C \eta.\epsilon_k$ and a corresponding path
of length $1$ to any pair
$\lambda, \rho \in \eta . \Lambda$, such that 
$\rho - \lambda = \eta \epsilon_k$, 
for some $k=1,..., N$. We let $V_{\rho,\lambda}=\emptyset$
for all other pairs of vertices. 
The vector space $V$ of the fundamental corepresentation is formed
by all paths of length $1$
\[
V=\bigoplus_{\lambda,\rho} V_{\rho,\lambda}\quad
\sim
\quad
\bigoplus_{\lambda,\rho}\;\,
\unitlength1mm
\begin{picture}(10,0)(0,-1)\scriptsize
\put(0,0){\line(1,0){10}}
\put(5,0){\vector(1,0){1}}
\put(0,0){\makebox(0,0)[cc]{$\times$}}
\put(10,0){\makebox(0,0)[cc]{$\times$}}
\put(0,1){\makebox(0,0)[cb]{$\rho$}}
\put(10,1){\makebox(0,0)[cb]{$\lambda$}}
\end{picture}
\quad = \quad\bigoplus_{\lambda,i}\quad
\begin{picture}(10,0)(0,-1)\scriptsize
\put(0,0){\line(1,0){10}}
\put(5,0){\vector(1,0){1}}
\put(0,0){\makebox(0,0)[cc]{$\times$}}
\put(10,0){\makebox(0,0)[cc]{$\times$}}
\put(0,1){\makebox(0,0)[cb]{$\lambda + \eta \epsilon_i$}}
\put(10,1){\makebox(0,0)[cb]{$\lambda$}}
\end{picture}
\]
As all spaces $V_{\rho,\lambda}$ are at most one-dimensional, we can
characterize the numerical $R$-matrix $R^{AC}_{BD}$ in the fundamental
corepresentation by just four indices
referring to the vertices of the ``square'' defined by paths 
$A,B,C,D$ in the case of nonzero matrix element $R^{AC}_{BD}$. Let us set
$\cdot B =\cdot C= \nu $, $\cdot D=B\cdot =\mu$, $D\cdot =A\cdot= \lambda$
and $\cdot A= C\cdot =\rho$ and 
also
\[ R^{AC}_{BD}=W\Big( {C {B \atop A} D} \Big)\equiv W\Big( {\nu \atop \rho}\,\,
{\mu\atop \lambda} \Big) \quad\sim\qquad
\unitlength 1.00mm
\linethickness{0.4pt}
\begin{picture}(11.00,5.00)(0,-1.00)
\put(0.00,-5.00){\line(0,1){10.00}}
\put(0.00,5.00){\line(1,0){10.00}}
\put(10.00,5.00){\line(0,-1){10.00}}
\put(10.00,-5.00){\line(-1,0){10.00}}
\put(5.00,5.00){\vector(1,0){1.00}}
\put(5.00,-5.00){\vector(1,0){1.00}}
\put(0.00,0.00){\vector(0,-1){1.00}}
\put(10.00,0.00){\vector(0,-1){1.00}}
\put(-1.00,5.00){\makebox(0,0)[rc]{$\nu$}}
\put(-1.00,-5.00){\makebox(0,0)[rc]{$\rho$}}
\put(11.00,5.00){\makebox(0,0)[lc]{$\mu$}}
\put(11.00,-5.00){\makebox(0,0)[lc]{$\lambda$}}
\end{picture}
\quad.\]
Then the non-zero Boltzmann weights as given by \cite{Jim}
are: ($i\neq j$)
\[ W\Big( {\lambda +2\eta \epsilon_i \atop \lambda +\eta \epsilon_i}\,\,
{\lambda +\eta \epsilon_i\atop \lambda}\Big| u \Big)= 1\quad\sim\quad
\unitlength 1.00mm
\begin{picture}(20.00,0.00)(0,-1.00)\small
\put(0.00,-0.50){\line(1,0){9.5}}
\put(0.00,0.00){\line(1,0){9.5}}
\put(4.00,0.00){\vector(1,0){1.00}}
\put(4.00,-0.50){\vector(1,0){1.00}}
\put(10.00,0.00){\line(1,0){9.5}}
\put(10.00,-0.50){\line(1,0){9.5}}
\put(14.00,0.00){\vector(1,0){1.00}}
\put(14.00,-0.50){\vector(1,0){1.00}}
\put(20.50,-0.25){\makebox(0,0)[lb]{$\lambda$}}
\put(19.50,-0.25){\makebox(0,0)[cc]{$\times$}}
\end{picture}
\quad ,\]
\[ W\Big( {\lambda + \eta (\epsilon_i+\epsilon_j) 
\atop \lambda +\eta \epsilon_i}\,\,
{\lambda +\eta \epsilon_i\atop \lambda} \Big| u \Big)
= \frac{\theta(\eta)\theta(-u
+ \lambda_{ij})}{\theta(u+\eta)\theta(\lambda_{ij})}\quad\sim
\quad
\unitlength 1.00mm
\begin{picture}(11.00,5.00)(0,-1.00)\small
\put(0.00,5.00){\line(0,-1){10.00}}
\put(0.00,-5.00){\line(1,0){10.00}}
\put(10,-4.75){\makebox(0,0)[cc]{$\times$}}
\put(0.50,5.00){\line(0,-1){9.50}}
\put(0.50,-4.50){\line(1,0){9.50}}
\put(0.00,0.50){\vector(0,-1){1.00}}
\put(0.50,0.50){\vector(0,-1){1.00}}
\put(5.50,-4.50){\vector(1,0){1.00}}
\put(5.50,-5.00){\vector(1,0){1.00}}
\put(11.00,-5.00){\makebox(0,0)[lb]{$\lambda$}}
\end{picture}
\quad ,\]
and
\[ W\Big( {\lambda + \eta (\epsilon_i+\epsilon_j) 
\atop \lambda +\eta \epsilon_i}\,\,
{\lambda +\eta \epsilon_j\atop \lambda} \Big| u \Big)
= \frac{\theta(u)\theta(\eta+ 
\lambda_{ij})}{\theta(u+\eta)\theta(\lambda_{ij})} 
\quad\sim\quad
\unitlength 1.00mm
\begin{picture}(11.00,5.00)(0,-1.00)\small
\put(0.00,5.00){\line(0,-1){10.00}}
\put(0.00,-5.00){\line(1,0){10.00}}
\put(10.00,-5.00){\line(0,1){10.00}}
\put(10.00,5.00){\line(-1,0){10.00}}
\put(0.00,0.00){\vector(0,-1){1.00}}
\put(5.00,-5.00){\vector(1,0){1.00}}
\put(10.00,0.00){\vector(0,-1){1.00}}
\put(5.00,5.00){\vector(1,0){1.00}}
\put(11.00,-5.00){\makebox(0,0)[lc]{$\lambda$}}
\put(10,-5){\makebox(0,0)[cc]{$\times$}}
\end{picture}
\quad.\]
Here  
$\lambda_{ij}\equiv \lambda_i - \lambda_j =
\langle\lambda,\epsilon_i-\epsilon_j\rangle$, \,
$\theta(u)$ is the Jacobi theta function
\eq
\theta(u)=\sum_{j\in\Z}e^{\pi i(j+\frac{1}{2})^2\tau 
+ 2\pi i(j+\frac{1}{2})
(u+\frac{1}{2})} , \label{theta}
\en
$u \in \C$ is the spectral parameter and $\tau$ is the 
elliptic modulus parameter.
The matrix element $T^A_B (\lambda|u)$ 
(now depending also on the spectral parameter $u$, which also enters
all previous expressions in the standard way) in the fundamental 
corepresentation is also uniquely determined by the value of the 
vertices that fix the length-$1$ paths $A$ and $B$.
Let us use the following notation for it ($i,j=1,\ldots,N$)
\[
\rule{0mm}{15mm} 
T^A_B (\lambda|u) = \sum_\mu  L^i_j(\mu, \lambda|u) e_\mu
\quad\sim\quad
\unitlength1mm
\begin{picture}(15,6)(-12,4)\scriptsize
\put(0,0){\line(1,0){10}}
\put(0,10){\line(1,0){10}}
\multiput(0,0)(0,1){10}{\line(0,1){0.5}}
\multiput(10,0)(0,1){10}{\line(0,1){0.5}}
\put(0,5){\vector(0,1){1}}
\put(10,5){\vector(0,1){1}}
\put(5,0){\vector(1,0){1}}
\put(5,10){\vector(1,0){1}}
\put(0,0){\makebox(0,0)[cc]{$\times$}}
\put(10,0){\makebox(0,0)[cc]{$\times$}}
\put(0,10){\makebox(0,0)[cc]{$\times$}}
\put(10,10){\makebox(0,0)[cc]{$\times$}}
\put(-1,0){\makebox(0,0)[rt]{$\mu + \eta \epsilon_j$}}
\put(11,0){\makebox(0,0)[lt]{$\mu$}}
\put(-1,10){\makebox(0,0)[rb]{$\lambda + \eta \epsilon_i$}}
\put(11,10){\makebox(0,0)[lb]{$\lambda$}}
\put(5,11){\makebox(0,0)[cb]{\small$A$}}
\put(5,-1){\makebox(0,0)[ct]{\small$B$}}
\end{picture}
\]\vspace{4mm}
where 
$\mu =B\cdot $, $\lambda
=A\cdot $, $\mu +\eta\epsilon_j=\cdot B$, 
$\lambda +\eta\epsilon_i=\cdot A$.

As the lattice $\eta . \Lambda$ that we consider is just one 
connected component
of a continuous family of disconnected graphs, the vertices
$\lambda$, $\mu$, \ldots are allowed to take any values in 
$\C^{(N-1)}$. This will be assumed implicitly
in the rest of this section.
Now we need to specify the appropriate representation of 
$F$ which can be read of from \cite{Has}.
We can characterize it by its path decomposition. 
To any pair of vertices
$\lambda, \mu$ we associate a one-dimensional vector space 
$\tilde V_{\lambda,\mu}\sim \C$ (path from $\mu$ to $\lambda$). 
The representation space $\tilde V$ is then
\[ 
\tilde V=\bigoplus_{\lambda,\mu \in h^*}\tilde V_{\lambda,\mu}.
\]
The matrix element $T^A_B (\lambda) \equiv L^i_j(\lambda,\mu |u)$ 
for fixed $A,B$ and hence also with fixed $i$, $j$, $\lambda$, $\mu$ 
is obviously non-zero
only if restricted to act from $\tilde V_{\lambda, \mu}$ 
to $\tilde V_{\lambda +\eta \epsilon_i, \mu +\eta \epsilon_j}$ 
in which case it acts as multiplication by
\begin{equation}
L^i_j(\lambda,\mu|u)
=\frac{\theta(\frac{c\eta}{N} +u +\lambda_i -\mu_j)}{\theta(u)}
\prod_{k\neq i}
\frac{\theta(\frac{c\eta}{N} +\lambda_{k} 
-\mu_j)}{\theta(\lambda_{k}-\lambda_i)}.\label{L}
\end{equation}
Here we used notation 
$\lambda_i=\langle \lambda , \epsilon_i \rangle$ 
for $\lambda \in h^*$. $c\in \C$ will play the role of coupling constant.

The Hamiltonian is chosen in accordance with Section 3 as $\ha=\sum_P
T^P_P$, i.e. the trace in the ``fundamental'' corepresentation. 
In accordance
with the discussion of the previous sections it is non-zero only when
acting on the diagonal subspace (closed paths)
\[ 
H=\bigoplus_{\lambda \in h^*}H_{\lambda}
\equiv\bigoplus_{\lambda \in h^*}\tilde V_{\lambda, \lambda}
\]
of $\tilde V$. So this is the actual state space of the integrable system
under consideration. The Lax operator $M^\pm$
acts from $\tilde V_{\lambda, \mu}$ 
to $\tilde V_{\lambda +\eta \epsilon_i, \mu +\eta \epsilon_j}$ 
as multiplication by $M^\pm{}^l_k(\lambda, \mu | u , v) =
(1 \ot \ha)^l_k(\lambda, \mu |v) - m_\pm{}^l_k(\lambda, \mu | u , v)$ with
\begin{equation}
(1 \ot \ha)^l_k(\lambda,\mu|v) 
= \delta_{\lambda,\mu} \delta^l_k 
\frac{\theta(\frac{c\eta}{N} + v)}{\theta(v)}
\prod_{j'\neq i}
\frac{\theta(\frac{c\eta}{N} 
+\lambda_{j',i})}{\theta(\lambda_{j',i})} , \label{h}
\end{equation}
\begin{eqnarray}
\lefteqn{m^+{}^l_k(\lambda,\mu|u , v)  = } \nonumber \\
&& \delta_{\epsilon_i + \epsilon_k , 
\epsilon_j + \epsilon_l}
\delta_{\lambda , \mu + \eta \epsilon_j}
\sum_{i,j} L^i_j(\mu + \eta \epsilon_k , \mu | v)
W\Big( {\lambda + \eta \epsilon_l 
\atop \mu +\eta \epsilon_k}\,\,
{\lambda \atop \mu} \Big| v - u \Big)\\
\rule{0mm}{2em} && \sim  
\unitlength1mm
\begin{picture}(13,6)(-12,4)\scriptsize
\put(0,0){\line(0,1){10}}
\put(10,0){\line(0,1){10}}
\put(10,0){\line(1,0){10}}
\put(10,10){\line(1,0){10}}
\put(20,0){\line(0,1){10}}
\multiput(0,0)(1,0){10}{\line(1,0){0.5}}
\multiput(0,10)(1,0){10}{\line(1,0){0.5}}
\put(0,5){\vector(0,-1){1}}
\put(10,5){\vector(0,-1){1}}
\put(20,5){\vector(0,-1){1}}
\put(5,10){\vector(1,0){1}}
\put(5,0){\vector(1,0){1}}
\put(15,10){\vector(1,0){1}}
\put(15,0){\vector(1,0){1}}
\put(20,0){\makebox(0,0)[cc]{$\times$}}
\put(20,10){\makebox(0,0)[cc]{$\times$}}
\put(0,0){\makebox(0,0)[cc]{$\times$}}
\put(0,10){\makebox(0,0)[cc]{$\times$}}
\put(10,0){\makebox(0,0)[cc]{$\bullet$}}
\put(10,10){\makebox(0,0)[cc]{$\bullet$}}
\put(21,0){\makebox(0,0)[lt]{$\mu$}}
\put(21,10){\makebox(0,0)[lb]{$\lambda$}}
\put(10,11){\makebox(0,0)[cb]{$\lambda + \eta \epsilon_l$}}
\put(10,-1){\makebox(0,0)[ct]{$\mu + \eta \epsilon_k$}}
\put(-1,0){\makebox(0,0)[rt]{$\mu$}}
\put(-1,10){\makebox(0,0)[rb]{$\lambda$}}
\end{picture} \nonumber 
\end{eqnarray}
\mbox{}\\[1ex]
$m^-$ is given by a similar formula with the inverse Boltzmann weight.
\paragraph{\it Dynamical Lax Equation}
$$
i\frac{d L^i_k(\lambda, \mu | u)}{d t} =
\sum_{j , \nu} M^\pm{}^i_j(\lambda, \nu | u, v) L^j_k(\nu, \mu |u)
- L^i_j(\lambda, \nu | u) M^\pm{}^j_k(\nu, \mu | u, v)
$$

The Hamiltonian $\ha$ maps 
the component $H_{\lambda}\sim \C.|\lambda \rangle$
into the component 
$H_{\lambda+\eta \epsilon_i}\sim \C.|\lambda +\eta\epsilon_i\rangle$.
Obviously $\tilde V$ can be understood as the 
complex vector space of all functions in $\lambda$, 
$\mu$ as well as $H$ can be understood as the
complex vector space of all functions in $\lambda$. 
In that case
the Hamiltonian $\ha$ given by (\ref{h}) is proportional 
to a difference operator 
in variable $\lambda \in h^*$
\begin{equation}
\ha\propto 
\sum_i t^{(\lambda)}_i\prod_{j\neq i }\frac{\theta(\frac{c\eta}{N} 
+\lambda_{j,i} )}{\theta(\lambda_{j,i})}, \label{Ruijsenaars}
\end{equation}
where $t^{(\lambda)}_i$ has an obvious meaning of the shift operator by 
$-\eta\epsilon_i$ in the variable $\lambda$. 
This is equivalent \cite{Die} to the Ruijsenaars Hamiltonian \cite{Rui}.
It follows from \cite{Has} that in the same way we 
can obtain the higher
order Hamiltonians concerning traces in properly fused 
``fundamental representations".

\chapter{{\sffamily Noncommutative gauge theory}}

A natural approach to gauge theory on noncommutative spaces can be
based on the simple observation that multiplication of a field by a
(coordinate) function is not a covariant concept if that function does 
not commute with gauge transformations~\cite{MSSW}.  
This problem can be cured by adding appropriate 
noncommutative gauge potentials and thus introducing covariant functions 
in complete analogy to the covariant derivatives of ordinary gauge theory.%
\footnote{From the phase space point of view ordinary gauge
theory is in fact a special case of this construction with gauge
potentials  for only half of the `coordinates' (momenta).}
This construction is of particular interest
because of its apparent relevance for the description of 
open strings in a
background $B$-field \cite{CLNY,CH,Scho}, where the D-brane world volume can be
interpreted as a noncommutative space whose fluctuations are governed by a noncommutative
version of Yang-Mills theory \cite{CDS,DH,MZ,Wati,SW}. 
It has been noticed (at least in the case of a constant $B$-field) that there
can be equivalent description of the effective theory both in terms of noncommutative gauge
theory and ordinary gauge
theory. From the physics perspective the two pictures are related by a choice of
regularization~\cite{SW,AD} which suggests a somewhat miraculous 
field redefinition that is usually
called Seiberg-Witten map~\cite{SW}. The inverse \mbox{$B$-field}, or more generally
the antisymmetric part of the inverse sum of $B$-field and metric, defines a
Poisson structure~$\theta$ whose quantization gives rise to the noncommutativity
on the D-brane world volume. Classically, the field strength $f$ describes
fluctuations of the $B$-field.
The Seiberg-Witten map expresses the noncommutative potential 
$\As$, 
noncommutative gauge 
parameter $\Lq$ and noncommutative field strength in terms of their 
classical counter parts $a$, $\lambda$, $f$ as formal power series in $\theta$, 
such that noncommutative gauge
transformations $\deltaq_{\Lq}$ are induced by ordinary gauge transformations
$\delta_\lambda$:
\eq
\As(a) + \deltaq_{\Lq}\As(a) = \As(a +\delta_\lambda a), \label{SWcond}
\en
where $\Lq$ is a function both of $a$ and $\lambda$. In section~\ref{swmapkont} we shall
focus on the rank one case and will explicitly 
construct the maps $\As(a)$ and $\Lq(a,\lambda)$ to all orders in theta for
the general case of an arbitrary Poisson manifold (which is relevant
for non-constant background fields)~\cite{JS,JSW}.\footnote{This is of
course neither restricted to magnetic fields -- $B^{0i}$ need not be zero --
nor to even dimensional manifolds.} The corresponding
star products can be computed with Kontsevich's formula~\cite{Kontsevich};
this formula continues to make sense even for non-closed $B$-field
although the corresponding star product will no longer be associative
(see also~\cite{Cornalbana}) but the non-associativity is still under control
by the formality (\ref{control}).
For a noncommutative gauge theory 
the rank one case does already include some information
about the nonabelian case, since it is always possible to include
a matrix factor in the definition of the 
underlying noncommutative space. We shall make this more
precise and will extend our results
to non-abelian gauge theories for \emph{any} Lie group.
The noncommutative gauge potential and field
strength in general take values in the universal enveloping algebra,
nevertheless thanks to the existence of the Seiberg-Witten map the theory 
can be consistently formulated in terms of only a finite number of fields;
this important observation has been discussed in 
~\cite{Jurco:2000ja}. A prerequisite for all this is an appropriate 
formulation of gauge theory on a more or less arbitrary noncommutative
space. (Here we are interested in the general case of an arbitrary
associative algebra of non-commuting variables,
important special examples with constant, linear and quadratic
commutation relations have been discussed in~\cite{MSSW}.) 
Particularly well-suited is the approach  based on the notion of 
covariant 
coordinates that we mentioned above%
\footnote{This is a somewhat nonstandard formulation of
noncommutative gauge theory that is not as intimately connected 
with a differential calculus
as it is in Connes approach. In particular examples, e.g. for the noncommutative
torus, both formulations may be used and give the same results.}
because it finds a natural interpretation in the framework of deformation 
quantization~\cite{BFFLS,Kontsevich,Sternheimer}. 
This is the natural setting since we are dealing with 
associative algebras and formal 
power series~-- it also allows rigorous statements by postponeing,
or rather circumventing difficult questions related to convergence and 
representation theory.
Deformation quantization for non-constant and possibly degenerate Poisson structures
goes far beyond the basic Weyl-Moyal product and the problem has only recently
found a general solution~\cite{Kontsevich}. To construct a Seiberg-Witten map 
we do in fact need the even
more general formality theorem of Kontsevich~\cite{Kontsevich}.
A link between Kontsevich quantization/formality  and quantum field theory
is given by the path integral approach~\cite{CattaneoFelder} which  relates the
graphs that determine the terms in the formality map to
Feynman diagrams. The relevant action~--
a Poisson sigma model~-- was originally studied
in~\cite{Schaller:1994es,Ikeda:1994fh};
see also \cite{Schaller:1994uj,Alekseev:1995py}.
Our discussion is entirely tree-level. Aspects of the quantization
of nonabelian noncommutative gauge theories have been discussed by several
authors~\cite{Bonora:2000ga,Armoni2,Armoni:2001br} (and references therein.)
Closest to the present discussion is the perturbative study of $\theta$-expanded
noncommutative gauge theories~\cite{Bichl:2001nf,Bichl2}.

We will see that the concept of covariant coordinates on noncommutative
spaces leads directly to 
gauge theories with generalized noncommutative gauge
fields of the type that arises in string theory with background $B$-fields.
I will argue that the theory is naturally expressed in terms of cochains
in an appropriate cohomology and will
discuss how it fits into the framework of projective
modules. We will use the equivalence of star products that arise
from the background field with and without fluctuations and
Kontsevich's formality theorem to explicitly construct a map that 
relates ordinary gauge theory and noncommutative gauge theory 
(Seiberg-Witten map.) As application we shall proof the exact equality of the
Dirac-Born-Infeld action
with $B$-field in the commutative setting and its semi-noncommutative
cousin in the intermediate picture. We will see that
a  consistent noncommutative gauge theory requires that the gauge
fields couple to gravity (except on flat manifolds, of course.)
Introducing noncommutative
extra dimensions we will extend the construction to
noncommutative nonabelian gauge theory for arbitrary gauge groups and will
in particular compute an explicit map between abelian and nonabelian gauge fields.
All constructions are also valid for non-constant $B$-field, Poisson
structure, metric and coupling constant.
 
This work was done in collaboration with B.\ Jur\v co and J.\ Wess
and will be published in ...
In the following we will briefly review how noncommutative gauge
theory appears in string theory.

\subsection*{Noncommutative gauge theory in string theory}
\label{sec:string}

\begin{figure}
$$
\includegraphics[width = .45\textwidth]{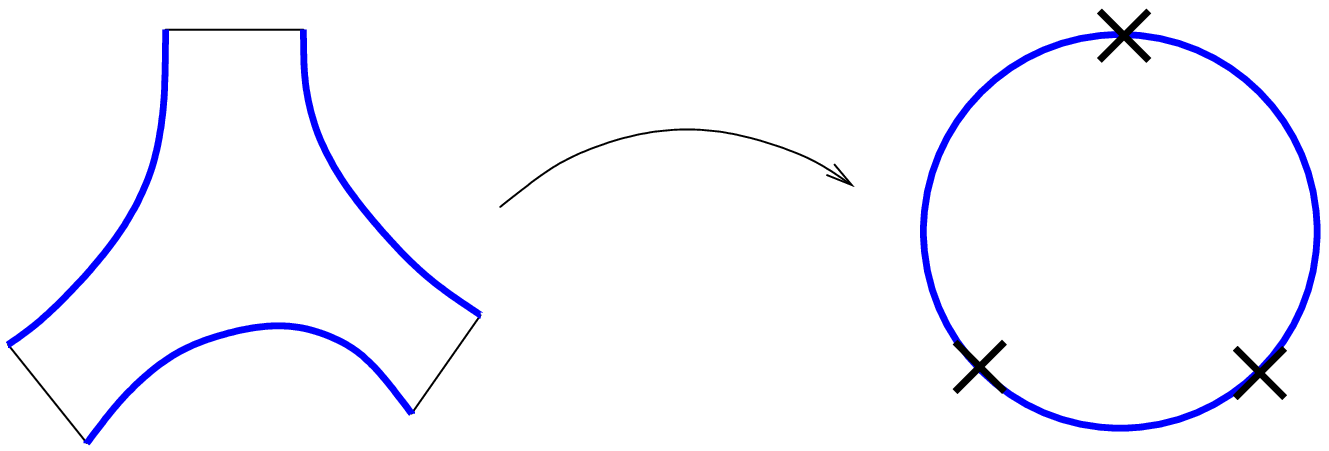}\qquad
\includegraphics[width = .45\textwidth]{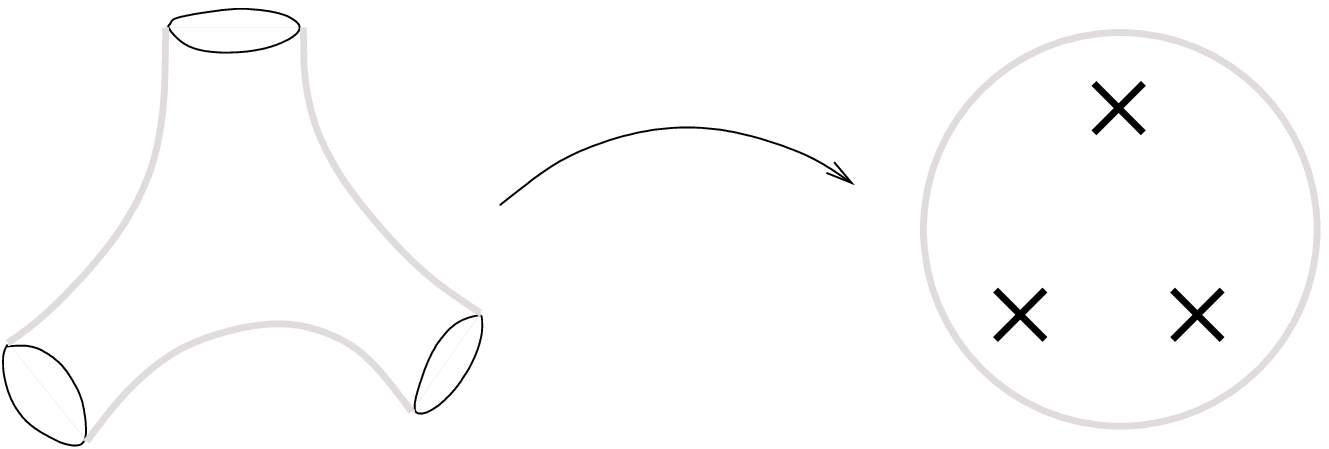}
$$
\caption{Open and closed strings conformally mapped to a disc and
a sphere respectively. The background $B$-field affects only
the open strings ending at the boundary of the disc and is not
felt by the closed strings.
Noncommutativity arises essentially since the open string vertex operators
at the boundary of
the disk cannot freely move past each other, while the closed string
vertex operators in the bulk are not affected.} \label{stringvertex}
\end{figure}

Let us briefly recall how star products and
noncommutative gauge theory arise
in string theory~\cite{CH,Scho,SW}:
Consider an open string $\sigma$-model with background
$B$-field
\eq
S_B = \int_\Sigma\left(\frac{1}{4\pi\alpha'} g_{ij}\pp_a x^i\pp^a x^j +
\frac{1}{2i}  B_{i\!j}\, \epsilon^{ab} \pp_a x^i \pp_b x^j\right)
\en
where the integral is over the string world-sheet and
$B$ is constant, nondegenerate and $dB = 0$.
The equations of motion give the boundary conditions
\eq
g_{ij} \pp_n x^i + 2\pi i \alpha' B_{ij} \pp_t x^j|_{\pp\Sigma} = 0.
\label{bry}
\en
The effect of the $B$-field is only felt by open strings and not by close
strings, since it is only relevant at the boundary and
not in the bulk (see figure~\ref{stringvertex}.)
The propagator at boundary points compatible with the boundary conditions
(\ref{bry})
is
\eq
\langle x^i(\tau) x^j(\tau')\rangle = -\alpha' {G^{ij}} \ln (\tau -\tau')^2
+ \frac{i}{2} {\theta^{ij}}
\varepsilon(\tau-\tau'),
\en
where $\varepsilon(\tau-\tau')$ is the step function and
\eq
\frac{1}{g + 2\pi\alpha' B} = \frac{{\theta}}{2\pi\alpha'}
+ \frac{1}{G (+ 2 \pi \alpha' \Phi)} .
\en
(The term $2 \pi \alpha' \Phi$ is relevant in the so-called
intermediate picture.)
The correlation functions
on the boundary of the disc in the decoupling limit ($g \rightarrow 0$,
$\alpha' \rightarrow 0$; $\theta = 1/B$) for constant $B$ are
\eq
\left\langle f_1(x(\tau_1)) \cdot\ldots\cdot f_n(x(\tau_n))\right\rangle_B
= \int dx \, f_1 \star \ldots
\star f_n , \quad (\tau_1 < \ldots < \tau_n) \label{corr}
\en
with the Weyl-Moyal star product
\eq
(f \star g ) (x)  = \left. e^{\frac{i\hbar}{2}\theta^{i\!j} \pp_i \pp_j'} f(x) g(x') \right|_{x'
\rightarrow x} ,
\en
which is the deformation quantization of the \emph{constant} Poisson structure $\theta$.
More generally a star product is an associative, $[[\hbar]]$-bilinear product
\eq
f \star g = f g + \sum_{n=1}^\infty (i\hbar)^n
\underbrace{B_n(f,g)}_{\mathrm{bilinear}} ,
\en
which is the deformation of a Poisson structure $\theta$:
\eq
\scom{f}{g} = i \hbar \pcom{f}{g} + \OO(\hbar^2), \quad \pcom{f}{g} =
\theta^{i\!j}(x) \pp_i f \, \pp_j g .
\en
We now perturb the constant $B$ field by adding a gauge potential $a_i(x)$:
$B \rightarrow B + da$, $S_B \rightarrow S_B + S_a$, with
\eq
S_a = -i \int_{\partial D} d\tau a_i(x(\tau))\partial_\tau x^i(\tau).
\en
Classically we have the naive gauge invariance
\eq
\delta a_i = \partial_i \lambda, \label{naive}
\en
but in the quantum theory this depends on the choice of regularization.
For Pauli-Villars (\ref{naive}) remains a symmetry but
if one expands $\exp S_a$ and employes a point-splitting regularization
then the functional integral is invariant under noncommutative gauge
transformations\footnote{In this form this formula is 
only valid for the Moyal-Weyl
star product.}
\eq
\hat\delta \hat A_i = \partial_i \hl + i \hl \star \hat A_i - i \hat A_i \star \hl.
\label{nctrans}
\en
Since a sensible quantum theory should be independent of the choice of regularization
there should be field redefinitions 
$\hat A(a)$, $\hl(a,\lambda)$ (Seiberg-Witten map) that relate (\ref{naive}) and
(\ref{nctrans}):
\eq
\hat A(a) +\hat\delta_{\hl} \hat A(a) = \hat A(a+\delta_\lambda a). \label{swcond1}
\en
It is instructive to study the effect of the extra factor $\exp S_a$ in
the correlation function (\ref{corr})
in more detail: It effectively shifts the
coordinates\footnote{Notation: $\DD$ should not be confused with a 
covariant
derivative (but it is related).}
\eq
x^i \rightarrow x^i + \theta^{ij}\hat A_j =: \DD x^i . \label{covco1}
\en
More generally, for a function $f$,
\eq
f \rightarrow f + f_A =: \DD f. \label{covfu}
\en
The mapping $\AA: \, f \mapsto f_A$ plays the 
role of a generalized gauge potential; it maps a function
to a new function that depends on the gauge potential.
The shifted coordinates and functions are covariant under noncommutative
gauge transformations:
\eq
\hat\delta (\DD x^i) = i\scom{\hl}{\DD x^i}, \qquad
\hat\delta (\DD f) = i\scom{\hl}{\DD f}.
\en
The first expression implies (\ref{nctrans}) (for $\theta$ constant and
nondegenerate).

The covariant coordinates (\ref{covco1}) are the
background independent operators of \cite{SW,Seiberg}; 
they and the covariant functions (\ref{covfu}) can also
be introduced abstractly in the general case of an
arbitrary noncommutative space as we shall discuss
in the next section.

\section{Gauge theory on noncommutative spaces}

\subsection{Covariant functions, covariant coordinates}

Take a more or less arbitrary noncommutative space, i.e.\ an
associative unital algebra $\Ax$ of noncommuting 
variables with multiplication~$\star$
and consider (matter) fields $\psi$ on this space.
The fields  can be taken to be elements of $\Ax$, or, more generally, 
a left module of it.
The notion of a gauge transformation is introduced 
as usual\footnote{We shall
often use the
infinitesimal version $\delta \psi = i\lambda\star\psi$ 
of~(\ref{gaugetrafo}) -- this is
purely for notational clarity. Other transformations, like, e.g.,
$\psi \mapsto \psi\star\Lambda$ or $\psi \mapsto \Lambda\star\psi\star\Lambda^{-1}$
can also be considered.}
\eq
\psi \mapsto \Lambda \star \psi , \label{gaugetrafo}
\en
where $\Lambda$ is an invertible element of $\Ax$.
In analogy to commutative geometry where a manifold can be described by the
commutative space of functions over it, we shall refer to the elements of $\Ax$
also as functions. Later we shall focus on the case where the noncommutative
multiplication is a star product; the elements of $\Ax$ are then in fact
ordinary functions in the usual sense of the word.
The left-multiplication  of a field with a function $f \in \Ax$ does in general
not result in a covariant object because of the noncommutativity of $\Ax$:
\eq
f\star\psi \mapsto f\star\Lambda\star\psi \neq \Lambda\star (f \star\psi) .
\label{noncov}
\en
(As in ordinary gauge theory the gauge transformation only acts on the
fields, i.e.\ on the elements of the left-module of $\Ax$ and not on 
the elements of $\Ax$ itself.) To cure (\ref{noncov}) we introduce covariant functions
\eq
\DD f = f + f_A ,
\label{covf}
\en
that transform under gauge transformations by conjugation
\eq
\DD f \mapsto \Lambda\star \DD f \star \Lambda^{-1} , \label{covft}
\en
by adding `gauge potentials' $f_A$ with appropriate transformation
property\footnote{Notation: $\scom{a}{b} \equiv a\star b - b \star a \equiv
[a,b]_\star$.}
\eq
f_A \mapsto \Lambda\star \scom{f}{\Lambda^{-1}} + \Lambda\star f_A \star
\Lambda^{-1} . \label{potentials}
\en
Further covariant objects can be constructed from covariant functions; 
the `2-tensor'
\eq
\FF(f,g) = \scom{\DD f}{\DD g} - \DD(\scom{f}{g}) , \label{tensor}
\en
for instance plays the role of covariant noncommutative field strength. 

\subsubsection{Canonical structure (constant $\mathbf \theta$) 
and noncommutative Yang-Mills}

Before we continue let
us illustrate all this in the particular simple case of an algebra $\Ax$ generated
by `coordinates' $x^i$ with canonical commutation relations
\eq
\scom{x^i}{x^j} = i\theta^{ij} , \quad \theta^{ij} \; \in \; \BC.
\label{canonical}
\en
This algebra arises in the decoupling limit of open strings in 
the presence of a \emph{constant} $B$-field. It
can be viewed as the quantization of a Poisson structure with
Poisson tensor $\theta^{ij}$ and the multiplication $\star$ is then the
Weyl-Moyal star product
\eq
f\star g = f e^{\frac{i}{2}\theta^{ij}
\stackrel{\leftarrow}{\partial_i} \otimes \stackrel{\rightarrow}{\partial_j}} g.
\en
(This formula holds only in the present example, where $\theta^{ij}$ is
constant and we shall also assume that it is non-degenerate. In the rest
of this chapter we drop both restrictions.) Let us focus on the coordinate functions
$x^i$. The corresponding covariant coordinates are
\eq
\DD x^i = x^i + x_A^i = x^i + \theta^{ij} \hat A_j, \label{covco}
\en
where we have used $\theta$ to lower the index on $\hat A_j$. 
Using (\ref{canonical}), we see that 
the transformation~(\ref{potentials}) of the 
noncommutative gauge potential $\hat A_j$ is 
\eq
\hat A_j \mapsto i\Lambda \star \pp_j(\Lambda^{-1}) + \Lambda \star \hat A_j
\star \Lambda^{-1},  \label{ncgt}
\en
or, infinitesimally
\begin{equation}
  \delta \hat A_j = \pp_j\lambda + i[\lambda,\hat A_j]_\star .
\end{equation}
The noncommutative field strength
\eq
\hat F_{kl} = \pp_k \hat A_l - \pp_l \hat A_k - i [\hat A_k,\hat A_l]_\star 
\label{ncfs}
\en
transforms covariantly 
\eq
\hat F_{kl} \mapsto \Lambda \star \hat F_{kl} \star \Lambda^{-1} .
\label{ncft}
\en
We have again used $\theta$ to lower indices to get
(\ref{ncfs}) from the definition (\ref{tensor})
\eq
i\hat F_{kl} \theta^{ik}\theta^{jl} \equiv
\FF(x^i,x^j) = [x_A^i,x^j]_\star + [x^i,x_A^j]_\star + [x_A^i,x_A^j]_\star . \label{covF}
\en
Note, that we should in general be more careful when using $\theta$ to 
lower indices as in (\ref{covco}) or (\ref{covF}) because this may spoil 
the covariance when $\theta$ is not constant as it was in this particular example. 
Relations (\ref{ncgt}), (\ref{ncfs}) and (\ref{ncft}) define what is usually called
Noncommutative Yang-Mills theory (NCYM) in the narrow sense: ordinary Yang-Mills
with all matrix products replaced by star products. This simple rule, however, only
really  works well for the Moyal-Weyl product, i.e.\ constant $\theta$.
In the general case it is wise to stick with the manifestly covariant and
coordinate-independent\footnote{We would like to thank Anton Alekseev for stressing 
the importance of this point.} objects defined in 
(\ref{covf}) and (\ref{tensor}). 
The fundamental objects  are really
the mappings (differential operators) $\DD$ and $\FF$ in these equations. 
The transformation of
$\AA = \DD - \id: f \mapsto f_A$
under gauge transformations is exactly
so that (\ref{covf})
transforms by conjugation.
The mappings $\AA \in \Hom(\Ax,\Ax)$ and $\FF \in \Hom(\Ax\wedge\Ax,\Ax)$ play the role
of generalized noncommutative gauge potential and noncommutative field strength.
There are several reasons, why one needs $\AA$ and $\DD$ and not just
$A^i \equiv \AA(x^i)$ (or $\hat A_i$, for $\theta$ constant):
If we perform a general coordinate transformation
$x^i \mapsto {x^i}'(x^j)$ and
transform $A^i$ (or $\hat A_i$) naively as its index structure suggests,
then we would obtain objects that are no longer covariant under
noncommutative gauge transformations. The correct transformation,
$\AA(x^i) \mapsto \AA'({x^i}')$, is more complicated and will be
discussed in section~\ref{sec:coordtransform}. 
Furthermore we may be interested
in covariant versions of scalar fields $\phi(x)$. These are given
by the corresponding covariant function $\DD(\phi(x))$.

In the next section we will propose an abstract definition of
the type of noncommutative gauge theory that is of present interest. Then we shall
proceed to give an interpretation in the framework of 
deformation quantization  
and will construct a particular important class of these operators.

\subsection{Cochains and projective modules}

Finite projective modules take the place of fiber bundles in the noncommutative
realm \cite{Connes}. This is also the case here, as we shall explain below,
but may not have been apparent 
since we have been working with component
fields as is customary in the physics literature.
We have argued in the previous section that $\AA \in C^1$, $\FF \in C^2$ with
$C^p = \Hom(\Ax^{\wedge p},\Ax)$, $C^0 \equiv \Ax$.
These $p$-cochains take the place of forms
on a noncommutative space $\Ax$, which for now is still an arbitrary associative 
algebra over a field $k$ with
multiplication~$\star$.
It is actually more convenient to start with the Hochschild complex
of $\Ax$, 
$H^p(\Ax,\Ax) = \Hom_k(\Ax^{\otimes p},\Ax)$,
with values in $\Ax$ considered as a left module of $\Ax$. 
(The formulas for $C^p$ can then be obtained by antisymmetrization.)
We have a coboundary operator
$\ds: H^p \rightarrow H^{p+1}$, $\ds^2 = 0$, $\ds 1 = 0$,
\eq
\ds \mathcal{C} = -\gcom{\mathcal{C}}{\star\,}, \label{dq}
\en
where $\gcom{}{}$ is the Gerstenhaber bracket (\ref{gerstenhaber}),
\eqa
\lefteqn{(\ds \mathcal{C})(f_1,\ldots,f_{p+1})\; = \;
f_1 \star \mathcal{C}(f_2,\ldots,f_{p+1}) 
-\mathcal{C}(f_1\star f_2, f_3, \ldots , f_{p+1}) +{}}&& \nn
 && {} + \mathcal{C}(f_1, f_2\star f_3, \ldots, f_{p+1}) 
 \mp \cdots 
   +(-)^{p} \mathcal{C}(f_1, f_2, \ldots,f_p\star f_{p+1}) \nn
  &&{} +(-)^{p+1} \mathcal{C}(f_1,\ldots,f_p) \star f_{p+1} , \label{hochschilddc}
\ena
and we have a cup product
$\star: H^{p_1} \otimes H^{p_2} \rightarrow H^{p_1 + p_2}$,
\eq
\left(\mathcal{C}_1\star\mathcal{C}_2\right)\left(f_1, 
\ldots,f_{p_1+p_2}\right)
= \mathcal{C}_1(  f_1,\ldots,f_{p_1} ) \star \mathcal{C}_2( f_{p_1+1},
    \ldots,f_{p_1+p_2}  ) .
\en
For a function $\lambda\in H^0 \equiv \Ax$ the coboundary operator is defined
as 
\eq
(\ds\lambda)(f) = f\star\lambda - \lambda\star f \label{hochschilddonf}
\en 
and the cup
product reduces to the multiplication $\star$ of $\Ax$ in the obvious way.
For this reason and since there seems to be little chance of confusion we have
used the same symbol $\star$ for the cup product and the
multiplication.
Let us apply the Hochschild formalism to the gauge transformation dependent
map $\DD \in H^1$ that we introduced in the definition of 
covariant functions (\ref{covf}) in the previous section. In view of the way that
$A^i \equiv \theta^{ij} \hat A_j$ appeared in the definition of covariant
coordinates~(\ref{covco}) we define an abstract noncommutative
gauge potential $\AA \in H^1$
\eq
\AA \equiv \DD - \id. \label{nca}
\en
Applying $\AA$ to coordinate functions in the setting of (\ref{covco}) 
we indeed 
recover $A^i$: 
\[
\AA(x^i) = \DD(x^i) - x^i =  A^i .
\]
Let us compute the behavior of $\AA$ under a gauge transformation. Using
(\ref{covft}) 
and the definitions of $\ds$ and
the cup product we find 
{\renewcommand\cup{\star}
\eq
\AA \mapsto \Lambda\star\ds\Lambda^{-1} + \Lambda\star\AA\star\Lambda^{-1},
\en
which gives (\ref{potentials}) when evaluated on a function.
The corresponding infinitesimal version is
\eq \label{deltaA}
\delta\AA =  i(-\ds\lambda + \lambda\cup\AA - \AA\cup\lambda).
\en
Next we introduce the ``Hochschild'' field strength $\FF_H \in H^2$ 
\eq  \label{FHS}
\FF_H \equiv \ds \AA + \AA\cup\AA 
\en
and compute
\eq
\FF_H(f,g) = \DD f \star \DD g - \DD(f\star g) \label{symF} 
\en
and find the Bianchi identity
\eq \label{dF}
\ds \FF_H + \AA\cup\FF_H - \FF_H\cup\AA = 0.
\en
Evaluated on three functions $f$, $g$, $h$ the latter reads
\eq
\DD((f\star g)\star h) - \DD(f\star(g\star h))
+(\DD f\star \DD g) \star \DD h -
\DD f \star (\DD g \star \DD h),
\en
which is zero by associativity of $\Ax$.
$\FF_H$ transforms covariantly under gauge transformations
\eq
\FF_H \mapsto \Lambda\star\FF_H\star\Lambda^{-1};
\en
infinitesimally
\eq \label{deltaF}
\delta\FF_H = i (\lambda\cup\FF_H - \FF_H\cup\lambda).
\en
}%
When we compare
(\ref{tensor}) and (\ref{symF}), we see (as expected) that our noncommutative field 
strength $\FF$ of the 
previous section  is an antisymmetric version of $\FF_H$. 
We can obtain $\FF$ directly by taking the antisymmetrized version of
Hochschild, where one considers $\Ax$ as a Lie algebra with
bracket $\scom{a}{b} = a\star b - b\star a$; this is the Chevalley
cohomology of $\Ax$ with values in $\Ax$: $C^p = \Hom_k(\Ax^{\wedge p},\Ax)$.
We find the relevant formulas in this setting by replacing $H^p$ with $C^p$
(whose elements are 
antisymmetric), and
by using corresponding antisymmetrized
formulas for the coboundary operator $\ds$ and the cup product
which we then denote by $\wedge$. The action of
$\mbox{Lie} \Ax$ on the module $\Ax$ is given by $\star$-multiplication as before.
We now see that equation (\ref{tensor}),
\eq
\FF(f,g) = \scom{\DD f}{\DD g} - \DD(\scom{f}{g}), \label{antisymF} 
\en
can be written
\eq
\FF \equiv \ds \AA + \AA\wedge\AA. \label{asymF}
\en
The remaining equations also do not change in form (as compared to the Hochschild case): (\ref{asymF}) implies
\eq
\ds \FF + \AA\wedge\FF - \FF\wedge\AA = 0 \label{dasymF}
\en
and the behavior under (infinitesimal )gauge transformations is
\eqa
\delta\AA &= & i(-\ds\lambda + \lambda\wedge\AA - \AA\wedge\lambda) , \label{deltaaA} \\
\delta\FF &= & i\left( \lambda\wedge\FF - \FF\wedge\lambda\right). \label{deltaaF}
\ena
Equations (\ref{asymF}), (\ref{dasymF}), (\ref{deltaaA}) and (\ref{deltaaF}) 
are reminiscent of
the corresponding equations of ordinary (nonabelian) gauge theory. The correspondence
is given by the following dictionary: One-forms become linear operators on $\Ax$
which
take one function as argument and yield a new function,
two-forms become bilinear operators on $\Ax$ which take two functions as arguments
and return one new function, and the Lie bracket is replaced
by the antisymmetrized cup product $\wedge$.\footnote{More educated:
$n$-forms become $n$-cochains. An even closer match with the usual
physics conventions is achieved by multiplying our $\ds$ and $\FF$ by $i$
(section~\ref{ordinary} and later: multiply by $i\hbar$).}
As in ordinary gauge theory, it may
not be possible to use one globally defined gauge potential $\AA$; we may need
to introduce several $\DD$ and corresponding gauge potentials $\AA$ for
functions defined on different ``patches''. We shall come back to this later.
 
\noindent \emph{Remark:}
One reason for going through the slightly more general Hochschild 
construction first is that
the symmetric part of $\FF_H$ may also contain interesting information
as we will see in section~\ref{metric}.
For invertible $\DD$ there is still another interesting object:
\eq
\widetilde\FF \equiv \DD^{-1} \circ \FF
\en
measures noncommutativity:
\eq
\widetilde\FF(f,g) = [f\stackrel{\star'}{,} g] - [f\stackrel{\star}{,} g], \label{ftilde}
\en
where the (associative) product $\star'$ is defined by
$f\star' g = \DD^{-1}\Big(\DD f \star \DD g\Big)$.
This ``field strength'' satisfies the Cartan-Maurer equation
\eq
\ds\widetilde\FF = \gcom{\widetilde\FF}{\widetilde\FF}.
\en

\subsubsection{Projective modules}

We shall now discuss how our formulae fit into the framework of finite
projective modules: 
The calculus of $p$-cochains in $C^p$ with the
coboundary operator $\ds$ uses only the algebraic structure of $\Ax$; it is
related to the standard universal calculus and one can obtain
other calculi by projection.
Consider a (finite) projective right $\Ax$-module $\EE$.
We introduce a connection on $\EE$ as a linear map
$\nabla : \EE \otimes_{\Ax} \!C^p \rightarrow \EE \otimes_{\Ax} C^{p+1}$
for $p \in \mathbb{N}_0$
which satisfies the Leibniz rule
\eq
\nabla(\eta \psi)
 =  (\tilde\nabla \eta) \psi
+ (-)^p \eta\,\dst\psi \nonumber
\en
for all $\eta\in\EE \otimes_{\Ax} \!C^p$, $\psi \in C^r$, and where
$\tilde\nabla\eta = \nabla\eta - (-)^p \eta\,\dst 1$,
\eq
\dst(a\wedge\psi) = (\ds a)\wedge\psi + (-)^q a \wedge(\dst\psi)
\en
for all $a \in C^q$, and $\dst 1$ is the identity operator on $\Ax$.
(The transformation of matter
fields $\deltaq \psi = i \hat\lambda\star\psi$ leads to a
slight complication here; for fields
that transform in the adjoint (by star-commutator) we would only need $\tilde\nabla$,
$\ds$ and not $\nabla$ and $\dst$.) 
Let $(\eta_a)$ be a  generating family for $\EE$;
any $\xi \in \EE$ can then be written as $\xi = \sum \eta_a \psi^a$ with $\psi^a \in \Ax$
(with only a finite number of terms different from zero). For a free module
the $\psi^a$ are unique, but we shall not assume that. Let the generalized
gauge potential be defined by the action of $\tilde\nabla$ on the elements
of the generating family: $\tilde\nabla \eta_a = \eta_b \AA^b_a$. In the
following we shall suppress indices and simply write $\xi = \eta.\psi$,
$\tilde\nabla \eta = \eta.\AA$ etc. We compute
\eq
\nabla\xi = \nabla(\eta.\psi) = \eta.(\AA\wedge\psi + \dst \psi)
= \eta.(\DD\wedge\psi).
\en
Evaluated on a function $f\in\Ax$ the component $\DD\wedge\psi$
yields a covariant function
times the matter field, $(\DD\wedge\psi) (f) = (\DD f)\star\psi$,
so in this framework covariant functions are related to the covariant
``derivative''
$\dst + \AA$:
\eq
[(\dst + \AA)\psi](f) = (f + \AA(f))\star\psi .
\en
The square of the connection gives
\eq
\nabla^2 \xi = \eta.(\AA\wedge\AA + \ds\AA).\psi = \eta.\FF.\psi
\en
with the field strength
\eq
\FF =  \ds\AA + \AA \wedge \AA.
\en

\section{Ordinary versus noncommutative gauge theory}
\label{ordinary}

We are particularly interested in the case where the algebra of our noncommutative
space $\Ax$ is given by a star product (via a quantization map).
A star product on a smooth $C^\infty$-Manifold $\MM$ is an associative
$\BC[[\hbar]]$-bilinear product
\eq
f \star g = f g + \sum_{n=1}^\infty \left(\frac{i\hbar}{2}\right)^n B_n(f,g) ,
\quad f,g \in C^\infty(\MM) ,
\en
where $B_n$ are bilinear operators and $\hbar$ is the formal deformation 
parameter; it is a deformation quantization of
the Poisson structure
\eq
\{ f, g\} \equiv \theta^{ij}(x) \pp_i f \, \pp_j g = B_1(f,g) - B_1(g,f) .
\en
Equivalent star products $\tilde\star$ can be constructed with the help
of invertible operators $D$
\eq
D(f \,\tilde\star\, g) = D f \star D g , \qquad D f \equiv f + \sum_{n=1}^\infty \hbar^n
D_n(f), \label{equiv}
\en
where $D_n$ are linear operators. This operation clearly does not spoil
associativity. There are also inner automorphisms for each invertible 
element~$\Lambda$
and their infinitesimal version, inner derivations,
\eq
f \mapsto \Lambda \star f \star \Lambda^{-1} , 
\qquad \delta f = \scom{i \lambda}{f}; \label{iauto}
\en
these operations do not change the star product.

The striking similarity between equations (\ref{covf}), (\ref{covft}) and equations
(\ref{equiv}), (\ref{iauto}) suggests the following interpretation of noncommutative gauge
theory in the star product formalism:
The covariance maps $\DD = \id + \AA$ are gauge equivalence maps $D$ for the
underlying star product,
combined with a change of coordinates $\rho^*$
\eq
\DD = D \circ \rho^*, \qquad \DD (f\star' g) = \DD f \star \DD g ;
\en
gauge transformations are inner automorphisms of the star
products.

\subsection*{Motivation from string theory} 

A Poisson tensor $\theta$ enters the discussion of Seiberg and 
Witten~\cite{SW} via a background 
$B$-field in the open string picture. In this setting
\eq
\theta^{ij} = 2\pi\alpha'\left(\frac{1}{g+2\pi\alpha' B}\right)^{ij}_A 
\en
appears in the propagator at boundary points of the string world sheet.
($g$ is the closed string metric and $A$ denotes the antisymmetric
part of a matrix.)
The 2-form
$\omega \equiv \frac{1}{2} B_{ij} dx^i\wedge dx^j$ is a symplectic form, provided
$B$ is nondegenerate and
$d\omega = 0$, which is e.g.\ obviously the case if $B$ is constant (but we
shall not require it to be constant.)
In the zero slope limit (or in the intermediate picture with $\Phi = -B$~\cite{SW},
see section~\ref{DBI})
\eq
\theta^{ij} = (B^{-1})^{ij}
\en
defines then a Poisson structure. It has been discussed by several authors how
the Moyal-Weyl star product enters the picture as a quantization of this
Poisson structure in the constant case~\cite{CH,Scho,SW}. 
A direct approach that is most suitable for our
purposes and that also works for non-constant $\theta$ is given by 
Cattaneo and Felder~\cite{CattaneoFelder}
in their QFT realization of Kontsevich's star product.

In the introduction we have discussed what happens if we add fluctuations $f$ (with
$df = 0$, i.e. locally $f = da$) to the background $B$ field: The action
is then naively invariant under ordinary gauge transformations $\delta a = d\lambda$,
but the invariance of the quantum theory depends on the choice of
regularization. A point splitting prescription \cite{SW,AD} leads in fact to
a noncommutative gauge invariance. Since in a consistent quantum theory the choice of 
regularization should not matter, Seiberg and Witten argued that there should
exist maps that relate ordinary and noncommutative gauge theory such that
(\ref{SWcond}) holds.
A more abstract argument that leads to the same conclusion, but gives the 
Seiberg-Witten maps more directly and also works for
non-constant $\theta$
can be based on a quantum version of Moser's lemma \cite{JS,JSW}.
Here is briefly the idea, in the next section we will review the details:
The addition of $f = da$ to $B$  defines
a new Poisson structure
\eq
\theta = \frac{1}{B} \quad\rightarrow\quad \theta' = \frac{1}{B + f}
\label{BplusF}
\en
which, according to Moser's lemma, is related to the original one by a change of
coordinates given by a flow $\rho^*_a$ that depends on the gauge potential $a$.
After quantization $\theta$ and $\theta'$ give rise to equivalent
star products $\star$ and $\star'$.
The equivalence map $D_a$, the full quantum flow
$\DD_a = D_a \circ \rho^*_a$
and the noncommutative gauge potential 
\eq
\AA_a = \DD_a - \id
\en
are also functions of $a$.
An additional infinitesimal gauge transformation 
\eq
\delta a = d\lambda
\en
does not change the Poisson structure (since $\delta f = 0$),
but it still induces an infinitesimal canonical transformation.
After quantization that transformation becomes an inner derivation of the
star product $\star$ and thus a noncommutative gauge
transformation
$\delta_{\hat\lambda}$  with $\hat\lambda = \hat\lambda(\lambda,a)$,
such that
\eq
\AA_{a + d\lambda} = \AA_a + \delta_{\hat\lambda}\AA_a . \label{swcond}
\en

\subsection{Seiberg-Witten map and Kontsevich formality}
\label{swmapkont}

Consider an abelian gauge theory on a manifold that also carries
a Poisson structure~$\theta$. The gauge potential, field strength
and infinitesimal gauge transformations are
\eq
a = a_i dx^i, \qquad f= \frac{1}{2}f_{ij}\,dx^i\wedge dx^j = da,
\qquad f_{ij} = \pp_i a_j - \pp_j a_i, \qquad \delta_\lambda a = d\lambda.
\en
We will first construct a semiclassical version of the Seiberg-Witten map,
where all star commutators are replaced by Poisson brackets. The construction
is essentially a formal generalization of Moser's lemma to Poisson manifolds.

\subsubsection{Semi-classical construction}
\label{sec:semiclass}

Let us consider the  nilpotent coboundary operator of
the Poisson cohomology (see \cite{Weinstein}) -- the semiclassical limit
of (\ref{dq}) -- 
\eq
\dpo = -\sncom{\,\cdot\,}{\theta},
\en
where $\sncom{}{}$ is the Schouten-Nijenhuis bracket (\ref{sncom})
and $\theta = \frac{1}{2} \theta^{ij} \partial_i\wedge\partial_j$ is
the Poisson bivector.
Acting with $\dpo$ on a function $f$ gives the Hamiltonian vector field corresponding
to $f$
\eq
\dpo f =  \{\,\cdot\,,f\} = \theta^{ij}(\pp_j f)\pp_i. \label{hamvec}
\en
It is natural to introduce a vector field 
\eq
\ap = a_i \dpo x^i = \theta^{ji} a_i \pp_j \label{vf}
\en
corresponding to the abelian gauge
potential $a$ and a bivector field
\eq
\fp = \dpo\ap = - \frac{1}{2}\theta^{ik} f_{kl} \theta^{lj}\, \pp_i\wedge\pp_j
\label{fs}
\en
corresponding to the abelian field strength $f = da$.
We have $\dpo\fp = 0$,
due to $\dpo^2 \propto \sncom{\theta}{\theta} = 0$
(Jacobi identity).

We are now ready to perturb the Poisson structure $\theta$ by introducing
a one-parameter deformation $\theta_t$ with $t \in [0,1]$:%
\footnote{In this
notation the equations resemble those of Moser's original lemma, which deals with
the symplectic 2-form $\omega$, the inverse of $\theta$ (provided it exists).
There, e.g., $\pp_t \omega_t = f$ for $\omega_t = \omega + t f$.}
\eq
\pp_t \theta_t = \fp[_t]  \label{evolution}
\en
with initial condition $\theta_0 =
\theta$. In local coordinates:
\eq
\pp_t \theta_t^{ij} = -(\theta_t f \theta_t)^{ij} ,\qquad \theta_0^{ij} = \theta^{ij},
\label{partialtf}
\en
with formal solution given by the geometric series
\eq
\theta_t = \theta - t \theta f \theta + t^2 \theta f \theta f \theta
- t^3 \theta f \theta f \theta f \theta \pm \cdots
= \theta \frac{1}{1 + t f \theta},
\en
if $f$ is not explicitly $\theta$-dependent.
(The differential equations (\ref{evolution}), (\ref{partialtf}) and the rest
of the construction
do make sense even if $f$ or $a$ are
$\theta$-dependent).
$\theta_t$ is a Poisson tensor for all $t$ because
$\sncom{\theta_t}{\theta_t}= 0$ at $t=0$ and
\[
\pp_t\sncom{\theta_t}{\theta_t}
= -2\dpo[_t]\fp[_t] \propto \sncom{\theta_t}{\theta_t} .
\]
The evolution (\ref{evolution}) of $\theta_t$ is generated by the vector field $\ap$:
\eq
\pp_t\theta_t = \dpo[_t]\ap[_t] = -\sncom{\ap[_t]}{\theta_t}.
\en
This Lie derivative can be integrated to a flow (see
appendix~\ref{s:t-evolution})
\eq
\rho^*_a = \left.\exp(\ap[_t] + \pp_t)\exp(-\pp_t)\right|_{t=0} \label{flow}
\en
that relates the Poisson structures $\theta' = \theta_1$ and $\theta = \theta_0$.
In analogy to (\ref{nca}) we define a semi-classical (semi-noncommutative)
generalized gauge potential
\eq
A_a = \rho^*_a - \id. \label{as}
\en
Under an infinitesimal gauge transformation $a \mapsto a + d\lambda$ the
vector field (\ref{vf}) changes by a Hamiltonian vector field $\dpo\lambda
= \theta^{ij} (\pp_j\lambda)\pp_i$:
\eq
\ap \mapsto \ap + \dpo\lambda. \label{astrafo}
\en
Let us compute the effect of this gauge transformation on the flow (\ref{flow}).
After some computation (see appendix~\ref{lambdatilde})
we find (infinitesimally: to first order in
$\lambda$)
\eq
\rho^*_{a+d\lambda}
=  (\id + \dpo\tilde\lambda)\circ\rho^*_a,
\quad\mbox{i.e.,}\quad
\rho^*_{a+d\lambda}(f)= \rho^*_a(f) + \{\rho^*_a(f),\tilde\lambda\}
\label{ctrafo}
\en
and
\eq
A_{a+d\lambda} = A_a + \, \dpo\tilde\lambda + \{A_a,\tilde\lambda\},
\label{cswcond}
\en
with
\eq
\tilde\lambda(\lambda,a) =
\sum_{n=0}^\infty \left.\frac{(\ap[_t] + \pp_t)^n(\lambda)}{(n+1)!}
\right|_{t=0}.
\label{SW2}
\en
Equations (\ref{SW2}) and (\ref{as}) with (\ref{flow}) are explicit
semi-classical versions of the Seiberg-Witten map.
The semi-classical (semi-noncommutative) generalized field strength
evaluated on two functions (e.g.\ coordinates) $f$, $g$  is
\eq
F_a(f,g) = \pcom{\rho^* f}{\rho^* g} - \rho^*\pcom{f}{g}
         = \rho^*\left(\pcom{f}{g}' - \pcom{f}{g}\right). \label{scgfs}
\en
Abstractly as 2-cochain:
\eq
F_a = \rho^* \circ \frac{1}{2}(\theta' -\theta)^{ik} \pp_i\wedge\pp_k
    = \rho^* \circ \frac{1}{2} (f')_{jl} \theta^{ij}\theta^{kl}\pp_i\wedge\pp_k
\en
with $\theta' f = \theta f'$, or
\eq
f' = \frac{1}{1 + f \theta} f,
\en
which we recognize as the  noncommutative field strength (with lower indices)
for constant $f$, $\theta$~\cite{SW}. The general result for non-constant $f$, $\theta$
is thus simply obtained by the application of the covariantizing map $\rho^*$
(after raising indices with $\theta$'s).

The Seiberg-Witten map in the semiclassical regime for constant $\theta$
has previously been discussed in~\cite{Cornalba1,Ishibashi}, where
it was understood as a coordinate
redefinition that eliminates fluctuations around a constant background.

We will now  use
Kontsevich's formality theorem to quantize everything. The goal is to
obtain (\ref{SWcond}) in the form (\ref{swcond})
of which (\ref{cswcond}) is the semi-classical limit.

\subsubsection{Kontsevich formality map}

Kontsevich's formality map 
is a collection  of skew-symmetric multilinear maps $U_n$ for $n=0 \ldots \infty$
that map tensor products of $n$ polyvector fields to differential operators.
More precisely
$U_n$ maps the tensor product
of $n$ $k_i$-vector fields to an $m$-differential operator, where $m$ is determined
by the matching condition
\eq
m = 2 - 2n + \sum_{i=1}^n k_i. \label{matching}
\en
$U_1$ in particular is the natural map from a $k$-vector field to a
$k$-differential operator
\eq
U_1(\xi_1\wedge\ldots\wedge\xi_k)(f_1,\ldots,f_k) = 
\frac{1}{k!}\sum_{\sigma\in\Sigma_k}
\mathrm{sgn}(\sigma) \prod_{i=1}^k \xi_{\sigma_i}(f_i),
\en
and $U_0$ is defined to be the ordinary multiplication of functions:
\eq
U_0(f,g) = fg.
\en
The $U_n$, $n \geq 1$, satisfy the formality condition~\cite{Kontsevich}
\eqa
\lefteqn{d_\mu U_n(\al_1,\ldots,\al_n) + \frac{1}{2}
\sum_{{}^{I\sqcup J = (1,\ldots,n)}_{I,J \neq \emptyset}} \pm
\gcom{U_{|I|}(\al_I)}{U_{|J|}(\al_J)}}&& \nn
&&=\sum_{i<j} \pm U_{n-1}\left(\sncom{\al_i}{\al_j},\al_1,\ldots,\widehat\al_i,
\ldots,\widehat\al_j,\ldots,\al_n\right),
\ena
where $d_\mu \CC \equiv -\gcom{\CC}{\mu}$, with the commutative multiplication $\mu(f,g)
= f\cdot g$
of functions; the hat  marks an omitted vector field.
See~\cite{Kontsevich,CattaneoFelder} for explicit constructions and more details
and~\cite{Arnal,Kontsevich} for the definition of the signs  in
this equation. In the following we collect the three
special cases that we actually use in this chapter.

Consider the formal series (see also~\cite{Manchon})
\eq
\Phi(\alpha) = \sum_{n=0}^\infty \frac{(i\hbar)^n}{n!}
U_{n+1}(\alpha,\theta,\ldots,\theta).
\en
According to the matching condition (\ref{matching}),
$U_{n+1}(\alpha,\theta,\ldots,\theta)$ is a
tridifferential operator for every $n$ if $\alpha$ is a trivector field,
it is a bidifferential operator if $\alpha$ is a bivector field,
it is a differential operator if $\alpha$ is a vector field and
it is a function if $\alpha$ is a function; in all cases $\theta$
is assumed to be a bidifferential operator.

\paragraph{Star products from Poisson tensors}

A Poisson bivector $\theta$ gives rise to a star product via the formality map:
According to the matching condition (\ref{matching}),
$U_n(\theta,\ldots,\theta)$ is a
bidifferential operator for every $n$ if $\theta$ is a bivector field.
This can be used to define a product
\eq
f \star g  =  \sum_{n=0}^\infty \frac{\left(i\hbar\right)^n}{n!} 
U_n(\theta,\ldots,\theta)(f,g) 
 =  f g + \frac{i\hbar}{2}\,\theta^{ij}\pp_i f\,\pp_j g + \cdots \; .
\label{kstar}
\en
The formality condition
implies
\eq
\ds \star = i\hbar\Phi(\dpo \theta), \label{control}
\en
or, $\gcom{\star}{\star} = i\hbar\Phi(\sncom{\theta}{\theta})$,
i.e., 
associativity of $\star$,
if $\theta$ is Poisson.
(If $\theta$ is not Poisson, i.e., has non-vanishing
Schouten-Nijenhuis bracket $\sncom{\theta}{\theta}$,
then the product $\star$ is not associative, but the
non-associativity is nevertheless under control
via the formality condition by (\ref{control}).)

\paragraph{Differential operators from vector fields}

We can define a
linear differential operator\footnotemark\addtocounter{footnote}{-1}
\eq \label{vecdiff}
\Phi(\xi)
= \xi + \frac{(i\hbar)^2}{2}
U_3(\xi,\theta,\theta) + \cdots
\en
for every vector field $\xi$.
For $\theta$ Poisson the formality condition gives
\eq
\ds \Phi(\xi) = i\hbar \Phi(\dpo\xi)
= i\hbar \,\dpo\xi + \cdots \; .
\label{dderiv}
\en
Vector fields $\xi$ that preserve the Poisson bracket, $\dpo\xi =
-\sncom{\theta}{\xi} = 0$, give rise to derivations of the star product
(\ref{kstar}): From (\ref{dderiv}) and the definition (\ref{hochschildd}),
(\ref{hochschilddc}) of $\ds$
\eq 
0 = [\ds\Phi(\xi)](f,g) = -[\Phi(\xi)](f\star g) + 
f \star [\Phi(\xi)](g) + [\Phi(\xi)](f)
\star g .
\en

\paragraph{Inner derivations from Hamiltonian vector fields}

Hamiltonian vector fields $\dpo f$ give rise to inner derivations of the
star product (\ref{kstar}):
We can define
a new function\footnote{$U_2(\xi,\theta) = 0$ and $U_2(f,\theta)=0$
by explicit computation of Kontsevich's formulas.}
\eq
\hat f \equiv \Phi(f) =
f + \frac{(i\hbar)^2}{2}
U_3(f,\theta,\theta) + \cdots  \label{fhat}
\en
for every function $f$.
For $\theta$ Poisson the formality condition gives
\eq
\ds \hat f = i\hbar \Phi(\dpo f)
\label{dfhat}
\en
Evaluated on a function $g$, this reads
\eq
[\Phi(\dpo f)](g) = \frac{1}{i\hbar}\scom{g}{\hat f} . 
\en
The Hamiltonian vector field $\dpo f$ is thus mapped to the inner derivation
$\frac{i}{\hbar}\scom{\hat f}{\,\cdot\,}$.

\subsubsection{Quantum construction}

The construction mirrors the semiclassical one, the exact correspondence
is given by the formality maps $U_n$ that are skew-symmetric 
multilinear  maps that take  $n$ polyvector fields
into a polydifferential operator.  We start with the differential
operator 
\begin{equation}
  \label{astar}
  \as = \sum_{n=0}^\infty \frac{\left(i\hbar\right)^{n}}{n!}
U_{n+1}(\ap,\theta,\ldots,\theta),
\end{equation}
which is
the image of the vector field $\ap$ under the formality map 
(\ref{vecdiff}); then we use the coboundary operator 
$\ds$ (\ref{dq}) to define
a bidifferential operator
\begin{equation}
  \label{fstar}
  \fs = \ds\as.
\end{equation}
This is the image of $\fp = \dpo\ap$ under
the formality map:
\eq 
  \label{fstarformal}
  \fs = \sum_{n=0}^\infty 
  \frac{(i\hbar)^{n+1}}{n!}U_{n+1}(\fp,\theta,\ldots,\theta).
\en
A $t$-dependent Poisson structure (\ref{evolution}) induces a $t$-dependent
star product via (\ref{kstar})
\begin{equation}
  \label{tstar}
  g \star_t h = \sum_{n=0}^\infty
  \frac{(i\hbar)^{n}}{n!} U_n(\theta_t,\ldots,\theta_t)(g,h).
\end{equation}
The $t$-derivative of this equation is
\begin{equation}
  \label{tstarder}
  \pp_t(g \star_t h) =  \sum_{n=0}^\infty
  \frac{(i\hbar)^{n+1}}{n!} 
  U_{n+1}(\fp[_t],\theta_t,\ldots,\theta_t)(g,h),
\end{equation}
where we have used (\ref{evolution}) 
and the skew-symmetry and multi-linearity
of $U_n$.
Comparing with (\ref{fstarformal}) we find
\begin{equation}
  \label{tstarf}
  \pp_t(g\star_t h) = \fs[_t](g,h),
\end{equation}
or, shorter, as an operator equation: $\pp_t(\star_t) = \fs[_t]$. But
$\fs[_t] = \ds[_t] \as[_t] = -\gcom{\as[_t]}{\star_t}$,
so the $t$-evolution is generated by the differential operator
$\as[_t]$
and can be integrated to a flow (see appendix~\ref{s:t-evolution})
\begin{equation}
  \label{Dq}
  \DD_a = \left.\exp(\as[_t] + \pp_t)\exp(-\pp_t)\right|_{t=0} ,
\end{equation}
that relates the star products $\star' = \star_1$
and $\star = \star_0$, and that defines the generalized noncommutative
gauge potential
\begin{equation}
  \label{ncgaugepot}
  \AA_a = \DD_a - \id .
\end{equation}
The transformation of $\as$ under 
an infinitesimal gauge transformation $a \mapsto a +  d\lambda$
can be computed from (\ref{astrafo}) with the help of (\ref{dfhat}),
see (\ref{hochschilddonf}):
\begin{equation}
  \label{deltaaq}
  \as \mapsto \as + \frac{1}{i\hbar} \ds\hat\lambda .
\end{equation}
The effect of this transformation on the quantum flow and on the noncommutative
gauge potential are (see appendix~\ref{lambdatilde})
\eq
\DD_{a+d\lambda} = (\id +\frac{1}{i\hbar} \ds\hat\Lambda)\circ\DD_a,
\quad\mbox{i.e.,}\quad\DD_{a+d\lambda}(f) = 
\DD_a f + \frac{i}{\hbar}\scom{\hat\Lambda}{\DD_a f}
\label{Dadl}
\en
\eq
\AA_{a+d\lambda} = \AA_a + \frac{1}{i\hbar}\left(\ds\hat\Lambda 
- \hat\Lambda\star\AA +\AA\star
\hat\Lambda\right).
\en
with
\eq
\hat\Lambda(\lambda,a) =
\sum_{n=0}^\infty \left.\frac{(\as[_t] + \partial_t)^n(\hat\lambda)}{(n+1)!}
\right|_{t=0}.
\label{tildehatlambda}
\en
Equations (\ref{ncgaugepot}) with (\ref{Dq}) and (\ref{tildehatlambda}) are
explicit versions of the abelian Seiberg-Witten map to all orders
in $\hbar$. They are unique up to
(noncommutative) gauge transformations.
Perhaps more importantly this construction provides us with
an explicit version of the ``covariantizer'' $\DD_a$ (the equivalence
map that sends coordinates and functions to their covariant
analogs) in terms of a finite number of (classical) fields
$a_i$.
The noncommutative gauge parameter (\ref{tildehatlambda}) also
satisfies the consistency condition
\eq
\delta_\alpha \hat\Lambda(\beta,a) - \delta_\beta \hat\Lambda(\alpha,a)
= \frac{i}{\hbar}\scom{\hat\Lambda(\alpha,a)}{\hat\Lambda(\beta,a)},
\en
with $\delta_\alpha (a_i) = \pp_i \alpha$, $\delta_\alpha(\beta) = 0$,
that follows from computing the commutator of abelian
gauge transformations on a covariant field~\cite{Jurco:2000ja}.

The generalized noncommutative field strength evaluated
on two functions (or coordinates) $f$, $g$ is
\eq
\FF_a(f,g) = \DD_a\left([f\stackrel{\star'}{,} g] - \scom{f}{g}\right).
\en

Up to order $\theta^2$ the series for $\AA_a$ and $\Lambda$
agree with the semiclassical results. In components:
\eq
\AA_a(x^i) = \theta^{ij} a_j + \frac{1}{2} \theta^{kl}a_l (\partial_k
(\theta^{ij} a_j) - \theta^{ij} f_{jk}) + \ldots ,
\en
\eq
\hat\Lambda = \lambda + \frac{1}{2} \theta^{ij} a_j \partial_i\lambda +
\frac{1}{6} \theta^{kl} a_l (\partial_k(\theta^{ij} a_j \partial_i \lambda)
-\theta^{ij} f_{jk} \partial_i \lambda) + \ldots .
\en

\noindent There are three major strategies for the computation of the Seiberg-Witten map:
\begin{enumerate}
\item[(i)] From the gauge equivalence condition (\ref{SWcond}) one can directly
obtain recursion relations for the terms in the Seiberg-Witten map.
For constant $\theta$ these can be cast in the form of differential
equations~\cite{SW}. Terms of low order in the gauge fields but all
orders in $\theta$ can be expressed in terms of
$\star_n$-products~\cite{Mehen:2000vs,LiuII}.\footnote{Some
motivation for the latter was provided
in~\cite{Das1}, and~\cite{Das2} provided some more concrete understanding of the
relationship of the generalized star product and Seiberg-Witten map.}

\item[(ii)] A path integral approach can be based on the relationship
between open Wilson lines in the
commutative and noncommutative picture~\cite{Okuyama,Okuyama:2001sw}.
\item[(iii)] The equivalence of the star products corresponding to
the perturbed and unperturbed Poisson structures leads to
our formulation in the framework of deformation quantization.
This allows a closed formula for the Seiberg-Witten map to
all orders in the gauge fields and in $\theta$.
\end{enumerate}

\subsection{Covariance and (non)uniqueness}
\label{sec:coordtransform}

The objects $\DD_a$ and $\hat\Lambda$ are not unique if all we ask is
that they satisfy the generalized Seiberg-Witten condition (\ref{Dadl})
with star product $\star$
and have the correct ``classical limit'' $\DD = \ap + \ldots$, $\hat\Lambda = \lambda
+ \ldots$.  The pair $\DD_2\circ\DD_a\circ\DD_1$, $\DD_2(\hat\Lambda)$,
where $\DD_2$ is an \mbox{$\star$-alge}\-bra automorphisms and $\DD_1$ is an equivalence map
(possibly combined with a change of coordinates of the
form $\id + o(\theta^2)$) is an equally valid solution. 
If we allow also a transformation to a new (but equivalent) star product 
$\overline\star$
then we can relax the condition on $\DD_2$: it may be any
fixed equivalence map possibly combined with a change of coordinates.
The maps (differential operators)
$\DD_1$ and $\DD_2$ may depend on the gauge potential $a_i$ via $f_{ij}$; it is 
important, however, that they are gauge-invariant.
The freedom in the choice of $\DD_1$ and $\DD_2$ represents the
freedom in the choice of coordinates and/or quantization scheme
in our construction; different $\DD_a$, $\hat\Lambda$
related by $\DD_1$, $\DD_2$ should be regarded as being equivalent.

\begin{figure}[htb]
\begin{center}
\vspace{1em}
\begin{picture}(0,0)%
\includegraphics{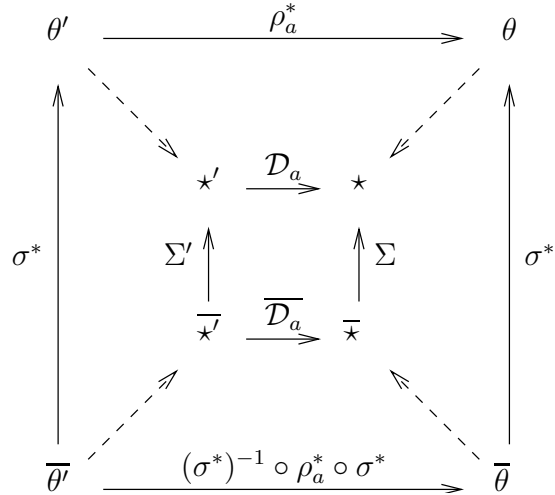}%
\end{picture}%
\setlength{\unitlength}{4144sp}%
\begingroup\makeatletter\ifx\SetFigFont\undefined%
\gdef\SetFigFont#1#2#3#4#5{%
  \reset@font\fontsize{#1}{#2pt}%
  \fontfamily{#3}\fontseries{#4}\fontshape{#5}%
  \selectfont}%
\fi\endgroup%
\begin{picture}(2880,2937)(811,-2773)
\put(3601,-61){\makebox(0,0)[c]{\smash{\SetFigFont{12}{14.4}{\rmdefault}{\mddefault}{\updefault}
$\theta$
}}}
\put(1801,-961){\makebox(0,0)[c]{\smash{\SetFigFont{12}{14.4}{\rmdefault}{\mddefault}{\updefault}
$\star'$
}}}
\put(2701,-961){\makebox(0,0)[c]{\smash{\SetFigFont{12}{14.4}{\rmdefault}{\mddefault}{\updefault}
$\star$
}}}
\put(1801,-1861){\makebox(0,0)[c]{\smash{\SetFigFont{12}{14.4}{\rmdefault}{\mddefault}{\updefault}
$\overline{\star'}$
}}}
\put(2656,-1861){\makebox(0,0)[c]{\smash{\SetFigFont{12}{14.4}{\rmdefault}{\mddefault}{\updefault}
$\overline{\star}$
}}}
\put(901,-2761){\makebox(0,0)[c]{\smash{\SetFigFont{12}{14.4}{\rmdefault}{\mddefault}{\updefault}
$\overline{\theta'}$
}}}
\put(3556,-2761){\makebox(0,0)[c]{\smash{\SetFigFont{12}{14.4}{\rmdefault}{\mddefault}{\updefault}
$\overline{\theta}$
}}}
\put(901,-61){\makebox(0,0)[c]{\smash{\SetFigFont{12}{14.4}{\rmdefault}{\mddefault}{\updefault}
$\theta'$
}}}
\put(2251, 29){\makebox(0,0)[b]{\smash{\SetFigFont{12}{14.4}{\rmdefault}{\mddefault}{\updefault}
$\rho^*_a$
}}}
\put(2251,-871){\makebox(0,0)[b]{\smash{\SetFigFont{12}{14.4}{\rmdefault}{\mddefault}{\updefault}
$\DD_a$
}}}
\put(2251,-1771){\makebox(0,0)[b]{\smash{\SetFigFont{12}{14.4}{\rmdefault}{\mddefault}{\updefault}
$\overline{\DD_a}$
}}}
\put(2251,-2671){\makebox(0,0)[b]{\smash{\SetFigFont{12}{14.4}{\rmdefault}{\mddefault}{\updefault}
$(\sigma^*)^{-1}\circ\rho_a^*\circ\sigma^*$
}}}
\put(1711,-1411){\makebox(0,0)[rt]{\smash{\SetFigFont{12}{14.4}{\rmdefault}{\mddefault}{\updefault}
$\Sigma'$
}}}
\put(2791,-1411){\makebox(0,0)[lt]{\smash{\SetFigFont{12}{14.4}{\rmdefault}{\mddefault}{\updefault}
$\Sigma$
}}}
\put(811,-1411){\makebox(0,0)[rt]{\smash{\SetFigFont{12}{14.4}{\rmdefault}{\mddefault}{\updefault}
$\sigma^*$
}}}
\put(3691,-1411){\makebox(0,0)[lt]{\smash{\SetFigFont{12}{14.4}{\rmdefault}{\mddefault}{\updefault}
$\sigma^*$
}}}
\end{picture}
\vspace{1ex}
\end{center}
\caption{Two nested commutative diagrams that illustrate
the covariance of the semiclassical and quantum constructions
under a change of coordinates given by $\sigma^*$.
The dashed lines indicate Kontsevich quantization.
$\Sigma$ and $\Sigma'$ are the equivalence maps (including $\sigma^*$)
that relate the star products that were computed in the new
coordinates with those computed in the old coordinates.
The top and bottom trapezia illustrate the construction of
the covariantizing equivalence maps in the old and new coordinates.
\vspace{1ex}} 
\label{pic:coordtransform}
\end{figure}
Figure~\ref{pic:coordtransform} illustrates how the semi-classical and
quantum  constructions are affected by a change of coordinates $\sigma^*$:
The quantization of $\overline{\theta}$ and $\overline{\theta'}$ in
the new coordinates leads to star products $\overline{\star}$ and
$\overline{\star'}$ that are related to the star products
$\star$ and $\star'$ in the old coordinates 
by equivalence maps $\Sigma$ and $\Sigma'$ respectively. (We have
included $\sigma^*$ in the definition of these maps.) Note that in general
$\Sigma \neq \Sigma'$.
The covariantizing
equivalence map, generalized gauge potential and field strength in the
new coordinates and old coordinates are related by:
\eqa
\overline{\DD_a} & = & \Sigma^{-1} \circ \DD_a \circ \Sigma' , \\
\overline{\AA_a} & = & \Sigma^{-1} \circ (\Sigma' - \Sigma) + \Sigma^{-1} \circ
\AA_a \circ \Sigma' , \\
\overline{\FF_a} & = & \Sigma^{-1} \circ \FF_a \circ (\Sigma')^{\otimes 2} .
\ena
Explicit (but complicated) expressions for
$\Sigma$ and $\Sigma'$ in terms of $\theta$, $\sigma^*$, and the
gauge potential $a$ can be computed with methods similar to the ones that
we have used to compute $\DD_a$ in the previous section.

\subsection[Born-Infeld action in the intermediate picture]{Born-Infeld
action in the intermediate picture\protect\footnote{To avoid confusion with the
matrices we will use bold face letters
for tensors and forms in this section
(e.g.: $\btheta = \frac{1}{2} \theta^{ij} \pp_i \wedge \pp_j$,
$\bomega = \frac{1}{2} \omega_{ij} dx^i \wedge dx^j$, $\theta = \omega^{-1}$);
for simplicity we shall assume that all matrices are nondegenerate when
needed.}}
\label{DBI}

Seiberg and Witten have argued that the open string theory effective
action in the presence of a background $B$-field can be expressed either in terms
of ordinary gauge theory written in terms of the combination $B+F$ or
in terms of noncommutative gauge theory with
gauge field $\widehat F$, where the $B$-dependence appears
only via the $\theta$-dependence of the star product and the open string
metric $G$ and effective coupling $G_s$. (Here we implicitly need to assume
that $\theta$ is Poisson, which is of course the case for constant $\theta$.)
There is also an intermediate picture 
with an effective noncommutative action which is a function of
$\widehat \Phi + \widehat F$, where $\widehat\Phi$ is a covariant version of
some antisymmetric matrix $\Phi$, with a $\theta$-dependent
star product and effective metric $G$ and string coupling $G_s$.
The proposed relations between
the new quantities and the background field $B$,
the given  closed string metric $g$ and the coupling $g_s$ are%
\newcommand*\deh{{\det}^{\frac{1}{2}}}
\newcommand*\dehi{{\det}^{-\frac{1}{2}}}
\eq
\frac{1}{G + \Phi} = \frac{1}{g + B} - \theta,
\qquad \frac{\deh(g + B)}{g_s} = \frac{\deh(G + \Phi)}{G_s}.
\en
The first relation can also be written more symmetrically:
\eq
[1+(G+\Phi)\theta][1-(g+B)\theta] = 1; \qquad G_s = g_s \dehi[1 - (g +
B)\theta]. \label{identities}
\en
Given $g$ and $B$ we can pick essentially \emph{any} antisymmetric matrix $\theta$
-- in particular one that satisfies the Jacobi identity
--
and find $G$ and $\Phi$ as symmetric and antisymmetric parts of the
following expression
\eq
G + \Phi = \frac{1}{1-(g+B)\theta} (g+B).
\en
For $\theta = 1/B$ (as in the zero slope limit):
$G = -B g^{-1} B$, $\Phi = -B$, $G_s = g_s \deh(-Bg^{-1})$.

For slowly varying but not necessarily small fields on a $D$-brane the
effective theory is given by the Dirac-Born-Infeld (DBI) action. In the following
we will show that the ordinary DBI action is exactly equal to the
semi-noncommutative DBI action in the intermediate picture. There are
no derivative corrections. By \emph{semi}-noncommutative we mean the
semiclassical limit of a noncommutative theory with star commutators,
e.g.\ in the noncommutative transformation law, replaced by Poisson brackets
as in section~\ref{sec:semiclass}.
Using (\ref{identities}) we can derive the following identity for scalar
densities
\eq
\frac{1}{g_s} \deh(g+ B + F) = \frac{1}{G_s}
\deh\!\left(\frac{\theta}{\theta'}\right) \deh(G+ \Phi + F'), \label{scalardens}
\en
where
\eq
\theta' = \theta\frac{1}{1 + F\theta},\qquad F' = \frac{1}{1+F\theta} F
\en
(understood as formal power series).
Raising indices on $F'$ with $\theta$ we get $\theta' - \theta
\equiv \pcom{x^i}{x^j}' - \pcom{x^i}{x^j}$ which we recognize as the
semiclassical version of
\eq
\tilde F^{ij} = [x^i\stackrel{\star'}{,} x^j] -
\scom{x^i}{x^j}, \label{ftild}
\en
compare equation (\ref{ftilde}). The semi-noncommutative field strength
changes by canonical transformation under gauge transformations (\ref{ctrafo}); 
it is obtained from the invariant $\theta' - \theta$ by action of the
covariantizing $\rho^*$  (see also (\ref{tensor}), (\ref{scgfs})): 
\eq
\rho^*(\theta' -\theta) = \pcom{\rho^* x^i}{\rho^* x^j} -
\rho^*\left(\pcom{x^i}{x^j}\right).
\en
The corresponding object with
lower indices is
\eq
\widehat F =\rho^* \left(F'\right).
\en 
The  Poisson structures $\btheta'$ and $\btheta$ are related by the change of
coordinates
$\rho^*$: $\rho^*\btheta' = \btheta$. The matrices $\theta'$, $\theta$
are consequently related to the Jacobian $\det(\pp \rho^*(x)/\pp x)$
of $\rho^*$:
\eq
\deh\!\left(\frac{\theta}{\rho^*\theta'}\right)\cdot\det\!\left(\frac{\pp
\rho^*(x)}{\pp x}\right) = 1.
\en
Using this we can derive the following exact equality for the DBI action with
background field $B$ and the semi-noncommutative DBI action without
$B$ (but with $\Phi$) in the intermediate picture,
\eq
\int d^px\,\frac{1}{g_s} \deh(g+ B + F) 
= \int d^px\,\frac{1}{\widehat G_s}
\frac{\deh(\rho^*\theta)}{\deh\theta}\deh(\widehat G + \widehat \Phi +
\widehat F), \label{sncdbi}
\en
with covariant $\widehat G_s \equiv \rho^* G_s$, $\widehat G \equiv \rho^* G$,
$\widehat \Phi \equiv \rho^* \Phi$, $\widehat F = \rho^* F'$ that transform
semi-classically under gauge transformations (\ref{ctrafo}).
The only object without a ``$\rho^*$'', $\dehi\theta$, is important
since it ensures that the semi-noncommutative action is invariant under
gauge transformations, i.e.\ canonical transformations.
The factor $\deh(\rho^*\theta)/\deh(\theta)$ can be absorbed in a redefinition
of $G_s$; it is equal to one in the case of constant $\theta$, since
$\rho^*$ does not change a constant;
\eq
\int d^px\,\frac{1}{g_s} \deh(g+ B + F) 
= \int d^px\,\frac{1}{\widehat G_s}
\deh(\widehat G + \widehat \Phi + \widehat F) \qquad\mbox{($\theta$ const.)} .
\en
General actions invariant under the semiclassical Seiberg-Witten map, which includes 
the Born-Infleld action as well as some actions with derivative terms,
have been discussed in~\cite{Cornalba2} in the case of constant
$\theta = B^{-1}$ and constant metric $g$.

One can consider a fully noncommutative version of
the DBI action with $\rho^*$ replaced by the equivalence map $\DD$ and with
star products
in the appropriate places.  That action no longer exactly equals
its commutative cousin but differs only by derivative terms as
can be seen~\cite{Paolo} by using the explicit form of the
equivalence map (\ref{Dq}) --
ordering ambiguities in the definition of the action
also contribute only derivative terms.
The equivalence up to derivative terms
of the commutative and noncommutative DBI actions was previously
shown  by direct computation~\cite{SW} in the case of constant $\theta$;
in this case an alternative derivation that is closer to 
our present discussion and is
based on a conjectured formula for the Seiberg-Witten map
was given in~\cite{LiuII}. Requireing equivalence of the
commutative and noncommutative descriptions one can compute
derivative corrections to the DBI action~\cite{Cornalba:2000ua}.

The semi-noncommutative actions have the general form
\eq
S_\Phi = \int d^px \, \frac{1}{\deh\theta}\,
\rho^*_{(\theta,a)}\Big(\mathcal{L}(G_s,G,\Phi,\theta',\theta)\Big),
\en
where $\mathcal{L}$ is a gauge invariant scalar function. The gauge potential
enters in two places: in $\theta'$
via the gauge invariant field strength $f$  and in
$\rho^*_{(\theta,a)}$. It is interesting to note that even the metric
and the coupling constants will in general transform under gauge
transformations since they depend on the gauge potential
via $\rho^*$.
Under gauge transformations $\rho^*\mathcal{L}$
transforms canonically:
\eq
\delta_\lambda (\rho^*\mathcal{L}) = \pcom{\rho^*\mathcal{L}}{\tilde\lambda}.
\en
Due to the special scalar density $\dehi\theta$, the action $S_\Phi$ is
gauge invariant and covariant under general coordinate transformations.

In the ``background independent'' gauge $\theta = B^{-1}$, $\Phi = -B$
\cite{Seiberg} the action (\ref{sncdbi}) becomes simply
\eq
 S_{\mathrm{DBI}} = \int d^px \, \frac{1}{\deh\theta}\: \rho^*\!\left(\frac{1}{g_s}
 \deh(1 + g \theta')\right),
\en
with $\theta' = (B+F)^{-1}$.
Expanding the determinant to lowest nontrivial order we find the following
semi-non\-com\-mu\-ta\-tive Yang-Mills action
\eq
\int d^px \, \frac{1}{\deh\theta}\: \rho^*\!\!\left(\frac{1}{4 g_s}  g_{ij}
\theta'^{jk} g_{kl} \theta'^{li}\right) =
\int d^px \, \frac{1}{\deh\theta}\, \frac{1}{4\hat g_s} \hat g_{ij}
\pcom{\hat x^j}{\hat x^k} \hat g_{kl} \pcom{\hat x^l}{\hat x^i},
\en
with covariant coupling constant $\hat g_s = \rho^* g_s$,
metric $\hat g_{ij} = \rho^* g_{ij}$ and coordinates $\hat x^i = \rho^* x^i$.
An analogous fully noncommutative version can be written with the
help of the covariantizing equivalence map $\DD$
\eq
S_{\mathrm{NC}} = \int d^px \, \deh\omega\: \DD\!\left(\frac{1}{4 g_s} \star' g_{ij} \star'
\tilde F^{jk} \star' g_{kl} \star' \tilde F^{li}\right),
\en
with an appropriate scalar
density $\deh\omega$ that ensures upon integration
a cyclic trace (this is important for the gauge invariance of the action.)
In the zero-slope limit $\tilde F$ is as given
in (\ref{ftild}), in the background independent gauge
$\tilde F^{ij} =
[x^i  \stackrel{\star'}{,}  x^j  ]$.
In the latter case
\eq
S_{\mathrm{NC}} = \int d^px \, \deh\omega\: \frac{1}{4 \hat g_s} \star \hat g_{ij} \star
\scom{\hat X^j}{\hat X^k} \star \hat g_{kl} \star \scom{\hat X^l}{\hat X^i}.
\en
This has the form of a matrix model potential (albeit with nonconstant
$g_s$, $g$) with covariant coordinates $\hat X^i = \DD x^i$ as dynamical variables.

\subsection{Some notes on symmetric tensors}
\label{metric}
The matrix
\eq
\wtheta^{ij} = \left(\frac{1}{B + g}\right)^{ij} = \theta^{ij} +
G^{ij} \label{matrix}
\en
plays a central role for strings in a background $B$-field (with $\Phi = 0$):
its symmetric part $\wtheta_S^{ij}$  is the effective open string
metric and its antisymmetric part $\wtheta_A^{ij}$
provides the Poisson structure (provided it is indeed Poisson)
that leads upon quantization to the noncommutativity
felt by the open strings. One may now ask out of pure curiosity whether
it is possible to quantize $\wtheta$ directly, i.e., whether it is
possible to find an associative
star product $\tilde\star$ which in lowest order in $\hbar$
is given by $\wtheta$ (which is not antisymmetric):
\eq
f \tilde\star g =  f g + \frac{i\hbar}{2} \wtheta^{ij} \pp_i f \pp_j g +
o(\hbar^2).
\en
This is indeed possible, provided $\wtheta_A$ is Poisson (i.e. satisfies
the Jacobi identity), since the symmetric part of the star product can
be gauged away by an equivalence map
\eq
\Xi(f\tilde\star g) = \Xi(f) \star \Xi(g)
\en
where
\eq
f \star g =  f g + \frac{i\hbar}{2} \wtheta_A^{ij} \pp_i f \pp_j g +
o(\hbar^2).
\en
An equivalence map that does the job can be given explicitly in terms
of the symmetric part of $\wtheta$:
\eq
\Xi = \exp(-\frac{i\hbar}{4} \wtheta_S^{ij} \pp_i \pp_j).
\en
It is enough to check  terms up to order $\hbar$
\eqa
\lefteqn{f g + \frac{i\hbar}{2}\wtheta^{ij}\pp_i f\pp_j g
-\frac{i\hbar}{4}\wtheta_S^{ij} \pp_i \pp_j (fg)}\nn
& = &f g + \frac{i\hbar}{2} \wtheta_A^{ij}\pp_i f\pp_j g
- \frac{i\hbar}{4} \wtheta_S^{ij}\left[(\pp_i\pp_j f) g + f (\pp_i\pp_j
g)\right]. \nonumber
\ena
To quantize $\wtheta$ we can thus proceed as follows: first one quantizes
$\wtheta_A$ e.g.\ with Kontsevich's formula and then one uses $\Xi$ to
get $\tilde\star$ from $\star$. In the previous sections we saw that
the full information about the noncommutative gauge fields is encoded
in the equivalence map $\DD_a$. Here we can similarly reconstruct the metric
field from $\Xi$ by evaluating its Hochschild field strength
\eq
\FF^\Xi_H(x^i,x^j) = \Xi(x^i) \star \Xi(x^j) - \Xi(x^i \star x^j)
= \frac{i\hbar}{2} G^{ij} + o(\hbar^2),
\en
or more directly: $G^{ij} = \frac{2i}{\hbar}\left(\Xi(x^i x^j) - x^i x^j\right)$.
Two more questions come up naturally: When is $\theta = (B + g)^{-1}_A$ Poisson?
Why is the relevant star product not a quantization of
$\tau \equiv B^{-1}$ as in the zero-slope limit? The answer to the second
question is of course that this is determined by the open string
propagator in the presence of a background $B$ and that happens to have
an antisymmetric part given by $\theta^{ij}$ and only in the zero-slope
limit this is equal to $B^{-1}$. Nevertheless the two
questions turn out to be related -- the star products based
on $\tau$ and $\theta$ are equivalent provided
that the 2-forms
\eq
\boldsymbol{B} = \frac{1}{2}(\tau^{-1})_{ij} dx^i\wedge dx^j, \qquad
\boldsymbol{\phi} = -\frac{1}{2}(g \tau g)_{ij} dx^i\wedge dx^j
\en
are closed, $\boldsymbol{\phi} = d\boldsymbol{\alpha}$ for
some 1-form $\boldsymbol{\alpha}$,  and $B(t) = B + t\phi$
is nondegenerate for $t\in[0,1]$; the closedness conditions
on $\boldsymbol{B}$ and $\boldsymbol{\phi}$
ensure in particular that $\alpha$ and $\theta$ are Poisson.
According to Moser's lemma the symplectic structures $B(1)$ and $B(0)$
are related by a change of coordinates generated by the vector
field $\chi_\alpha = \tau^{ij} \alpha_j \pp_i$
and, moreover, according to Kontsevich the star products resulting from
quantization of $B(0)$ and $B(1)$ are equivalent. Since $B(0) = \tau^{-1}$
and $B(1) = \theta^{-1}$ we have demonstrated our claim.
Let us remark that $\boldsymbol{B}$ and $\boldsymbol{\phi}$ are
closed if $g$ and $B$ are derived from a K\"ahler metric,
since then $\boldsymbol{B}$ is closed and $B$, $g B^{-1} g$ are proportional.
Instead of using Moser's lemma we could also drop some assumptions
and work directly with Poisson structures as in previous sections.
We would then require that $\tau$ is Poisson
and $g \tau g$ is closed and 
introduce a 1-parameter deformation $\tau(t)$, $t \in [0,1]$, with
\eq
\tau(0) = \tau, \qquad \pp_t \tau(t) = \tau(t)\cdot (g \tau g)\cdot \tau(t)
\en
and solution
\eq
\tau(t) = \tau + t \tau g \tau g \tau + t^2 \tau g \tau g \tau g \tau g \tau + \ldots = \tau\frac{1}{1 - t(g\tau)^2}.
\en
This is the antisymmetric part of
\eq
\wtheta(t) = \frac{1}{B + t^{\frac{1}{2}} g}.
\en
The symmetric part is $t^{\frac{1}{2}} G^{ij}(t) = -t^{\frac{1}{2}}
 [\tau(t) g \tau]^{ij}$
with $G(0) = -B^{-1} g B^{-1}$ and $G(1) = G$,
while $\tau(0) = B^{-1}$ and $\tau(1) = \theta$ with
$G$ and $\theta$ as given in (\ref{matrix}).
This suggest that $\phi = -g \tau g$ represents ``metric fluctuations''
around the background $B$ that can be gauged away by an equivalence
transformation that curiously leads to the zero-slope values $G(0)$, $\tau(0)$
of the metric and Poisson structure.

\section{Nonabelian noncommutative gauge fields}

We will now extend the discussion to nonabelian gauge theories, i.e.,
Lie algebra-valued gauge potentials and gauge fields. We will argue
that a Seiberg-Witten map can be explicitly constructed for \emph{any}
gauge group by treating both the space-time noncommutativity and
the noncommutativity of the nonabelian gauge group on equal
footing. Both structures are obtained from appropriate Poisson structures
by deformation quantization. This construction
generalizes to fairly arbitrary noncommutative
internal spaces.

Let us mention that  it is possible to  absorb
a matrix factor (e.g. GL(n) or U(n) in the defining representation)
 directly into the definition of the noncommutative space $\Ax$
and then work with the abelian results of the previous sections,
however, for other gauge groups it is not a priory clear how to do this consistently.
In any case even for GL(n) and U(n) that approach would not give a very
detailed description of the nonabelian Seiberg-Witten map.

\subsection{Nonabelian setting}

In this section we shall establish notation and will
give a precise definition of the problem that
we would like to solve. Consider a manifold ``(noncommutative) space-time'' with
a noncommutative structure provided by a star product that is
derived from a Poisson structure $\Theta^{\mu\nu}$. On this space consider
a nonabelian gauge theory with gauge group G,
field strength $F_{\mu\nu}$, that can be locally expressed as
\eq
F_{\mu\nu} = \pp_\mu A_\nu - \pp_\nu A_\mu - i [A_\mu,A_\nu] 
\en
with nonabelian gauge potential $A_\mu = A_{\mu b} T^b$
where $T^b \in \mathrm{Lie}(G)$ are generators with
commutation relations
$-i[T^a,T^b] = C^{ab}_c T^c$, and nonabelian gauge transformations
\eq
\delta_\Lambda A_\mu = \pp_\mu \Lambda + i [\Lambda, A_\mu] .
\en
Our main goal is to find a noncommutative gauge potential $\As = \As(A_\mu)$ and
a noncommutative gauge parameter $\LLq(A_\mu,\Lambda)$  such that
a nonabelian gauge transformation $\delta_\Lambda$ of $A_\mu$ induces
a noncommutative gauge transformation $\deltaq_{\LLq}$ of $\As$:
\eq
\As(A_\mu + \delta_\Lambda A_\mu) = \As(A_\mu) + \deltaq_{\LLq} \As(A_\mu) .
\en
$\As(A_\mu)$ should be a universal enveloping algebra-valued
formal power series in $\Theta^{\mu\nu}$,
starting with $\Theta^{\mu\nu} A_\nu \pp_\mu$, that
contains polynomials of  $A_\mu$ and it's derivatives.
Similarly $\LLq(A_\mu,\Lambda)$ should be a universal enveloping algebra-valued
formal power series in $\Theta^{\mu\nu}$,
starting with $\Lambda$, that
contains polynomials of  $A_\mu$, $\Lambda$ and their derivatives.
The product in the definition of the noncommutative gauge transformation is
a combination of the star product on space-time and the matrix product of
the $T^a$.
We expect that it should be possible to find expressions, where the
structure constants $C^{ab}_c$ do not appear explicitly, except via commutators of
the Lie algebra-valued $A_\mu$, $\Lambda$.

A secondary goal is to find a construction that stays as close as possible to
the method that we used in the abelian case. There, we used a generalization of
Moser's lemma to relate Poisson structures $\btheta$ and $\btheta'$ (and, after
quantization, star products $\star$ and $\star'$). The motivation for this
and some of the complications of the nonabelian case can be most easily
understood in the special case of invertible, i.e. symplectic, Poisson structures.
The inverses of $\theta$ and $\theta'$ define closed 2-forms $\boldsymbol B$ and
$\boldsymbol B'$ that
differ by the addition of a (closed) gauge field $\boldsymbol f$ (\ref{BplusF}). Physically
$\boldsymbol B'$ is the background $\boldsymbol B$-field plus fluctuations
$\boldsymbol f$. In the nonabelian
case we would like to keep this picture but with $\boldsymbol f$ replaced by the
nonabelian field strength $\boldsymbol F$:
\[
{\boldsymbol B'} = {\boldsymbol B} + {\boldsymbol F} ,
\]
where $B = \Theta^{-1}$.
The trouble with this is that $d{\boldsymbol F} = -{\boldsymbol A}\wedge {\boldsymbol A}
\neq 0$ in the nonabelian case
so $\boldsymbol B$ and $\boldsymbol B'$ cannot both be closed 2-forms,
which they should be
if we want to interpret their inverses as Poisson structures. Ignoring this,
we could then look for a ``vector field'' $\chi$
that generates a coordinate transformation
that relates $\boldsymbol B$ and $\boldsymbol B'$. A natural generalization from the abelian
case (\ref{vf}) is
$\chi = \Theta^{\mu\nu} A_\nu D_\mu$,
where we have replaced the abelian gauge potential by the nonabelian one
and have also switched to a covariant derivative
$D_\mu = \pp_\mu + i[A_\mu , \,\cdot\,]$. The trouble here is that
it is not clear how to act with the matrix-valued $\chi$ on ``coordinates''.
$\chi$ is certainly no vector field; it is not even a derivation. (It's action
turns out to involve complete symmetrization over the constituent matrices
$T^a$.)

The solution to both problems is to consider a larger space that is spanned
by the space-time coordinates $x^\mu$ and by symbols $t^a$ for the
generators $T^a$ of the Lie Group. $\chi$~is then the projection onto
space-time of a true vector field and $\Theta$ and $\Theta'$ are 
the space-time components of true Poisson structures on the enlarged space.
We obtain the desired nonabelian noncommutative gauge theory by quantizing
both the external and internal part of the enlarged space at the same time.
To use the method of the previous sections we need to encode the
nonabelian data in an abelian gauge theory on the enlarged space. This program
is successful, if the  ``commutative'' nonabelian gauge theory can be recovered at
an intermediate step.

\subsection{Noncommutative extra dimensions}

\emph{Notation:} Greek indices $\mu$, $\nu$, $\xi$, \ldots\ belong to
the external space, indices from the beginning of the alphabet $a$, $b$, $c$,
\ldots\ belong to the internal space and indices $i$, $j$, $k$,~\ldots\
run over the whole space (internal and external). We shall use
capital letters ($A_\mu$, $F_{\mu\nu}$, $\Theta^{\mu\nu}$) for things
related to the nonabelian theory on the external space and
small letters ($a_i$, $f_{ij}$, $\theta^{ij}$) for objects related to
the abelian theory on the enlarged space or the internal space
($a_b$, $\vartheta^{bc}$).

The $t^a$ are commutating coordinate functions on the internal space (``Lie algebra'') 
just like the $x^\mu$ are commuting coordinate functions on the external space
(``space-time''). We later recover the Lie algebra in the form of star-commutators
on the internal space and the matrices $T^a$ by taking a representation of that
algebra. The star product on the internal space is a quantization of
its natural Poisson structure
\eq
\Liecom{t^a}{t^b}  = C^{ab}_c t^c =: \vartheta^{ab}.
\en
In the new language
\eq
F_{\mu\nu} = \pp_\mu A_\nu - \pp_\nu A_\mu + \Liecom{A_\mu}{A_\nu} 
\en
with $A_\nu(x,t) = A_{\nu b}(x) t^b$ and
\eq
\delta_\Lambda A_\mu = \pp_\mu \Lambda + \Liecom{A_\mu}{\Lambda}
\en
with $\Lambda(x,t) = \Lambda_b(x) t^b$. (``Lie algebra-valued'' translates
into ``linear in $t$''.)
We equip the enlarged space with
a Poisson structure $\theta^{ij}$ which is the direct sum of the
external $\Theta^{\mu\nu}$ and internal $\vartheta^{ab}$ Poisson structures
\eq
\theta=
\left(\begin{array}{c|c} \Theta &  0 \\
\hline  0 & \vartheta \end{array}\right).
\en
Only for $t=0$ is the Poisson structure block-diagonal.
$\theta(t)$ and in particular $\theta' = \theta(1)$ acquire off-diagonal
terms through the $t$-evolution
\eq
\theta(0) = \theta,\; \pp_t \theta(t) = -\theta(t) f \theta(t),\mbox{ i.e., }
\theta(t) = \theta - t \theta f(t) \theta + t^2 \theta f(t) \theta f(t) \theta
\mp \cdots
\en
generated by an abelian gauge
field $f$ that is itself not block-diagonal (but whose internal
components $f_{ab}$ are zero as we shall argue below.)
The space-time components of $\theta(t)$ can be re-summed in a series
in $\Theta$ and we miraculously obtain an expression
that looks like the series for $\theta(t)$ but with
$f$ replaced by $F_{\mu\nu}(t) \equiv f_{\mu\nu} - t f_{\mu a} \theta^{ab}
f_{b\nu}$, which at $t=1$ (and $a_b = 0$, see below) becomes
the nonabelian field strength $F \equiv F(1) = 
\pp_\mu a_\nu - \pp_\nu a_\mu + \Liecom{a_\mu}{a_\nu}$:
\eqa
\Theta^{\mu\nu}(t) \, \equiv \, \theta^{\mu\nu}(t) &  = & \theta^{\mu\nu} - t\theta^{\mu i}f_{ij}\theta^{j\nu}
+ t^2 \theta^{\mu j} f_{jk} \theta^{kl} f_{lm} \theta^{m\nu} \mp \cdots \nn
& = & \theta^{\mu\nu} - t \theta^{\mu\kappa} \left(f_{\kappa\sigma} -
t f_{\kappa a} \theta^{ab} f_{b\sigma}\right) \theta^{\sigma\nu} + \cdots .
\ena
To all orders in $\Theta$:
\eq
\Theta(t)\, =\, \Theta - t \Theta F(t) \Theta + t^2 \Theta F(t) \Theta F(t) \Theta
\mp \cdots \,= \,\Theta\frac{1}{1 + t F(t) \Theta}.
\en
In the case of invertible $\Theta$, $\Theta'$ (with $\Theta' \equiv \Theta(1)$) we have
\eq
\frac{1}{\Theta'} = \frac{1}{\Theta} + F.
\en
This resembles the relation $B' = B + F$ that one would have naively
expected, but we should note that $\Theta$, $\Theta'$ are not necessarily
Poisson (they are just the space-time components of the Poisson
structures $\theta$, $\theta'$) and $F$ is not exactly a non-abelian
field strength (it is in fact a gauge-invariant expression in
abelian gauge fields that coincides with the nonabelian
field strength in the special gauge $a_b = 0$; see next section.) The other components of $\theta(t)$ are
computed similarly (again using $f_{ab} = 0$):
\eq
\theta^{\mu b}(t) = -\theta^{b
\mu}(t) = -t \Theta^{\mu\nu}(t) f_{\nu a} \vartheta^{ab},\qquad
\theta^{ab}(t) = \vartheta^{ab} + t^2 [\vartheta f \Theta(t) f
\vartheta]^{ab}.
\en
$\Theta(t)$ is not the only object that acquires a non-abelian look at $t = 1$;
this is also the case for Moser's vector field
\eq
\ap['] = (\theta')^{ij} a_j \pp_i = (\Theta')^{\mu\nu} \bar A_\nu \bar D_\mu +
\vartheta^{ab} a_b \pp_a,
\en
where $\bar A_\nu = a_\nu - f_{\nu a} \vartheta^{ab} a_b$ and
$\bar D_\mu = \pp_\mu - f_{\mu a} \vartheta^{ab} \pp_b$ are
the first terms in the expansions for the nonabelian gauge potential
and covariant derivative, valid around the special gauge $a_b = 0$ (see next
section.)

Now we need to identify appropriate abelian gauge fields and gauge
transformations on the
enlarged space that upon quantization give the desired
nonabelian noncommutative gauge fields and noncommutative gauge transformations.
For this we consider the terms of lowest order in $\Theta$ of
the Seiberg-Witten condition, where we expect
to see a purely nonabelian gauge transformation. Up to this order it is in fact enough to
work with the semiclassical condition (\ref{cswcond})
\[
A_{a+d\lambda} = A_a + \dpo\tilde\lambda + \{A_a,\tilde\lambda\}.
\]
Evaluating this on $x^\mu$ and collecting terms of order $\Theta$ we
get
\eq
A_\mu(a + d\lambda) = A_\mu(a) + \pp_\mu\Lambda + \Liecom{A_\mu(a)}{\Lambda},
\en
with $A_\mu(a)$ and $\Lambda(\lambda,a)$ defined
by
\eq
A_a(x^\nu) = \Theta^{\nu\mu} A_\mu(a) + o(\Theta^2),
\qquad \tilde\lambda = \Lambda(\lambda,a) + o(\Theta). \label{nadata}
\en
An abelian gauge transformation $\delta a_i = \pp_i \lambda$ thus results
in a nonabelian gauge transformation of $A_\mu$ with gauge parameter
$\Lambda$. Since we would like to identify $A_\mu$ and $\Lambda$
with the gauge potential and parameter of the ordinary nonabelian gauge theory
that we started with, they should both be linear in the coordinates $t^a$
of the internal space. Studying gauge transformations and
the explicit expression (\ref{flow}), (\ref{as})
for $A_a(x^\mu)$ (see next section) we find that this implies that the internal components
of the abelian gauge potential are independent of the $t^a$,
while the external components are linear in the $t^a$. This is preserved by
abelian gauge transformations with gauge parameters that are linear
in the $t^a$:
\eq
a_\mu = a_{\mu b}(x) t^b, \quad a_b = a_b(x),\quad \lambda
= \lambda_{b} (x) t^b; \quad \delta a_\mu = i(\pp_\mu \lambda_{b}) t^b,
\quad \delta a_b = i \lambda_{b}. \label{abdata}
\en
The gauge invariant characterization of the desired abelian
gauge fields is $f_{ab} = 0$, $f_{\mu b} = - f_{b\mu}$
independent of the~$t^a$, $f_{\mu\nu}$ linear in the $t^a$.
By a gauge transformation with parameter $-a_b(x) t^b$ we
can always go to a special gauge with vanishing internal gauge potential
$a_b' = 0$ (and $a'_\mu = a_\mu - \pp_\mu(a_b) t^b$.)

We can now apply the method that we developed for the abelian case
in the previous sections to obtain the desired nonabelian
noncommutative gauge fields in terms of the abelian data (\ref{abdata}).
These are $\theta$-expanded noncommutative gauge fields
that become ordinary nonabelian gauge fields at lowest nontrivial
order in $\Theta$ (but all orders in $\vartheta$.)
We claim that a re-summation of the $\theta$-series gives
in fact $\Theta$-expanded noncommutative gauge fields in terms
of nonabelian gauge fields. This is a much stronger statement
and can be checked by inspection using the special gauge $a_b = 0$.
A rigorous formal proof is however missing.

\subsection{Mini Seiberg-Witten map}

Nonabelian gauge theory is of course also a type of noncommutative
gauge theory and one may thus wonder whether a Seiberg-Witten map
exists from abelian to nonabelian gauge fields. Computing
$A_\mu(a)$ and $\Lambda(\lambda,a)$ (\ref{nadata}) using
the results from the previous sections does in fact provide
such maps:
\eq
A_\mu(a)  =  \left(e^{\api}\right)(a_\mu - \pp_\mu \alpha) +
\left(\frac{e^{\api} -1}{\api}\right)(\pp_\mu\alpha),
\en
\eq
\Lambda(\lambda, a)  =  \left(\frac{e^{\api} -1}{\api}\right)(\lambda)
\en
with $\alpha(x,t) = a_b(x) t^b$, the parameter of the 
gauge transformation that gives $a_b = \pp_b \alpha$ starting
from the special gauge $a_b = 0$. Note that 
\[
\api = \vartheta^{ab} a_b \pp_a =
\Liecom{\cdot}{\alpha}.
\]
In components
\eq
A_\mu(a) = \sum_{n=0}^\infty \frac{1}{(n+1)!} t^a (M^n)^b_a(a_{\mu b} - n f_{\mu b}),
\en
\eq
\Lambda(\lambda,a) = \sum_{n=0}^\infty \frac{1}{(n+1)!} t^a (M^n)^b_a \lambda_b,
\en
with the matrix $M_a^b = C^{bc}_a a_c$ and
$a_{\mu b}t^b = a_\mu$, $\lambda_b t^b = \lambda$.
Under an abelian gauge transformation $\delta_\lambda a_i = \pp_i \lambda$,
\eq
\delta_\lambda A_\mu(a) = \pp_\mu \Lambda(\lambda, a)
+ \Liecom{A_\mu(a)}{\Lambda(\lambda, a)}.
\en
In the special gauge of vanishing internal gauge potential $a_b = 0$ the
maps
becomes simply $A_\mu(a) = a_\mu$, $\Lambda(\lambda,a) = \lambda$.

\appendix

\chapter{{\sffamily Brackets, evolution and para\-meters}}

\section[Schouten-Nijenhuis and Gerstenhaber brackets]{Schouten-Nijenhuis
bracket\protect\footnote{%
A good reference for the material in this section 
is~\cite{Weinstein}.}}
\label{App1}

The Schouten-Nijenhuis bracket
of two polyvector fields is defined
by
\eqa
\lefteqn{\sncom{\xi_1\wedge\ldots\wedge\xi_k}{\eta_1\wedge\ldots\wedge\eta_l}}
\phantom{\sncom{f}{\xi_1\wedge\ldots\wedge\xi_k}}&& \nn[2pt]
& = & \sum_{i,j}(-)^{i+j} \,
[\xi_i,\eta_j]\wedge\xi_1\wedge\ldots\wedge\hat{\xi}_i      
\wedge\ldots\wedge
\hat{\eta}_j\wedge\ldots\wedge\eta_l , \nn[5pt]
\sncom{f}{\xi_1\wedge\ldots\wedge\xi_k} &=&
\sum_{i}(-)^{i-1}\, \xi_i(f)\xi_1\wedge\ldots\wedge\hat{\xi}_i
\wedge\ldots\wedge\xi_k, \label{sncom}
\ena
if all $\xi$'s and $\eta$'s are vector fields and $f$ is a function.
The hat marks omitted vector fields.
A Poison tensor is a bivector field $\theta = 
\frac{1}{2}\theta^{ij}\pp_i\wedge\pp_j$
that satisfies the Jacobi identity
\eq
0 = \sncom{\theta}{\theta} \equiv 
\frac{1}{3}\Big(\theta^{il}\pp_l(\theta^{jk})
+\theta^{jl}\pp_l(\theta^{ki})
+\theta^{kl}\pp_l(\theta^{ij}) \Big)\pp_i\wedge\pp_j\wedge\pp_k .
\label{jacobi}
\en
In terms of the coboundary operator
\eq
\dpo = -\sncom{\cdot}{\theta},
\en
this can also be expressed as $\dpo\theta = 0$ or $\dpo^2 = 0$.

\section*{Gerstenhaber bracket}
\label{App2}
The Gerstenhaber bracket is
given by
\eq
\gcom{\CC_1}{\CC_2} = \CC_1 \circ \CC_2 - (-)^{(p_2+1)(p_1+1)} \CC_2 \circ \CC_1,
\label{gerstenhaber}
\en
where composition $\circ$
for $\CC_1 \in C^{p_1}$ and $\CC_2 \in C^{p_2}$ is defined as
\eqa
\lefteqn{(\CC_1 \circ \CC_2)(f_1,f_2,\ldots,f_{p_1+p_2-1})
= \CC_1\Big(\CC_2(f_1,\ldots,f_{p_2}),f_{p_2+1},\ldots,f_{p_2+p_1-1}\Big)}&&\nn
&& {}-(-)^{p_2} \CC_1\Big(f_1,
   \CC_2(f_2,\ldots,f_{p_2+1}),f_{p_2+2},\ldots,f_{p_2+p_1-1}\Big) \pm \ldots +\nn
&&   {} (-)^{(p_2+1)(p_1+1)}
   \CC_1\Big(f_1,\ldots,f_{p_1-1},\CC_2(f_{p_1},\ldots,f_{p_1+p_2-1})\Big) ;
\ena
$C^p$ may be either $\Hom_k(\Ax^{\otimes p},\Ax)$
or the space of $p$-differential operators $D_\mathrm{poly}^p$. 
In analogy to (\ref{jacobi}) we can express the 
associativity of a product $\star \in C^2$ as
\eq
\gcom{\star}{\star} = 0 , \qquad 
\gcom{\star}{\star}(f,g,h) \equiv 2\Big((f\star g)\star h - f\star(g\star
h)\Big).
\en
In terms of the coboundary operator (see also (\ref{hochschilddc}))
\eq
\ds: C^p \rightarrow C^{p+1}, \qquad \ds\CC = - \gcom{\CC}{\star},
\label{hochschildd}
\en
this can also be written as $\ds\star = 0$ or $\ds^2 = 0$.

\section[$\Lambda$-symmetry and $t$-evolution]{$\Lambda$-symmetry and t-evolution}
\label{s:t-evolution}

Under $\Lambda$-symmetry transformations in string theory the terms in
$B+ F$ get rearranged
as follows: $B \mapsto B + \Lambda$, $F \mapsto F - \Lambda$. The only
$\Lambda$-gauge invariant
combination of $B$ and $F$ is thus $B + F$. We usually interpret $B$ as a
background field and $F$ as fluctuations. The continuous mapping between
expressions with and without fluctuations is given by the ``$t$-evolution''.

Consider a $t$-dependent function $f(t)$ whose $t$-evolution is governed
by
\eq
\left(\pp_t + A(t)\right) f(t) = 0, \label{tevolution}
\en
where $A(t)$ is an operator (vector field or differential operator of arbitrary
degree) whose $t$-dependence is given. We are interested to relate $f(1)$ to
$f(0)$. There is a simple way to integrate (\ref{tevolution}) without having
to resort to $t$-ordered exponentials: By Taylor expansion
\eq
e^{-\pp_t} f(t) = f(t-1).
\en
Due to (\ref{tevolution}) we can insert $\exp(\pp_t + A(t))$ without
changing anything
\eq
e^{-\pp_t} e^{\pp_t + A(t)} f(t) = f(t-1).
\en
The trick hereby is that due to the Baker-Campbell-Hausdorff formula
all $\pp_t$ are saturated in the
product of the exponentials; there are no free $t$-derivatives
acting on $f(t)$, so we can evaluate at $t=1$ and get
\eq
\left.e^{-\pp_t} e^{\pp_t + A(t)}\right|_{t=1} f(1) = f(0) ,
\en
or, slightly rearranged
\eq
\left.e^{\pp_t + A(t)} e^{-\pp_t} \right|_{t=0} f(1) = f(0) .
\en
The first few terms in the expansion of the exponentials are
\[
1 + A + \frac{1}{2}(A^2 + \dot A) + \frac{1}{6}(A^3 + \dot A A + 2 A \dot A + \ddot A)
+ \cdots \,.
\]

\section{Semi-classical and quantum gauge para\-meters}
\label{lambdatilde}

Let $A$ and $B$ be two operators (vector fields or differential operators) and
\eq
B_0 \equiv B, \qquad B_{n+1} = [A , B_n],
\en
then
\eq
e^{A + \epsilon B} - e^A
= \epsilon\sum_{n=0}^\infty \frac{B_n }{(n+1)!} \: e^A
+ o(\epsilon^2) . \label{useful}
\en
\paragraph{Semi-classical:}
We would like to proof (\ref{flow}) and (\ref{SW2})
\eq
\rho^*_{a+d\lambda} - \rho^*_a = (\dpo\tilde\lambda)\circ\rho^*_a + o(\lambda^2),
\qquad \tilde\lambda(\lambda,a) =
\sum_{n=0}^\infty \frac{\left.(\ap[_t] + \pp_t)^n(\lambda)\right|_{t=0}}{(n+1)!} .
\en
It is helpful to first evaluate
\eq
[\pp_t + \ap[_t], \dpo[_t] \lambda] = \dpo[_t] [(\ap[_t]+\partial_t)(\lambda)].
\en
(Note that both $\dpo[_t] \lambda$ and $\dpo[_t] \ap[_t](\lambda)$ are Hamiltonian
vector fields.)\\[5pt]
\emph{Proof}: (we suppress the $t$-subscripts and the explicit $t$-dependence
of $\lambda$)
\eqa
[\pp_t + \ap, \dpo \lambda](f)
& = & \pp_t( \{ f , \lambda \} ) + \ap( \{ f , \lambda \} ) - \{ \ap(f) ,
      \lambda \}  \nn
& = & \fp(f,\lambda) - (\dpo\ap)(f,\lambda) + \{ f , \ap(\lambda) \}\nn
& = & \dpo (\ap(\lambda))(f).
\ena
We have repeatedly used the definition of $\dpo$
(\ref{hamvec}) and, in the last step, (\ref{fs}).
Now we can use (\ref{useful}) to evaluate
\eqa
\rho^*_{a+d\lambda} - \rho^*_a
 & = & \left.\left(e^{\pp_t + \ap[_t] + \dpo[_t] \lambda} - e^{\pp_t + \ap[_t]}\right)
 e^{-\pp_t}\right|_{t=0}\nn
 & = &
 \sum_{n=0}^\infty \left.\frac{\dpo[_t](\ap[_t] + \pp_t)^n(\lambda)}{(n+1)!} 
 e^{\pp_t + \ap[_t]}e^{-\pp_t}\right|_{t=0} + o(\lambda^2). 
\ena
\paragraph{Quantum:}
We would like to proof (\ref{Dadl}) and (\ref{tildehatlambda})
\eq
\DD_{a+d\lambda} - \DD_a= (\frac{1}{i\hbar} \ds\Lambda)\circ\DD_a
+ o(\lambda^2),
\qquad
\Lambda(\lambda,a) =
\sum_{n=0}^\infty \frac{(\as[_t] + \pp_t)^n(\hat\lambda)|_{t=0}}{(n+1)!}.
\label{toshow}
\en
First we evaluate
\eq
[\pp_t + \as[_t], \ds[_t] \hat\lambda] = \ds[_t] [(\as[_t] +
\partial_t)(\hat\lambda)].
\en
(Note that both $\ds[_t] \hat\lambda$ and $\ds[_t] \as[_t](\hat\lambda)$
are inner derivations. Also note that the proof of this equation is in principle much harder than
its semi-classical counterpart, since we are now dealing with differential
operators of arbitrary degree.)
\\[5pt]
\emph{Proof}: (we suppress the $t$-subscripts and the explicit $t$-dependence
of $\lambda$)
\eqa
\lefteqn{[\pp_t + \as, \ds \hat\lambda](f)}\nn
& = & \pp_t[f,\hat\lambda]_\star +
 	\as (\ds(\hat\lambda) f)
	- \ds(\hat\lambda)(\as(f)) \nn
	& = & \fs(f,\hat\lambda) - \fs(\hat\lambda,f) +
	\as ([f,\hat\lambda]_{\star})
	-[\as (f),\hat\lambda]_{\star} \nn
	& = & \fs(f,\hat\lambda) - \fs(\hat\lambda,f)
	-(\ds\as)(f,\hat\lambda) + [f,\as \hat\lambda]_\star
	+ (\ds\as)(\hat\lambda,f) \nn
	& = & \ds(\as(\hat\lambda))(f).
\ena
The desired result (\ref{toshow}) follows now from (\ref{useful}) as in
the semi-classical case.

\chapter[{\sffamily The main theorem
for Poisson-Lie groups}]{{\sffamily The main theorem for factorizable
Poisson-Lie groups}}

\label{AppB}

The so-called main theorem for factorizable Poisson-Lie groups
concerns the solution by factorization of the Hamilton equations
of motion. It is the classical limit of corresponding quantum
theorem  presented
in chapter~\ref{TQLE}.

For the convenience of the interested reader 
we briefly recall an elegant proof of the main theorem following
a lecture of N.Y.\ Reshetikhin~\cite{Reshetikhin}, 
see also~\cite{STSDress,ReySTS}.

\section{The main theorem}

Let $I(G) \subset C^\infty(G)$ be the subspace of Ad${}_G$-invariant
functions on a factorizable Poisson-Lie group $(G,p)$.

\begin{thm} \hfill
\begin{enumerate}
\item[i)] $I(G)$ is a commutative Poisson algebra in $C^\infty(G)$.
\item[ii)] The flow lines of the Hamiltonian  $H\in I(G)$ 
have the form 
\[ x(t)=g_\pm(t)^{-1}x g_\pm(t), \]
where the mappings $g_\pm(t)$ are determined by 
\[ g_+(t)g_-(t)^{-1}=\exp\left(tI\left(\nabla H(x)\right)\right), \]
and $I:\gtg^*\longrightarrow\gtg$ is the factorization isomorphism.
\end{enumerate}
\end{thm}

\noindent {\sc Proof:} \
Firstly, let us verify:
\[ Ad_{g_+(t)}(x)=Ad_{g_-(t)}(x). \]
Indeed, this is equivalent to
\eq  \label{Ad-invariance}
\exp\left(tI\left(\nabla H(x)\right)\right)x=x\exp\left(tI\left(\nabla H(x)\right)\right), 
\en
which follows from the $G$-invariance of $I$ and $H$.\\
Secondly, let us verify that the transformation 
\eq \label{transformation}
x(t)=g_\pm(t)^{-1}x g_\pm(t)
\en 
never leaves the symplectic leaf in $G$ passing through $x\in G$. For this it suffices to prove that (\ref{transformation}) is a dressing transformation. Indeed, by using equation (\ref{Ad-invariance}) we obtain
\begin{eqnarray*}
x^{g(t)} &=& \left(x g(t) x^{-1}\right)^{-1}_+xg_+(t) \\
&=& g(t)^{-1}_+xg_+(t).
\end{eqnarray*} 
Now, let us prove formula (\ref{transformation}).
For that, we consider the Heisenberg double 
$\DD_+\simeq G\times G\simeq G\times G^*$:
\begin{center}
\unitlength 1mm
\linethickness{0.4pt}
\begin{picture}(55.00,25.00)
\put(25.00,20.00){\vector(-1,-1){15.00}}
\put(30.00,20.00){\vector(1,-1){15.00}}
\put(10.00,4.00){\makebox(0,0)[ct]{$G^* \simeq \DD_+/G$}}
\put(45.00,4.00){\makebox(0,0)[ct]{$G \simeq \DD_+/G^*$}}
\put(27.00,21.00){\makebox(0,0)[cb]{$\mathcal{D}_+$}}
\put(16.00,12.00){\makebox(0,0)[rb]{$p$}}
\put(39.00,12.00){\makebox(0,0)[lb]{$\pi$}}
\end{picture}
\end{center}
$\DD_+/G$ comes from the right action of $\DD$ on $\DD_+\simeq\DD$: 
\[ (x,y)\cdot h=(xh,yh), \]
and $\DD_+/G^*$ comes from the dressing action of $G^*\subset\DD^*$ on $\DD_+\simeq\DD$:
\[ (x,y)\cdot\xi=(\xi_+^{-1}x\xi_-,\xi_+^{-1}y\xi_-). \]
Here and in the following, we denote points of $G\times G$ by $(x,y)$ and points of $G\times G^*$ by $(g,\xi)$, such that we can write $(x,y)=(g\xi_+,g\xi_-)$ under the identification $G\simeq G^*$.\\
The projections $p$ and $\pi$ on the corresponding cosets are therefore given by
\begin{eqnarray*}
p(x,y) &=& xy^{-1}, \\
\pi(x,y) &=& y_+^{-1}xy_-
\end{eqnarray*}

\begin{lemma}
The projections $p$ and $\pi$ are Poisson maps.
\end{lemma}
The idea of proof of the identity (\ref{transformation}) is illustrated
in the following figure:
\begin{center}
\unitlength 1.00mm
\linethickness{0.4pt}
\begin{picture}(52.55,29.27)
\put(25.00,20.00){\vector(-1,-1){15.00}}
\put(30.00,20.00){\vector(1,-1){15.00}}
\put(27.00,21.00){\makebox(0,0)[cb]{$\mathcal{D}_+$}}
\put(16.00,12.00){\makebox(0,0)[rb]{$p$}}
\put(39.00,12.00){\makebox(0,0)[lb]{$\pi$}}
\put(9.00,4.00){\makebox(0,0)[rt]{$G^*$}}
\put(46.00,4.00){\makebox(0,0)[lt]{$G$}}
\put(-1.01,5.98){\makebox(0,0)[rc]{{\footnotesize constant evolution}}}
\put(-1.01,2.99){\makebox(0,0)[rc]{{\footnotesize with $H$}}}
\put(31.02,29.04){\makebox(0,0)[lc]{{\footnotesize linear evolution}}}
\put(31.02,25.82){\makebox(0,0)[lc]{{\footnotesize with $\pi^* H = p^* H$}}}
\put(49.00,5.00){\oval(4.00,4.00)[rt]}
\put(48.50,4.50){\oval(3.00,3.00)[rt]}
\put(48.72,4.72){\oval(3.44,3.44)[rt]}
\put(52.55,8.00){\makebox(0,0)[lc]{{\footnotesize Hamiltonian evolution}}}
\put(52.55,5.00){\makebox(0,0)[lc]{{\footnotesize with $H$}}}
\put(24.44,26.44){\line(6,1){5.00}}
\put(24.22,27.11){\line(6,1){5.00}}
\put(24.00,27.78){\line(6,1){5.00}}
\put(23.78,28.44){\line(6,1){5.00}}
\put(49.22,5.22){\oval(4.44,4.44)[rt]}
\put(1.00,4.44){\circle*{0.89}}
\end{picture}
\end{center}
We complete the proof in three steps:\\[1ex]
{\bf Step 1:} Identifying $G$ with $G^*$ we consider the evolution induced by $H\in I(G)$ on $G^*$. (Note, that $G\simeq G^*$ as a manifold, but not with respect to the Poisson structure.)
\begin{lemma} 
The space of $Ad_G$-invariant functions $I(G)$ is the central subalgebra in the Poisson algebra $C^\infty(G^*)$.
\end{lemma}
Since $\{H,C^\infty(G^*)\}=0$, we conclude
\begin{cor}
The flow lines induced by $H$ in $G^*$ are points.
\end{cor}

\noindent
{\bf Step 2:} Let us lift the Hamiltonian dynamics generated by $H$ on $G$ to $\DD_+$:
\[ (p^*H)(x,y)=H(p(x,y))=H(xy^{-1}),\qquad p^*H\in C^\infty(\DD_+). \]
\begin{lemma} \label{flow lines}
The flow lines of $p^*H$ in $\DD_+\simeq G\times G$ have the following form:
\[ (x(t),y(t))=\left(g(t)\xi_+(t),g(t)\xi_-(t)\right), \]
where
\[ g(t)=g\exp\left(tI(\nabla H(\xi))\right),\qquad\xi(t)=\xi. \]
\end{lemma}
Noting that $\pi^*H(x,y)=H(y_+^{-1}xy_-)=H(xy^{-1})=p^*H(x,y)$, we have
\begin{lemma}
$p^*H=\pi^*H$.
\end{lemma}
Therefore we conclude
\begin{cor}
$\pi$ projects the flow lines of the Hamiltonian $p^*H$ in $\DD_+$ (described in Lemma {\rm \ref{flow lines}}) to the flow lines of the Hamiltonian $H$ in $G$.
\end{cor}

\noindent
{\bf Step 3:} \ Let us compute $\pi(x(t),y(t))$: with  
$X(\xi) \equiv I(\nabla H(\xi))$ we have
\newcommand{\e}{\mbox{e}}
\begin{eqnarray*}
\lefteqn{\pi(x(t),y(t))=} \\
&=& \pi\left(g\e^{tX(\xi)}\xi_+,g\e^{tX(t)}\xi_-\right) \\
&=& \left(g\e^{tX(\xi)}\xi_-\right)^{-1}_+g\e^{tX(\xi)}\xi_+\left(g\e^{tX(\xi)}\xi_-\right)_- \\
&=& \left(g_+\e^{tX(\xi_{g_-})}g_-^{-1}\xi_-\right)_+^{-1}g_+\e^{tX(\xi_{g_-})}g_-^{-1}\xi_+\left(g_+\e^{tX(\xi_{g_-})}g_-^{-1}\xi_-\right)_- \\
&=& g_+(t)^{-1}g_+^{-1}g_+g(t)g_-^{-1}\xi_+\xi_-^{-1}g_-g_-(t) \\
&=& g_-(t)^{-1}\xi_{g_-}g_-(t),  
\end{eqnarray*}
where $\xi_{g_-}=g_-^{-1}\xi\,g_-=\pi(x,y)$ and 
$g_+(t)g_-(t)^{-1}=\exp(tX(\xi_{g_-}))$. \hfill $\mybox$

\vspace{1em}

\noindent\emph{Remark:} \
The algebra of functions $I(G)$ restricted to the symplectic leaf $\mathcal{L}_x$ in $G$ passing through $x\in G$ determines a completely integrable system on $\mathcal{L}_x$, if it has $\dim(\mathcal{L}_x)/2$ independent functions in it.

\section{Twisted version of the main theorem}

Let $(G,p)$ be a Poisson-Lie group 
and let $\theta:G\longrightarrow G$ be an automorphism 
of the Poisson-Lie group $G$.\\[1ex]
Consider the space of functions $I_\theta(G)\subset C^\infty(G)$ invariant under twisted conjugations
\eq
g:h\longmapsto g^{-1}h\theta(g).
\en

\begin{thm}[Twisted version of the main theorem] \hfill
\begin{enumerate}
\item[i)] $I_\theta(G)$ is a commutative Poisson subalgebra in $C^\infty(G)$.
\item[ii)] If the Hamiltonian $H\in I_\theta(G)$, the integral flow is given by
\[ x(t)=g_\pm(t)^{-1}x\theta(g_\pm(t)), \]
where $g_+(t)g_-(t)^{-1}=\exp(tI(\nabla H(x)))$.
\end{enumerate}
\end{thm}

\noindent
The proof parallels the one of the nontwisted version; 
it involves the twisted double construction instead of 
the normal one~\cite{STS}.\\[1ex]
\emph{Remark:} \
Let us verify that $x(t)=g_\pm(t)^{-1}x\theta(g_\pm(t))$ 
is a dressing transformation: Since $H\in I_\theta(G)$, 
$H\circ\theta=H$.
By the same reason, we have for $g(t)=\exp(tI(\nabla H(x)))$: 
\[ xg(t)x^{-1}=\theta^{-1}(g(t)). \] 
Therefore
\[
x^{g(t)} = (xg(t)x^{-1})_+^{-1}xg_+(t) 
= \theta^{-1}(g_+(t))^{-1}xg_+(t).
\]


\backmatter
\end{document}